# Non-Markovian quantum Brownian motion: a non-Hamiltonian approach[1]

A. O. Bolivar[2]

Instituto Mario Schönberg de Física-Matemática-Filosofia, CLSW 100, Bloco A, 141, 70670-051, Sudoeste, Brasília, D.F, Brazil.

## Abstract

We generalize the classical theory of Brownian motion so as to reckon with non-Markovian effects on both Klein-Kramers and Smoluchowski equations. For a free particle and a harmonic oscillator, it is shown that such non-Markovian effects account for the differentiability of the Brownian trajectories as well as the breakdown of the energy equipartition of statistical mechanics at short times in some physical situations. This non-Markovian approach is also extended to look at anomalous diffusion. Next, we bring in the dynamical-quantization method for investigating open quantum systems, which does consist in quantizing the classical Brownian motion starting directly from our non-Markovian Klein-Kramers and Smoluchowski equations, without alluding to any model Hamiltonian. Accordingly, quantizing our non-Markovian Klein-Kramers in phase space gives rise to a non-Markovian quantum master equation in configuration space, whereas quantizing our non-Markovian Smoluchowski equation in configuration space leads to a non-Markovian quantum Smoluchowski equation in phase space. In addition, it is worth noticing that non-Markovian quantum Brownian motion takes place in presence of a generic environment (e.g. a non-thermal quantum fluid). As far as the special case of a heat bath comprising of quantum harmonic oscillators is concerned, a non-Markovian Caldeira-Leggett master equation and a thermal quantum Smoluchowski equation are derived and extended to bosonic and fermionic heat baths valid for all temperatures $T \geq 0$. For the cases of a free particle and a harmonic oscillator, it is shown that quantum Brownian trajectories are also differentiable. According to our Caldeira-Leggett master equation for a free particle, it is predicted that the energy equipartition theorem is violated in the following cases: at zero temperature for all times $t \geq 0$; and at high temperatures for times of the order of the quantum time $t_q = \hbar/2k_B T$, where $\hbar$ is the Planck constant, $k_B$ the Boltzmann constant, and $T$ the temperature of the thermal bath. For a free particle and a harmonic oscillator, our quantum Smoluchowski equation in turn leads to the violation of the equipartition theorem for all times $t \geq 0$ and temperatures $T \geq 0$. Quantum anomalous diffusion is investigated, too. Further, we address the phenomenon of tunneling of a quantum Brownian particle over a potential barrier. By regarding our quantum Smoluchowski equation, we predict a kind of dissipationless quantum tunneling for all temperatures as well as in the low-temperature regime, including $T = 0$.

    By way of internal consistency, both the deterministic and classical limits of the non-Markovian quantum Brownian motion are examined. Furthermore, we wish to point out that

---

our theoretical predictions uphold the view that our non-Hamiltonian quantum mechanics is able to fathom novel features inherent in quantum Brownian motion, thereby overcoming some shortcomings underlying the usual Hamiltonian approach to open quantum systems. Lastly, some ontological implications of the status of the open-system concept on the foundations of quantum mechanics are taken up, as well.

# Table of Contents

# 1. Introduction: Going far beyond the Schrödinger equation

## 1.1. Historical background

Realistic physical phenomena or processes are never found isolated, for they are immersed in the surrounding environment and interact continuously with it. This fact had been recognized by great physicists such as Aristotle, Plato, Galileo, Newton, Descartes, Einstein, Bohr, Schrödinger, and Heisenberg [1,2]. Yet, since Galileo physical phenomena turned out to be mathematically idealized as isolated systems. Accordingly, within such a Galilean paradigm non-isolated physical processes existing in nature, such as in experimental conditions, should be explained on the ground of isolated systems which in turn do not exist in it, but in a mathematical world [3-5].

The main goal of this Introduction is historically to draw a continuous path starting from Galileo's inertia equation to Schrödinger's equation concerning the concept of isolated system. In addition, because of the conceptually quite controversial role played by the measuring apparatus in the heart of quantum mechanics, we nevertheless bring out the need to abandon such a concept of isolated system. This is achieved by idealizing mathematically a quantum system as an open system, thereby leading to theoretical predictions that go far beyond those predictions based on the Schrödinger equation.

### 1.1.1. Isolated systems in classical physics

Galileo's main contribution to setting the concept of isolated system was to image mathematically a physical system thoroughly in isolation from external influences (eliminating, for instance, the friction which always accompanies motion) [3-6]. To bring out such a mathematization processes we could, following Heidegger [3], literally quote Galileo: *"Mobile mente concipio omni secluso impedimento."*[3] This Galilean mathematization leaded to the highly idealized concept of isolated system characterized by a particle (a material point) undergoing a special movement, called inertial motion. The trajectory of such a Galilean particle is described by the linear equation

$$x(t) = x_0 + v_0 t, \qquad (1.1)$$

---

[3] *"I think in my mind of something moveable that is left entirely to itself"*.

where $x_0 \equiv x(t = t_0)$ and $v_0 \equiv v(t = t_0)$ are initial conditions: its initial position and velocity at time $t = t_0$, respectively. It follows from the linear kinematical law (1.1) that a Galilean particle is given by a constant velocity for all time $t$, i.e., $v(t) = v_0$, and, accordingly, by a vanishing acceleration. Idealized experiments involving a Galilean particle can never be actually performed, although they lead to a profound understanding of real experiments [3,6].

In the wake of Galileo's achievements, Newton introduced the concept of force as deviation from the Galilean inertial motion [3,6,7]. A special kind of force is one that is derived from the potential energy $V(x)$ inherent in the particle. Furthermore, because the mechanical energy (kinetic energy plus potential energy), $E = mv^2/2 + V(x)$, is conserved, such a type of force $K(x) = -dV(x)/dx$ is called conservative. The gravitational force between any two bodies with mass $m_1$ and $m_2$ is the most prominent example of conservative force [7,8]. Another example of conservative system is the case of a free fall of a Newtonian particle of mass $m$ under the action of a force derived from the potential energy $V(x) = mgx$, $g$ being the gravitational acceleration, which is described by the nonlinear trajectory

$$x(t) = x_0 + v_0 t - \frac{g}{2} t^2. \qquad (1.2)$$

A Newtonian particle undergoes therefore a non-inertial motion in view of the quadratic term $-gt^2/2$ in Eq. (1.2). As far as $g = 0$ is concerned the Galilean inertial motion (1.1) is retrieved.

With Euler, Lagrange, Hamilton, Jacobi, Poisson and many others, the physics of conservative systems reached a high degree of mathematical abstraction and generality due mainly to the introduction of the concept of generalized coordinates in terms of which both Lagrangian and Hamiltonian functions turn out to be expressed. In the Hamiltonian formalism of classical mechanics of isolated systems, the Hamilton equations replace Newton equations, whereas in the Lagrangian formalism the dynamics is given by the Lagrange equations. The procedure of obtaining both Lagrange and Hamilton equations is the variational principle of least action [2,9,10].

On the basis of the classical dynamics, given by Newton's or Hamilton's and Lagrange's equations, a classical system should be then idealized as an isolated system:

$$\boxed{\text{Classical System}} = \boxed{\text{Isolated System}}$$

A characteristic feature inherent in isolated systems is its deterministic nature. This implies that, ontologically, nothing prevents to predict theoretically with absolute precision the trajectory of an isolated system. A Newtonian particle, for instance, is assumed to have a well defined trajectory, existing as an unobserved objective reality. Epistemologically, a measurement process simply reveals the preexisting value of a physical quantity within a limited accuracy, since the accuracy of the initial conditions is always diminished by the ubiquitous errors of measurement, by our ignorance of the forces acting on the particle, and by our limited computational capability.

Although physical systems are not in isolation, the concept of isolated system turned out to be central in physics, underpinning both Einstein's theories of relativity [6,11] and reverberating in quantum physics too. Yet, in quantum mechanics such a concept of isolated system seems to be highly problematic due to the issue of the interpretative status of the Schrödinger function as well as the controversial role played by the measurement process.

## 1.1.2. Isolated systems in quantum physics?

Historically, quantum physics had emerged from the experiments dealing with the stability problem of atoms and molecules, diffraction of electrons, and diffraction of light under the theoretical influence of ideas coming from the Hamiltonian and Lagrangian frameworks for conservative systems [9,12-19]. During the period of the so-called old quantum theory (1900-1925), eminent physicists such as Planck, Bohr, Sommerfeld, Einstein, Epstein, Wilson, and Schwarzschild [20,21] introduced ad hoc rules for quantizing physical quantities such as the energy and angular momentum of the electron in a hydrogen atom. The basic mathematical tool was the action integral.

In 1925, starting from Newton's equation and replacing the physical quantities $x$ (position) and $p$ (linear momentum) with operators $\hat{x}$ and $\hat{p}$ complying with a noncommutative algebra, Heisenberg [22] obtained equations of motion describing isolated quantum systems (Heisenberg quantization). One year later, from the Hamilton-Jacobi formulation of classical mechanics, Schrödinger [23] arrived at the so-termed Schrödinger equation (Schrödinger quantization). In the same year of 1926, Dirac [24] unified both Heisenberg and Schrödinger approaches by quantizing via the Hamiltonian formalism[4] (Dirac quantization). It is worth highlighting that such quantization methods presuppose the use of classical concepts in the quantum realm.

---

[4]Also, in 1948 Feynman [25] derived the Schrödinger equation from the Lagrangian formalism of classical mechanics (Feynman quantization).

From a formal point of view an isolated quantum system, that is, a Schrödinger particle with mass $m$, is then mathematically described by the Schrödinger equation [23]

$$i\hbar\frac{\partial\psi(x,t)}{\partial t} = V(x)\,\psi(x,t) - \frac{\hbar^2}{2m}\frac{\partial^2\psi(x,t)}{\partial x^2}, \quad (1.3a)$$

where the phenomenological parameter $\hbar$, dubbed the Planck constant $h$ divided by $2\pi$, accounts for the signature of the quantum world. It is worth stressing that in the Schrödinger equation (1.3a) it is assumed that the mass $m$, the time $t$, as well as the coordinate $x$ show up as $\hbar$-independent physical quantities. In addition, because the Schrödinger equation (1.3a) relies on the imaginary number $i = \sqrt{-1}$, a quantum system may be also imagined as evolving backward in time[5] according to the complex conjugate of Eq. (1.3a), i.e.,

$$-i\hbar\frac{\partial\psi^*(x,t)}{\partial t} = V(x)\,\psi^*(x,t) - \frac{\hbar^2}{2m}\frac{\partial^2\psi^*(x,t)}{\partial x^2}. \quad (1.3b)$$

Before a measurement, therefore, an isolated quantum system could be formally imagined as an object flowing simultaneously forward and backward in time. Physically, the connection of both Schrödinger functions $\psi(x,t)$ and $\psi^*(x,t)$ with experiment is performed by interpreting the function $P(x,t) = \psi^*(x,t)\psi(x,t) = |\psi(x,t)|^2$, the so-called Born rule [29-31], as a probability density function from which average values of the physical quantity $x(t)$ can be calculated. This probabilistic character is in sharp contrast to the deterministic values found through the classical equations (1.1) and (1.2), for example. In brief, measurement accounts for breaking down the time symmetry of the time evolution of $\psi(x,t)$ and $\psi^*(x,t)$, thereby choosing one-way direction of time through Born's probabilistic rule:

$$\begin{array}{c}\psi(x,t)\\ \psi^*(x,t)\end{array} \quad \Rightarrow \quad \boxed{MEASUREMENT} \quad \Rightarrow \quad |\psi(x,t)|^2.$$

It is worth highlighting that before a measurement there is no probability concept because the quantum system is isolated and hence deterministically described by the Schrödinger equation. Probability arises in the theoretical predictions only insofar as the quantum system is considered as a non-isolated system, i.e., as measurement outcomes are interpreted according to the Born rule [32].

---

[5]This interpretation is originally due to Eddington [26]. See also the Schwinger interpretation of quantum mechanics [27,28].

Although Born's calculation rule is responsible for the enormous empirical success of the probabilistic predictions of quantum mechanics, serious conceptual difficulties arise on attempting to comprehend its foundations [33]. The origin of such difficulties lies in the existence of the following dualistic structure underlying the mathematical formalism [20,33-37]: On the one hand, an isolated world without probability (a "complex" world) ruled by $\psi(x,t)$ and $\psi^*(x,t)$ and, on the other hand, a non-isolated world governed by measurements (a real world) providing probabilities through $\psi(x,t)\psi^*(x,t)$. Accordingly, the following question still holds unanswered: Could probability emerge from a world without probability or does it display an irreducible character in the quantum realm?

Answering this issue requires to come up with an interpretation of quantum mechanics. The earliest attempt to interpret quantum mechanics aimed just to break down the concept of isolated system, whereby causality is often identified with determinism. According to Heisenberg (see Ref. [16], pp. 63-64):

*In fact, our ordinary description of nature, and the idea of exact laws, rests on the assumption that it is possible to observe the phenomena without appreciably influencing them. To co-ordinate a definite cause to a definite effect has sense only when both can be observed without introducing a foreign element disturbing their interrelation. The law of causality, because of its very nature, can only be defined for isolated systems, and in atomic physics even approximately isolated systems cannot be observed.*

Quantum systems should be viewed as non-isolated physical systems because, in contrast to a passive role in classical physics, the measurement apparatus turned out to play a relevant role for establishing the very physical nature of the quantum phenomena. As pointed out by Heisenberg [38], only after observing the physical property "position in time" through a measuring device constructed for this purpose can we speak of the trajectory concept: *"Die Bahn entsteht erst dadurch, dass wir sie beobachten."*[6]

According to Bohr [39], in turn, isolated systems are abstractions without any physical content[7]:

---

[6] *"The trajectory concept arises provided that we observe it."*
[7] Later, Bohr [40] brought out the *"impossibility of any sharp separation between the behavior of atomic objects and the interaction with the measuring instruments which serve to define the very conditions under which the phenomena appear."*

> *It must be kept in mind that (...) radiation in free space as well as isolated material particles are abstractions, their properties on the quantum theory being definable and observable only through their interaction with other systems.*

This implies that there is no independent reality in the quantum realm [39]:

> *... the quantum postulate [symbolised by Planck's quantum of action] implies that any observation of atomic phenomena will involve an interaction with the agency of observation not to be neglected. Accordingly, an independent reality in the ordinary physical sense can neither be ascribed to the phenomena nor to the agencies of observation.*

The breakdown of the concept of isolated system implies that the Schrödinger function does not describe ontological or intrinsic features underlying a given quantum system[8], such as the trajectory of an electron. That is, before a measurement the wave function is epistemologically interpreted as a mathematical symbol without any physical meaning. Bohm, for instance, had summarized such an epistemic interpretation of quantum mechanics as follows [42]:

> *... quantum theory requires us to give up the idea that the electron, or any other object has, by itself, any intrinsic properties at all. Instead, each object should be regarded as something containing only incompletely defined potentialities that are developed when the object interacts with an appropriate system.*

More specifically, a quantum system does behave as either a particle or a wave depending on how it is treated by the surrounding environment [42]. That is, only after being measured a quantum system manifests itself in a complementary way either as a wave-like or as a particle-like.

In summary, for both Bohr and Heisenberg a quantum system cannot be imagined as being in isolation from its surroundings. In truth, it only comes into being as far as its interaction with a certain environment (e.g., a measuring apparatus) is concerned. The definition of a quantum system (or a *quantum phenomenon* after Bohr) could be then schematized as displaying the following dualistic structure:

---

[8] *"In our description of nature the purpose is not to disclose the real essence of phenomena but only to track down as far as possible relations between the multifold aspects of our experience"* [41].

$$\boxed{\begin{array}{c}\text{Quantum System}\end{array}} = \boxed{\begin{array}{c}\text{Isolated System}\\ \psi\end{array}} + \boxed{\begin{array}{c}\text{Measuring Apparatus}\\ |\psi(x,t)|^2\end{array}}$$

where the symbol "+" denotes an interaction process between an isolated system, formally represented by the Schrödinger function before being measured, and a measuring apparatus that could be assumed to be of classical nature (*à la* Bohr and Heisenberg) or of quantum nature (*à la* von Neumann [43]). Quantum-mechanical probabilistic predictions can be calculated, but cannot be derived, from the Schrödinger equation. Probability shows up therefore as an epistemic concept because there is no ontological quantum world[9], i.e., an isolated quantum system.

While von Neumann [43] brought out the dualistic structure of the mathematical formalism of quantum mechanics by resorting to a non-isolated process that in turn cannot be described by the Schrödinger equation— the so-called collapse[10] (or reduction) of the wave function—, pointing out the controversial role of the observer's consciousness, Bohr [46] attempted to surpass such a dualism within the framework of his principle of complementarity [47,48] and the assumption of classicality of the measurement apparatus [33,49]. Like the existence of quantization methods mentioned above, it is worth stressing that both Bohr's classicality requirement and von Neumann's collapse suggest the non-universality of the Schrödinger equation.

Another way to try to surpass such a dualistic framework underlying the quantum-mechanical formalism is to restore the concept of isolated system, that is, to question what a quantum system (e.g., an electron) really *is* in the absence of any measuring devices. So from this ontological perspective a quantum system is to be viewed indeed as an isolated system:

$$\boxed{\begin{array}{c}\text{Quantum System}\end{array}} = \boxed{\begin{array}{c}\text{Isolated System}\\ \psi\end{array}}$$

Such an ontological stance on the foundations of quantum mechanics had prompted and guided the various criticisms raised by Einstein (e.g., [36, 50-52]) against the consequences of the epistemic interpretations of quantum mechanics advanced by Bohr, Heisenberg, Born, and others. In essence, for Einstein — following the Galilean tradition— no measurement processes can account for establishing the physical reality of a given isolated system. It is this view that underlies the EPR criterion of physical reality of any isolated systems [50]:

> *If, without in any way disturbing a system, we can predict with certainty (i.e., with probability equal to unity) the value of a physical quantity, then there exists an element of physical reality corresponding to this physical quantity.*

---

[9]As reported by Aage Petersen [44] when asked *"whether the algorithm of quantum mechanics could be considered as somehow mirroring an underlying quantum world"*, Bohr would answer: *"There is no quantum world. There is only an abstract quantum physical description. It is wrong to think that the task of physics is to find out how nature is. Physics concerns what we can say about nature"*. See also [34].

[10]Surprisingly, such a collapse process cannot be described by the Schrödinger equation itself (understanding the gist of the collapse constitutes the problem of the measurement in quantum mechanics [45]).

However, Einstein was not able to come up with an ontic interpretation according to which the probabilistic feature could be derived from a physics of isolated quantum systems [53,54]. Instead, he claimed that the Schrödinger function *per se* fails to unveil ontological properties underlying quantum systems [50-52]. It should be indeed interpreted as an ensemble of similar systems, and not as referring to individual systems [52, 55-58]. Despite that Einstein's epistemic interpretation of the wave function, some physicists [59] (e. g., de Broglie [60-62], Schrödinger [32], Bialobrzeski [63], Bohm [64], and Takabayasi [65]) had strived to interpret ontologically the Schrödinger function with or without additional variables (hidden variables) [20,33-37,59,66-75]. In this context, the wave function is viewed as a physical symbol having as much relationship to material systems as other symbols have [59]. Nevertheless, the physical meaning of the wave function $\psi(x,t)$ still holds one quite controversial issue [33,37,76-105]. Moreover, it has been argued that the various existing ontic interpretations of quantum mechanics (e.g., the de Broglie-Bohm theory) also lead to serious conceptual difficulties [33,37,91-93,106,107].

As already pointed out above, from an epistemic standpoint the isolated-system concept is a useful mathematical abstraction, whereas from an ontological point of view there remains a need to reveal properties inherent in a physical processes without resorting to any measuring apparatuses (or collapse of the wave function). Nevertheless, according to some researchers the most effective way of surpassing such conceptual difficulties is to go beyond the Schrödinger equation: Prigogine and coworkers [108,109], Blokhintsev [110], Cini [111-114], Balian [115], Bohm and Hiley [116], and Fain [117], for instance, has put forward that the von Neumann function or matrix density (or its Fourier transform, i.e., the so-called Wigner function), and not the Schrödinger function, is the fundamental quantity of quantum mechanics. According to Prigogine and coworkers, the so-termed Large Poincaré Systems (LPS) requires an extension of quantum mechanics whereby the basic quantity is the density matrix, and no longer the wave function [108,109]. In Blokhintsev's approach to quantum mechanics [110], which is regarded as a generalization of classical statistical mechanics, the basic concept is not the wave function, but the von Neumann function. For Cini [111-114] in turn the whole structure of quantum mechanics in phase space, expressed in terms of Wigner function, can be deduced from a single quantum postulate without also bringing in wave functions or probability amplitudes. Balian [115] has pointed out some shortcomings of the wave function concept in the study of certain physical systems, such as the description of the measurement process in quantum mechanics. For Bohm and Hiley [116] as well as Fain [117] the density matrix is regarded as the fundamental description of the state of a system, while the wave function is taken as an abstraction.

In order to eschew conceptual difficulties underlying both epistemic and ontic interpretations of quantum mechanics based on the Schrödinger function, in the present work we have as a *leitmotiv* the intention of unifying the different viewpoints of Bohr and Heisenberg, on the one hand, and Einstein, on the other. More specifically, from Bohr's and Heisenberg's physical views we reckon with their suggestion of abandoning the concept of isolated system and from Einstein's philosophical viewpoint we take into account his stance of unveiling the ontological status of physical systems without resorting to any measurement apparatus. To this end, we generalize the physics of isolated system through the concept of open quantum system whose description is based on the von Neumann function.

## 1.2. Open quantum systems

Let us consider a quantum system $A$ immersed in an environment or reservoir $B$. $A$ is then said to be an open system [118]. In the absence of any environment, $A$ is called an isolated system. That is a general definition of open system, whereby isolated systems arise as a special case.

The system as a whole, i.e., $A + B$, is deemed to be described by the von Neumann function $\varrho(q_1, q_2, x_1, x_2, t)$, where $x_1$ and $x_2$ denote the coordinates of the $A$ system and $q_1$ and $q_2$ the coordinates of the environment $B$. Generalizing Landau's and Lifschitz's definition of density matrix [119] (see also [120], p. 426), the reduced density matrix[11] of the quantum system $A$ may be defined as

$$\rho(x_1, x_2, t) = \iint \varrho(q_1, q_2, x_1, x_2, t) dq_1 dq_2. \qquad (1.4)$$

---

[11]Strictly speaking, the density matrix refers to the matrix representation of the density operator in a particular basis. The concept of density matrix was introduced by von Neumann [121], Landau [122] and Bloch [119].

In general, $\varrho(q_1, q_2, x_1, x_2, t) \neq \varrho(q_1, q_2, t)\varrho(x_1, x_2, t)$, meaning that both systems $A$ and $B$ cannot be separated from one another, hence they are said to be entangled. If the total quantum system $A + B$ is supposed to be isolated, then the joint von Neumann function $\varrho(q_1, q_2, x_1, x_2, t)$ can be factorized as the pure case $\varrho(q_1, q_2, x_1, x_2, t) = \Psi(q_1, x_1, t)\Psi^*(q_2, x_2, t)$, where $\Psi(q_1, x_1, t)$ is the joint wave function at $x_1$ and $q_1$ and $\Psi^*(q_2, x_2, t)$ its complex conjugate at points $x_2$ and $q_2$, so that the reduced density matrix (1.4) associated with the system $A$ also represents a pure case, i.e., $\rho(x_1, x_2, t) = \psi(x_1, t)\psi^*(x_2, t)$, with $\psi(x_1, t) = \int \Psi(q_1, x_1, t)dq_1$ and $\psi^*(x_2, t) = \int \Psi^*(q_2, x_2, t)dq_2$. In this case, starting from the Schrödinger equation (1.3a) at point $x_1$ and its complex conjugate (1.3b) at $x_2$, it is readily to show that the equation of motion for pure states reads[12]

$$i\hbar \frac{\partial \rho(x_1, x_2, t)}{\partial t} = \left[V(x_1, t) - V(x_2, t) - \frac{\hbar^2}{2m}\left(\frac{\partial^2}{\partial x_1^2} - \frac{\partial^2}{\partial x_2^2}\right)\right]\rho(x_1, x_2, t). \quad (1.5)$$

The Schrödinger equation in the density matrix representation, given by Eq. (1.5), is dubbed the Liouville-von Neumann equation or simply von Neumann equation.

However, if in general the total quantum system $A + B$ is assumed to be non-isolated, then $\varrho(q_1, q_2, x_1, x_2, t)$ represents a non-pure state called *improper mixture* [35], characterized by $\varrho(q_1, q_2, x_1, x_2, t) \neq \Psi(q_1, x_1, t)\Psi^*(q_2, x_2, t)$ or $\varrho(x_1, x_2, t) \neq \sum_n \omega_n \Psi_n(q_1, x_1, t)\Psi_n^*(q_2, x_2, t)$, thereby implying that $\rho(x_1, x_2, t)$ cannot be factorized as well, i.e., $\rho(x_1, x_2, t) \neq \psi(x_1, t)\psi^*(x_2, t)$, or cannot be expressed as a proper mixture $\rho(x_1, x_2, t) \neq \sum_n \omega_n \psi_n(x_1, t)\psi_n^*(x_2, t)$. So, underlying an open quantum system there exists no Schrödinger wave function [123]. In this context, a quantum system should be indeed idealized as an open system described by the von Neumann function $\rho$, i.e.,

$$\boxed{\text{Quantum System}} = \boxed{\begin{array}{c}\text{Open System}\\ \rho\end{array}}$$

In general, an open quantum system is mathematically described by master equations of the form

$$i\hbar \frac{\partial \rho(x_1, x_2, t)}{\partial t} = \mathcal{L}\rho(x_1, x_2, t), \quad (1.6)$$

---

[12]Nonpure states, called statistical mixtures or *proper mixtures* [35], given by $\rho(x_1, x_2, t) = \sum_n \omega_n \psi_n(x_1, t)\psi_n^*(x_2, t)$, the positive weights $\omega_n$ being time independent, are also described by the von Neumann equation (1.5). However, if $\omega_n$ rely on time, then there exists no wave function underlying $\rho(x_1, x_2, t)$.

where the so-called superoperator $\mathcal{L}$ acting on the von Neumann function $\rho(x_1, x_2, t)$ bears the Planck constant $\hbar$, some environmental features such as coupling constants (a sort of friction constant) and a kind of fluctuation energy as well as some properties inherent in the tagged particle such as its mass $m$ and its position. In addition, it is expected that in the absence of environment the quantum master equation (1.6) reduces to the von Neumann equation (1.5). The main issue in the quantum theory of open systems is therefore to find out master equations of the form (1.6)[13].

Under the mathematical criterion of complete positivity of the density matrix, a general form for Markovian quantum master equations, known as Lindblad master equations, has been put forward by the semigroup approach to open quantum systems [126-136]. Recently, such an approach has been generalized for non-Markovian quantum master equations [137].

Qantum master equations of the form (1.6) have physical applications in the following areas [138,139]: Nuclear Magnetic Resonance (NMR) [124,140-143], quantum optics [117,125,144-156], condensed matter physics [117,133], condensed phase chemical physics [117,157,158], and macroscopic quantum mechanics [159].

As a prototype of the movement undergone by a quantum open system, one considers quantum Brownian motion. In quantum optics, for example, the environment is represented by a quantized radiation field while the Brownian particle is deemed to be an atom or a molecule [125]. The corresponding master equation is called quantum optical master equation [133,151].

The predominant paradigm for deriving quantum Brownian master equations could be termed Hamiltonian, for it is based on the concept of isolated system, and can be summarized as follows[14] [118,133,164]: Imagine a generic environment coupled to a Brownian particle so that a quantum Hamiltonian function of the isolated system (particle plus environment) can be built up. Having established this Hamiltonian picture, the Nakajima-Zwanzig projection operator techniques [165,166] are employed to obtain quantum master equations, after getting rid of the environment's variables. In the specific case of a medium consisting of a set of harmonic oscillators it is usual to enter into quantum world via the canonical quantization procedure (Dirac or Heisenberg quantization) (see e.g. [159,167]). Here, the Feynman path integral formalism [163] is invocated for deriving the reduced dynamics of the quantum Brownian particle, after eliminating the environmental variables.

---

[13]The first derivation of a quantum master equation of the type (1.6) describing the relaxation of nuclear spin orientation seems to have been performed by Wangsness and Bloch [124] (see also Fano [123] and Haken [125]).

[14]From a historical point of view, the first attempt to set up a theory of quantum Brownian motion within a Hamiltonian framework seems to have been made by Claude George [160] on the basis of the canonical quantization of Prigogine's classical theory of Brownian movement [161]. Alternatively, Schwinger [162] as well as Feynman and Vernon [163] developed a theory of quantum Brownian motion under the Lagrangian structure: Schwinger's theory [162] was based on his quantum action principle (a differential approach) whereas Feynman and Vernon [163] employed path integral techniques.

In contrast to the Hamiltonian approach, an alternative methodology for deriving quantum master equations is to start from on the outset with a non-isolated system, thereby obtaining the isolated systems as special case. This non-Hamiltonian approach to quantum Brownian motion can be split into two ways of investigation: one way is to build up kinetical models leading to Lindbladian quantum master equations as a limiting case of the so-termed quantum linear Boltzmann equation [168-180]. Here, no quantization process is used. Alternatively, making use of a Hamiltonian-independent quantization procedure, called *dynamical quantization*, we have derived various Markovian and non-Markovian quantum master equations in Refs. [181-190].

The subject of the present book is to feature the foundations of our non-Hamiltonian approach to Brownian motion. First, on the ground of previous papers [191-192] we generalize the classical theory of Brownian motion by reckon with non-Markovian effects. Next, this non-Markovian generalization of Brownian motion is quantized by means of a non-Hamiltonian quantization method [181-190]. By way of consistency, the classical limit $\hbar \to 0$ of our quantum Brownian motion is calculated too. In addition, as far as thermal heat baths are concerned the quantum limit $T = 0$ is looked at.

## 1.3. Organization of the book

In Chapter 2 we present our generalization of the classical theory of (normal) Brownian motion by reckoning with non-Markovian effects on both the Klein-Kramers equation (in presence of inertial forces) and the Smoluchowski equation (in absence of inertial forces). Our non-Markovian Klein-Kramers equation is solved for a free particle whereas our non-Markovian Smoluchowski equation is solved for a free particle and a harmonic oscillator. Besides, our non-Markovian approach is applied to the phenomenon of anomalous diffusion as much in the presence as in the absence of inertial forces as well as being used to calculate a time-dependent Kramers escape rate.

Chapters 3 through 7 are devoted to investigate both non-Markovian and quantum effects on the motion of a Brownian particle immersed in thermal and non-thermal environments. Thermal environments, which may be a heat bath of quantum harmonic oscillators, a fermionic heat bath or a bosonic heat bath, are valid for all temperatures $T \geq 0$. The classical limit of our non-Markovian quantum Brownian motion is also examined.

In Chapter 3 a non-Markovian quantum master equation in the presence of a generic environment is derived. As far as a harmonic oscillators heat bath is concerned our quantum master equation becomes the non-Markovian Caldeira-Leggett equation. In addition, we generalize our non-Markovian quantum master equation to both bosonic and fermionic heat baths.

In Chapter 4 by making use of the Wigner representation of quantum mechanics we solve our non-Markovian quantum master equation describing a free Brownian particle in a general medium.

In Chapter 5 we obtain a non-Markovian quantum Smoluchowski equation in phase space as the quantization of a non-Markovian classical Smoluchowski equation derived in Chapter 2. Our quantum-mechanical Smoluchowski equation is then solved for a free particle and a harmonic oscillator in the presence of both thermal and non-thermal environments. Emphasis is on thermal systems at zero and high temperatures.

Chapter 6 in turn deals with the quantum anomalous diffusion in presence and absence of inertial force for thermal and non-thermal environments. Thermal systems are analyzed at zero and high temperatures.

Lastly, Chapter 7 addresses the problem of the quantum tunneling in presence as well as in absence of inertial force. Specifically, both Markovian and non-Markovian effects on the quantum tunneling for thermal and non-thermal open systems are investigated too.

Our conclusions are presented in Chapter 8, whereby our main theoretical predictions are summed up and conceptual implications of our non-Hamiltonian quantum mechanics are also stressed. Eventual generalizations of our approach are mentioned as well.

Eight appendices are included. Appendices A and B deal with the Einstein and Langevin approaches to Markovian Brownian motion, respectively. The former is based on the time evolution for the probability density function (the diffusion equation), while the latter starts from the stochastic differential equation for the random variable. Appendix C addresses the problem of deriving general equations of motion for the probability density function, which we have dubbed them Kolmogorov equations, from stochastic differential equations. In the Gaussian approximation our Kolmogorov equations reduce to Fokker-Planck equations the diffusion coefficient of which turn out to rely on a time-dependent function $I(t)$, called correlational function. This function, which is responsible for non-Markovian effects, is examined in Appendix D, and in Appendix E we find out the general solution to a non-Markovian classical Smoluchowski equation for a free Brownian particle so as to look at the differentiability property of the Brownian trajectories.

In Appendix F we show how the Schrödinger equation can be derived as a special case within the framework of our non-Hamiltonian quantum mechanics. Appendix G is devoted to the environment's physics, whereas the last Appendix H aims at to derive a non-Markovian quantum Smoluchowski equation in the Gaussian approximation for nonlinear potential energy function $V(x)$.

# 2. Classical Brownian motion: Non-Markovian effects

Brownian motion is physically interpreted as a diffusion phenomenon brought about by a myriad of collisions between the fluid's particles and a tagged particle[15]. Such a diffusion comes into being owing the interplay between two closely entangled processes: fluctuation and dissipation. On the one hand, fluctuations are responsible for activating the Brownian movement through correlational effects that in general hold both particle and environment statistically correlated during a span of time $t_c$, the so-called correlation timescale. For the $t_c > 0$ case, the interaction of the Brownian particle with its environment is said to be a non-Markovian one, while the $t_c \to 0$ limiting case makes it Markovian, that is, uncorrelated. Dissipation processes, on the other hand, account for damping the motion of the Brownian particle at a given relaxation timescale $t_r$.

This Chapter is organized as follows. Section 2.1 aims at presenting a brief historical overview of the Markovian approach to Brownian motion[16] as well as pointing out the need to go beyond such Markovian assumption. Our non-Markovian approach to (normal) Brownian motion is the subject of Sect. 2.2 and next extended to the phenomena of anomalous diffusion in Sect. 2.3. Non-Markov effects on Kramers escape rate are investigated in Sect. 2.4. Lastly, in Sect. 2.5 we summarize the main features of our non-Markovian approach to Brownian motion and put it in perspective by comparing our findings with other non-Markovian accounts existing in literature.

## 2.1. Is Brownian motion actually Markovian?

Historically, the physics of Brownian movement had been mathematically addressed by Einstein [194] and Langevin [195] at the beginning of the past century on the basis of the concept of probability. Nevertheless, it is worth pointing out that the mathematical formalization of the probability concept was achieved by Kolmogorov only in 1933 on the ground of measure theoretic probability [196,197]. In the Kolmogorov probabilistic framework physical quantities, such as the displacement and the velocity of a Brownian particle, are represented by random variables whose realizations are distributed according to a certain probability distribution function. The physical-mathematical contributions of Einstein [194, 198-

---

[15]Under normal conditions, in a liquid, a Brownian particle suffers about $10^{21}$ collisions per second [193].
[16]Appendices A and B are devoted to the Einstein and Langevin approaches to Markovian Brownian motion, respectively.

201], Langevin [195], and Kolmogorov [196,197,202,203], we dub them the Einstein-Langevin-Kolmogorov picture of Brownian motion.

Thanks to Einstein [194], the random motion of a free Brownian particle (in the absence of inertial force) could be described by the diffusion equation for the probability distribution function $\mathcal{F}(x,t)$, i.e.,

$$\frac{\partial \mathcal{F}(x,t)}{\partial t} = D \frac{\partial^2 \mathcal{F}(x,t)}{\partial x^2}. \tag{2.1}$$

This Eq. (2.1), which is a sort of Fokker-Planck equation, may be derived from the Bachelier-Einstein integral equation[17] [194,205]

$$\mathcal{F}(x, t+\tau) = \int_{-\infty}^{\infty} \mathcal{F}(x+\Delta x, t)\varphi(\Delta x, \tau)d(\Delta x) \tag{2.2}$$

in the Markovian limit $\tau \to 0$ (see Appendix A). The function $\varphi(\Delta x, \tau)$ in Eq. (2.2) accounts for the transition from the distribution function $\mathcal{F}(x + \Delta x, t)$ to $\mathcal{F}(x, t + \tau)$. The diffusion constant $D$ in Eq. (2.1) reads

$$D \equiv \frac{1}{2} \lim_{\tau \to 0} \frac{1}{\tau} \int_{-\infty}^{\infty} (\Delta x)^2 \varphi(\Delta x, \tau) d(\Delta x). \tag{2.3}$$

As far as a thermal reservoir in thermodynamic equilibrium is concerned, one can show that the Brownian motion of a particle of mass $m$ is characterized by the Sutherland-Einstein diffusion constant [194,206]

$$D = \frac{k_B T}{\beta m}, \tag{2.4}$$

where $k_B$ is the Boltzmann constant and $T$ the (absolute) temperature of the thermal reservoir. The diffusion constant (2.4), or expression (2.3), connects dissipation processes through the viscosity coefficient $\alpha \equiv \beta m$, characterized by the relaxation timescale $t_r \equiv \beta^{-1}$, with fluctuations coming from the thermal energy $\mathcal{E} \equiv k_B T$ of the heat bath defined in the steady regime, $t \to \infty$, i.e., $t \gg \tau$. In addition, $\beta$ conveys information on some mechanical properties of the environment (e.g., its viscosity) as well as some geometrical features inherent in the Brownian particle (its size, for instance)[18]. Thus, the Sutherland-Einstein diffusion constant (2.4) is considered as the first example of fluctuation-dissipation relation [208-212].

---

[17]The integral equation (2.2) is commonly known as the Chapman-Kolmogorov equation or the Smoluchowski equation (see, e.g., [120,204]).

On the other hand, Langevin [195] set out to address the problem of Brownian motion by focusing on the concept of random variable $X \equiv X(t)$ whose time evolution is given by a stochastic differential equation, the so-called Langevin equation. According to that approach, a free Brownian particle (in the presence of inertial force) is described by the following Langevin equation (see Appendix B)

$$m \frac{d^2 X}{dt^2} = -\beta m \frac{dX}{dt} + L(t), \qquad (2.5)$$

where the inertial force $m d^2 X/dt^2$ offsets two kinds of environmental forces: A linearly velocity-dependent dissipative force $F_\mathrm{d} = -\beta m dX/dt$, accounting for stopping the particle's motion, as well as an anti-dissipative force $L(t)$, dubbed Langevin's force, responsible for activating the particle's movement through fluctuations brought about by collisions with the environmental particles [120].

In the Langevin approach to Brownian motion, Markovianity assumption turns up in the following statistical property to be satisfied by the Langevin force[19] $L(t)$

$$\langle L(t) L(t') \rangle = 2D\delta(t - t'), \qquad (2.6)$$

where $D$ is the diffusion constant (2.4) and $\langle ... \rangle$ denotes the average value evaluated over the probability distribution function associated with the stochastic process $L(t)$. The delta correlated function (2.6) stands for that each collision occurs instantaneously and that successive collisions are uncorrelated so that $L(t)$ and $L(t')$ are totally independent for arbitrarily small time $|t - t'|$ [120].

While Einstein [194] focused on the time evolution of the probability distribution function, Langevin [195] pinpointed the dynamics of random variable governed by stochastic differential equation. The connection between both Einstein and Langevin approaches to Brownian movement had been carried out by Ornstein [214,215] in velocity space, and then by Klein [216] (and independently by Kramers [217] ) in phase space.  Later, Uhlenbeck and Ornstein [218] investigated new features inherent in such connection as much in configuration space as in velocity space. Finally, Wang and Uhlenbeck [219] were able to provide a concise way to derive Fokker-Planck equations from Langevin equations. In this context Stratonovich's book [204] is a highly recommended reference to the construction of Fokker-Planck equations from Langevin equations.

---

[18]If the Brownian particle is deemed to be spherical and the fluid is treated as a continuous medium (that is, if the mean free particle path of the fluid particles is small compared with the size of the Brownian particle), then the viscosity coefficient $\alpha$ may be calculated by Stokes' law in hydrodynamics as $\alpha = 6\pi\theta a$ or $\alpha = 4\pi\theta a$, where $a$ denotes the radius of the Brownian particle and $\theta$ the viscosity of the fluid [207].

[19]According to Naqvi [213] the first glimpses of the delta function had turn up in Ornstein's 1919 paper on Brownian motion [214].

The existence of the Bachelier-Einstein integral equation (2.2) at $\tau \to 0$, or the statistical property (2.6), is deemed to be the pivotal assumption underpinning the Markovian theory of Brownian motion [193,219]. Hence, van Kampen has claimed that *"[t]he art of the physicist is to find those variables that are needed to make the description (approximately) Markovian"* [120]. The consequence of the Markovian assumption is that the (stochastic) Brownian trajectories are nondifferentiable and, accordingly, the Langevin equation (2.5) cannot be mathematically interpreted as a genuine differential equation. This result is in sharp contrast to the (deterministic) Newtonian trajectories that are assumed to be differentiable functions. In other words, the actual path of a Brownian particle displays no physical reality (see Appendices A and B).

Nevertheless, it has been argued that the Markov assumption (2.6) is a highly idealized feature since any physical interaction between the Brownian particle and ambient's particles actually comes about during a finite correlation time $t_c \neq 0$ [120,192,204,220-224][20]. Furthermore, the mere existence of the Bachelier-Einstein equation (2.2) does not imply Markovianity [120,225]. In short, Markovianity assumption does not fathom the gist of Brownian movement. This conclusion has been recognized by van Kampen himself who laconically asserted [224]: *"Non-Markov is the rule, Markov is the exception."*

In what follows we intend to present our approach to Brownian motion [192] in which we uphold the view that this sort of erratic movement is actually non-Markovian. Hence, the random trajectories of a Brownian particle are to be described by differentiable functions, thereby exhibiting physical reality.

## 2.2. Non-Markovian Brownian motion

*Stochastic differential equations in phase space.* In the presence of a generic environment, it is assumed that the phase-space dynamics of a Brownian particle of mass $m$, position $X \equiv X(t)$, and linear momentum $P \equiv P(t) = mdX(t)/dt$, moving in an external potential $V \equiv V(X)$, may be probabilistically described by the following system of stochastic differential equations

---
[20]According to van Kampen [120], for instance, the use of the delta function in Eq. (2.6) is a mere convenience issue.

$$\frac{dP(t)}{dt} = -\frac{dV(X)}{dX} - \beta P(t) + b\Psi(t), \tag{2.7a}$$

$$\frac{dX(t)}{dt} = \frac{P(t)}{m}. \tag{2.7b}$$

The term $dP(t)/dt$, called inertial force and proportional to the acceleration $d^2X(t)/dt^2$, offsets two sorts of force: A conservative force $F_c(X) = -dV(X)/dX$, derived from the Brownian particle's potential energy $V(X)$, and an environmental force $F_{\text{env}}(P,\Psi) = -\beta P + b\Psi(t)$, made up by the dissipative force, $F_d(P) = -\beta P$, and the fluctuating force $L(t) = b\Psi(t)$. If $\Psi(t)$ has dimensions of [time$^{-1/2}$], then $b$ displays dimensions of [mass × length × time$^{-3/2}$].

From a mathematical point of view, both Eqs. (2.7) are defined within the Kolmogorov probabilistic framework in which $X(t), P(t)$, and $\Psi(t)$ are interpreted as stochastic processes (or random variables) in the sense that there is a probability distribution function, $\mathcal{F}_{XP\Psi}(x,p,\psi,t)$, expressed in terms of the possible values (realizations) $x = \{x_i(t)\}$, $p = \{p_i(t)\}$, and $\psi = \{\psi_i(t)\}$, with $i \geq 1$, distributed about the sharp values $x'$, $p'$, and $\psi'$ of $X(t), P(t)$, and $\Psi(t)$, respectively. By contrast, the parameters $t$ (time), $\beta$ (frictional constant), $m$ (mass), and $b$ (fluctuation strength) in the stochastic differential equations above show up as non-random quantities so having sharp values.

In the context of the theory of stochastic processes, the average value of any random function $A(X,P,\Psi,t)$ is deemed to be expressed as

$$\langle A(X,P,\Psi,t)\rangle = \int_{-\infty}^{\infty}\int_{-\infty}^{\infty}\int_{-\infty}^{\infty} a(x,p,\psi,t)\mathcal{F}_{XP\Psi}(x,p,\psi,t)dxdpd\psi \tag{2.8}$$

which in turn is a time-dependent non-random quantity. It is straightforward to check from Eq. (2.8) that the normalization condition of $\mathcal{F}_{XP\Psi}(x,p,\psi,t)$ reads

$$\langle 1 \rangle = \int_{-\infty}^{\infty}\int_{-\infty}^{\infty}\int_{-\infty}^{\infty} \mathcal{F}_{XP\Psi}(x,p,\psi,t)dxdpd\psi = 1. \tag{2.9}$$

The deterministic limit follows from the Brownian dynamics (2.7) as far as $\mathcal{F}_{XP\Psi}(x,p,\psi,t) = \delta(x-x')\delta(p-p')\delta(\psi-\psi')$ is concerned. So the stochastic differential equations (2.7) change into the following deterministic differential equations

$$\frac{dp(t)}{dt} = -\frac{dV(x)}{dx} - \beta p(t) + f(t), \quad (2.10a)$$

$$\frac{dx(t)}{dt} = \frac{p(t)}{m}, \quad (2.10b)$$

which describe a dissipative Newtonian system in terms of the sharp variables $x(t) = \langle X(t) \rangle$ and $p(t) = \langle P(t) \rangle$ in the presence of a time-dependent external force (a driving force) given by $f(t) \equiv b\psi(t) = b\langle \Psi(t) \rangle$ (see, e.g., [226]).

*Kolmogorov equation in phase space.* The stochastic differential equations in phase space (2.7) give rise to the following Kolmogorov equation in phase space (see Appendix C)

$$\frac{\partial \mathcal{F}(x, p, t)}{\partial t} = \mathbb{K}\mathcal{F}(x, p, t) \quad (2.11)$$

for the (marginal) probability distribution function (the Kolmogorov function)

$$\mathcal{F}(x, p, t) = \int_{-\infty}^{\infty} \mathcal{F}_{XP\Psi}(x, p, \psi, t)\, d\psi. \quad (2.12)$$

Equation of motion (2.11) is an exact equation in which the Kolmogorovian operator $\mathbb{K}$ acts on the function $\mathcal{F}(x, p, t)$ according to the prescription (see Appendix C)

$$\mathbb{K}\mathcal{F}(x, p, t) = \sum_{k=1}^{\infty} \sum_{r=0}^{k} \frac{(-1)^k}{r!\,(k-r)!} \frac{\partial^k}{\partial x^{k-r} \partial p^r} \left[ A^{(k-r,r)}(x, p, t) \mathcal{F}(x, p, t) \right], \quad (2.13)$$

the time-dependent coefficients $A^{(k-r,r)}(x, p, t)$ being given by[21]

$$A^{(k-r,r)}(x, p, t) \equiv \lim_{\epsilon \to 0} \left[ \frac{\langle (\Delta X)^{k-r} \rangle \langle (\Delta P)^r \rangle}{\epsilon} \right], \quad (2.14)$$

where the average quantities, $\langle (\Delta X)^{k-r} \rangle$ and $\langle (\Delta P)^r \rangle$, are deemed to be calculated about their sharp values $x'$ and $p'$, respectively, i.e.,

$$\mathcal{F}_{XP\Psi}(x, p, \psi, t) = \delta(x - x')\delta(p - p')\mathcal{F}_\Psi(\psi, t). \quad (2.15)$$

Both increments $\Delta X \equiv X(t + \epsilon) - X(t)$ and $\Delta P \equiv P(t + \epsilon) - P(t)$ in Eq. (2.14) are evaluated from Eqs. (2.7) and hence expressed respectively in the integral form

---

[21]The coefficients (2.14) are time dependent in accordance with Pawula [227,228] and Coffey, Kalmykov, and Waldron [229], but in sharp contrast to the derivations performed by some authors (e.g., [120,159]).

$$\Delta X = \frac{1}{m}\int_t^{t+\epsilon} P(t)dt, \qquad (2.16)$$

$$\Delta P = -\int_t^{t+\epsilon}\left[\frac{dV(X)}{dX}+\beta P(t)\right]dt + b\int_t^{t+\epsilon}\Psi(t)dt. \qquad (2.17)$$

Since Eq. (2.11) relies on an infinite number of autocorrelation functions of the random force $\langle L(t_1)L(t_2)\dots L(t_n)\rangle$, with $n=2,\dots,\infty$, both Eqs. (2.7) and (2.11) describe the time evolution in phase space of a Brownian particle immersed in a generic non-Gaussian environment.

It is interesting to notice that both the root mean square displacement, $\mathbb{X}(t)\equiv\sqrt{\langle X^2(t)\rangle-\langle X(t)\rangle^2}$, and the root mean square momentum, $\mathbb{P}(t)\equiv\sqrt{\langle P^2(t)\rangle-\langle P(t)\rangle^2}$, satisfy the following fluctuation relationship in phase space

$$\mathbb{X}(t)\mathbb{P}(t)\geq 0. \qquad (2.18)$$

The $\mathbb{X}(t)\mathbb{P}(t)=0$ case is valid for either $\mathbb{X}(t)=0$ or $\mathbb{P}(t)=0$. For isolated systems, it is expected that $\mathbb{X}(t)=0$ and $\mathbb{P}(t)=0$.

In addition, the non-Gaussian equation of motion (2.11) is to be solved starting from a given initial condition $\mathcal{F}(x,p,t=0)=\mathcal{F}_0(x,p)$. If the solution $\mathcal{F}(x,p,t)$ renders steady in the long-time regime

$$\lim_{t\to\infty}\mathcal{F}(x,p,t)=\mathcal{F}(x,p), \qquad (2.19)$$

then it is said that the Brownian particle has reached the same stationary state of the environment. Otherwise, i.e., if there exists the asymptotic

$$\lim_{t\to\infty}\mathcal{F}(x,p,t)\approx\tilde{\mathcal{F}}(x,p,t), \qquad (2.20)$$

the Brownian particle holds in a non-stationary state even at long times.

### 2.2.1. Non-Markovian Fokker-Planck equations

*Generalized Fokker-Planck equations in phase space.* As far as the central limit theorem or Pawula's theorem [227,228] is concerned, the statistical properties of the random force $L(t)$ in Eq. (2.11) turn out to be characterized only by its mean $\langle L(t_1)\rangle$ and the autocorrelation function $\langle L(t_1)L(t_2)\rangle$. Accordingly, in this Gaussian approximation the stochastic differential equations (2.7) are dubbed Langevin equations, whereas the Kolmogorov equation (2.11) reduces to the so-called Fokker-Planck equation, i.e.,

$$\frac{\partial \mathcal{F}}{\partial t} = -\frac{p}{m}\frac{\partial \mathcal{F}}{\partial x} + \frac{\partial}{\partial p}\left[\frac{\partial \mathcal{V}_{\text{eff}}(x,t)}{\partial x} + \beta p\right]\mathcal{F} + \mathcal{D}_p(t)\frac{\partial^2 \mathcal{F}}{\partial p^2}, \qquad (2.21)$$

albeit its solutions $\mathcal{F} \equiv \mathcal{F}(x, p, t)$ are in general non-Gaussian functions due to the nonlinear character of the potential energy $V(x)$ present in the effective potential $\mathcal{V}_{\text{eff}}(x, t)$, given by

$$\mathcal{V}_{\text{eff}}(x,t) = V(x) - xb\langle\Psi(t)\rangle, \qquad (2.22a)$$

$b\langle\Psi(t)\rangle$ being the average of the Langevin random force $L(t) = b\Psi(t)$, i. e.,

$$b\langle\Psi(t)\rangle = b\int_{-\infty}^{\infty}\psi \mathcal{F}_\Psi(\psi,t)d\psi, \qquad (2.22b)$$

where we have used the fact that $\langle\Psi(t)\rangle = \lim_{\epsilon\to 0}(1/\epsilon)\int_t^{t+\epsilon}\langle\Psi(t')\rangle dt'$. The time-dependent diffusion coefficient $\mathcal{D}_p(t)$ in Eq. (2.21) is given by

$$\mathcal{D}_p(t) = \beta m \mathcal{E}(t), \qquad (2.23a)$$

which is deemed to be defined in terms of the time-dependent diffusion energy

$$\mathcal{E}(t) = \mathcal{E}I(t), \qquad (2.23b)$$

where $I(t)$ is the dimensionless function

$$I(t) = \lim_{\epsilon\to 0}\frac{1}{\epsilon}\int_t^{t+\epsilon}\int_t^{t+\epsilon}\langle\Psi(t')\Psi(t'')\rangle dt' dt'', \qquad (2.24)$$

and $\mathcal{E}$ the time-independent diffusion energy

$$\mathcal{E} = \frac{b^2}{2\beta m}, \qquad (2.25a)$$

yielding in turn the fluctuation-dissipation relation in the form

$$b = \sqrt{2\beta m \mathcal{E}}. \qquad (2.25b)$$

So the parameter $b$ measuring the fluctuation strength in the Langevin equation (2.7a) is determined by the friction constant $\beta$, the particle's mass $m$, as well as the diffusion energy $\mathcal{E}$.

The characteristic feature underlying the concept of time-dependent diffusion energy (2.23b), $\mathcal{E}(t) = b^2 I(t)/2\beta m$, is that it sets up a general relationship between fluctuation and dissipation phenomena: While the term $b^2 I(t)$ comes from the autocorrelation of the Langevin fluctuating force $L(t) = b\Psi(t)$, the frictional constant $\beta$ gives rise to dissipation processes.

Even though the average value of the Langevin force accounts for the appearance of both the external force $f(t)$ in the deterministic limit (2.10a) and the effective potential (2.22a), if for convenience we could define the random variable $\Psi(t)$ as $\Psi(t) = \Phi(t) - \langle \Phi(t) \rangle$, then we obtain the following statistical property[22]

$$\langle \Psi(t) \rangle = 0, \tag{2.26}$$

where the average is taken over the environmental distribution $\mathcal{F}_\Psi(\psi, t)$ according to Eq. (2.22b). Thus, taking condition (2.26) into account and making use of Eqs. (2.23) the Fokker-Planck equation (2.21) reads

$$\frac{\partial \mathcal{F}}{\partial t} = -\frac{p}{m}\frac{\partial \mathcal{F}}{\partial x} + \frac{\partial}{\partial p}\left[\frac{dV(x)}{dx} + \beta p\right]\mathcal{F} + \beta m \mathcal{E} I(t) \frac{\partial^2 \mathcal{F}}{\partial p^2}. \tag{2.27}$$

Underlying this Gaussian Fokker-Planck equation (2.27), there exist the Langevin equations (2.7) satisfying the dissipation-fluctuation relation (2.25b) and exhibiting the following Gaussian statistical properties: The zero mean (2.26) and the unspecified autocorrelation function $\langle \Psi(t')\Psi(t'') \rangle$ present in $I(t)$.

*The correlational function $I(t)$.* For long times $t \to \infty$, if the correlational function (2.24) displays the following steady behavior

$$\lim_{t\to\infty} I(t) = 1, \tag{2.28}$$

then the fluctuations render Markovian in the sense that the time-dependence for the statistical autocorrelations of the Langevin force is "forgotten" in the steady regime, whereby the Brownian particle's diffusion energy (2.23b) becomes stationary:

---

[22]Without employing assumption (2.26), we have recently looked at both classical and quantum Brownian motion in Refs. [189,190,192], whereby it has been shown that averaging effects of the Langevin force $L(t) = b\Psi(t)$ are unobservable. Therefore, taking $\langle L(t) \rangle = 0$ seems to be an unnecessary assumption in the theory of Brownian motion. Furthermore, it is worth emphasizing that the average in Eq. (2.26) should be taken over $\mathcal{F}_\Psi(\psi, t)$, and not over test-particles of velocity $v(t)$ [230], or over a large number of equal test-particles having the same initial velocity $v(t = 0)$ [219], for instance. Moreover, the definition of $\Psi(t)$ as $\Psi(t) = \Phi(t) - \langle \Phi(t) \rangle$ so as to obtain the statistical property (2.26) reveals to be more general than the arguments constructed by Ferrari [230] and by Naqvi [213] for justifying $\langle \Psi(t) \rangle = 0$ on the ground of the Brownian motion of a free particle. After all, assuming $\langle L(t) \rangle = 0$ as a statistical property to be satisfied by the Langevin force is a mere notational convenience!

$\lim_{t\to\infty} \mathcal{E}(t) = \mathcal{E}(\infty) \equiv \mathcal{E}$. Accordingly, the particle-environment interaction is said to be non-Markovian in the non-steady range $0 < t < \infty$.

The non-Markovian character is also manifest if $I(t)$ does feature the following asymptotic behavior

$$\lim_{t\to\infty} I(t) \approx I'(t). \tag{2.29}$$

In that case the stochastic process remains ever non-Markovian.

In Appendix D, we have built up a non-Markovian correlational function $I(t)$ exhibiting the Markovian behavior (2.28) as

$$I(t) = 1 - e^{\frac{-t}{t_c}}, \tag{2.30}$$

where the correlation time $t_c$ explicitly represents the non-Markovianity parameter of the Brownian motion. At short times $t \to 0$, Eq. (2.30) approaches $I(t) \sim t/t_c$ so implying that $t_c > 0$, since the Markovian limit $t_c \to 0$ would lead to the following unphysical result: $\mathcal{D}_p(t) \to \infty$. On the other hand, Eq. (2.30) in the overcorrelated case $t_c \to \infty$ predicts no diffusion phenomenon for all time $t$: $I(t) = 0$. Accordingly, the correlation time $t_c$ in Eq. (2.30) is to be held within the range $0 < t_c < \infty$.

One example of $I(t)$ exhibiting the non-steady behavior (2.29) has also been put forward in Appendix D as

$$I(t) = \lambda \frac{t^{\lambda-1}}{t_c^{\lambda-1}} + 1 - e^{\frac{-t}{t_c}}, \tag{2.31}$$

for $\lambda > 1$. Section 2.3 below is devoted to the study of the anomalous diffusion in which $\lambda$ turns out to be interpreted as an anomaly parameter or deviation from the normal diffusion $\lambda = 0$.

*The non-Markovian Klein-Kramers equation.* Let us consider an environment in thermodynamic equilibrium at temperature $T$ and characterized by the thermal energy $k_B T$, where $k_B$ denotes the Boltzmann constant. Upon identifying the Brownian particle's diffusion energy $\mathcal{E}$ with the reservoir's thermal energy $k_B T$, i.e.,

$$\mathcal{E} \equiv k_B T, \tag{2.32}$$

the non-Markovian Fokker-Planck equation (2.27) reads

$$\frac{\partial \mathcal{F}}{\partial t} = -\frac{p}{m}\frac{\partial \mathcal{F}}{\partial x} + \frac{\partial}{\partial p}\left[\frac{dV(x)}{dx} + \beta p\right]\mathcal{F} + \beta m k_B T I(t)\frac{\partial^2 \mathcal{F}}{\partial p^2}. \tag{2.33a}$$

Under condition $I(t) = 1$, and using $\beta = 2\gamma$, the non-Markovian Fokker-Planck equation (2.33) is called the Markovian Klein-Kramers equation [216,217]

$$\frac{\partial \mathcal{F}}{\partial t} = -\frac{p}{m}\frac{\partial \mathcal{F}}{\partial x} + \frac{\partial}{\partial p}\left[\frac{dV(x)}{dx} + 2\gamma p\right]\mathcal{F} + 2\gamma m k_B T \frac{\partial^2 \mathcal{F}}{\partial p^2}. \quad (2.33b)$$

Hence, we have dubbed Eq. (2.33a) the non-Markovian Klein-Kramers equation [192].

It is worth noticing that due to the behavior (2.28) both equations of motion (2.33) do possess the same steady motion as $t \to \infty$, i.e.,

$$-\frac{p}{m}\frac{\partial \mathcal{F}}{\partial x} + \frac{\partial}{\partial p}\left[\frac{dV(x)}{dx} + 2\gamma p\right]\mathcal{F} + 2\gamma m k_B T \frac{\partial^2 \mathcal{F}}{\partial p^2} = 0. \quad (2.34)$$

We now proceed to investigate non-Markov effects on the Brownian movement of an inertial free particle according to both Fokker-Planck and Langevin descriptions.

*Fokker-Planck description of an inertial free particle.* Let us consider the inertial Brownian motion of a free particle, $V(x) = 0$, described by the non-Markovian Klein-Kramers equation (2.33a), with the correlational function (2.30),

$$\frac{\partial \mathcal{F}}{\partial t} = -\frac{p}{m}\frac{\partial \mathcal{F}}{\partial x} + 2\gamma\frac{\partial}{\partial p}(p\mathcal{F}) + 2\gamma m k_B T \left(1 - e^{\frac{-t}{t_c}}\right)\frac{\partial^2 \mathcal{F}}{\partial p^2}. \quad (2.35)$$

We seek for a solution of that equation of motion in the factorized form

$$\mathcal{F}(x,p,t) = f(x,t)\mathcal{F}(p,t), \quad (2.36)$$

such that on integrating over $x$, i.e., $\mathcal{F}(p,t) = \int_{-\infty}^{\infty} \mathcal{F}(x,p,t)dx$, we obtain from Eq. (2.35) the non-Markovian Rayleigh equation[23]

$$\frac{\partial \mathcal{F}(p,t)}{\partial t} = 2\gamma\frac{\partial}{\partial p}[p\mathcal{F}(p,t)] + 2\gamma m k_B T \left(1 - e^{\frac{-t}{t_c}}\right)\frac{\partial^2 \mathcal{F}(p,t)}{\partial p^2}. \quad (2.37)$$

Inserting Eq. (2.37) into Eq. (2.35) yields the following equation of motion for the function $f(x,t)$

---

[23] An alternative manner of deriving the non-Markovian Rayleigh equation (2.37) directly from the Langevin equation $dP(t)/dt = -2\gamma P(t) + \sqrt{4\gamma m k_B T}\Psi(t)$ can be found in Appendix C.

$$\frac{\partial f(x,t)}{\partial t} = -\frac{p}{m}\frac{\partial f(x,t)}{\partial x}, \qquad (2.38)$$

which is deterministic in the sense that it contains no diffusion term like Eq. (2.37). To solve Eq. (2.38) we make use of the strategy of adding to it the diffusion term $D\partial^2 f(x,t)/\partial x^2$, under the condition $D \to 0$. In so doing, we find the solution $f(x,t) = \delta(x)$, implying that the position of the Brownian particle has no fluctuation:

$$\mathbb{X}(t) \equiv \sqrt{\langle X^2(t)\rangle - \langle X(t)\rangle^2} = 0. \qquad (2.39)$$

In order to solve the non-Markovian Rayleigh equation (2.37), on the other hand, we assume the initial condition to be $\mathcal{F}(p, t=0) = \delta(p)$. So its solution reads

$$\mathcal{F}(p,t) = \frac{1}{\sqrt{4\pi \mathcal{G}(t)}} e^{\frac{-p^2}{4\mathcal{G}(t)}}, \qquad (2.40)$$

with

$$\mathcal{G}(t) = \frac{mk_B T}{2}\left[1 - e^{-4\gamma t} + \frac{4\gamma t_c}{4\gamma t_c - 1}\left(e^{-4\gamma t} - e^{\frac{-t}{t_c}}\right)\right]. \qquad (2.40a)$$

The probability distribution function (2.40) does bring the following timescales: The evolution time $t$, the relaxation time $t_r = (4\gamma)^{-1}$ as well as the correlation time $t_c$. Furthermore, it yields both the mean momentum $\langle P(t)\rangle = 0$ and the mean square momentum

$$\langle P^2(t)\rangle = 2\mathcal{G}(t) = mk_B T\left[1 - e^{-4\gamma t} + \frac{4\gamma t_c}{4\gamma t_c - 1}\left(e^{-4\gamma t} - e^{\frac{-t}{t_c}}\right)\right]. \qquad (2.41)$$

The mean mechanical energy, $\langle E(t)\rangle \equiv \langle P^2(t)\rangle/2m$, then reads

$$\langle E(t)\rangle = \frac{k_B T}{2}\left[1 - e^{-4\gamma t} + \frac{4\gamma t_c}{4\gamma t_c - 1}\left(e^{-4\gamma t} - e^{\frac{-t}{t_c}}\right)\right], \qquad (2.42)$$

whereas the root mean square momentum, $\mathbb{P}(t) = \sqrt{2m\langle E(t)\rangle}$, becomes

$$\mathbb{P}(t) = \sqrt{mk_B T\left[1 - e^{-4\gamma t} + \frac{4\gamma t_c}{4\gamma t_c - 1}\left(e^{-4\gamma t} - e^{\frac{-t}{t_c}}\right)\right]}, \qquad (2.43)$$

in which non-Markovian effects account for diminishing the strength of the momentum fluctuations.

Taking into account the solution $\delta(x)$ to Eq. (2.38) and solution (2.40) to Eq. (2.37), the Fokker-Planck description of Brownian free motion yields therefore as solution (2.36) the function

$$\mathcal{F}(x,p,t) = \delta(x)\frac{1}{\sqrt{4\pi\mathcal{G}(t)}}e^{\frac{-p^2}{4\mathcal{G}(t)}}, \qquad (2.44)$$

thereby implying that an inertial free Brownian particle in phase space fulfills the fluctuation relation (2.18) in the form $\mathbb{X}(t)\mathbb{P}(t) = 0$.

*Steady regime.* In the steady regime $t \to \infty$, physically interpreted as $t \gg (4\gamma)^{-1}, t_c$, the non-equilibrium solution (2.40) leads rightly to the Maxwell probability distribution function

$$\mathcal{F}(p) = \frac{1}{\sqrt{2\pi m k_B T}}e^{\frac{-p^2}{2mk_B T}}, \qquad t \to \infty, \qquad (2.45)$$

thus implying that a classical Brownian free particle satisfies the energy equipartition theorem, $\langle E(\infty)\rangle = k_B T/2$.

*Ballistic regime.* At short times $t \to 0$, i.e., as $t^3 \ll 1$, the momentum fluctuation (2.43) does display the following ballistic behavior

$$\mathbb{P}(t) \sim t\sqrt{\frac{2\gamma m k_B T}{t_c}}, \qquad t \to 0, \qquad (2.46)$$

which is differentiable, thereby giving rise to the thermal force $\mathbb{F}(t) \equiv d\mathbb{P}(t)/dt$, i.e.,

$$\mathbb{F}(t) \sim \sqrt{\frac{2\gamma m k_B T}{t_c}}, \qquad t \to 0. \qquad (2.47)$$

Notice that the differentiability of $\mathbb{P}(t)$, or the existence of $\mathbb{F}(t)$, is due to the presence of non-Markov effects encapsulated in the correlation time $t_c$. Since the momentum fluctuation $\mathbb{P}(t)$, Eq. (2.46), leads to the violation of the energy equipartition theorem at $t = 0$, i.e., $\langle E(0)\rangle \neq k_B T/2$, the non-Markovian thermal force (2.47) can measure the violation of such a theorem at short time scales.

By way of example [192], on considering $t_c \sim 10^{-12}$s, $\gamma \sim 10^{12}$s$^{-1}$, $m \sim 10^{-3}$kg, $k_B \sim 10^{-23}$m$^2$ kg s$^{-2}$ K$^{-1}$, and $T \sim 300$K, we obtain $\mathbb{F}(t) \sim 1$N, which is the magnitude of the thermal force (2.47) exerted by a heat bath at room temperature 27°C on a Brownian free particle of mass 1g.

The property of differentiability of the momentum fluctuation (2.43) implies therefore the existence of the thermal force

$$\mathbb{F}(t) = \sqrt{\frac{2\gamma m k_B T}{t_c}} \mathrm{B}(t) \qquad (2.48)$$

for all time $t \geq 0$. The dimensionless function $\mathrm{B}(t)$ is given by

$$\mathrm{B}(t) = \frac{\sqrt{2\gamma t_c}\left(e^{\frac{-t}{t_c}} - e^{-4\gamma t}\right)}{\sqrt{(4\gamma t_c - 1)\left[e^{-4\gamma t} - 1 + 4\gamma t_c\left(1 - e^{\frac{-t}{t_c}}\right)\right]}}, \qquad (2.48a)$$

valid for $4\gamma t_c \neq 1$. If $\mathrm{B}(t) > 0$, then the thermal force (2.48) is repulsive: $\mathbb{F}(t) > 0$. On the contrary, if $\mathrm{B}(t) < 0$, then it renders repulsive, i.e., $\mathbb{F}(t) < 0$. It is worth stressing that the dimensionless factor $\mathrm{B}(t)$ above is expressed in terms of the following experimentally accessible time scales: The evolution time $t$, the correlation time $t_c$, as well as the relaxation time $t_r = (4\gamma)^{-1}$. Moreover, notice that the force (2.48) does vanish at the thermal equilibrium $t \to \infty$, thus meaning that it can only be measured in the non-equilibrium regime $t < \infty$.

*Langevin description of an inertial free particle.* The Brownian motion of a free particle immersed in a thermal reservoir is described by the Langevin equations

$$\frac{dP(t)}{dt} = -2\gamma P(t) + \sqrt{4\gamma m k_B T}\Psi(t), \qquad (2.49)$$

$$\frac{dX(t)}{dt} = \frac{P(t)}{m}, \qquad (2.50)$$

the formal solutions of which are respectively

$$P(t) = P(0)e^{-2\gamma t} + \sqrt{4\gamma m k_B T}\int_0^t e^{-2\gamma(t-s)}\Psi(s)ds, \qquad (2.49a)$$

and

$$X(t) = X(0) + \frac{P(0)}{2\gamma m}(1 - e^{-2\gamma t}) + \sqrt{\frac{k_B T}{\gamma m}} \int_0^t [1 - e^{-2\gamma(t-s)}]\Psi(s)ds. \quad (2.50a)$$

Assuming the autocorrelation function $\langle \Psi(t')\Psi(t'')\rangle$ to be given by (see Appendix D)

$$\langle \Psi(t')\Psi(t'')\rangle = \left[1 - e^{\frac{-(t'+t'')}{2t_c}}\right]\delta(t' - t''), \quad (2.51)$$

and $\langle P(0)\Psi(t)\rangle = 0$, one can derive from Eq. (2.49a) the momentum autocorrelation function

$$\langle P(t)P(t')\rangle = [\langle P^2(0)\rangle - mk_B T]e^{-2\gamma(t+t')} + mk_B T e^{-2\gamma|t-t'|}$$
$$- mk_B T \frac{4\gamma t_c}{4\gamma t_c - 1}\left[e^{-2\gamma|t-t'|}e^{\frac{-t}{t_c}} - e^{-2\gamma(t+t')}\right] \quad (2.52)$$

and from Eq. (2.50a), under the condition $\langle X(0)P(0)\rangle = \langle X(0)\Psi(t)\rangle = 0$, the displacement autocorrelation function

$$\langle X(t)X(t')\rangle = \langle X^2(0)\rangle + \frac{\langle P^2(0)\rangle}{(2\gamma m)^2}(1 - e^{-2\gamma t})^2 + \frac{k_B T}{\gamma m}C(t,t') \quad (2.53)$$

with

$$C(t,t') = t - \frac{1}{2\gamma}(1 + e^{-2\gamma|t-t'|} - e^{-2\gamma t'} - e^{-2\gamma t}) + \frac{1}{4\gamma}[e^{-2\gamma|t-t'|} - e^{-2\gamma(t+t')}]$$
$$+ t_c\left(e^{\frac{-t}{t_c}} - 1\right) + \frac{t_c}{2\gamma t_c - 1}\left[e^{2\gamma(t-t') - \frac{t}{t_c}} - e^{-2\gamma t'}\right]$$
$$+ \frac{t_c}{2\gamma t_c - 1}\left[e^{2\gamma(t-t') - \frac{t}{t_c}} - e^{-2\gamma t'}\right]$$
$$- \frac{t_c}{4\gamma t_c - 1}\left[e^{2\gamma(t-t') - \frac{t}{t_c}} - e^{-2\gamma(t+t')}\right]. \quad (2.53a)$$

In addition, the correlation between the variables $P(t)$ and $X(t)$ is

$$\langle P(t)X(t)\rangle = 2k_B T\left[\frac{1}{2\gamma}(1 - e^{-2\gamma t}) + \frac{1}{4\gamma}(1 - e^{-4\gamma t}) + \frac{t_c}{(2\gamma t_c - 1)}\left(e^{\frac{-t}{t_c}} - e^{-2\gamma t}\right)\right.$$
$$\left. + \frac{t_c}{(4\gamma t_c - 1)}\left(e^{\frac{-t}{t_c}} - e^{-4\gamma t}\right)\right]. \quad (2.54)$$

Supposing the validity of the energy equipartition at $t = 0$, i.e., $\langle P^2(0)\rangle = mk_B T$, and taking into account $\langle P(0)\rangle = 0$ as well as the statistical property (2.26), $\langle \Psi(t) \rangle = 0$, the momentum autocorrelation function (2.52) at $t = t'$ gives rise to the non-Markovian root mean square momentum

$$\mathbb{P}(t) = \sqrt{mk_BT\left[1 - \frac{4\gamma t_c}{4\gamma t_c - 1}\left(e^{\frac{-t}{t_c}} - e^{-4\gamma t}\right)\right]}, \qquad (2.55)$$

the differentiation of which provides the thermal force

$$\mathbb{F}(t) = -2\gamma\sqrt{mk_BT}\frac{\left(4\gamma t_c e^{-4\gamma t} - e^{\frac{-t}{t_c}}\right)}{\sqrt{(4\gamma t_c - 1)\left[4\gamma t_c\left(1 - e^{\frac{-t}{t_c}} + e^{-4\gamma t}\right) - 1\right]}} \qquad (2.56)$$

which vanishes at long times $t \to \infty$, i.e., $\mathbb{F}(\infty) = 0$, whereas at short times $t \to 0$ it becomes

$$\mathbb{F}(t) = -2\gamma\sqrt{mk_BT}, \quad t \to 0. \qquad (2.56a)$$

In contrast to the non-Markovian repulsive force (2.47) which predicts the violation of the energy equipartition theorem, the Markovian attractive force (2.56a) is compatible with such a theorem at $t = 0$ because of the assumption $\langle P^2(0) \rangle = mk_BT$. So taking, for example, $\gamma \sim 10^{12}\text{s}^{-1}$, $m \sim 10^{-3}\text{kg}$, $k_B \sim 10^{-23}\text{m}^2 \text{ kg s}^{-2} \text{ K}^{-1}$, and $T \sim 300\text{K}$, we find $\mathbb{F}(t) \sim -3{,}5\text{N}$.

On the other hand, starting from the deterministic condition $\langle X^2(0) \rangle = \langle X(0) \rangle = 0$, Eq. (2.53) leads to the non-Markovian root mean square displacement (RMSD)

$$\mathbb{X}(t) = \sqrt{\frac{k_BT}{2m\gamma^2}[2\gamma t - 1 + e^{-2\gamma t} - 2\gamma t_c \mathcal{N}(t)]}. \qquad (2.57)$$

On taking the Markovian limit $t_c \to 0$, quantity (2.57) reduces to the Ornstein-Fürth RMSD (see Eq. (B.10), with $\beta = 2\gamma$, in Appendix B), i.e.,

$$\mathbb{X}(t) = \sqrt{\frac{k_BT}{2m\gamma^2}(2\gamma t - 1 + e^{-2\gamma t})}. \qquad (2.57a)$$

Differentiating Eq. (2.57) yields the following instantaneous velocity

$$\mathbb{V}(t) = \sqrt{\frac{k_BT}{2m}}\frac{\left[1 - e^{-2\gamma t} - t_c \frac{d\mathcal{N}(t)}{dt}\right]}{\sqrt{2\gamma t - 1 + e^{-2\gamma t} - 2\gamma t_c \mathcal{N}(t)}} \qquad (2.58)$$

and the non-Markovian Ornstein-Fürth diffusion coefficient, $\mathbb{D}(t) \equiv \mathbb{X}(t)\mathbb{V}(t)$,

$$\mathbb{D}(t) = \frac{k_B T}{2\gamma m}\left[1 - e^{-2\gamma t} - t_c \frac{d\mathcal{N}(t)}{dt}\right]. \tag{2.59}$$

The dimensionless function $\mathcal{N}(t)$ in Eq. (2.57) and its time derivative $d\mathcal{N}(t)/dt$ present in both Eqs. (2.58) and (2.59) read, respectively,

$$\mathcal{N}(t) = 1 - e^{\frac{-t}{t_c}} + \frac{2}{1 - 2\gamma t_c}\left(e^{\frac{-t}{t_c}} - e^{-2\gamma t}\right) - \frac{1}{1 - 4\gamma t_c}\left(e^{\frac{-t}{t_c}} - e^{-4\gamma t}\right), \tag{2.60a}$$

and

$$\frac{d\mathcal{N}(t)}{dt} = e^{\frac{-t}{t_c}} + \frac{2}{1 - 2\gamma t_c}\left(2\gamma t_c e^{-2\gamma t} - e^{\frac{-t}{t_c}}\right) - \frac{1}{1 - 4\gamma t_c}\left(4\gamma t_c e^{-4\gamma t} - e^{\frac{-t}{t_c}}\right) \tag{2.60b}$$

For long times, $t \to \infty$, our non-Markovian Brownian particle described by Eq. (2.57) enters the diffusive realm characterized by the Markovian Einstein RMSD $\mathbb{X}(t) \sim \sqrt{(k_B T/\gamma m)t}$, the instantaneous velocity $\mathbb{V}(t) \sim \sqrt{(k_B T/4\gamma mt)}$, and the Sutherland-Einstein diffusion constant $\mathbb{D}(\infty) = k_B T/2\gamma m$. On the other hand, for short times $t \to 0$ it attains the Markovian ballistic regime $\mathbb{X}(t) \sim \sqrt{(k_B T/m)}t$, from which we obtain the result $\mathbb{V}(t) \sim \sqrt{(k_B T/m)}$ that is in accordance with the validity assumption $\langle P^2(0)\rangle = mk_B T$ of the energy equipartition at $t = 0$. In other words, the existence of the instantaneous velocity (2.58) suggests the verification of the validity of the energy equipartition theorem at short times. In fact, this theoretical prediction made by Fürth [231] and Ornstein [214] in the second decade of the past century has been only borne out by recent experiments [232].

From a mathematical viewpoint, the differentiability property of both the root mean square momentum (2.55) and the root mean square displacement (2.57) implies that the Langevin equations (2.49) and (2.50) should be interpreted as a genuine differential equation and not as an integral equation *à la* Doob, for instance [229,233] (see also Appendix B).

### 2.2.2. The Kolmogorov equation in configuration space

In the absence of inertial force, that is, as the inertial force is too small in comparison with the friction force, $|dP(t)/dt| \ll |\beta P(t)|$, the set of stochastic differential equations (2.7) satisfying the fluctuation-dissipation relation (2.25b) does approximate to

$$\frac{dX(t)}{dt} = -\frac{1}{\beta m}\frac{dV(X)}{dX} + \sqrt{\frac{2\mathcal{E}}{\beta m}}\Psi(t) \tag{2.61}$$

which in turn gives rise to the Kolmogorov stochastic equation in configuration space (see Appendix C)

$$\frac{\partial \mathcal{F}(x,t)}{\partial t} = \mathcal{K}\mathcal{F}(x,t), \tag{2.62}$$

where the Kolmogorovian operator $\mathcal{K}$ acts on the function $\mathcal{F}(x,t)$ according to

$$\mathcal{K}\mathcal{F}(x,t) = \sum_{k=1}^{\infty} \frac{(-1)^k}{k!} \frac{\partial^k}{\partial x^k}[A_k(x,t)\mathcal{F}(x,t)], \tag{2.63}$$

the coefficients $A_k(x,t)$ being given by

$$A_k(x,t) = \lim_{\epsilon \to 0}\left[\frac{\langle \Delta X^k \rangle}{\epsilon}\right], \tag{2.64}$$

where the average values $\langle \Delta X^k \rangle$ are to be calculated about the sharp values $x'$, i.e.,

$$\mathcal{F}_{X\Psi}(x,\psi,t) = \delta(x-x')\mathcal{F}_\Psi(\psi,t). \tag{2.65}$$

*The non-Markovian Fokker-Planck equation in configuration space.* The Kolmogorov equation (2.62) in the Gaussian approximation reduces to the so-called non-Markovian Fokker-Planck equation in configuration space

$$\frac{\partial \mathcal{F}(x,t)}{\partial t} = \frac{1}{\beta m}\frac{\partial}{\partial x}\left[\frac{dV(x)}{dx}\mathcal{F}(x,t)\right] + \mathcal{D}_x(t)\frac{\partial^2 \mathcal{F}(x,t)}{\partial x^2}, \tag{2.66}$$

where have used the statistical property (2.26), i.e., $\langle \Psi(t) \rangle = 0$. $\mathcal{D}_x(t)$ denotes the time-dependent diffusion coefficient

$$\mathcal{D}_x(t) = \frac{\mathcal{E}}{\beta m}I(t), \tag{2.67}$$

associated with the motion of $\mathcal{F}(x,t)$. The diffusion energy $\mathcal{E}$ and the correlational function $I(t)$ in Eq. (2.67) are given by Eqs. (2.25a) and (2.24), respectively.

*The non-Markovian Smoluchowski equation.* As far as $\mathcal{E} \equiv k_B T$ and $\beta \equiv \gamma$ are concerned, Eq. (2.66) reads

$$\frac{\partial \mathcal{F}(x,t)}{\partial t} = \frac{1}{\gamma m}\frac{\partial}{\partial x}\left[\frac{dV(x)}{dx}\mathcal{F}(x,t)\right] + \frac{k_B T}{\gamma m}I(t)\frac{\partial^2 \mathcal{F}(x,t)}{\partial x^2} \tag{2.68}$$

which in turn reduces for $I(t) = 1$ to the Markovian Smoluchowski equation [234]

$$\frac{\partial \mathcal{F}(x,t)}{\partial t} = \frac{1}{\gamma m}\frac{\partial}{\partial x}\left[\frac{dV(x)}{dx}\mathcal{F}(x,t)\right] + \frac{k_B T}{\gamma m}\frac{\partial^2 \mathcal{F}(x,t)}{\partial x^2}. \quad (2.69)$$

For this reason, we have termed the Fokker-Planck equation (2.68) the non-Markovian Smoluchowski equation [189].

(a) *Free particle.* For a free Brownian particle, $V(x) = 0$, the non-Markovian Smoluchowski equation (2.68), with[24] $I(t) = 1 - e^{-t/t_c}$, reads

$$\frac{\partial \mathcal{F}(x,t)}{\partial t} = \frac{k_B T}{m\gamma}\left(1 - e^{\frac{-t}{t_c}}\right)\frac{\partial^2 \mathcal{F}(x,t)}{\partial x^2}. \quad (2.70)$$

Starting from the deterministic initial condition $\mathcal{F}(x, t = 0) = \delta(x)$, we obtain the following time solution to Eq. (2.70)

$$f(x,t) = \sqrt{\frac{m\gamma}{4\pi A(t)}}\, e^{\frac{-m\gamma x^2}{4A(t)}}, \quad (2.71)$$

where

$$A(t) = k_B T \int_0^t I(t)dt = k_B T\left[t + t_c\left(e^{\frac{-t}{t_c}} - 1\right)\right]. \quad (2.72)$$

Thereby, solution (2.71) yields both the mean $\langle X(t)\rangle = 0$ and the mean square displacement

$$\langle X^2(t)\rangle = \frac{2k_B T}{m\gamma}\left[t + t_c\left(e^{\frac{-t}{t_c}} - 1\right)\right]. \quad (2.73)$$

Hence, the root mean square displacement, $\mathbb{X}(t) = \sqrt{\langle X^2(t)\rangle}$, reads

$$\mathbb{X}(t) = \sqrt{\frac{2k_B T}{m\gamma}\left[t - t_c\left(1 - e^{\frac{-t}{t_c}}\right)\right]} \quad (2.74)$$

the differentiation of which leads to the non-Markovian thermal velocity

---

[24]In Appendix E we have solved the non-Markovian Smoluchowski equation (2.70) for a free particle reckoning with a general expression for the correlational function $I(t)$.

$$\mathbb{V}(t) = \sqrt{\frac{k_B T}{m}} \frac{\left(1 - e^{\frac{-t}{t_c}}\right)}{\sqrt{2\gamma \left[t - t_c\left(1 - e^{\frac{-t}{t_c}}\right)\right]}}. \tag{2.75}$$

The non-Markovian diffusion coefficient $\mathbb{D}(t) = \mathbb{X}(t)\mathbb{V}(t)$ in turn reads

$$\mathbb{D}(t) = \frac{k_B T}{m\gamma}\left(1 - e^{\frac{-t}{t_c}}\right), \tag{2.76}$$

which coincides with the diffusion coefficient $\mathcal{D}_x(t) = (k_B T/m\gamma)(1 - e^{-t/t_c})$ in Eq. (2.69).

For long times $t \to \infty$, i.e., $t \gg t_c$, our result (2.74) leads to the Einstein's diffusive regime, $\mathbb{X}(t) \sim \sqrt{(2k_B T/m\gamma)t}$, while Eq. (2.75) goes to $\mathbb{V}(t) \sim \sqrt{(k_B T/2m\gamma t)}$, and Eq. (2.76) provides the steady diffusion constant $\mathbb{D}(\infty) = k_B T/m\gamma$. On the other hand, for short times $t \to 0$, i.e., $t \ll t_c$, Eq. (2.74) predicts the ballistic regime, given by the displacement fluctuation

$$\mathbb{X}(t) \sim t\sqrt{\frac{k_B T}{m\gamma t_c}}, \tag{2.77}$$

whose derivative yields the instantaneous velocity

$$\mathbb{V}(t) \sim \sqrt{\frac{k_B T}{m\gamma t_c}}. \tag{2.78}$$

In addition, the Brownian movement at short times is characterized by the time-dependent diffusion coefficient

$$\mathbb{D}(t) \sim \frac{k_B T}{m\gamma t_c} t. \tag{2.79}$$

If $t_c = 1/\gamma$, then the instantaneous velocity (2.78) leads to the validity of the energy equipartition theorem of statistical mechanics at short times. Yet, in case of $t_c \neq 1/\gamma$, on account of non-Markovian effects on the instantaneous velocity (2.78) account for the violation of the energy equipartition at $t = 0$.

From a mathematical viewpoint, the differentiability of Eq. (2.74) implies that the Langevin equation (2.61), given by

$$\frac{dX(t)}{dt} = \sqrt{\frac{2k_B T}{\gamma m}} \Psi(t), \qquad (2.80)$$

should be interpreted as a genuine differential equation and not as an integral one à la Doob [229,233] (see Appendix B).

The formal solution of Eq. (2.80) reads

$$X(t) = X(0) + \sqrt{\frac{2k_B T}{\gamma m}} \int_0^t \Psi(s)\, ds. \qquad (2.81)$$

Because $\langle X(0)\Psi(t)\rangle = 0$, Eq. (2.81) leads to the displacement autocorrelation function

$$\langle X(t)X(t')\rangle = \langle X^2(0)\rangle + \frac{2k_B T}{\gamma m} \int_0^t \int_0^{t'} \langle \Psi(s)\Psi(s')\rangle\, ds ds', \qquad (2.82)$$

which in turn yields at $t = t'$ the root mean square displacement (2.74), after using the property (2.51) for the autocorrelation $\langle \Psi(s)\Psi(s')\rangle$ and $\langle X^2(0)\rangle = \langle X(0)\rangle = 0$.

(b) *Harmonic oscillator.* We reckon with a Brownian particle moving with potential energy $V(X) = m\omega^2 X^2(t)/2$, where $\omega$ is its oscillation frequency,

$$\frac{dX(t)}{dt} = -\frac{\omega^2}{\gamma} X(t) + \sqrt{\frac{2k_B T}{\gamma m}} \Psi(t). \qquad (2.83)$$

So the non-Markovian Smoluchowski equation (2.68) turns out to be given by

$$\frac{\partial f(x,t)}{\partial t} = \frac{\omega^2}{\gamma} \frac{\partial}{\partial x}[xf(x,t)] + \frac{k_B T}{\gamma m}\left(1 - e^{\frac{-t}{t_c}}\right)\frac{\partial^2 f(x,t)}{\partial x^2}. \qquad (2.84)$$

Starting from the initial condition $f(x, t = 0) = \delta(x)$, a solution to Eq. (2.84) reads

$$f(x,t) = \frac{1}{\sqrt{4\pi G(t)}} e^{\frac{-x^2}{4G(t)}}, \qquad (2.85)$$

the function $G(t)$ being the expression

$$G(t) = \frac{k_B T}{2m\omega^2}\left[1 - e^{\frac{-t}{\tau_r}} - \frac{t_c}{t_c - \tau_r}\left(e^{\frac{-t}{t_c}} - e^{\frac{-t}{\tau_r}}\right)\right], \qquad (2.86)$$

where $\tau_r$ is the relaxation time $\tau_r = \gamma/2\omega^2$. Solution (2.85) yields the mean $\langle X(t) \rangle = 0$ and the mean square displacement

$$\langle X^2(t) \rangle = \frac{k_B T}{m\omega^2}\left[1 - e^{\frac{-t}{\tau_r}} - \frac{t_c}{t_c - \tau_r}\left(e^{\frac{-t}{t_c}} - e^{\frac{-t}{\tau_r}}\right)\right]. \quad (2.87)$$

The mean mechanical energy (the mean energy potential $m\omega^2 \langle X^2(t) \rangle/2$) is given by

$$\langle E(t) \rangle = \frac{k_B T}{2}\left[1 - e^{\frac{-t}{\tau_r}} - \frac{t_c}{t_c - \tau_r}\left(e^{\frac{-t}{t_c}} - e^{\frac{-t}{\tau_r}}\right)\right], \quad (2.88)$$

obeying the equipartition theorem at long times, i.e., $\langle E(\infty) \rangle = k_B T/2$.

Making use of Eq. (2.87) and $\langle X(t) \rangle = 0$, the root mean square displacement reads

$$\mathbb{X}(t) = \sqrt{\frac{k_B T}{m\omega^2}\left[1 - e^{\frac{-t}{\tau_r}} - \frac{t_c}{t_c - \tau_r}\left(e^{\frac{-t}{t_c}} - e^{\frac{-t}{\tau_r}}\right)\right]}, \quad (2.89)$$

which in turn leads to the instantaneous velocity

$$\mathbb{V}(t) = \sqrt{\frac{k_B T}{m}}\frac{1}{2\omega}\frac{\left(e^{\frac{-t}{\tau_r}} - e^{\frac{-t}{t_c}}\right)}{\sqrt{(t_c - \tau_r)\left[t_c\left(1 - e^{\frac{-t}{t_c}}\right) - \tau_r\left(1 - e^{\frac{-t}{\tau_r}}\right)\right]}} \quad (2.90)$$

as well as the diffusion coefficient

$$\mathbb{D}(t) = \frac{k_B T}{2m\omega^2}\left(\frac{e^{\frac{-t}{t_c}} - e^{\frac{-t}{\tau_r}}}{t_c - \tau_r}\right). \quad (2.91)$$

It is worth noticing that the diffusion coefficient (2.91) is different from the diffusion coefficient $\mathcal{D}_x(t) = (k_B T/m\gamma)(1 - e^{-t/t_c})$ in Eq. (2.84).

At long times, solution (2.85) yields the Boltzmann distribution

$$f(x, \infty) = \sqrt{\frac{m\omega^2}{2\pi k_B T}} e^{\frac{-m\omega^2 x^2}{4 k_B T}}, \quad (2.92)$$

and the RMSD (2.89) becomes $\mathbb{X}(\infty) = \sqrt{k_B T/m\omega^2}$. In addition, at $t \to \infty$ both the instantaneous velocity (2.90) and the diffusion coefficient (2.91) vanish, while $\mathcal{D}_x(\infty) = k_B T/m\gamma$.

At short times $t \to 0$, on the other hand, we obtain from Eqs. (2.89 − 2.91), with $\tau_r = \gamma/2\omega^2$, respectively, the displacement fluctuation

$$\mathbb{X}(t) \sim t\sqrt{\frac{k_B T}{m\gamma t_c}}, \quad t \to 0, \qquad (2.93)$$

the instantaneous velocity

$$\mathbb{V}(t) \sim \sqrt{\frac{k_B T}{m\gamma t_c}}, \quad t \to 0, \qquad (2.94)$$

and the diffusion coefficient

$$\mathbb{D}(t) \sim \frac{k_B T}{m\gamma t_c} t, \quad t \to 0, \qquad (2.95)$$

which in turn is the same that $\mathcal{D}_x(t) = (k_B T/m\gamma)(1 - e^{-t/t_c})$ in the Smoluchowski equation (2.84) at short times, i.e., $\mathcal{D}_x(t) \sim (k_B T/m\gamma t_c)t$. As long as $\gamma t_c \neq 1$ the instantaneous velocity (2.94) of a harmonic oscillator does suggest the violation of energy equipartition at short times.

As a consequence of the differentiability of the displacement fluctuation (2.89) at $t = 0$, the Langevin equation (2.84) is to be interpreted actually as a stochastic differential equation.

## 2.3. Anomalous diffusion

So far, we have addressed the phenomenon of Brownian motion characterized by the normal diffusion law $\langle X^2(t) \rangle \sim Dt$ at long times, $t \to \infty$. Nevertheless, as far as a disordered environment is concerned, it has been reported that the long-time behavior the mean square displacement (MSD) turns out to be given by [235]

$$\langle X^2(t) \rangle \sim C^{(\lambda)} t^\lambda, \quad t \to \infty, \qquad (2.96)$$

where $C^{(\lambda)}$ is a constant labeled by the index $\lambda \neq 1$. The disordered environment containing impurities, defects, or some sort of intrinsic disorder, gives rise through Eq. (2.96) to two different domains of anomalous diffusion according to the values of $\lambda$ [236]: (a) subdiffusion for $0 < \lambda < 1$; and (b) superdiffusion for $\lambda > 1$. Specific cases of superdiffusion are the following: For $\lambda = 2$, the diffusion is said to be ballistic; the $\lambda = 3$ case is called turbulent diffusion [237]. Examples of such disordered media are disordered lattice, porous media, and dopped conductors and semiconductors [235-237].

Our aim is to show how our correlational function $I(t)$, Eq. (2.31), can be viewed as the physical mechanism giving rise to the anomalous diffusion (2.96). Furthermore, we look at how the anomaly parameter $\lambda$ affects the differentiability property of the Brownian trajectories at short times taking into account two physical situations: anomalous diffusion in the absence as well as in the presence of inertial effects on Brownian motion.

### 2.3.1. Anomalous diffusion in the absence of inertial force

*Free particle*. We now proceed to show how our non-Markovian approach to Brownian motion can explain diffusion processes exhibiting the anomalous behavior given by Eq. (2.96). As a first example, we start with the Brownian motion of a free particle in the absence of inertial force described by the Langevin equation (2.61), with $V(X) = 0$, i. e.,

$$\frac{dX(t)}{dt} = \sqrt{\frac{2\mathcal{E}}{\gamma m}} \Psi(t). \tag{2.97}$$

The corresponding diffusion equation (2.66), with Eq. (2.31), reads

$$\frac{\partial f(x,t)}{\partial t} = D_x^{(\lambda)}(t) \frac{\partial^2 f(x,t)}{\partial x^2}, \tag{2.98}$$

where $D_x^{(\lambda)}(t)$ is the time-dependent diffusion coefficient

$$D_x^{(\lambda)}(t) = \frac{\mathcal{E}}{\gamma m}\left(\lambda \frac{t^{\lambda-1}}{t_c^{\lambda-1}} + 1 - e^{\frac{-t}{t_c}}\right), \tag{2.99}$$

indexed by the parameter $\lambda > 1$ (see Appendix D). Starting from the initial condition $f(x, t = 0) = \delta(x)$, solution to Eq. (2.98) is given by the Gaussian function

$$f(x,t) = \frac{1}{\sqrt{2\pi\langle X^2(t)\rangle}} e^{\frac{-x^2}{2\langle X^2(t)\rangle}}, \tag{2.100}$$

expressed in terms of the mean square displacement

$$\langle X^2(t) \rangle = \frac{2\mathcal{E}}{\gamma m} \left[ \frac{t^\lambda}{t_c^{\lambda-1}} + t + t_c \left( e^{\frac{-t}{t_c}} - 1 \right) \right], \qquad (2.101)$$

which displays the following long-time behavior

$$\langle X^2(t) \rangle \sim \frac{2\mathcal{E}}{\gamma m t_c^{\lambda-1}} t^\lambda, \qquad t \to \infty. \qquad (2.102)$$

Because $\lambda > 1$ the anomalous diffusion (2.102) is termed superdiffusion.

To investigate the effects of the anomaly parameter $\lambda$ on the differentiability property of the anomalous Brownian paths, we take into account the square root of Eq. (2.101) at short times $t \to 0$, such that $t^n \ll 1$, for $n > 2$, i.e.,

$$\mathbb{X}(t) \sim \sqrt{\frac{2\mathcal{E}}{\gamma m} \left( \frac{t^\lambda}{t_c^{\lambda-1}} + \frac{t^2}{2t_c} \right)}, \qquad t \to 0, \qquad (2.103)$$

which is nondifferentiable for $1 < \lambda < 2$. For $\lambda = 2$, the quantity (2.103), given by $\mathbb{X}(t) \sim t\sqrt{3\mathcal{E}/\gamma m t_c}$, renders differentiable, i.e.,

$$\mathbb{V}(t) \sim \sqrt{\frac{3\mathcal{E}}{\gamma m t_c}}, \qquad t \to 0. \qquad (2.104a)$$

while for $\lambda > 2$ we obtain $\mathbb{X}(t) \sim t\sqrt{\mathcal{E}/\gamma m t_c}$ whose time derivative yields

$$\mathbb{V}(t) \sim \sqrt{\frac{\mathcal{E}}{\gamma m t_c}}, \qquad t \to 0. \qquad (2.104b)$$

For thermal systems, i.e., $\mathcal{E} \equiv k_B T$, both results (2.104) imply the violation of the energy equipartition law $\mathbb{V}(t) \sim \sqrt{k_B T/m}$. Mathematically, the consequence of the existence of the instantaneous velocities (2.104), or the differentiability of the square root of Eq. (2.101), $\mathbb{X}(t) = \sqrt{\langle X^2(t) \rangle}$, is that the Langevin equation (2.97) underlying the Fokker-Planck (2.98) can be interpreted as a true random differential equation as long as the anomalous case $\lambda \geq 2$ is concerned.

From the definition of instantaneous velocity, it follows that the diffusion coefficient $\mathbb{D}(t) = \mathbb{V}(t)\mathbb{X}(t)$ is identical with Eq. (2.99), i.e., $\mathbb{D}(t) \equiv D_x^{(\lambda)}(t) = (\mathcal{E}/\gamma m)[\lambda(t/t_c)^{\lambda-1} + 1 - e^{-t/t_c}]$, exhibiting the asymptotic behavior $\mathbb{D}(t) \sim (\mathcal{E}/\gamma m \lambda t/t_c \lambda - 1$, at long times $t \to \infty$, and $\mathbb{D}t \sim 0$, at short times $t \to 0$.

*Harmonic oscillator.* The non-Markovian Smoluchowski equation (2.68), with Eq. (2.31) and $\mathcal{E} = k_B T$, reads

$$\frac{\partial f(x,t)}{\partial t} = \frac{\omega^2}{\gamma}\frac{\partial}{\partial x}[xf(x,t)] + \frac{\mathcal{E}}{\gamma m}\left(\lambda\frac{t^{\lambda-1}}{t_c^{\lambda-1}} + 1 - e^{\frac{-t}{t_c}}\right)\frac{\partial^2 f(x,t)}{\partial x^2}, \quad (2.105)$$

(i) For $\lambda = 2$, we find

$$\langle X^2(t)\rangle = \frac{\mathcal{E}}{m\omega^2}\left\{\frac{2}{t_c}\left[t + \tau\left(e^{\frac{-t}{\tau}} - 1\right)\right] + 1 - e^{\frac{-t}{\tau}} - \frac{t_c}{(t_c - \tau)}\left(e^{\frac{-t}{t_c}} - e^{\frac{-t}{\tau}}\right)\right\} \quad (2.106)$$

with

$$\langle X^2(t)\rangle \sim \frac{2\mathcal{E}}{m\omega^2 t_c}t, \quad t \to \infty, \quad (2.106a)$$

and

$$\langle X^2(t)\rangle \sim \frac{3\mathcal{E}}{m\gamma t_c}t^2, \quad t \to 0; \quad (2.106b)$$

(ii) for $\lambda = 3$, we have

$$\langle X^2(t)\rangle = \frac{\mathcal{E}}{m\omega^2}\left\{\frac{3}{t_c^2}\left[t^2 - 2t\tau + 2\tau^2\left(1 - e^{\frac{-t}{\tau}}\right)\right] + 1 - e^{\frac{-t}{\tau}}\right.$$
$$\left. - \frac{t_c}{(t_c - \tau)}\left(e^{\frac{-t}{t_c}} - e^{\frac{-t}{\tau}}\right)\right\} \quad (2.107)$$

with

$$\langle X^2(t)\rangle \sim \frac{3\mathcal{E}}{m\omega^2 t_c^2}t^2, \quad t \to \infty, \quad (2.107a)$$

and

$$\langle X^2(t)\rangle \sim \frac{\mathcal{E}}{m\gamma t_c}t^2, \quad t \to 0; \quad (2.107b)$$

(iii) for $\lambda = 4$, Eq. (2.105) leads to

$$\langle X^2(t) \rangle = \frac{\mathcal{E}}{m\omega^2} \left\{ \frac{4}{t_c^3} \left[ t^3 - 3\tau t^2 + 6\tau^2 t - 6\tau^3 \left( 1 - e^{\frac{-t}{\tau}} \right) \right] + 1 - e^{\frac{-t}{\tau}} \right.$$
$$\left. - \frac{t_c}{(t_c - \tau)} \left( e^{\frac{-t}{t_c}} - e^{\frac{-t}{\tau}} \right) \right\} \tag{2.108}$$

with

$$\langle X^2(t) \rangle \sim \frac{4\mathcal{E}}{m\omega^2 t_c^3} t^3, \quad t \to \infty, \tag{2.108a}$$

and

$$\langle X^2(t) \rangle = \frac{\mathcal{E}}{m\gamma t_c} t^2, \quad t \to 0. \tag{2.108b}$$

In Eqs. (2.106-2.108), $\tau$ denotes the relaxation time $\tau = \gamma/2\omega^2$ and the limit $t \to 0$ is meant to be $t^n \ll 1$, for $n > 2$. By generalizing results (2.106a-2.108a), the anomalous behavior reads

$$\langle X^2(t) \rangle \sim \lambda \frac{\mathcal{E}}{m\omega^2} \frac{t^{\lambda-1}}{t_c^{\lambda-1}}, \quad t \to \infty. \tag{2.109}$$

It is readily to verify that starting from the short-time MSD (2.106b) one can derive the instantaneous velocity $\mathbb{V}(t) \sim \sqrt{3\mathcal{E}/m\gamma t_c}$, whereas both Eqs. (2.107b) and (2.108b) provide $\mathbb{V}(t) \sim \sqrt{\mathcal{E}/m\gamma t_c}$, so leading to the breakdown of the energy equipartition law $\mathbb{V}(t) \sim \sqrt{k_B T/m}$.

In brief, we have shown that Brownian paths are differentiable at short times in the anomalous cases $\lambda = 2, 3, 4$. Consequently, the Langevin equation underlying the Fokker-Planck equation (2.105) is in fact a differential random equation.

## 2.3.2. Anomalous diffusion in the presence of inertial force

*Anomalous Brownian motion in momentum space*[25]. We turn to examine anomalous Brownian motion in momentum space. We start with the Langevin equation

$$\frac{dP(t)}{dt} = -2\gamma P(t) + \sqrt{4\gamma m \mathcal{E}} \Psi(t) \tag{2.110}$$

---

[25] The interested reader is invited to compare our present approach with the anomalous diffusion in momentum space currently studied in Ref. [238].

giving rise to the non-Markovian Rayleigh equation, with the correlational function (2.31),

$$\frac{\partial \mathcal{F}(p,t)}{\partial t} = 2\gamma \frac{\partial}{\partial p}[p\mathcal{F}(p,t)] + 2\gamma m\mathcal{E}\left(\lambda \frac{t^{\lambda-1}}{t_c^{\lambda-1}} + 1 - e^{\frac{-t}{t_c}}\right)\frac{\partial^2 \mathcal{F}(p,t)}{\partial p^2}. \quad (2.111)$$

We take into account the following cases:

(i) for $\lambda = 2$, we find

$$\langle P^2(t) \rangle = m\mathcal{E}\left\{\frac{2}{t_c}\left[t + \frac{1}{4\gamma}(e^{-4\gamma t} - 1)\right] + 1 - e^{-4\gamma t} + \frac{4\gamma t_c}{4\gamma t_c - 1}\left(e^{-4\gamma t} - e^{\frac{-t}{t_c}}\right)\right\},$$

(2.112)

exhibiting the non-Markovian diffusive regime

$$\langle P^2(t) \rangle \sim \frac{2m\mathcal{E}}{t_c} t, \qquad t \to \infty, \quad (2.112a)$$

and the ballistic one

$$\langle P^2(t) \rangle \sim \frac{4\gamma m\mathcal{E}}{t_c} t^2, \qquad t \to 0; \quad (2.112b)$$

(ii) for $\lambda = 3$, we have

$$\langle P^2(t) \rangle = m\mathcal{E}\left\{\frac{3}{t_c^2}\left[t^2 - \frac{t}{2\gamma} + \frac{2}{(4\gamma)^2}(1 - e^{-4\gamma t})\right] + 1 - e^{-4\gamma t}\right.$$
$$\left. + \frac{4\gamma t_c}{4\gamma t_c - 1}\left(e^{-4\gamma t} - e^{\frac{-t}{t_c}}\right)\right\} \quad (2.113)$$

displaying the following ballistic behaviors

$$\langle P^2(t) \rangle \sim \frac{3m\mathcal{E}}{t_c^2} t^2, \qquad t \to \infty; \quad (2.113a)$$

$$\langle P^2(t) \rangle \sim \frac{2\gamma m\mathcal{E}}{t_c} t^2, \qquad t \to 0; \quad (2.113b)$$

(iii) for $\lambda = 4$, Eq. (2.111) yields

$$\langle P^2(t)\rangle = m\mathcal{E}\left\{\frac{16\gamma}{t_c^3}\left[\frac{t^3}{4\gamma} - \frac{3t^2}{(4\gamma)^2} + \frac{6t}{(4\gamma)^3} - \frac{6(1-e^{-4\gamma t})}{(4\gamma)^4}\right] + 1 - e^{-4\gamma t}\right.$$
$$\left. + \frac{4\gamma t_c}{4\gamma t_c - 1}\left(e^{-4\gamma t} - e^{\frac{-t}{t_c}}\right)\right\} \tag{2.114}$$

leading to the turbulent regime

$$\langle P^2(t)\rangle \sim \frac{4m\mathcal{E}}{t_c^3}t^3, \quad t \to \infty, \tag{2.114a}$$

and the ballistic one

$$\langle P^2(t)\rangle \sim \frac{2\gamma m\mathcal{E}}{t_c}t^2, \quad t \to 0. \tag{2.114b}$$

Generalizing the long-time results (2.112a-2.114a) leads to the anomalous diffusion in momentum space

$$\langle P^2(t)\rangle \sim \lambda m\mathcal{E}\frac{t^{\lambda-1}}{t_c^{\lambda-1}}, \quad \text{for } \lambda > 1. \tag{2.115}$$

The ballistic behavior of $\langle P^2(t)\rangle$, given by Eqs. (2.112b-2.114b), leads to the differentiability of $\mathbb{P}(t) \equiv \sqrt{\langle P^2(t)\rangle}$, i.e., the existence of the force $\mathbb{F}(t) \equiv d\mathbb{P}(t)/dt$,

$$\mathbb{F}(t) \sim \sqrt{\frac{4\gamma m\mathcal{E}}{t_c}}, \quad \text{for } \lambda = 2; \tag{2.116a}$$

$$\mathbb{F}(t) \sim \sqrt{\frac{2\gamma m\mathcal{E}}{t_c}}, \quad \text{for } \lambda = 3 \text{ and } 4. \tag{2.116b}$$

Considering $\mathcal{E} \equiv k_B T$, both Eqs. (2.116) could verify the violation of the energy equipartition at short times.

Moreover, from the existence of the forces (2.116) it follows that the Langevin equation (2.110) should be viewed as a genuine stochastic differential equation because the inertial term $dP/dt = md^2X/dt^2$ can mathematically be well defined. Physically, this stands for that particles undergoing anomalous Brownian movement in general do have a finite acceleration.

*Anomalous Brownian motion in configuration space.* Assuming $\langle X^2(0)\rangle = \langle X(0)P(0)\rangle = \langle P(0)\Psi(s)\rangle = 0$ and the validity of the equipartition energy at $t = 0$ through the relation $\langle P^2(0)\rangle = mk_BT$, the Langevin equation (2.110) leads to the mean square displacement

$$\langle X^2(t)\rangle = \frac{k_BT}{4\gamma^2 m}(1 - e^{-2\gamma t})^2$$
$$+ \frac{k_BT}{\gamma m}\int_0^t\int_0^t [1 - e^{-2\gamma(t-s)}][1 - e^{-2\gamma(t-r)}]\langle\Psi(s)\Psi(r)\rangle drds. \quad (2.117)$$

Supposing now the autocorrelation function $\langle\Psi(s)\Psi(r)\rangle$ in Eq. (2.117) to be given by (see Appendix D)

$$\langle\Psi(s)\Psi(r)\rangle = \left[\lambda\frac{r^{\lambda-1}}{t_c^{\lambda-1}} + 1 - e^{\frac{-(r+s)}{2t_c}}\right]\delta(s-r), \quad (2.118)$$

with $\lambda > 1$, it follows that

$$\langle X^2(t)\rangle = \frac{k_BT}{4\gamma^2 m}(1 - e^{-2\gamma t})^2 + a(t) + a^{(\lambda)}(t), \quad (2.119)$$

where $a(t)$ is given by

$$a(t) = \frac{-k_BTt_c}{\gamma m}\left[1 - e^{\frac{-t}{t_c}} + \frac{2}{1-2\gamma t_c}\left(e^{\frac{-t}{t_c}} - e^{-2\gamma t}\right) - \frac{1}{1-4\gamma t_c}\left(e^{\frac{-t}{t_c}} - e^{-4\gamma t}\right)\right].$$

$$(2.120)$$

For the case $\lambda = 2$, the term $a^{(\lambda)}(t)$ in Eq. (2.119) reads

$$a^{(\lambda=2)}(t) = \frac{2k_BT}{\gamma m t_c}\left\{\frac{t^2}{2} - \left[\frac{t}{\gamma} + \frac{2}{(2\gamma)^2}(e^{-2\gamma t} - 1)\right] + \left[\frac{t}{4\gamma} + \frac{1}{(4\gamma)^2}(e^{-4\gamma t} - 1)\right]\right\}.$$

$$(2.121)$$

So, at long times we have

$$\langle X^2(t)\rangle \sim \frac{k_BT}{\gamma m t_c}t^2, \quad t \to \infty. \quad (2.122)$$

For the $\lambda = 3$ case, the term $a^{(\lambda)}(t)$ in Eq. (2.119) is written down as

$$a^{(\lambda=3)}(t) = \frac{3k_BT}{\gamma m t_c^2}\left\{\frac{t^3}{3} - 2\left[\frac{t^2}{2\gamma} - \frac{2t}{(2\gamma)^2} + \frac{2}{(2\gamma)^3}(1-e^{-2\gamma t})\right]\right.$$
$$\left. + \left[\frac{t^2}{4\gamma} - \frac{2t}{(4\gamma)^2} + \frac{2}{(4\gamma)^3}(1-e^{-4\gamma t})\right]\right\}. \tag{2.123}$$

Thus, we obtain at long times the turbulent diffusion

$$\langle X^2(t)\rangle \sim \frac{k_BT}{\gamma m t_c^2} t^3, \quad t\to\infty. \tag{2.124}$$

Moreover, the term $a^{(\lambda)}(t)$, for $\lambda = 4$, in Eq. (2.119) reads

$$a^{(\lambda=4)}(t) = \frac{4k_BT}{\gamma m t_c^3}\left\{\frac{t^4}{4} - 2\left[\frac{t^3}{2\gamma} - \frac{3t^2}{(2\gamma)^2} + \frac{6t}{(2\gamma)^3} - \frac{6}{(2\gamma)^4} + \frac{6e^{-2\gamma t}}{(2\gamma)^4}\right]\right.$$
$$\left. + \left[\frac{t^3}{4\gamma} - \frac{3t^2}{(4\gamma)^2} + \frac{6t}{(4\gamma)^3} - \frac{6}{(4\gamma)^4} + \frac{6e^{-4\gamma t}}{(4\gamma)^4}\right]\right\}. \tag{2.125}$$

At long times, inserting Eq. (2.125) into Eq. (2.119) yields

$$\langle X^2(t)\rangle \sim \frac{k_BT}{\gamma m t_c^3} t^4. \tag{2.126}$$

Generalizing results (2.122), (2.124) and (2.126) leads to the anomalous diffusion in configuration space

$$\langle X^2(t)\rangle \sim \frac{k_BT}{\gamma m t_c^{\lambda-1}} t^\lambda. \tag{2.127}$$

It is readily to check that from the MSD (2.119), for $\lambda = 2, 3, 4$, at short times, we can derive the instantaneous velocity $\mathbb{V}(t)\sim\sqrt{k_BT/m}$, thereby implying no violation of the energy equipartition. This result is consistent with the initial condition $\langle P^2(0)\rangle = mk_BT$ assumed in Eq. (2.117).

## 2.4. Non-Markovian Kramers escape rate

Due to environmental fluctuations, the escape rate of a Brownian particle over a barrier separating two metastable states— in a double-well potential, for instance— is known as the Kramers problem [120,191,193,207,217,220,229,239-241]. In the classical realm, the transition of such a particle over a potential barrier from a metastable state towards another state is said to be an activation phenomenon [239].

In 1940, on the ground of the steady Fokker-Planck equation (2.34) Kramers [217] investigated the issue of metastability of a Brownian particle in phase space and came up with the following steady escape rate characterized by Markovian effects

$$\Gamma(\infty) = \frac{\omega_a}{2\pi\omega_b}\left(\sqrt{\gamma^2 + \omega_b^2} - \gamma\right)e^{\frac{-V(x_b)}{k_B T}}, \qquad (2.128)$$

where $\omega_b$ denotes the particle's oscillation frequency over the potential barrier and $\omega_a$ the oscillation frequency coming from the number of particles in the metastable state around the well at $x_a$.

In this section we wish to extend Kramers' technique [217] to our non-Markovian Klein-Kramers equation (2.33a), with $\beta = 2\gamma$ and Eq. (2.31), on assuming that during a finite time $t < \infty$ for observing the Brownian particle the non-steady function $\mathcal{F}(x, p, t)$ could be factorized as

$$\mathcal{F}(x, p, t) = e^{2\gamma t}\mathcal{F}(x, p), \quad t < \infty. \qquad (2.129)$$

So, we derive from Eq. (2.33) the following time-independent equation of motion for $\mathcal{F}(x, p)$

$$-\frac{p}{m}\frac{\partial \mathcal{F}(x,p)}{\partial x} + \left[\frac{dV(x)}{dx} + 2\gamma p\right]\frac{\partial \mathcal{F}(x,p)}{\partial p} + 2\gamma m k_B T I(\alpha)\frac{\partial^2 \mathcal{F}(x,p)}{\partial p^2} = 0, \qquad (2.130)$$

where $I(\alpha)$ should be interpreted as a constant given by

$$I(\alpha) = \lambda \alpha^{\lambda-1} + 1 - e^{-\alpha}, \qquad (2.130a)$$

obtained from Eq. (2.31) at $t = \alpha t_c$, $\alpha$ being a dimensionless parameter. Notice that the correlational function (2.130a) also depends on the parameter $\lambda$: For $\lambda = 0$, the diffusion is called normal; for $\lambda > 1$, it is anomalous.

Assuming the barrier top to be located at point $x_b$, while the two bottom wells are at $x_a$ and $x_c$, such that $V(x_c) = 0 = V(x_a)$, with $x_a < x_b$, we perform an expansion of the potential function $V(x) = -m\omega_b^2 x^2/2$ in a Taylor series around the point $x_b$, i.e., $V(x) \sim V(x_b) - (m\omega_b^2/2)(x - x_b)^2$, where the quantity $\omega_b$ denotes the particle's oscillation frequency over the potential barrier. Next, we introduce the new variable given by

$$\xi = p - a(x - x_b). \qquad (2.131)$$

On making use of the fact that

$$\frac{\partial}{\partial x} = \frac{\partial \xi}{\partial x}\frac{\partial}{\partial \xi} = -a\frac{\partial}{\partial \xi} \tag{2.132a}$$

and

$$\frac{\partial}{\partial p} = \frac{\partial \xi}{\partial p}\frac{\partial}{\partial \xi} = \frac{\partial}{\partial \xi}, \tag{2.132b}$$

Eq. (2.130) changes into

$$\frac{d^2\mathcal{F}(\xi)}{d\xi^2} = -A\xi\frac{d\mathcal{F}(\xi)}{d\xi}, \tag{2.133}$$

where

$$A = \frac{2\gamma m + a}{2\gamma m^2 k_B T I(\alpha)}, \tag{2.133a}$$

with

$$a = -\gamma m \pm m\sqrt{\gamma^2 + \omega_b^2}. \tag{2.133b}$$

Under the condition that $\mathcal{F}(\xi \to \infty) = 1$, the solution of Eq. (2.133) reads

$$\mathcal{F}(\xi) = \sqrt{\frac{A}{2\pi}} \int_{-\infty}^{\xi} e^{\frac{-A}{2}\xi^2} d\xi. \tag{2.134}$$

Because $A > 0$, it follows that

$$a = -\gamma m + m\sqrt{\gamma^2 + \omega_b^2}. \tag{2.135}$$

Unlike Kramers [217], we make use of the non-steady function (2.129), with Eq. (2.134), to build up the following time-dependent probability distribution function

$$W(x,p,t) = e^{2\gamma t}\sqrt{\frac{A}{2\pi}}e^{\frac{-1}{k_B T}\left[\frac{p^2}{2m}+V(x_b)-\frac{m\omega_b^2}{2}(x-x_b)^2\right]}\int_{-\infty}^{\xi=p-a(x-x_b)} e^{\frac{-A}{2}\xi^2} d\xi, \tag{2.136}$$

where the Maxwell-Boltzmann distribution shows up at $t = 0$ as long as $\xi \to \infty$. On the basis of function (2.136) we can calculate the non-steady probability current $J_b$ over the potential barrier located at $x = x_b$, i.e.,

$$J_b(t) = \int_{-\infty}^{\infty} \frac{p}{m} W(x = x_b, p, t) dp = e^{2\gamma t} k_B T e^{\frac{-V(x_b)}{k_B T}} \sqrt{\frac{m k_B T A}{1 + m k_B T A}}, \quad (2.137)$$

where we have used the result

$$\int_{-\infty}^{\infty} e^{-up^2} p \, dp \int_{-\infty}^{\xi=p} e^{-v\xi^2} d\xi = \frac{1}{2u} \sqrt{\frac{\pi}{u+v}} \quad (2.138)$$

with $u = 1/2m k_B T$ and $v = A/2$.

On the other hand, the number of particles $n_a$ in the metastable state around $x_a$ can be calculated from the Maxwell-Boltzmann distribution, which is obtained from Eq. (2.136) at $t = 0$ on replacing $x_b$ with $x_a$ and taking into account the positive concavity of the potential, $k = m\omega_a^2$, as well as the limit $\xi \to \infty$. The result is

$$n_a = e^{\frac{-V(x_a)}{k_B T}} \int_{-\infty}^{\infty} \int_{-\infty}^{\infty} e^{\frac{-1}{k_B T} \left[\frac{p^2}{2m} + \frac{m\omega_a^2}{2}(x-x_a)^2\right]} dx dp = \frac{2\pi k_B T}{\omega_a} e^{\frac{-V(x_a)}{k_B T}}. \quad (2.139)$$

Using Eq. (2.137), $A$ being given by Eq. (2.133a), as well as Eq. (2.139), the time-dependent escape rate $\Gamma(t) = J_b(t)/n_a$ reads

$$\Gamma(t) = e^{2\gamma t} \frac{\omega_a}{2\pi} \sqrt{\frac{\gamma + \sqrt{\gamma^2 + \omega_b^2}}{2\gamma I(\alpha) + \gamma + \sqrt{\gamma^2 + \omega_b^2}}} e^{\frac{-[V(x_b)-V(x_a)]}{k_B T}}, \quad t < \infty, \quad (2.140)$$

where the constant $I(\alpha)$ accounts for non-Markov effects through Eq. (2.130a), which is valid for normal ($\lambda = 0$) and anomalous ($\lambda > 1$) diffusion. For instance, if $t \sim t_c$, then $I(\alpha)$ is determined by $I(1) = \lambda + 1 - e^{-1}$. For $t \gg t_c$, i.e., $\alpha \gg 1$, we find $I(\alpha) \sim \lambda \alpha^{\lambda-1}$, with $\lambda > 1$. At $t = 0$, the constant $I(\alpha = 0)$ vanishes implying that our escape rate (2.140) is Markovian and takes place in the absence of dissipation, i.e.,

$$\Gamma(0) = \frac{\omega_a}{2\pi} e^{\frac{-[V(x_b)-V(x_a)]}{k_B T}}, \quad t = 0. \quad (2.141)$$

## 2.5. Summary and Discussion

From a mathematical standpoint Brownian motion is described by two sorts of non-Gaussian equations of motion: A stochastic differential equation for random variables, e.g., Eqs. (2.7), and the corresponding time evolution for the probability distribution function (the Kolmogorov equation (2.11), for instance). Nevertheless, both equations of motion turn out to have some physical significance as long as a Gaussian approximation is taken into account. Accordingly, the Kolmogorov equation reduces to the Fokker-Planck equation whereas its underlying stochastic differential equation turns out to be dubbed the Langevin equation. The gist of our approach lies in the existence of the dimensionless correlational function $I(t)$, defined by Eq. (2.24), but not introduced *ad hoc*. On the contrary, it arises naturally by building up the Fokker-Planck equation from a given Langevin equation, giving rise to time-dependent diffusion coefficients. According to our approach, therefore, the Brownian dynamics is only Markovian in the steady regime $t \to \infty$, as $I(\infty) = 1$. Non-Markovianity features turn up in the non-steady regime $t < \infty$, as $I(t) \neq 1$. Surprisingly, this feature seems to have been overlooked in the centennial literature on Markovian Brownian motion, whereby it is assumed that $I(t) = 1$ for all time $t$.

On the ground of our correlational function $I(t)$ we have examined in the Gaussian approximation non-Markov effects on normal and anomalous Brownian movement in the presence and absence of inertial forces. In the case of normal diffusion we have shown that non-Markov effects account for the differentiability of the erratic trajectories of a Brownian particle, whereas anomalous diffusion may give rise to differentiable or nondifferentiable paths. As to the phenomenon of escape rate our main upshot is the time-dependent non-Markovian Kramers rate, Eq. (2.140), which is valid for both normal and anomalous diffusion.

The differentiability property of non-Markovian Brownian motion implies the breakdown of the energy equipartition theorem of statistical mechanics at short times in four physical situations:

(i) In the presence of inertial forces in momentum space via the existence of the concept of thermal force, Eq. (2.47);

(ii) In the absence of inertial forces in configuration space through our instantaneous velocity $\mathbb{V}(t) \sim \sqrt{k_B T/m\gamma t_c}$, which is valid for a free particle, Eq. (2.78), and a harmonic oscillator, Eq. (2.94);

(iii) In the non-inertial anomalous diffusion of a free particle for the cases in which $\lambda \geq 2$, on account of the existence of the instantaneous velocities (2.104), as well as in the anomalous diffusion of a harmonic oscillator as described by the mean square displacements (2.106b), (2.107b), and (2.108b);

(iv) In the inertial anomalous diffusion for $\lambda = 2, 3, 4$ due to the existence of the forces (2.116).

A mathematical implication of the differentiability of Brownian trajectories is that the underlying Langevin equations should be interpreted as genuine differential equations. Otherwise, they should be interpreted as integral equations according to a determined interpretation (Doob's rule, for instance).

Conceptually, the cases in which the Brownian paths are analytic reveals that stochastic trajectory is a concept intrinsic to the motion of a Brownian particle just as the concept of deterministic trajectory is assumed to be a feature inherent in the motion of a Newtonian particle. In short, our main conceptual finding is that a Brownian particle actually follows a physical trajectory that is a mathematically well-defined concept owing to non-Markovian effects. In contrast, non-differentiability implies that the Brownian motion is not a mathematically and physically well-defined phenomenon at short times, albeit non-differentiable paths make up a prolific abstraction in the strictly mathematical realm.

Also, our approach to Brownian motion brings out that anomalous diffusion is compatible with the central limit theorem. This result is in contrast to accounts based on fractional Fokker-Planck equations, for instance, which claim that a revision of such a theorem should be required to explain the phenomenon of anomalous diffusion [236].

By way of discussion, we would like to clarify some points concerning the differences between the present study and van Kampen's, Mori's approaches as well as the Hamiltonian approach to Brownian motion. In his authoritative monograph van Kampen states [120]:

> *The Fokker-Planck equation is a special type of master equation, which is often used as an approximation to the actual equation or as a model for more general Markov process.*

This van Kampen's statement could raise doubts concerning to the correctness of dealing with non-Markovian phenomena under the framework of Fokker–Planck equations. By contrast, our present study (see also [189]) can shed some light on the non-Markovian character of Brownian motion within a structure of generalized Fokker-Planck equations carrying our correlational function $I(t)$, given by Eq. (2.24).

In our approach to Brownian motion we have postulated Langevin equations as phenomenological equations of motion and investigated the mathematical consequences. According to this phenomenological approach, the parameters and the functional form of both frictional and random forces have to be determined from experimental data [242]. It remains to compare the resulting predictions with experiments in order to (in)validate our theory [242].

Alternatively, in order to justify the form of the environmental force on the Brownian particle (e.g., the splitting into a friction force and a random force), one believes that both Langevin and Fokker-Planck equations could be derived from microscopic models[26] which in turn may be built up on the basis of the Boltzmann equation (kinetic models) [243-245] or model Hamiltonians. In the case of Hamiltonian microscopic models, a general scheme of deriving Brownian dynamics (both generalized Langevin and Fokker-Planck equations) is based on the framework of nonequilibrium statistical mechanics [132,149,166,207,222,246-250] in which both the system (the Brownian particle) and the environment are regarded as an isolated whole deemed to be ruled by a total Hamiltonian of the form

$$H_{\mathrm{T}} = H_{\mathrm{S}} + H_{\mathrm{E}} + H_{\mathrm{SE}}, \qquad (2.142)$$

where $H_{\mathrm{S}}$ is the system Hamiltonian, $H_{\mathrm{E}}$ the environment Hamiltonian, and $H_{\mathrm{SE}}$ the interaction Hamiltonian. On eliminating the environmental variables via a coarse-graining procedure [250-252] with the help of the technique of projection operators [166,207,222,253] or the multiple-time method [245,254,255], for instance, as well as making use of simplifications and approximations (e.g., expansion in powers of the square root of the mass ratio $\sqrt{m/M}$ [245,254,256-260], where $M$ is the mass of the Brownian particle and $m$ the mass of the environment particles, one believes that the deterministic Hamilton equations (or Newton's equations) generated by the Hamiltonian function (2.142) are able to lead to generalized Langevin equations while the Liouville equation associated with Eq. (2.142) yields generalized Fokker-Planck equations [255]. The source of ramdomness shows up on assuming a specific initial condition to the statistical behavior of the medium at thermal equilibrium [132,222,261,262]. So, in this context non-Markovian Langevin and/or Fokker-Planck equations in the presence of arbitrary thermal environments can be found, for example, in Refs. [132,149,166,207,222,225,248,251,252,255-260,263-276]. More specifically, assuming the environment to be a bath of harmonic oscillators,

---

[26]Microscopic models are also known as a first-principles approach according to which the concept of probability should be derived from deterministic models. In other words, the statistical hypotheses used in open systems undergoing Brownian motion must be justified on the basis of the dynamics of isolated systems (Newton's laws or Schrödinger equation).

generalized Langevin equations and/or generalized Fokker-Planck equations had been derived in Refs. [149,178,222,223,261,262,278-289].

Following the Hamiltonian approach to Brownian motion, Zwanzig [222,262] derived the so-called generalized Langevin equation

$$m\frac{d^2X}{dt^2} = -\frac{dV(X)}{dX} - \int_0^t \beta(t'-t'')\frac{dX(t')}{dt}dt' + L(t) \qquad (2.143)$$

put originally forward by Mori [208,253]. Such a derivation shows that both memory effects present in the friction kernel $\beta(t'-t'')$ and non-Markovian fluctuation effects ingrained in the autocorrelation function $\langle L(t')L(t'')\rangle$ satisfy the Mori-Kubo dissipation-fluctuation relationship [208,250,253]

$$\langle L(t')L(t'')\rangle = k_B T \beta(t'-t''), \qquad (2.144)$$

where the thermal energy $k_B T$ comes from the assumption that the oscillator bath initial conditions are taken from a distribution, given by

$$f_{eq}(p,x) \propto e^{\frac{-H_E}{k_B T}}, \qquad (2.145)$$

in which the bath is at thermodynamic equilibrium. As pointed out by Zwanzig himself [222], the dissipation-fluctuation relation (2.144) is only obtained for a specific kind of initial distribution. Morita [290] in turn recalled that the generalized Langevin equation (2.143) contradicts one of the most fundamental requirements in non-equilibrium statistical mechanics, namely, the fact that the Langevin force should not be correlated to the initial position in the long-time regime: $\lim_{t\to\infty}\langle X(0)L(t)\rangle =$ *X0Lt=0*. On the other hand, Costa et al. [291] have disclosed some inconsistency between Eq. (2.144) and the Mori-Kubo relation (2.144) in the case of superdiffusion $\langle X^2(t)\rangle \sim C^{(\lambda)}t^\lambda$, with $\lambda \geq 2$, whereas Porrá, Wang, and Masoliver [292] as well as Wang and Tokuyama [293] have shown that the dissipation-fluctuation relation (2.144) may be an unnecessary constraint to investigating anomalous diffusion.

Although one believes that such "first principles"-based approaches (both Hamiltonian and kinetic approaches) can clarify the range of validity of both Langevin and Fokker-Planck equations, looked upon as phenomenological equations, a strong criticism raised against such studies is based on the fact that the heuristic simplifications and assumptions used are not reliable than the phenomenological assumptions [294]. Such assumptions are heuristic because they are not able to be justified or verified with complete rigor before the results can be accepted [207].

Lastly, it should be emphasized that the existence of the correlational function $I(t)$ has been overlooked by the extensive literature concerning non-Markov effects in both Hamiltonian [132,149,166,178,207,222,223,225,248,251,252,256,258-289,295,296] and non-Hamiltonian [191,207,221,223,224,232,267,290,297-331] approaches to Brownian motion, including studies on anomalous diffusion [223,235-237,291,332-349].

# 3. Quantum Brownian motion: Non-Markovian effects

The classical theory of Brownian motion predicts no random movement as the diffusion energy $\mathcal{E}$, defined by Eq. (2.25a), vanishes, e.g., for thermal systems $\mathcal{E} = k_B T$ at zero temperature. More specifically, in the escape rate theory the steady Kramers rate (2.128) or our non-steady one (2.140) leads to no thermal activation at zero temperature. Yet, it is expected that non-classical effects on both Brownian particle and environment account for the appearance of novel phenomena in the quantum realm at low temperatures, e. g., the quantum mechanical tunneling. Such quantum effects on the Brownian motion of a particle we call Quantum Brownian Motion[27]. In this context, a remarkable upshot is to predict quantum Brownian movement at zero temperature.

Our contribution to a general theory of quantum Brownian motion consists in deriving quantum master equations for both non-thermal and thermal systems via a non-Hamiltonian method of quantizing Fokker-Planck equations. For thermal systems, in Sect. 3.1 our main finding is the derivation of the non-Markovian Caldeira-Leggett equation for a particle in the presence of a heat bath of harmonic oscillators and its generalization to bosonic and fermionic baths. Sect. 3.2 in turn is devoted to discussing our Hamiltonian-independent method of quantizing Brownian motion pointing out the need of going beyond the Caldeira-Leggett Hamiltonian model.

## 3.1. Deriving non-Markovian quantum master equations

### 3.1.1. Non-thermal systems

We consider a Brownian particle immersed in a non-Gaussian environment described by the generalized Langevin equations (2.7) and its corresponding Kolmogorov equation in phase space, Eq. (2.11). We now wish to quantize such generalized Brownian movement by means of a non-Hamiltonian quantization process called dynamical quantization [181-190].

Following closely our previous works [189,190], we first obtain the equation of motion

---

[27]To the author's knowledge, the first glances of an eventual *quantum Brownian motion* arose in the paper by Einstein and Stern [350] on arguing in favor of the existence of molecular agitation at zero temperature. Yet, Einstein accomplished no attempt towards quantizing the diffusion equation derived by him in his 1905 paper [194]. In Sect. 5 we show how we can quantize directly a non-Markovian generalization of the Einstein's diffusion equation.

$$\frac{\partial \chi(x,\eta,t)}{\partial t} = \int_{-\infty}^{\infty} \mathbb{K}\mathcal{F}(x,p,t)e^{ip\eta}dp, \tag{3.1}$$

after performing on the Kolmogorov equation (2.11) the Fourier transform

$$\chi(x,\eta,t) = \frac{1}{2\pi}\int_{-\infty}^{\infty} \mathcal{F}(x,p,t)e^{ip\eta}dp, \tag{3.2}$$

where the exponential $e^{ip\eta}$ is deemed to be a dimensionless term. Notice that in the case $\eta = 0$ Eq. (3.2) is simply the marginal probability distribution function $\chi(x,t) = (1/2\pi)\int_{-\infty}^{\infty} \mathcal{F}(x,p,t)dp$.

Once obtained Eq. (3.1), the Kolmogorov stochastic dynamics (2.11) is said to be quantized by introducing the following quantization conditions

$$x_1 = x + \frac{\eta\hbar}{2} \tag{3.3a}$$

and

$$x_2 = x - \frac{\eta\hbar}{2} \tag{3.3b}$$

so as to derive the non-Gaussian quantum master equation

$$\frac{\partial \rho(x_1,x_2,t)}{\partial t} = \mathbb{B}\rho(x_1,x_2,t), \tag{3.4}$$

with

$$\mathbb{B}\rho(x_1,x_2,t) = \int_{-\infty}^{\infty} \mathbb{K}\mathcal{F}\left(\frac{x_1+x_2}{2},p,t\right)e^{ip\frac{(x_1-x_2)}{\hbar}}dp. \tag{3.4a}$$

Equation of motion (3.4) describes a quantum Brownian particle in the presence of a generic quantum fluid.

The parameter $\hbar$ characterizing the change of variables $(x,\eta) \mapsto (x_1,x_2)$, given by Eqs. (3.3), exhibits dimensions of angular momentum, i.e., [mass × length$^2$ × time$^{-1}$], and hence it is assumed to be identified numerically and conceptually with Planck's constant (divided by $2\pi$), whereas the variable $\eta$ displays dimensions of [time × mass$^{-1}$ × length$^{-1}$], as expected from the Fourier transformation (3.2). The geometric meaning of the quantization conditions (3.3) has to do with the existence of a minimal distance between the points $x_1$ and $x_2$, i.e., $|x_2 - x_1| = |\eta\hbar|$, by virtue of the quantum nature of space, so that in the classical limit $\hbar \to 0$, physically interpreted as $|\eta\hbar| \ll |x_2 - x_1|$, the mathematical point $x_2 = x_1 = x$ can be readily recovered.

On going from the classical equation of motion (3.1) to the quantum dynamics (3.4) via both quantization conditions (3.3), we have replaced the classical function $\chi \equiv \chi(x, \eta, t)$ with the quantum function $\rho \equiv \rho(x_1, x_2, t)$, for $\rho$ turns out to depend on the Planck constant, $\hbar$. Moreover, we dub $\rho(x_1, x_2, t)$ the von Neumann function because our Eq. (3.4) changes into the von Neumann equation in the absence of environment (see Appendix F). Physically, our quantization conditions (3.3) stand for that a quantum Brownian particle does display two coordinates $x_1$ and $x_2$ moving at the same time in the quantum configuration space in contrast to the classical Brownian particle described by only one coordinate $x$ in the classical configuration space given by Eq. (3.1), with $\chi(x, t) = (1/2\pi) \int_{-\infty}^{\infty} \mathcal{F}(x, p, t) dp$. So, it is the difference of the two motions that gives rise to quantum effects[28].

By assuming $\langle \Psi(t) \rangle = 0$ and taking into consideration the Gaussianity condition $|x_2 - x_1|^3 \ll 1$, our non-Gaussian quantum master equation (3.4) reduces to

$$i\hbar \frac{\partial \rho}{\partial t} = [V(x_1) - V(x_2)]\rho - \frac{\hbar^2}{2m}\left(\frac{\partial^2 \rho}{\partial x_1^2} - \frac{\partial^2 \rho}{\partial x_2^2}\right)$$
$$- i\hbar\beta \frac{(x_1 - x_2)}{2}\left(\frac{\partial \rho}{\partial x_1} - \frac{\partial \rho}{\partial x_2}\right) - \frac{i}{\hbar}\beta m \mathcal{E}_\hbar I(t)(x_1 - x_2)^2 \rho, \quad (3.5)$$

which corresponds to the quantization of the non-Markovian Fokker-Planck equation (2.27). In the Gaussian quantum master equation (3.5) the time evolution parameter $t$, the mass $m$, the frictional constant $\beta$, as well as the correlational function $I(t)$, given by Eq. (2.24), have been deemed to be non-quantized quantities, that is, they are $\hbar$-independent quantities, in contrast to the classical diffusion energy $\mathcal{E}$ that has been quantized via $\mathcal{E} \to \mathcal{E}_\hbar$. In brief, the quantum nature of Brownian motion shows up through both the variable change (3.3) and the quantization of the environment encapsulated in the quantum diffusion energy $\mathcal{E}_\hbar$.

---

[28]This interpretation is also valid for isolated quantum systems (see Appendix F). In Refs. [184,187,190] we have shown that the Fourier transform (3.2) endows classical mechanics with a commutative operator algebra which in turn generates two quantum-mechanical algebras via the quantization conditions (3.3).

## 3.1.2. Thermal systems

The physics of an environment made up of a myriad of moving quantum particles is assumed to be determined by the statistical thermodynamics (see Appendix G) according to which thermodynamic quantities turn up in conjunction with statistical properties [351]. For instance, the temperature —a $\hbar$-independent physical quantity— is looked upon as the fundamental thermodynamic concept characterizing the environment as a whole in thermal equilibrium situations, whereas the environmental particles may be probabilistically described by means of three kinds of $\hbar$-dependent statistics: The Maxwell-Boltzmann statistics, the Bose-Einstein statistics or the Fermi-Dirac statistics. On the basis of each of these statistics, one can calculate the thermodynamic internal energy $U \equiv U(T; \hbar)$ that in turn turns out to be statistically interpreted as the mean energy of the environmental particles. For this reason, under the assumption that it is such a kind of thermo-statistical energy $U$ which actually accounts for the quantum Brownian motion, the diffusion energy $\mathcal{E}_\hbar$ present in our quantum master equation (3.5) is to be identified with the reservoir's internal energy per particle, i.e.,

$$\mathcal{E}_\hbar = \frac{U(T;\hbar)}{N}. \tag{3.6}$$

*Heat bath of quantum harmonic oscillators.* On assuming the environment to be a heat bath consisting of $N$ quantum harmonic oscillators with angular frequency $\omega$, the Brownian particle's quantum diffusion energy (3.6) reads (see Appendix G)

$$\mathcal{E}_\hbar = \frac{\omega\hbar}{2}\coth\left(\frac{\omega\hbar}{2k_B T}\right), \tag{3.7}$$

where $k_B T$ is the thermal energy of the heat bath at high temperatures, $T \gg \omega\hbar/2k_B$, whereas $\omega\hbar/2$ corresponds to its quantum zero-point energy at $T = 0$:

$$\mathcal{E}_\hbar = \frac{\omega\hbar}{2}, \quad T = 0. \tag{3.7a}$$

Notice that in expression (3.7) besides the temperature $T$, the angular frequency $\omega$, as well as the Boltzmann constant $k_B$ do not depend on Planck's constant $\hbar$. They are then looked upon as classical quantities even in the quantum realm.

*Heat bath of fermions.* On the condition that the quantum Brownian motion takes place in presence of a fermionic heat bath of noninteracting particles contained in a volume $V$ and displaying spin degeneracy $g = 2s + 1$, the diffusion energy (3.6) turns out to be (see Appendix G)

$$\mathcal{E}_\hbar = \frac{gVm^{3/2}}{N\sqrt{2}\pi^2\hbar^3} \int_0^\infty \frac{\varepsilon^{3/2} d\varepsilon}{z^{-1}e^{\frac{\varepsilon}{k_B T}} + 1}, \qquad (3.8)$$

where $z = e^{\mu/k_B T}$ is the fugacity of the quantum gas expressed in terms of its chemical potential $\mu$. At low temperatures $T \to 0$ the quantum diffusion energy (3.8) becomes

$$\mathcal{E}_\hbar \sim \frac{3}{5} k_B T_F \left(1 + \frac{5\pi^2}{12} \frac{T^2}{T_F^2}\right), \quad T \to 0, \qquad (3.8a)$$

where $T_F$ is the Fermi temperature defined as

$$T_F = \frac{\epsilon_F}{k_B}, \qquad (3.9)$$

$\epsilon_F$ being the Fermi energy

$$\epsilon_F = \frac{\hbar^2}{2m} \left(\frac{6\pi^2}{g}\right)^{2/3} \left(\frac{N}{V}\right)^{2/3}, \qquad (3.9a)$$

expressed in terms of the thermodynamic quantities $N$ and $V$. From Eq. (3.8a) and using Eq. (3.9), it follows that the diffusion energy of a quantum Brownian particle moving in a fermion gas at $T = 0$ is determined only by the Fermi energy, i.e.,

$$\mathcal{E}_\hbar = \frac{3}{5} \epsilon_F, \qquad T = 0. \qquad (3.10)$$

*Heat bath of bosons.* If the quantum Brownian particle is immersed in an ideal boson gas, consisting of $N$ free noninteracting particles in a volume $V$, whose internal energy is given by $U_{BE}$, then its bosonic diffusion energy reads (see Appendix G)

$$\mathcal{E}_\hbar = \frac{gVm^{3/2}}{N\sqrt{2}\pi^2\hbar^3} \int_0^\infty \frac{\varepsilon^{3/2} d\varepsilon}{z^{-1}e^{\frac{\varepsilon}{k_B T}} - 1}, \quad T > T_{BE}, \qquad (3.11)$$

$z = e^{\mu/k_B T}$ being the fugacity and the quantity $g = 2s + 1$, which assumes integral values, the degeneracy factor of the spin states. $T_{BE}$ in turn is dubbed the Bose-Einstein temperature defined as

$$T_{BE} = g \frac{2\pi \hbar^2}{m k_B} \left(\frac{1}{2.612}\right)^{2/3} \left(\frac{N}{V}\right)^{2/3}. \qquad (3.11a)$$

At temperatures below or equal to $T_{BE}$, the quantum diffusion energy of a Brownian particle due to the internal energy of the bosonic gas reads [352,353]

$$\mathcal{E}_\hbar = 0.77 k_B \left( \sqrt{\frac{T^5}{T_{\mathrm{BE}}^3}} + T_{\mathrm{BE}} \sqrt{1 - \frac{T^3}{T_{\mathrm{BE}}^3}} \right), \quad T \leq T_{\mathrm{BE}}, \tag{3.12}$$

leading to the zero-point diffusion energy

$$\mathcal{E}_\hbar = 0.77 k_B T_{\mathrm{BE}}, \quad T = 0, \tag{3.12a}$$

which is equal to the diffusion energy (3.12) at $T = T_{\mathrm{BE}}$.

Both quantum diffusion energies (3.8) and (3.11) at high temperatures, $T \gg T_{\mathrm{F}}, T_{\mathrm{BE}}$, tend to the classical thermal energy $\mathcal{E} \propto k_B T$ for the case of an one-dimensional gas of fermions or bosons.

## 3.2. Discussion: Going beyond the Caldeira-Leggett model

We have stressed in recent articles [189,190] that the non-Markovian master equation (3.5), with $\beta \equiv 2\gamma$, $\mathcal{E}_\hbar$ given by Eq. (3.7), and $I(t) = 1$, becomes the Markovian Caldeira-Leggett equation

$$i\hbar \frac{\partial \rho}{\partial t} = [V(x_1, t) - V(x_2, t)] \rho - \frac{\hbar^2}{2m} \left( \frac{\partial^2 \rho}{\partial x_1^2} - \frac{\partial^2 \rho}{\partial x_2^2} \right) - i\hbar\gamma (x_1 - x_2) \left( \frac{\partial \rho}{\partial x_1} - \frac{\partial \rho}{\partial x_2} \right)$$
$$- i\gamma m \omega \coth\left( \frac{\omega \hbar}{2 k_B T} \right) (x_1 - x_2)^2 \rho, \tag{3.13a}$$

which in turn reduces at high temperatures, i.e., $\mathcal{E}_\hbar \sim k_B T$, to

$$i\hbar \frac{\partial \rho}{\partial t} = [V(x_1, t) - V(x_2, t)] \rho - \frac{\hbar^2}{2m} \left( \frac{\partial^2 \rho}{\partial x_1^2} - \frac{\partial^2 \rho}{\partial x_2^2} \right) - i\hbar\gamma (x_1 - x_2) \left( \frac{\partial \rho}{\partial x_1} - \frac{\partial \rho}{\partial x_2} \right)$$
$$- \frac{2 i \gamma m k_B T}{\hbar} (x_1 - x_2)^2 \rho, \tag{3.13b}$$

Both Markovian Caldeira-Leggett quantum master equations (3.13) had been originally derived on the ground of the path-integral-based Feynman quantization procedure[29] [25,163,354,355] as long as the Brownian particle and heat bath are initially uncorrelated[30]. The master equation (3.13a) was found by Caldeira,

---

[29]Indeed, Eq. (3.13a) had been first derived by Agarwal [144] (see also Ref. [138]) without making use of path integrals.
[30]The uncorrelated initial condition assumption between the Brownian particle and thermal bath is criticized as being thoroughly non-realistic, since they give rise to non-physical results [356-363].

Cerdeira, and Ramaswami [364] under the condition of weakness of the damping[31], i.e., $\gamma \ll \omega$ (for any temperature $T$), whereas Eq. (3.13b) was first derived by Caldeira and Leggett [365] at high temperatures $T \gg \hbar\omega/2k_B$ (for any $\gamma$). In contrast, it is worth highlighting that according to our approach to quantum Brownian motion the non-Markovian equation (3.13a) has been derived for any initial condition $\rho(x_1, x_2, t = 0)$, friction constant $\gamma$, and temperature $T$.

Further, because both Markovian Caldeira-Leggett equations (3.13) are not of the Lindblad form [128,129,366,367], they are plagued with the problem of positivity of the von Neumann function $\rho(x_1, x_2, t)$, so giving rise to unphysical results [133,138,139,151,240,368,369]. In brief, it has claimed that the derivation of both Markovian Caldeira-Leggett equations (3.13), as achieved in Refs. [364,365] on the basis of a model Hamiltonian, cannot be looked upon as a *bona fide* description of quantum Brownian motion [139,189,190,356-363,370,371].

As a last comment, we would like to stress that because our non-Markovian master equation (3.5) leads to Markovian Caldeira-Leggett equations we have dubbed it the non-Markovian Caldeira-Leggett equation. Moreover, as far as the quantum diffusion energy (3.8) and (3.11) are concerned, we say that our non-Markovian Caldeira-Leggett equation (3.5) is in presence of a heat bath of fermions and of bosons, respectively. Such an upshot has been reached without resorting to any Hamiltonian (or Lagrangian) model as performed in Refs. [139,173,372-375]. This fact suggests that we should go beyond the Caldeira-Leggett model so as to fathom the gist of quantum Brownian motion. Hence, we intend in the next Chapter to provide physical significance to our non-Hamiltonian approach by investigating the quantum master equation (3.5) in presence of three kinds of thermal baths.

---

[31] Recently, making use of the influence functional path-integral method of Feynman and Vernon [163], Fleming et al. [370] have shown that the Caldeira-Leggett equation found in Ref. [364] cannot be considered the correct quantum master equation in the weak coupling regime.

# 4. Quantum Brownian free motion

In the sequel, we scrutinize the quantum Brownian motion of a free particle described by the non-Markovian quantum master equation (3.5) in a general environment $\mathcal{E}_\hbar$. First, we take up a description in terms of quantum Fokker-Planck equations in the Wigner representation of quantum mechanics. Next, a description in terms of quantum Langevin equations is advanced.

## 4.1. Description in terms of quantum Fokker-Planck equations

### 4.1.1. The Fokker-Planck equation in quantum phase space

In the absence of external potential, i.e., $V(x_1) = V(x_2) = 0$, the non-Markovian quantum master equation (3.5) reads

$$i\hbar \frac{\partial \rho}{\partial t} = -\frac{\hbar^2}{2m}\left(\frac{\partial^2 \rho}{\partial x_1^2} - \frac{\partial^2 \rho}{\partial x_2^2}\right) - i\hbar\gamma(x_1 - x_2)\left(\frac{\partial \rho}{\partial x_1} - \frac{\partial \rho}{\partial x_2}\right)$$
$$- \frac{2i\gamma m \mathcal{E}_\hbar}{\hbar}\left(1 - e^{\frac{-t}{t_c}}\right)(x_1 - x_2)^2 \rho, \tag{4.1}$$

where we have used $\beta \equiv 2\gamma$ and $I(t) \equiv 1 - e^{-t/t_c}$. In the Wigner representation of quantum phenomena [376] the equation of motion in quantum configuration space (4.1) changes into the following dynamics in quantum phase space[32] (a sort of quantum Fokker-Planck equation)

$$\frac{\partial W}{\partial t} = -\frac{p}{m}\frac{\partial W}{\partial x} + 2\gamma \frac{\partial}{\partial p}pW + 2\gamma m \mathcal{E}_\hbar \left(1 - e^{\frac{-t}{t_c}}\right)\frac{\partial^2 W}{\partial p^2}, \tag{4.2}$$

where $W \equiv W(x, p, t)$. The quantum Fokker-Planck equation (4.2) has been obtained after performing upon Eq. (4.1) the Fourier transform, the so-called Wigner function [376],

$$W(x, p, t) = \frac{1}{2\pi\hbar}\int_{-\infty}^{\infty} \rho(x_1, x_2, t)e^{-i\frac{p}{\hbar}(x_1 - x_2)}d(x_1 - x_2). \tag{4.3}$$

---

[32]The quantum Fokker-Planck equation (4.2) can be viewed as the quantization of the classical Fokker-Planck equation (2.27), with $\beta \equiv 2\gamma$ and $V(x) = 0$, on performing the notational change $\mathcal{E} \to \mathcal{E}_\hbar$ and $\mathcal{F} \to W$.

For the case of a free particle, it is straightforward to show that underlying the quantum Fokker-Planck equation (4.2) there exist the quantum Langevin equations (see Appendix C)

$$\frac{dP(t)}{dt} = -2\gamma P(t) + \sqrt{4\gamma m \mathcal{E}_\hbar}\Psi(t), \tag{4.4a}$$

$$\frac{dX(t)}{dt} = \frac{P(t)}{m}, \tag{4.4b}$$

the Gaussian random function $\Psi(t)$ having the following statistical properties

$$\langle\Psi(t)\rangle = 0 \tag{4.5a}$$

and

$$\langle\Psi(t')\Psi(t'')\rangle = \left[1 - e^{\frac{-(t'+t'')}{2t_c}}\right]\delta(t' - t''). \tag{4.5b}$$

Notice that the mass $m$, the frictional parameter $\gamma$, the correlation time $t_c$, as well as both statistical properties (4.5) are not quantized. Non-classical effects enter only through the quantum diffusion energy $\mathcal{E}_\hbar$ in Eq. (4.4a). While Eq. (4.2) sets up a description of a free particle in terms of quantum Fokker-Planck equation, the differential equations (4.4), along with properties (4.5), describe it in terms of quantum Langevin equations to be studied in Sect. 4.2.

### 4.1.2. The initial condition

To solve Eq. (4.2), we start with the Gaussian initial condition

$$W(x, p, t = 0) = \frac{1}{\pi\hbar} e^{-\left(\frac{ap^2}{\hbar m} + \frac{mx^2}{\hbar a}\right)}, \tag{4.6}$$

which couples the Brownian free particle of mass $m$ with the environment through the constant $a$ displaying dimension of time. It could therefore be identified with the relaxation time $t_r = (2\gamma)^{-1}$ or the correlation time $t_c$, since such time scales are present in the quantum Fokker-Planck equation (4.2). In the first case the initial condition (4.6) would be Markovian, while in the second one it would be non-Markovian. Further, since the function (4.6) gives rise to $\langle P^2(0)\rangle = \hbar m/2a$, $\langle X^2(0)\rangle = a\hbar/2m$, and $\langle X(0)\rangle = \langle P(0)\rangle = 0$, the range of $a$ is $0 < a < \infty$. So, it is readily to verify the Heisenberg fluctuation relationship $\mathbb{X}(0)\mathbb{P}(0) = \hbar/2$, where $\mathbb{X}(0) \equiv \sqrt{\langle X^2(0)\rangle - \langle X(0)\rangle^2}$ and $\mathbb{P}(0) \equiv \sqrt{\langle P^2(0)\rangle - \langle P(0)\rangle^2}$.

### 4.1.3. The time-dependent solution

We intend to find a time-dependent solution $W(x,p,t)$ to the quantum Fokker-Planck equation (4.2) in the factored form

$$W(x,p,t) = \mathcal{W}(p,t)\mathcal{G}(x,t) \qquad (4.7)$$

but satisfying the Heisenberg minimal fluctuation relationship

$$\mathbb{X}(t)\mathbb{P}(t) = \frac{\hbar}{2}, \qquad (4.8)$$

which measures the localization of any quantum particle in phase space. To find out the function $\mathcal{W}(p,t)$ in Eq. (4.7), we perform the integral transformation $\mathcal{W}(p,t) = \int_{-\infty}^{\infty} W(x,p,t)dx$ on Eq. (4.2) and obtain the so-called quantum Rayleigh equation in momentum space [189,190]

$$\frac{\partial \mathcal{W}(p,t)}{\partial t} = 2\gamma \frac{\partial}{\partial p}[p\mathcal{W}(p,t)] + 2\gamma m \mathcal{E}_\hbar \left(1 - e^{\frac{-t}{t_c}}\right) \frac{\partial^2 \mathcal{W}(p,t)}{\partial p^2} \qquad (4.9)$$

whose solution is the Gaussian function

$$\mathcal{W}(p,t) = \frac{1}{\sqrt{4\pi A(t)}} e^{\frac{-p^2}{4A(t)}}, \qquad (4.10)$$

with

$$A(t) = \frac{m\hbar}{4a}\left\{\left(1 - \frac{2a\mathcal{E}_\hbar}{\hbar}\right)e^{-4\gamma t} + \frac{2a\mathcal{E}_\hbar}{\hbar}\left[1 + \frac{4\gamma t_c}{(4\gamma t_c - 1)}\left(e^{-4\gamma t} - e^{\frac{-t}{t_c}}\right)\right]\right\}. \qquad (4.10a)$$

Notice that function (4.10) is mathematically and physically well-defined as long as $A(t) > 0$, implying $4\gamma t_c \neq 1$. Moreover, it bears the following four parameters related to time scales: The evolution time $t$ (e.g., the observation time), the parameter $a$ (due to the initial condition), the relaxation time $t_r = (2\gamma)^{-1}$ (due to Markovian dissipation process), and the correlation time $t_c$ (due to non-Markovian fluctuation process).

Solution (4.10) leads to $\langle P(t) \rangle = 0$ and $\langle P^2(t) \rangle = 2A(t)$. Hence, the root mean square momentum $\mathbb{P}(t)$ reads

$$\mathbb{P}(t) = \sqrt{2A(t)}. \qquad (4.11)$$

From the Heisenberg constraint (4.8), it follows that the position fluctuation is

$$\mathbb{X}(t) = \sqrt{\frac{\hbar^2}{8A(t)}}. \tag{4.12}$$

It is readily to check that this result (4.12) is generated by the Gaussian function

$$\mathcal{G}(x,t) = \sqrt{\frac{4A(t)}{\pi\hbar^2}} e^{-\frac{4A(t)x^2}{\hbar^2}}. \tag{4.13}$$

Accordingly, using (4.13) and (4.10) our solution (4.7) reads

$$W(x,p,t) = \frac{1}{\pi\hbar} e^{-\left[\frac{p^2}{4A(t)} + \frac{4A(t)x^2}{\hbar^2}\right]}. \tag{4.14}$$

### 4.1.4. The steady regime

In the stationary regime $t \to \infty$, physically interpreted as the evolution time $t$ is too large in comparison with both the relaxation time $t_r = (2\gamma)^{-1}$ and the correlation time $t_c$, from Eq. (4.10a) we have $A(\infty) = m\mathcal{E}_\hbar/2$. The probability distribution function (4.14) then becomes

$$W(x,p) = \frac{1}{\pi\hbar} e^{-\left(\frac{p^2}{2m\mathcal{E}_\hbar} + \frac{2m\mathcal{E}_\hbar x^2}{\hbar^2}\right)}, \qquad t \to \infty. \tag{4.15}$$

The momentum (4.11) and position (4.12) fluctuations become, respectively,

$$\mathbb{P}(\infty) = \sqrt{m\mathcal{E}_\hbar}, \qquad t \to \infty, \tag{4.16}$$

and

$$\mathbb{X}(\infty) = \frac{\hbar}{\sqrt{4m\mathcal{E}_\hbar}}, \qquad t \to \infty. \tag{4.17}$$

Since the Heisenberg constraint reads $\mathbb{X}(\infty)\mathbb{P}(\infty) = \hbar/2$ the localization of quantum particles relies only on the Planck constant, as expected from Eq. (4.8).

For thermal systems, the following temperature regimes may be taken into account:

(a) *High-temperature regime.* For the case of thermal environments (a bath of quantum harmonic oscillators, and both fermionic and bosonic baths), the diffusion energy $\mathcal{E}_\hbar$ tends to the classical thermal energy $\mathcal{E} = k_B T$ as far as the high-temperature regime $T \to \infty$, i.e., $T \gg (\omega\hbar/2k_B), T_F, T_{BE}$, is concerned. So quantities (4.16) and (4.17) turn out to be written, respectively, as

$$\mathbb{P}(\infty) = \sqrt{mk_B T}, \qquad T \to \infty, \qquad (4.18)$$

and

$$\mathbb{X}(\infty) = \frac{\hbar}{\sqrt{4mk_B T}}, \qquad T \to \infty, \qquad (4.19)$$

so implying that a quantum Brownian free particle at high temperatures obeys the energy equipartition through the momentum fluctuation (4.18), despite quantum effects on its displacement fluctuation (4.19)[33]. In the classical limit $\hbar \to 0$ the position turns out to be a deterministic variable.

(b) *Zero-temperature regime.* A quantum Brownian free particle immersed in a heat bath of harmonic oscillators with angular frequency $\omega$ is characterized by the diffusion energy $\mathcal{E}_\hbar = (\omega\hbar/2)\coth(\omega\hbar/2k_B T)$ valid for all temperatures $T \geq 0$. The low-temperature regime is specified by the limit $T \to 0$, physically interpreted as $T \ll \omega\hbar/2k_B$. Thus, at $T = 0$ we obtain from Eqs. (4.16) and (4.17), with $\mathcal{E}_\hbar = \omega\hbar/2$, both results

$$\mathbb{P}(\infty) = \sqrt{\frac{m\omega\hbar}{2}}, \qquad T = 0, \qquad (4.20)$$

and

$$\mathbb{X}(\infty) = \frac{\hbar}{\sqrt{2m\omega\hbar}}, \qquad T = 0. \qquad (4.21)$$

In the case of a Brownian free particle in a fermionic bath at low temperatures, $\mathcal{E}_\hbar$ is given by Eq. (3.8a). So we find at $T = 0$

$$\mathbb{P}(\infty) = \sqrt{\frac{3}{5}m\epsilon_F}, \qquad T = 0, \qquad (4.22)$$

and

---

[33]Result (4.19) suggests the existence of quantum effects at high temperatures, thereby implying that the high-temperature regime may not be a sufficient condition for the classical limit of quantum-mechanical quantities.

$$\mathbb{X}(\infty) = \sqrt{\frac{5}{12}\frac{\hbar^2}{m\epsilon_F}}, \qquad T = 0. \qquad (4.23)$$

For a bosonic environment the quantum diffusion energy of a Brownian free particle at temperatures below or equal to the Bose-Einstein temperature $T_{BE}$ is given by Eq. (3.12). Thus, at $T = 0$ we find

$$\mathbb{P}(\infty) = \sqrt{0.77 m k_B T_{BE}}, \qquad T = 0, \qquad (4.24)$$

and

$$\mathbb{X}(\infty) = \frac{\hbar}{\sqrt{3.08 m k_B T_{BE}}}, \qquad T = 0, \qquad (4.25)$$

Therefore, a quantum Brownian free particle in the steady regime obeys the Heisenberg relation $\mathbb{X}(\infty)\mathbb{P}(\infty) = \hbar/2$ in the three cases of thermal reservoirs at zero temperature. That result bears out the assumption (4.8) according to which the localization of particles in quantum phase space is independent of their mass, the time, the temperature and the thermal nature of the medium.

As to the postulate of the energy equipartition, it is satisfied at high temperatures via Eq. (4.18) but it is violated at $T = 0$ in the cases (4.20), (4.22), and (4.24), as expected.

### 4.1.4. The differentiability property

From Eq. (4.10a) at short times $t \to 0$, i.e., $t \ll t_c, t_r = (2\gamma)^{-1}$, we obtain $A(t) \sim (m\hbar/4a)(1 - 4\gamma t)$, hence $\mathbb{P}(t) \sim \sqrt{(m\hbar/2a)(1 - 4\gamma t)}$ and $\mathbb{X}(t) \sim \hbar/\sqrt{2(m\hbar/a)(1 - 4\gamma t)}$. Accordingly, both $\mathbb{P}(t)$ and $\mathbb{X}(t)$ are differentiable or analytic functions at $t = 0$, i.e.,

$$\left.\frac{d\mathbb{P}(t)}{dt}\right|_{t=0} = -\gamma\sqrt{\frac{2m\hbar}{a}}, \qquad (4.26)$$

$$\left.\frac{d\mathbb{X}(t)}{dt}\right|_{t=0} = \gamma\sqrt{\frac{2a\hbar}{m}}, \qquad (4.27)$$

where $a$ comes from the initial condition (4.6) and $\gamma$ is the dissipation parameter.

From a mathematical viewpoint, the differentiability property of both $\mathbb{X}(t)$ and $\mathbb{P}(t)$ implies that the quantum Langevin equations (4.4) should be interpreted as true differential equations and not as integral ones according to a given interpretation. On the other hand, physically such analyticity property leads to the existence of two physical quantities for all time $t \geq 0$: The quantum force $\mathbb{F}(t) \equiv d\mathbb{P}(t)/dt$ and the instantaneous velocity $\mathbb{V}(t) \equiv d\mathbb{X}(t)/dt$. The time-dependent quantum force, also expressed as $\mathbb{F}(t) = \left(1/\sqrt{2A(t)}\right)[dA(t)/dt]$, acting on the free Brownian particle is given by

$$\mathbb{F}(t) = -\gamma \sqrt{\frac{2m\hbar}{a}} B(t), \qquad (4.28)$$

where $B(t)$ is the dimensionless function

$$B(t) = \frac{\left(1 - \frac{2a\mathcal{E}_\hbar}{\hbar}\right) e^{-4\gamma t} + \frac{2a\mathcal{E}_\hbar}{\hbar(4\gamma t_c - 1)}\left(4\gamma t_c e^{-4\gamma t} - e^{\frac{-t}{t_c}}\right)}{\left\{\left(1 - \frac{2a\mathcal{E}_\hbar}{\hbar}\right) e^{-4\gamma t} + \frac{2a\mathcal{E}_\hbar}{\hbar}\left[1 + \frac{4\gamma t_c}{(4\gamma t_c - 1)}\left(e^{-4\gamma t} - e^{\frac{-t}{t_c}}\right)\right]\right\}^{1/2}}, \qquad (4.28a)$$

which is defined for $4\gamma t_c > 1$ and ranges from $B(0) = 1$ to $B(\infty) = 0$. For $B(t) > 0$, $\mathbb{F}(t)$ is attractive. At the instant of time when $B(t) = 0$, $\mathbb{F}(t)$ vanishes. In addition, $\mathbb{F}(t)$ becomes a repulsive force for $B(t) < 0$ towards the long-time regime when its influence decays to zero. Thus, the environmental force (4.28) is a non-steady effect.

The differentiability of $\mathbb{X}(t)$ in turn leads to the instantaneous velocity, $\mathbb{V}(t) = -\sqrt{[\hbar^2/32A^3(t)]}[dA(t)/dt]$,

$$\mathbb{V}(t) = \gamma \sqrt{\frac{2a\hbar}{m}} C(t), \qquad (4.29)$$

$C(t)$ being the following dimensionless function

$$C(t) = \frac{B(t)}{\left(1 - \frac{2a\mathcal{E}_\hbar}{\hbar}\right) e^{-4\gamma t} + \frac{2a\mathcal{E}_\hbar}{\hbar}\left[1 + \frac{4\gamma t_c}{(4\gamma t_c - 1)}\left(e^{-4\gamma t} - e^{\frac{-t}{t_c}}\right)\right]}, \qquad (4.29a)$$

where $B(t)$ is given by Eq. (4.28a). Due to Eq. (4.29a), both quantities $\mathbb{F}(t)$ and $\mathbb{V}(t)$ are connected through the quite formally suggestive equation

$$\mathbb{F}(t) = m\mathbb{A}(t), \tag{4.30}$$

$\mathbb{A}(t)$ being the quantum-mechanical acceleration

$$\mathbb{A}(t) = -\frac{\mathbb{V}(t)}{a}\left\{\left(1-\frac{2a\mathcal{E}_\hbar}{\hbar}\right)e^{-4\gamma t} + \frac{2a\mathcal{E}_\hbar}{\hbar}\left[1+\frac{4\gamma t_c}{(4\gamma t_c - 1)}\left(e^{-4\gamma t} - e^{\frac{-t}{t_c}}\right)\right]\right\}.$$

$$\tag{4.31}$$

A consequence of the existence of the instantaneous velocity (4.29) is the concept of time-dependent quantum diffusion coefficient $\mathbb{D}(t) \equiv \mathbb{X}(t)\mathbb{V}(t) = (1/2)(d\langle X^2(t)\rangle/dt)$, i.e.,

$$\mathbb{D}(t) = \frac{\gamma \hbar a}{m} D(t), \tag{4.32}$$

where the dimensionless function $D(t)$ is given by

$$D(t) = \frac{B(t)}{\left\{\left(1-\frac{2a\mathcal{E}_\hbar}{\hbar}\right)e^{-4\gamma t} + \frac{2a\mathcal{E}_\hbar}{\hbar}\left[1+\frac{4\gamma t_c}{(4\gamma t_c - 1)}\left(e^{-4\gamma t} - e^{\frac{-t}{t_c}}\right)\right]\right\}^{3/2}} \tag{4.32a}$$

which should be defined for $B(t) \geq 0$. In the steady regime $t \to \infty$, Eq. (4.32) vanishes, whereas at $t = 0$ it is given by $\mathbb{D}(0) = \gamma \hbar a/m$.

It is worth pointing out that both the instantaneous velocity (4.29) and the quantum diffusion coefficient (4.32) show up in our description of quantum Brownian free motion as a consequence of the Heisenberg relation (4.8).

### 4.1.6. The classical limit

Formally, the limit $\hbar \to 0$ of the quantum state (4.14) yields the classical probability distribution function $\mathcal{F}(x,p,t)$, solution of Eq. (2.35), i.e.,

$$\lim_{\hbar \to 0} W(x,p,t) = \mathcal{F}(x,p,t) = e^{\frac{-p^2}{4\mathcal{A}(t)}}\delta(x), \tag{4.33}$$

provided that $\lim_{\hbar \to 0} \mathcal{E}_\hbar = \mathcal{E}$. In the quantum-to-classical transition the parameters $m, a, \gamma,$ and $t_c$ are deemed to be $\hbar$-independent. So, it is said that the quantum Fokker-Planck equation (4.2) tends to the classical Fokker-Planck equation (2.35) in the limit $\hbar \to 0$. In addition, the quantum Langevin equations (4.4) go to the classical Langevin equations (2.7).

The physical significance of the classical limiting process (4.33) is the following: A classical free Brownian particle is characterized only by its momentum fluctuations, whereas its position $x$ appears as a deterministic variable, thereby breaking down Heisenberg's relationship (4.8), i.e., $\mathbb{X}(t)\mathbb{P}(t) = 0$. In other words, Eq. (4.8) is an intrinsically quantum feature without classical analogous. Alternatively, such a violation of Heisenberg's relation can be viewed directly from the limit $\hbar \to 0$ of Eq. (4.12).

Also, as far as the classical limit $\hbar \to 0$ is concerned the Gaussian initial condition (4.6) tends to $\delta(x)\delta(p)$, meaning that both $x$ and $p$ are sharp variables in the classical domain.

The classical limit of the quantum force (4.28) is

$$\mathbb{F}(t) = -2\gamma\sqrt{m\mathcal{E}}\,\frac{\left(e^{-4\gamma t} - e^{\frac{-t}{t_c}}\right)}{\sqrt{(4\gamma t_c - 1)\left[e^{-4\gamma t} - 1 + 4\gamma t_c\left(1 - e^{\frac{-t}{t_c}}\right)\right]}}, \quad (4.34)$$

which is identical with Eq. (2.48) for thermal systems, i.e., as $\mathcal{E} = k_B T$. From result (4.34) and relation (4.30), it follows that the corresponding classical acceleration reads

$$\mathbb{A}(t) = -2\gamma\sqrt{\frac{\mathcal{E}}{m}}\,\frac{\left(e^{-4\gamma t} - e^{\frac{-t}{t_c}}\right)}{\sqrt{(4\gamma t_c - 1)\left[e^{-4\gamma t} - 1 + 4\gamma t_c\left(1 - e^{\frac{-t}{t_c}}\right)\right]}}. \quad (4.35)$$

Moreover, assuming again the parameters $a, \gamma$, and $t_c$ to be $\hbar$-independent both the quantum instantaneous velocity (4.29) and the quantum diffusion coefficient (4.32) vanish in the classical limit, standing for that such physical quantities can exist only in the quantum realm.

## 4.1.7. Determining the parameters $a$, $\gamma$, and $t_c$

The quantum Brownian motion of a free particle in phase space, described by the Wigner function (4.14), exhibits three independent time-scales related to the parameters $a, \gamma$, and $t_c$ which in their turn do not depend on both the Planck constant $\hbar$ and the quantum energy $\mathcal{E}_\hbar$. Nevertheless, under the assumptions that the three time scales $a$, $t_r = (2\gamma)^{-1}$, and $t_c$ are identical with each other, i.e.,

$$a = \frac{1}{2\gamma} = t_c, \qquad (4.36)$$

and both the steady solution $W(x,p,t \to \infty)$ and the initial condition $W(x,p,t=0)$ are the same, i.e.,

$$W(x,p,t=0) = W(x,p,t \to \infty), \qquad (4.37)$$

it follows that $a$, $\gamma$, and $t_c$ turn out to be determined in terms of both $\hbar$ and $\mathcal{E}_\hbar$ as

$$a = \frac{\hbar}{2\mathcal{E}_\hbar}, \qquad (4.38)$$

$$\gamma = \frac{\mathcal{E}_\hbar}{\hbar}, \qquad (4.39)$$

and

$$t_c = \frac{\hbar}{2\mathcal{E}_\hbar}. \qquad (4.40)$$

As far as thermal systems are concerned, the parameters $a = t_c$ and $\gamma$ rely on the temperature of the heat bath after identifying the Brownian particle's quantum diffusion energy $\mathcal{E}_\hbar$ with Eqs. (3.7), (3.8), or (3.11). In the case of a heat bath of quantum harmonic oscillators, whereby $\mathcal{E}_\hbar = (\omega\hbar/2)\coth(\omega\hbar/2k_BT)$, we have

$$t_c = \frac{1}{\omega \coth\left(\frac{\omega\hbar}{2k_BT}\right)} \qquad (4.41\text{a})$$

and

$$\gamma = \frac{\omega}{2}\coth\left(\frac{\omega\hbar}{2k_BT}\right). \qquad (4.41\text{b})$$

At high temperatures $T \gg \omega\hbar/2k_B$, although the diffusion energy $\mathcal{E}_\hbar \sim \mathcal{E} = k_BT$ does not depend on $\hbar$, the parameters $a = t_c \sim \hbar/2k_BT$ and $\gamma \sim k_BT/\hbar$ hold $\hbar$-dependent. In contrast, in the quantum limit at $T = 0$, both $a = t_c$ and $\gamma$ render $\hbar$-independent, thereby being determined only by the angular frequency of the bath oscillators, i.e., $a = t_c = \omega^{-1}$ and $\gamma = \omega/2$, while the diffusion energy is $\hbar$-dependent: $\mathcal{E}_\hbar = \omega\hbar/2$.

For a fermionic thermal reservoir at low temperatures, characterized by Eq. (3.8a), the correlation time (4.41) reads

$$t_c = \frac{5\hbar}{6k_B T_F \left[1 + \frac{5\pi^2}{12}\left(\frac{T}{T_F}\right)^2\right]}, \qquad (4.42a)$$

whereas the frictional parameter (4.41b) becomes

$$\gamma = \frac{3}{5}\frac{k_B T_F}{\hbar}\left[1 + \frac{5\pi^2}{12}\left(\frac{T}{T_F}\right)^2\right]. \qquad (4.42b)$$

Both quantities (4.42a) and (4.42b) are also defined at zero temperature as $t_c = (5/6)(\hbar/k_B T_F)$ and $\gamma = (3/5)(k_B T_F/\hbar)$, respectively.

Lastly, in a bosonic heat bath at temperatures below or equal to $T_{BE}$, given by Eq. (3.12), we find

$$t_c = \frac{\hbar}{1.54 k_B \left[\sqrt{\frac{T^5}{T_{BE}^3}} + T_{BE}\sqrt{1 - \left(\frac{T}{T_{BE}}\right)^3}\right]} \qquad (4.43a)$$

and

$$\gamma = \frac{0.77 k_B}{\hbar}\left[\sqrt{\frac{T^5}{T_{BE}^3}} + T_{BE}\sqrt{1 - \left(\frac{T}{T_{BE}}\right)^3}\right]. \qquad (4.43b)$$

leading respectively to $t_c = \hbar/1.54 k_B T_{BE}$ and $\gamma = 0.77 k_B T_{BE}/\hbar$ at $T = 0$.

### 4.1.8. The quantum force

On inserting the parameters (4.38), (4.39), and (4.40) into expression (4.28), the quantum force reads

$$\mathbb{F}(t) = -\sqrt{\frac{4m\mathcal{E}_\hbar^3}{\hbar^2}} \frac{e^{-\frac{2\mathcal{E}_\hbar t}{\hbar}}\left(2e^{-\frac{2\mathcal{E}_\hbar t}{\hbar}} - 1\right)}{\left[1 + 2e^{-\frac{2\mathcal{E}_\hbar t}{\hbar}}\left(e^{-\frac{2\mathcal{E}_\hbar t}{\hbar}} - 1\right)\right]^{1/2}}, \qquad (4.44)$$

expressed now in terms of only two time scales: The evolution time $t$ (e.g., the observation time) and the quantum time $t_q = \hbar/2\mathcal{E}_\hbar$, interpreted physically as a correlation time or a relaxation time by virtue of our assumption (4.36).

In the time window $0 \leq t < t_q = \hbar/2\mathcal{E}_\hbar$, the force (4.44) renders attractive. At the instant $t = t_q \ln 2$ it vanishes, becoming a repulsive force for $t > t_q = \hbar/2\mathcal{E}_\hbar$, and going towards the long-time regime $t \gg t_q = \hbar/2\mathcal{E}_\hbar$ when its influence decays to zero. It is worth noting that the existence of the non-equilibrium thermal force (4.44) for all time $t \geq 0$ accounts for the analyticity property of the root mean square momentum

$$\mathbb{P}(t) = \sqrt{m\mathcal{E}_\hbar \left[1 + 2e^{\frac{-2\mathcal{E}_\hbar t}{\hbar}}\left(e^{\frac{-2\mathcal{E}_\hbar t}{\hbar}} - 1\right)\right]}. \qquad (4.45)$$

For thermal systems the quantum diffusion energy $\mathcal{E}_\hbar$ in Eq. (4.44) at high temperatures is proportional to $k_B T$. Making use of the diffusion energy (3.7) at high temperatures $T \gg \omega\hbar/2k_B$, the quantum force (4.44) reads

$$\mathbb{F}(t) = \sqrt{\frac{4m(k_B T)^3}{\hbar^2}} \frac{e^{\frac{-2k_B T}{\hbar}t}\left(1 - 2e^{\frac{-2k_B T}{\hbar}t}\right)}{\left[1 + 2e^{\frac{-2k_B T}{\hbar}t}\left(e^{\frac{-2k_B T}{\hbar}t} - 1\right)\right]^{1/2}}, \quad T \gg \frac{\omega\hbar}{2k_B}. \qquad (4.46a)$$

Note that the thermal force (4.46a) is an intrinsically quantum effect, for it is not defined as $\hbar \to 0$. Besides, it does depend on the evolution time $t$ and the thermal quantum time scale $t_q = \hbar/2k_B$. At $t = 0$, the quantum force $\mathbb{F}(0) = \sqrt{4m(k_B T)^3/\hbar^2}$ corresponds to the momentum fluctuation $\mathbb{P}(0) = \sqrt{mk_B T}$ which in turn implies the validity of the energy equipartition theorem of statistical mechanics in the high-temperature regime. At $t \sim t_q$, we have $\mathbb{F}(t_q) \sim (1/4)\mathbb{F}(0)$, while in the cases when $t = 10t_q$ and $t = 100t_q$, for example, the magnitude of the thermal force (4.46a) is given by $\mathbb{F}(10t_q) \sim 10^{-5}\mathbb{F}(0)$ and $\mathbb{F}(100t_q) \sim 10^{-44}\mathbb{F}(0)$, respectively. Therefore, measuring the thermal force (4.46a) can indicate deviations from the equipartition theorem at times of the order of $t_q = \hbar/2k_B T$. For example, employing $k_B \sim 10^{-23}$ m² kg s⁻² K⁻¹ and $\hbar \sim 10^{-34}$ m² kg s⁻¹ for a particle of mass $m \sim 10^{-15}$ kg immersed in a thermal bath at room temperature $T \sim 10^2$ K, Eq. (4.46a) leads to $\mathbb{F}(t) \sim 10^{-6}$ N at $t \sim 10^{-14}$ s.

*The quantum force brought about by a thermal reservoir of quantum harmonic oscillators at* $T = 0$. Inserting Eq. (3.7a), i.e., $\mathcal{E}_\hbar = \omega\hbar/2$, into the quantum force (4.44) provides

$$\mathbb{F}(t) = \sqrt{\frac{m\omega^3\hbar}{2}} \frac{e^{-\omega t}(1 - 2e^{-\omega t})}{\sqrt{1 + 2e^{-\omega t}(e^{-\omega t} - 1)}}, \qquad T = 0, \qquad (4.46b)$$

generated by the momentum fluctuation $\mathbb{P}(t) = \sqrt{(m\omega\hbar/2)[1 + 2e^{-\omega t}(e^{-\omega t} - 1)]}$, breaking down the energy equipartition at zero temperature for all times $t \geq 0$. For instance, Eq. (4.46b) with $\omega \sim 10^{10}$ s$^{-1}$ at $t \sim 10^{-10}$ s yields the quantum force $\mathbb{F}(t) \sim 10^{-11}$ N acting on the particle of mass $m \sim 10^{-15}$ kg.

*The quantum force brought about by a thermal reservoir of fermions at* $T = 0$. The use of the fermionic diffusion energy at low-temperatures, Eq. (3.8a), into Eq. (4.44) leads to the following quantum force brought about by a heat bath of fermions

$$\mathbb{F}(\lambda) = \frac{6}{5}\sqrt{\frac{3m(k_B T_F)^3}{5\hbar^2}}\left(1 + \frac{5\pi^2}{12}\frac{T^2}{T_F^2}\right)^{3/2} \frac{e^{-\lambda}(1 - 2e^{-\lambda})}{\sqrt{1 + 2e^{-\lambda}(e^{-\lambda} - 1)}}, T \leq T_F, \quad (4.47)$$

where we have introduced the dimensionless parameter

$$\lambda = \frac{6k_B T_F}{5\hbar}\left(1 + \frac{5\pi^2}{12}\frac{T^2}{T_F^2}\right)t. \qquad (4.48)$$

Such a fermionic force (4.47) corresponds to the momentum fluctuation $\mathbb{P}(t) = \sqrt{(3/5)mk_B T_F[1 + (5\pi^2/12)(T/T_F)^2][1 + 2e^{-\lambda}(e^{-\lambda} - 1)]}$ which also predicts the violation of the energy equipartition for all $\lambda \geq 0$, i.e., for all times $t \geq 0$.

At $T = T_F$, Eq. (4.47) reads

$$\mathbb{F}(\lambda) = \frac{6}{5}\sqrt{\frac{3}{5}\left(1 + \frac{5\pi^2}{12}\right)^3}\sqrt{\frac{m(k_B T_F)^3}{\hbar^2}} \frac{e^{-\lambda}(1 - 2e^{-\lambda})}{\sqrt{1 + 2e^{-\lambda}(e^{-\lambda} - 1)}}, \quad T = T_F, \quad (4.47a)$$

with

$$\lambda \equiv \lambda(T_F) = \frac{6k_B T_F}{5\hbar}\left(1 + \frac{5}{12}\pi^2\right)t, \qquad (4.48a)$$

while at zero temperature it goes to

$$\mathbb{F}(\lambda) = \frac{6}{5}\sqrt{\frac{3m(k_B T_F)^3}{5\hbar^2}} \frac{e^{-\lambda}(1-2e^{-\lambda})}{\sqrt{1+2e^{-\lambda}(e^{-\lambda}-1)}}, \qquad T=0, \quad (4.47b)$$

with

$$\lambda \equiv \lambda(0) = \frac{6k_B T_F}{5\hbar} t. \qquad (4.48b)$$

Assuming a metal to be a Fermi gas of electrons at $T_F \sim 10^4$ K contained in a box, with a density $n = (N/V) \sim 10^{16}$ m$^{-3}$ [377], the zero-point fermionic force (4.47b) on a quantum Brownian particle of mass $m \sim 10^{-15}$ kg is predicted to be of order $10^{-4}$ N for short times $t \sim 10^{-15}$ s.

*The quantum force brought about by a thermal reservoir of bosons at $T = 0$.* Using Eq. (3.12), the quantum force brought about by a bosonic heat bath at both temperatures $T = 0$ and $T = T_{BE}$ reads

$$\mathbb{F}(\zeta) = \sqrt{\frac{4m}{\hbar^2}} (0.77 k_B T_{BE})^{3/2} \frac{e^{-\zeta}(1-2e^{-\zeta})}{\sqrt{1+2e^{-\zeta}(e^{-\zeta}-1)}}, \qquad (4.49)$$

where the dimensionless quantity $\zeta$ is given by

$$\zeta = 1.54 \frac{k_B T_{BE}}{\hbar} t. \qquad (4.50)$$

The bosonic quantum force (4.49) is associated with the momentum fluctuation $\mathbb{P}(t) = \sqrt{0.77 m k_B T_{BE}[1+2e^{-\zeta}(e^{-\zeta}-1)]}$. Again, it is predicted no energy equipartition for bosonic system at zero temperature for all $\zeta \geq 0$, i.e., $t \geq 0$. Our Eq. (4.49) can therefore be employed for bearing out such a theoretical prediction. For example, at the Boson-Einstein temperature $T_{BE} \sim 3$ K in an ideal gas of $^4$He Eq. (4.49) predicts a bosonic force of order $10^{-11}$ N on a particle of mass $m \sim 10^{-15}$ kg to occur at short time $t \sim 10^{-11}$ s.

The simple numerical examples above suggest that the strength of the quantum forces (4.46), (4.47), and (4.49) brought about by thermal reservoirs at short times could be measured in experiences, for instance, using trapped ions in which measurement of forces of the order of yoctonewton, i.e., $10^{-24}$ N, has been recently reported [378].

### 4.1.9. The quantum instantaneous velocity

Under conditions (4.38), (4.39), and (4.40) the instantaneous velocity (4.29) reads

$$\mathbb{V}(t) = \sqrt{\frac{\mathcal{E}_\hbar}{m}} \frac{e^{\frac{-2\mathcal{E}_\hbar t}{\hbar}}\left(2e^{\frac{-2\mathcal{E}_\hbar t}{\hbar}} - 1\right)}{\left[1 + 2e^{\frac{-2\mathcal{E}_\hbar t}{\hbar}}\left(e^{\frac{-2\mathcal{E}_\hbar t}{\hbar}} - 1\right)\right]^{3/2}}. \quad (4.51)$$

Notice that this instantaneous velocity (4.51) also displays two time scales: The evolution time $t$ and the quantum time $t_q = \hbar/2\mathcal{E}_\hbar$. At $t = 0$, it becomes $\mathbb{V}(0) = \sqrt{\mathcal{E}_\hbar/m}$. At $t \sim t_q$, we find $\mathbb{V}(t_q) \sim (1/4)\mathbb{V}(0)$. In the cases when $t \sim 10 t_q$ and $t \sim 100 t_q$, for instance, the magnitude of the velocity is given by $|\mathbb{V}(10 t_q)| \sim 10^{-5}\mathbb{V}(0)$ and $|\mathbb{V}(100 t_q)| \sim 10^{-44}\mathbb{V}(0)$, respectively.

*The quantum instantaneous velocity brought about by a thermal reservoir of quantum harmonic oscillators.* At high temperatures, such that $\mathcal{E}_\hbar \to \mathcal{E} = k_B T$, Eq. (4.51) becomes

$$\mathbb{V}(t) = \sqrt{\frac{k_B T}{m}} \frac{e^{\frac{-2k_B T t}{\hbar}}\left(2e^{\frac{-2k_B T t}{\hbar}} - 1\right)}{\left[1 + 2e^{\frac{-2k_B T t}{\hbar}}\left(e^{\frac{-2k_B T t}{\hbar}} - 1\right)\right]^{3/2}}, \quad T \gg \frac{\omega\hbar}{2k_B}. \quad (4.52a)$$

At $t = 0$, Eq. (4.52a) yields simply the classical instantaneous velocity: $\mathbb{V}(0) = \sqrt{k_B T/m}$, while for long times $t \to \infty$ it vanishes. Hence, the appearance of quantum effects occurs at $0 < t < \infty$. At $t \sim t_q = \hbar/2k_B T$, for example, we find $\mathbb{V}(t_q) \sim (1/4 k_B T/m)$. Therefore, measuring the quantum instantaneous velocity at high-temperatures, Eq. (4.52a), could suggest experimentally the validity of the equipartition energy theorem at $t = 0$ and its violation at time scales of the order of the thermal time $t_q = \hbar/2k_B T$.

As a numerical example, let us consider a particle of mass $m \sim 10^{-15}$ kg immersed in a thermal bath at $T \sim 10^2$ K. Its initial instantaneous velocity is then about $10^{-3}$ m s$^{-1}$ according to the energy equipartition theorem [232]. Yet, during a span of time $t \sim 10^{-13}$ s we find from Eq. (4.52a) the instantaneous velocity $\mathbb{V}(t \sim 10^{-13}) \sim 10^{-4}$ m s$^{-1}$, thus standing for that quantum effects account for diminishing the initial instantaneous velocity $\mathbb{V}(0) \sim 10^{-3}$ m s$^{-1}$ of the Brownian particle.

As far as $\mathcal{E}_\hbar = \omega\hbar/2$ is concerned the instantaneous velocity (4.51) at zero temperature becomes

$$\mathbb{V}(t) = \sqrt{\frac{\omega\hbar}{2m}} \frac{e^{-\omega t}(2e^{-\omega t} - 1)}{[1 + 2e^{-\omega t}(e^{-\omega t} - 1)]^{3/2}}, \quad T = 0, \qquad (4.52b)$$

predicting the violation of the energy equipartition for all $t \geq 0$. By way of illustration, taking $\omega \sim 10^{10}$ s$^{-1}$ the instantaneous velocity (4.52b) at $t = 0$ of a particle of mass $m \sim 10^{-15}$ kg is about $10^{-5}$ m s$^{-1}$, reducing to about $10^{-10}$ m s$^{-1}$ at $t \sim 10^{-9}$ s.

*The quantum instantaneous velocity brought about by a thermal reservoir of fermions.* The instantaneous velocity $\mathbb{V}(t)$ of a free Brownian particle immersed in a heat bath of $N$ fermions at low temperatures is given by Eq. (4.51), with $\mathcal{E}_\hbar \sim (3/5 k_B T_F 1 + 5/12\pi 2T/T_F 2$. At zero temperature, it reads

$$\mathbb{V}(t) = \sqrt{\frac{3k_B T_F}{5m}} \frac{e^{-\frac{6k_B T_F}{5\hbar}t}\left(2e^{-\frac{6k_B T_F}{5\hbar}t} - 1\right)}{\left[1 + 2e^{-\frac{6k_B T_F}{5\hbar}t}\left(e^{-\frac{6k_B T_F}{5\hbar}t} - 1\right)\right]^{3/2}}, \quad T = 0, \qquad (4.53)$$

ranging from $\mathbb{V}(0) = (3/5m)k_B T_F$ to $\mathbb{V}(\infty) = 0$. Accordingly, the instantaneous velocity (4.53) also predicts the violation of the energy equipartition for all $t \geq 0$.

In a Fermi gas of electrons at $T_F \sim 10^4$ K, for instance, the zero-point fermionic velocity (4.53) at $t = 0$ of a particle of mass $m \sim 10^{-15}$ kg is about $10^{-3}$ m s$^{-1}$, about $10^{-4}$ m s$^{-1}$ at $t \sim 10^{-15}$ s, and about $10^{-47}$ m s$^{-1}$ at $t \sim 10^{-17}$ s.

*The quantum instantaneous velocity brought about by a thermal reservoir of bosons.* In this case, the instantaneous velocity $\mathbb{V}(t)$ of a free Brownian particle in a bosonic heat bath at and below the Bose-Einstein temperature is given by expression (4.51), with $\mathcal{E}_\hbar = 0.77 k_B \left[\sqrt{T^5/T_{BE}^3} + T_{BE}\sqrt{1 - (T/T_{BE})^3}\right]$. At $T = 0$ and $T = T_{BE}$, the instantaneous velocity $\mathbb{V}(t)$ reads

$$\mathbb{V}(t) = \sqrt{\frac{0.77 k_B T_{BE}}{m}} \frac{e^{\frac{-1.54 k_B T_{BE}}{\hbar}t}\left(2e^{\frac{-1.54 k_B T_{BE}}{\hbar}t} - 1\right)}{\left[1 + 2e^{\frac{-1.54 k_B T_{BE}}{\hbar}t}\left(e^{\frac{-1.54 k_B T_{BE}}{\hbar}t} - 1\right)\right]^{3/2}}, \quad (4.54)$$

ranging from $\mathbb{V}(0) = \sqrt{0.77 k_B T_{BE}/m}$ to $\mathbb{V}(\infty) = 0$. Thus, the instantaneous velocity (4.54) also leads to the violation of the energy equipartition for all $t \geq 0$.

In the case, for example, of an ideal gas of $^4$He at which the Boson-Einstein temperature is $T_{BE} \sim 3$ K, we find that the bosonic velocity (4.54) at $t = 0$ is $\mathbb{V}(0) \sim 10^{-4}$ m s$^{-1}$ for a particle of mass $m \sim 10^{-15}$ kg and about $10^{-9}$ m s$^{-1}$ at $t \sim 10^{-9}$ s.

### 4.1.10. The quantum diffusion coefficient

Making use of assumptions (4.38), (4.39), and (4.40) the quantum time-dependent diffusion coefficient (4.32) turns out to be written down as

$$\mathbb{D}(t) = \frac{\hbar}{2m} \frac{e^{\frac{-2\mathcal{E}_\hbar t}{\hbar}}\left(2e^{\frac{-2\mathcal{E}_\hbar t}{\hbar}} - 1\right)}{\left[1 + 2e^{\frac{-2\mathcal{E}_\hbar t}{\hbar}}\left(e^{\frac{-2\mathcal{E}_\hbar t}{\hbar}} - 1\right)\right]^2}, \quad (4.55)$$

At $t = 0$, the Brownian motion of a free particle is characterized by the quantum diffusion constant

$$\mathbb{D}(0) = \frac{\hbar}{2m}, \quad (4.55a)$$

which depends only on the Planck constant $\hbar$ and the mass $m$. It is readily to see that this is so because in the expression $\mathbb{D}(0) = \mathbb{X}(0)\mathbb{V}(0)$ both the initial position fluctuation $\mathbb{X}(0)$ and the initial instantaneous velocity $\mathbb{V}(0)$ do rely on the medium properties in the following way: $\mathbb{X}(0) = (\hbar/2)(1/\sqrt{m\mathcal{E}_\hbar})$ and $\mathbb{V}(0) = \sqrt{\mathcal{E}_\hbar/m}$. Our particle of mass $m \sim 10^{-15}$ kg moves owing to the quantum diffusion constant $\mathbb{D}(0) \sim 10^{-20}$ m$^2$ s$^{-1}$.

Environmental effects present in the time-dependent quantum diffusion coefficient (4.55) show up at the time window $0 < t < (\hbar/2\mathcal{E}_\hbar)\ln 2$ and are responsible for diminishing the initial diffusion constant (4.55a). At $t = (\hbar/2\mathcal{E}_\hbar)\ln 2$, Eq. (4.55) is null, rendering negative for $t > (\hbar/2\mathcal{E}_\hbar)\ln 2$. Hence, the quantum diffusion coefficient (4.55) should be defined in the time range $0 \leq t \leq (\hbar/2\mathcal{E}_\hbar)\ln 2$.

*The quantum diffusion coefficient brought about by a thermal reservoir of quantum harmonic oscillators.* At high temperatures, Eq. (4.55) reads

$$\mathbb{D}(t) = \frac{\hbar}{2m} \frac{e^{\frac{-2k_B T t}{\hbar}} \left(2 e^{\frac{-2k_B T t}{\hbar}} - 1\right)}{\left[1 + 2 e^{\frac{-2k_B T t}{\hbar}} \left(e^{\frac{-2k_B T t}{\hbar}} - 1\right)\right]^2}, \quad T \gg \frac{\omega \hbar}{2k_B}. \quad (4.56a)$$

At room temperature, for example, environmental effects then emerge in the time interval $0 < t < 10^{-14}$ s.

On the other hand, the quantum diffusion coefficient (4.55) at $T = 0$ becomes

$$\mathbb{D}(t) = \frac{\hbar}{2m} \frac{e^{-\omega t}(2 e^{-\omega t} - 1)}{[1 + 2 e^{-\omega t}(e^{-\omega t} - 1)]^2}, \quad T = 0. \quad (4.56b)$$

At zero temperature environmental effects arise therefore owing to the angular frequency $\omega$ at the time window $0 < t < \omega^{-1}$.

*The quantum diffusion coefficient brought about by a thermal reservoir of fermions.* Using the fermionic diffusion energy (4.15) at zero temperature, Eq. (4.55) changes into

$$\mathbb{D}(t) = \frac{\hbar}{2m} \frac{e^{\frac{-6k_B T_F}{5\hbar}t} \left(2 e^{\frac{-6k_B T_F}{5\hbar}t} - 1\right)}{\left[1 + 2 e^{\frac{-6k_B T_F}{5\hbar}t} \left(e^{\frac{-6k_B T_F}{5\hbar}t} - 1\right)\right]^2}, \quad T = 0. \quad (4.57)$$

In a Fermi gas of electrons at $T_F \sim 10^4$ K, for example, environmental effects on the initial diffusion constant $\mathbb{D}(0) = \hbar/2m$ appear for $t < 10^{-16}$ s.

*The quantum diffusion coefficient brought about by a thermal reservoir of bosons.* The quantum diffusion coefficient (4.55) in a bosonic heat bath at both temperatures $T = 0$ and $T = T_{BE}$ is given by

$$\mathbb{D}(t) = \frac{\hbar}{2m} \frac{e^{\frac{-1.54 k_B T_{BE}}{\hbar}t} \left(2 e^{\frac{-1.54 k_B T_{BE}}{\hbar}t} - 1\right)}{\left[1 + 2 e^{\frac{-1.54 k_B T_{BE}}{\hbar}t} \left(e^{\frac{-1.54 k_B T_{BE}}{\hbar}t} - 1\right)\right]^2}. \quad (4.58)$$

In an ideal gas of $^4$He, for instance, at which the Boson-Einstein temperature is $T_{BE} \sim 3$ K, environmental effects show up at very short time scale $t < 10^{-10}$ s, so diminishing the initial diffusion constant $\mathbb{D}(0) = \hbar/2m$.

## 4.2. Description in terms of Langevin equations

### 4.2.1. Non-thermal systems

From the quantum Langevin equations (4.4), we can derive the following formal solutions

$$P(t) = P(0)e^{-2\gamma t} + \sqrt{4\gamma m \mathcal{E}_\hbar} \int_0^t e^{-2\gamma(t-s)} \Psi(s) ds \qquad (4.59a)$$

and

$$X(t) = X(0) + \frac{P(0)}{2\gamma m}(1 - e^{-2\gamma t}) + \sqrt{\frac{\mathcal{E}_\hbar}{\gamma m}} \int_0^t [1 - e^{-2\gamma(t-s)}] \Psi(s) ds. \qquad (4.59b)$$

Starting from the initial condition (4.6) complying with assumption (4.37), such that $\langle P^2(0) \rangle = m\mathcal{E}_\hbar$ and $\langle X^2(0) \rangle = \hbar^2/4m\mathcal{E}_\hbar$, and making use of the statistical properties (4.5), solution (4.59a) yields $\langle P(t) \rangle = 0$ and the following momentum autocorrelation function

$$\langle P(t)P(t') \rangle = m\mathcal{E}_\hbar e^{-2\gamma|t-t'|} - m\mathcal{E}_\hbar \frac{4\gamma t_c}{4\gamma t_c - 1}\left[e^{2\gamma(t-t') - \frac{t}{t_c}} - e^{-2\gamma(t+t')}\right], \qquad (4.60)$$

valid for $4\gamma t_c - 1 \neq 0$, while solution (4.59b) provides $\langle X(t) \rangle = 0$ and the displacement autocorrelation function

$$\langle X(t)X(t') \rangle = \frac{\hbar^2}{4m\mathcal{E}_\hbar} + \frac{\mathcal{E}_\hbar}{4m\gamma^2}(1 - e^{-2\gamma t})^2 + \frac{\mathcal{E}_\hbar}{\gamma m} C(t, t'), \qquad (4.61)$$

with

$$C(t, t') = t - \frac{1}{2\gamma}\left(1 + e^{-2\gamma|t-t'|} - e^{-2\gamma t'} - e^{-2\gamma t}\right) + \frac{1}{4\gamma}\left[e^{-2\gamma|t-t'|} - e^{-2\gamma(t+t')}\right]$$
$$+ t_c\left(e^{\frac{-t}{t_c}} - 1\right) + \frac{2t_c}{2\gamma t_c - 1}\left[e^{2\gamma(t-t') - \frac{t}{t_c}} - e^{-2\gamma t'}\right]$$
$$- \frac{t_c}{4\gamma t_c - 1}\left[e^{2\gamma(t-t') - \frac{t}{t_c}} - e^{-2\gamma(t+t')}\right].$$

(4.61a)

Notice that Eq. (4.61) should satisfy both conditions $2\gamma t_c - 1 \neq 0$ and $4\gamma t_c - 1 \neq 0$.

For $t = t'$, the momentum autocorrelation function (4.60) rightly yields the root mean square momentum (4.11). Yet, the displacement autocorrelation function (4.61) at $t = t'$ is

$$\mathbb{X}(t) = \sqrt{\frac{\hbar^2}{4m\mathcal{E}_\hbar} + \frac{\mathcal{E}_\hbar}{2m\gamma^2}(2\gamma t + e^{-2\gamma t} - 1) + \mathcal{N}(t)}, \tag{4.62}$$

which is different from the root mean square displacement (4.12) for $t > 0$. The function $\mathcal{N}(t)$ in Eq. (4.62) accounting for non-Markovian effects is given by the expression

$$\mathcal{N}(t) = \frac{\mathcal{E}_\hbar}{\gamma m}\left[t_c\left(e^{\frac{-t}{t_c}} - 1\right) + \frac{2t_c}{2\gamma t_c - 1}\left(e^{\frac{-t}{t_c}} - e^{-2\gamma t}\right) - \frac{t_c}{4\gamma t_c - 1}\left(e^{\frac{-t}{t_c}} - e^{-4\gamma t}\right)\right]. \tag{4.62a}$$

Furthermore, it is worth pointing out that while both quantities (4.12) and (4.11) satisfy the Heisenberg minimal fluctuation constraint (4.8), Eq. (4.62) along with Eq. (4.11) obey the relationship

$$\mathbb{X}(t)\mathbb{P}(t) > \frac{\hbar}{2}, \tag{4.63}$$

for $t > 0$. Moreover, in contrast to solution (4.14) of the quantum Fokker-Planck equation (4.2), solutions (4.59) in the Langevin description do predict that there exists the following correlation between $P(t)$ and $X(t)$

$$\langle P(t)X(t)\rangle = 2\mathcal{E}_\hbar\left[\frac{1}{2\gamma}(1 - e^{-2\gamma t}) + \frac{1}{4\gamma}(1 - e^{-4\gamma t}) + \frac{t_c}{(2\gamma t_c - 1)}\left(e^{\frac{-t}{t_c}} - e^{-2\gamma t}\right) \right.$$
$$\left. + \frac{t_c}{(4\gamma t_c - 1)}\left(e^{\frac{-t}{t_c}} - e^{-4\gamma t}\right)\right], \tag{4.64}$$

starting from the uncorrelated initial condition $\langle P(0)X(0)\rangle = 0$ toward the steady correlation $\langle P(\infty)X(\infty)\rangle = (3/2)(\mathcal{E}_\hbar/\gamma)$.

Differentiating Eq. (4.62) with respect to time leads to the instantaneous velocity $\mathbb{V}(t)$, which may be written down as

$$\mathbb{V}(t) = \frac{\mathbb{D}(t)}{\mathbb{X}(t)}, \tag{4.65}$$

where the quantum diffusion coefficient $\mathbb{D}(t)$ is given by

$$\mathbb{D}(t) = \frac{\mathcal{E}_\hbar}{2\gamma m}\left[1 - e^{-2\gamma t} - e^{\frac{-t}{t_c}} + \frac{2}{2\gamma t_c - 1}\left(2\gamma t_c e^{-2\gamma t} - e^{\frac{-t}{t_c}}\right)\right.$$
$$\left. - \frac{1}{4\gamma t_c - 1}\left(4\gamma t_c e^{-4\gamma t} - e^{\frac{-t}{t_c}}\right)\right]. \tag{4.66}$$

Note that both $\mathbb{V}(t)$ and $\mathbb{D}(t)$ above are mathematically well-defined functions for all $t \geq 0$: While $\mathbb{V}(t)$ ranges from $\mathbb{V}(0) = 0$ to $\mathbb{V}(\infty) = 0$, $\mathbb{D}(t)$ goes from $\mathbb{D}(0) = 0$ to the steady value $\mathbb{D}(\infty) = \mathcal{E}_\hbar/2\gamma m$.

Summing up, in the Langevin description of the quantum Brownian motion the position fluctuation, given by Eq. (4.62), is a differentiable quantity so implying that the quantum trajectory or path of a free particle is governed by the time-dependent diffusion coefficient $\mathbb{D}(t)$, Eq. (4.66).

We now proceed to examine in details the behavior of $\mathbb{X}(t)$, $\mathbb{V}(t)$, and $\mathbb{D}(t)$ in both short- and long-time regimes.

*The long-time regime.* For long times $t \to \infty$, i.e., $t \gg t_c, (2\gamma)^{-1}$, the quantum Brownian motion of our non-Markovian free particle (4.62) attains the diffusive behavior

$$\mathbb{X}(t) \sim \sqrt{\frac{\mathcal{E}_\hbar t}{\gamma m}}, \quad t \to \infty, \tag{4.67}$$

while its instantaneous velocity (4.65) tends to

$$\mathbb{V}(t) \sim \sqrt{\frac{\mathcal{E}_\hbar}{4\gamma m t}}, \quad t \to \infty. \tag{4.68}$$

Both quantum Brownian trajectories (4.67) and (4.68) are characterized by the stationary diffusion coefficient

$$\mathbb{D}(\infty) = \frac{\mathcal{E}_\hbar}{2\gamma m}, \quad t \to \infty. \tag{4.69}$$

Equations (4.67), (4.68), and (4.69) depend on environmental features through the quantum diffusion energy $\mathcal{E}_\hbar$ and the coupling constant $\gamma$. If $\mathcal{E}_\hbar = \hbar/2t_c$, for example, the diffusive regime (4.67), (4.68), and (4.69) has the following explicitly non-Markovian nature

$$\mathbb{X}(t) \sim \sqrt{\frac{\hbar t}{2\gamma m t_c}}, \qquad t \to \infty, \tag{4.70a}$$

$$\mathbb{V}(t) \sim \sqrt{\frac{\hbar}{8 t_c \gamma m t}}, \qquad t \to \infty, \tag{4.71a}$$

and

$$\mathbb{D}(\infty) = \frac{\hbar}{4\gamma m t_c}, \qquad t \to \infty. \tag{4.72a}$$

Yet, assuming $\mathcal{E}_\hbar$ to be given by the non-thermal energy $\mathcal{E}_\hbar = \gamma \hbar$, we obtain the Markovian diffusive behavior

$$\mathbb{X}(t) \sim \sqrt{\frac{\hbar t}{m}}, \qquad t \to \infty, \tag{4.70b}$$

$$\mathbb{V}(t) \sim \sqrt{\frac{\hbar}{4mt}}, \qquad t \to \infty, \tag{4.71b}$$

and

$$\mathbb{D}(\infty) = \frac{\hbar}{2m}, \qquad t \to \infty. \tag{4.72b}$$

Inserting expression (4.72b) into Eq. (4.70b), the diffusive regime of our quantum free Brownian particle turns out to be written down as $\mathbb{X}(t) \sim \sqrt{2\mathbb{D}(\infty)t}$ which is formally identical with Einstein's law of diffusion of a classical Brownian particle $\mathbb{X}(t) \sim \sqrt{2D(\infty)t}$. Yet, in contrast to the classical diffusion constant $D(\infty) = 2k_B T/m\gamma$, the quantum diffusion constant $\mathbb{D}(\infty) = \hbar/2m$ is independent of the properties of the environment. The Markovian diffusive regime (4.70b), (4.71b), and (4.72b) therefore stands out as a universal feature inherent in the quantum Brownian free motion, without any classical correspondent.

*The short-time regime.* In the short-time regime $t \to 0$, i.e., $t \ll t_c, (2\gamma)^{-1}$, such that $t^3 \ll 1$, we obtain from Eq. (4.62) the non-ballistic behavior

$$\mathbb{X}(t) \sim \sqrt{\frac{\hbar^2}{4m\mathcal{E}_\hbar} + \frac{\mathcal{E}_\hbar}{m} t^2}, \qquad t \to 0, \qquad (4.73)$$

while the instantaneous velocity (4.65) turns out to be

$$\mathbb{V}(t) \sim \left(\frac{\hbar^2}{4m\mathcal{E}_\hbar} + \frac{\mathcal{E}_\hbar}{m} t^2\right)^{-1/2} \frac{\mathcal{E}_\hbar}{m} t, \qquad t \to 0. \qquad (4.74)$$

For times $t \ll \hbar/2\mathcal{E}_\hbar$, the fluctuation (4.73) becomes a constant given by $\mathbb{X}(t) \sim \sqrt{\hbar^2/4m\mathcal{E}_\hbar}$ and the instantaneous velocity (4.74) renders linear in time: $\mathbb{V}(t) \sim t\sqrt{4\mathcal{E}_\hbar^3/m\hbar^2}$. On the other hand, for times $t \gg \hbar/2\mathcal{E}_\hbar$ the non-ballistic behavior (4.73) turns out to be ballistic

$$\mathbb{X}(t) \sim t\sqrt{\frac{\mathcal{E}_\hbar}{m}}, \qquad t \to 0, \qquad (4.73a)$$

and the instantaneous velocity (4.74) approaches a constant

$$\mathbb{V}(t) \sim \sqrt{\frac{\mathcal{E}_\hbar}{m}}, \qquad t \to 0. \qquad (4.74b)$$

In both non-ballistic and ballistic regimes the diffusion coefficient is linear in time

$$\mathbb{D}(t) \sim \frac{\mathcal{E}_\hbar}{m} t, \qquad t \to 0. \qquad (4.75)$$

If $\mathcal{E}_\hbar \propto \gamma \hbar$, then the upshoots (4.73), (4.74), and (4.75) are said to be Markovian effects and reliant on the environment via the friction constant $\gamma$. Otherwise, if $\mathcal{E}_\hbar \propto \hbar/t_c$, then Eqs. (4.73), (4.74), and (4.75) render non-Markovian.

In the classical limit $\hbar \to 0$, i.e., as $\mathcal{E}_\hbar \to \mathcal{E}$, Eqs. (4.73a) and (4.74b) read $\mathbb{X}(t) \sim t\sqrt{\mathcal{E}/m}$ and $\mathbb{V}(t) \sim \sqrt{\mathcal{E}/m}$, respectively. The quantum diffusion coefficient (4.75) in turn becomes $\mathbb{D}(t) \sim (\mathcal{E}/m)t$.

### 4.2.2. Thermal systems

*A free Brownian particle in a thermal reservoir of quantum harmonic oscillators.* The quantum Brownian motion of a free particle in the presence of a heat bath of harmonic oscillators is described for all times by the Markovian properties (4.62), (4.65), and (4.66), with $\mathcal{E}_\hbar = (\omega\hbar/2)\coth(\omega\hbar/2k_B T)$. As far as the long-time scale is concerned the diffusive regime at zero temperature, that is, Eqs. (4.67), (4.68), and (4.69) with $\mathcal{E}_\hbar = \omega\hbar/2$, is characterized by

$$\mathbb{X}(t) \sim \sqrt{\frac{\omega\hbar}{2\gamma m}}\, t, \quad t \to \infty, \quad T = 0, \quad (4.76a)$$

$$\mathbb{V}(t) \sim \sqrt{\frac{\omega\hbar}{8\gamma m t}}, \quad t \to \infty, \quad T = 0, \quad (4.77a)$$

and

$$\mathbb{D}(\infty) = \frac{\hbar\omega}{4\gamma m}, \quad t \to \infty, \quad T = 0. \quad (4.78a)$$

On the other hand, at short time scales the free particle attains the non-ballistic regime determined by Eqs. (4.73) and (4.74) that at zero temperature read, respectively,

$$\mathbb{X}(t) \sim \sqrt{\frac{\hbar}{2m\omega} + \frac{\omega\hbar}{2m} t^2}, \quad t \to 0, T = 0, \quad (4.76b)$$

$$\mathbb{V}(t) \sim \left(\frac{\hbar}{2m\omega} + \frac{\omega\hbar}{2m} t^2\right)^{-1/2} \frac{\omega\hbar}{2m} t, \quad t \to 0, T = 0, \quad (4.77b)$$

The non-ballistic regime (4.76b) and (4.77b) for times $t \ll \omega^{-1}$ turns out to be given by $\mathbb{X}(t) \sim \sqrt{\hbar/2m\omega}$ and $\mathbb{V}(t) \sim t\sqrt{\hbar\omega^3/2m}$. It becomes ballistic for $t \gg \omega^{-1}$, i.e.,

$$\mathbb{X}(t) \sim t\sqrt{\frac{\omega\hbar}{2m}}, \quad t \to 0, T = 0, \quad (4.76c)$$

with constant instantaneous velocity

$$\mathbb{V}(t) \sim \sqrt{\frac{\omega\hbar}{2m}}, \quad t \to 0, T = 0. \quad (4.77c)$$

Both non-ballistic and ballistic regimes given above are characterized by the time-dependent quantum diffusion coefficient

$$\mathbb{D}(t) \sim \frac{\omega \hbar}{2m} t, \qquad t \to 0, T = 0. \tag{4.78b}$$

Moreover, notice that at short times Eqs. (4.76b, c), (4.77b, c), and (4.78b) are independent of frictional parameter, in contrast to Eqs. (4.76a), (4.77a), and (4.78a) at long times.

In the case of a Brownian particle of mass $m \sim 10^{-15}$ kg in a heat bath of oscillators with $\omega \sim 10^{15}$ s$^{-1}$, the non-ballistic regime (4.76b), (4.77b), and (4.78b) at $t \sim 10^{-15}$ s provides: $\mathbb{X}(t) \sim 10^{-17}$ m, $\mathbb{V}(t) \sim 10^{-3}$ m s$^{-1}$, and $\mathbb{D}(t) \sim 10^{-20}$ m$^2$ s$^{-1}$. At $t \sim 10^{-10}$ s and $\omega \sim 10^{15}$ s$^{-1}$, the ballistic regime (4.76c), (4.77c), and (4.78b) yields: $\mathbb{X}(t) \sim 10^{-12}$ m, $\mathbb{V}(t) \sim 10^{-2}$ m s$^{-1}$, and $\mathbb{D}(\infty) \sim 10^{-14}$ m$^2$ s$^{-1}$. At long times $t \sim 10$ s and $\gamma \sim 10^{11}$ s$^{-1}$, the diffusive regime (4.76a), (4.77a), and (4.78a) provides $\mathbb{X}(t) \sim 10^{-7}$ m, $\mathbb{V}(t) \sim 10^{-9}$ m s$^{-1}$, and $\mathbb{D}(\infty) \sim 10^{-16}$ m$^2$ s$^{-1}$, respectively.

*A free Brownian particle in a thermal reservoir of fermions.* At long times a free Brownian particle immersed in a heat bath of $N$ fermions at low temperatures is described by Eqs. (4.67), (4.68), and (4.69), the diffusion energy being $\mathcal{E}_\hbar \sim (3/5 k_B T_F \left[1 + 5/12 \pi^2 T/T_F\right]^2$. Accordingly, at zero temperature we obtain the position fluctuation

$$\mathbb{X}(t) \sim \sqrt{\frac{3t}{5\gamma m} k_B T_F}, \qquad t \to \infty, T = 0, \tag{4.79a}$$

the instantaneous velocity

$$\mathbb{V}(t) \sim \sqrt{\frac{3}{20\gamma m t} k_B T_F}, \qquad t \to \infty, T = 0, \tag{4.80a}$$

and the steady diffusion coefficient

$$\mathbb{D}(\infty) = \frac{3}{10\gamma m} k_B T_F, \qquad t \to \infty, T = 0. \tag{4.81a}$$

At short time scales $t^3 \ll 1$ and zero temperature, we obtain from Eqs. (4.73) and (4.74) the non-ballistic regime

$$\mathbb{X}(t) \sim \sqrt{\frac{5 \hbar^2}{12 m k_B T_F} + \frac{k_B T_F}{m} t^2}, \qquad t \to 0, T = 0, \tag{4.79b}$$

and

$$\mathbb{V}(t) \sim \left(\frac{5\hbar^2}{12mk_BT_F} + \frac{k_BT_F}{m}t^2\right)^{-1/2} \frac{k_BT_F}{m}t, \qquad t \to 0, T = 0. \qquad (4.80b)$$

At very short times $t \ll (\hbar/k_BT_F)\sqrt{5/12}$, Eq. (4.79b) reduces to the constant $\mathbb{X}(t) \sim \sqrt{5\hbar^2/12mk_BT_F}$ and Eq. (4.80b) to $\mathbb{V}(t) \sim t\sqrt{12(k_BT_F)^3/5m\hbar^2}$. In the case of ballistic behavior, $t \gg (\hbar/k_BT_F)\sqrt{5/12}$, we obtain from Eqs. (4.79b) and (4.80b) respectively

$$\mathbb{X}(t) \sim t\sqrt{\frac{k_BT_F}{m}}, \qquad t \to 0, T = 0, \qquad (4.79c)$$

and

$$\mathbb{V}(t) \sim \sqrt{\frac{k_BT_F}{m}}, \qquad t \to 0, T = 0. \qquad (4.80c)$$

In both ballistic and non-ballistic regimes at short times, the time-dependent diffusion coefficient is given by the same expression:

$$\mathbb{D}(t) \sim \frac{k_BT_F}{m}t, \qquad t \to 0, T = 0. \qquad (4.81b)$$

For a Brownian particle of mass $m \sim 10^{-15}$ kg in a Fermi gas of electrons at $T_F \sim 10^4$ K, at short times the non-ballistic regime occurs at $t \sim 10^{-16}$ s and the ballistic one at $t \gg 10^{-16}$ s. So, at $t \sim 10^{-16}$ s we obtain the non-ballistic behavior $\mathbb{X}(t) \sim 10^{-18}$ m, $\mathbb{V}(t) \sim 10^{-3}$ m s$^{-1}$, and $\mathbb{D}(t) \sim 10^{-21}$ m$^2$ s$^{-1}$, while at $t \sim 10^{-10}$ s we find the ballistic one $\mathbb{X}(t) \sim 10^{-12}$ m, $\mathbb{V}(t) \sim 10^{-2}$ m s$^{-1}$, and $\mathbb{D}(t) \sim 10^{-14}$ m$^2$ s$^{-1}$. On the other hand, at $t \sim 10$ s the particle reaches the diffusive regime characterized by $\mathbb{X}(t) \sim 10^{-8}$ m, $\mathbb{V}(t) \sim 10^{-9}$ m s$^{-1}$, and $\mathbb{D}(t) \sim 10^{-17}$ m$^2$ s$^{-1}$. Here, we have used $\gamma \sim 10^{11}$ s$^{-1}$.

*A free Brownian particle in a thermal reservoir of bosons.* Quantities (4.62), (4.65), and (4.66), with $\mathcal{E}_\hbar = 0.77k_B\left[\sqrt{T^5/T_{BE}^3} + T_{BE}\sqrt{1-(T/T_{BE})^3}\right]$, describe a free Brownian particle in a bosonic heat bath at and below the Bose-Einstein temperature. At $T = 0$ and $T = T_{BE}$, specifically, we obtain at long times

$$\mathbb{X}(t) \sim \sqrt{\frac{0.77 k_B T_{BE}}{\gamma m} t}, \quad t \to \infty, T = 0 \text{ and } T = T_{BE}, \qquad (4.82a)$$

$$\mathbb{V}(t) \sim \sqrt{\frac{0.77 k_B T_{BE}}{4 \gamma m t}}, \quad t \to \infty, T = 0 \text{ and } T = T_{BE}, \qquad (4.83a)$$

and

$$\mathbb{D}(\infty) \sim \frac{0.77 k_B T_{BE}}{2 \gamma m}, \quad t \to \infty, T = 0 \text{ and } T = T_{BE}, \qquad (4.84a)$$

while at short time scales $t \to 0$ we have the non-ballistic regime

$$\mathbb{X}(t) \sim \sqrt{\frac{\hbar^2}{3.08 m k_B T_{BE}} + \frac{0.77 k_B T_{BE}}{m} t^2}, \quad t \to 0, T = 0, T = T_{BE}, \qquad (4.82b)$$

$$\mathbb{V}(t) \sim \left( \frac{\hbar^2}{3.08 m k_B T_{BE}} + \frac{0.77 k_B T_{BE}}{m} t^2 \right)^{-1/2} \frac{0.77 k_B T_{BE}}{m} t, \quad t \to 0, T = 0, T = T_{BE}. $$

$$(4.83b)$$

At very short times $t \ll \hbar/1.54 k_B T_{BE}$, Eqs. (4.82b) and (4.83b) become $\mathbb{X}(t) \sim \sqrt{\hbar^2/3.08 m k_B T_{BE}}$ and $\mathbb{V}(t) \sim 1.35 t \sqrt{(k_B T_{BE})^3/\hbar^2 m}$, respectively. For $t \gg \hbar/1.54 k_B T_{BE}$, the following ballistic regime is attained

$$\mathbb{X}(t) \sim t \sqrt{\frac{0.77 k_B T_{BE}}{m}}, \quad t \to 0, T = 0, T = T_{BE}, \qquad (4.82c)$$

$$\mathbb{V}(t) \sim \sqrt{\frac{0.77 k_B T_{BE}}{m}}, \quad t \to 0, T = 0, T = T_{BE}. \qquad (4.83c)$$

Both ballistic and non-ballistic regimes are governed by the time-dependent diffusion coefficient

$$\mathbb{D}(t) \sim \frac{0.77 k_B T_{BE}}{m} t, \quad t \to 0, T = 0, T = T_{BE}. \qquad (4.84b)$$

In the case of a particle of mass $m \sim 10^{-15}$ kg in an ideal gas of $^4$He at $T_{BE} \sim 3$ K, at short time scales $t \ll t_c, (2\gamma)^{-1}$ the non-ballistic regime occurs for times of the order of $10^{-12}$ s, whereas the ballistic one is found to arise at times $t \gg 10^{-12}$ s. At $t \sim 10^{-12}$ s, we obtain $\mathbb{X}(t) \sim 10^{-16}$ m, $\mathbb{V}(t) \sim 10^{-4}$ m s$^{-1}$, and $\mathbb{D}(t) \sim 10^{-20}$ m$^2$ s$^{-1}$. At $t \sim 10^{-6}$ s, we find $\mathbb{X}(t) \sim 10^{-10}$ m, $\mathbb{V}(t) \sim 10^{-4}$ m s$^{-1}$, and $\mathbb{D}(t) \sim 10^{-14}$ m$^2$ s$^{-1}$. At $\sim 100$ s, we have the diffusive regime: $\mathbb{X}(t) \sim 10^{-9}$ m, $\mathbb{V}(t) \sim 10^{-11}$ m s$^{-1}$, and $\mathbb{D}(t) \sim 10^{-20}$ m$^2$ s$^{-1}$.

*A free Brownian particle in quantum thermal reservoirs at high temperatures.* At high temperatures $T \to \infty$, i.e., $T \gg T_F, T_{BE}, \omega\hbar/2k_B$, the diffusion energy of the three quantum thermal systems renders $\hbar$-independent: $\mathcal{E}_\hbar \to \mathcal{E} \propto k_B T$. Accordingly, we obtain from Eqs. (4.67), (4.68), and (4.69) the classical Brownian motion in the diffusive regime

$$\mathbb{X}(t) \sim \sqrt{\frac{k_B T t}{\gamma m}}, \quad t \to \infty, \quad T \to \infty, \qquad (4.84a)$$

$$\mathbb{V}(t) \sim \sqrt{\frac{k_B T}{4\gamma m t}}, \quad t \to \infty, \quad T \to \infty, \qquad (4.85a)$$

and

$$\mathbb{D}(\infty) \sim \frac{k_B T}{2\gamma m}, \quad t \to \infty, \quad T \to \infty. \qquad (4.86a)$$

Therefore, a quantum free Brownian particle at high temperatures behaves classically in the diffusive regime, that is, without any influence of Planck's constant on its movement. However, at short time scales $t \to 0$, such that $t^3 \ll 1$, we find the following quantum effects on $\mathbb{X}(t)$ and $\mathbb{V}(t)$ in the non-ballistic regime

$$\mathbb{X}(t) \sim \sqrt{\frac{\hbar^2}{4m k_B T} + \frac{k_B T}{m} t^2}, \quad t \to 0, \quad T \to \infty, \qquad (4.84b)$$

and

$$\mathbb{V}(t) \sim \left(\frac{\hbar^2}{4m k_B T} + \frac{k_B T}{m} t^2\right)^{-1/2} \frac{k_B T}{m} t, \quad t \to 0, \quad T \to \infty, \qquad (4.85b)$$

although the diffusion coefficient $\mathbb{D}(t) = \mathbb{X}(t)\mathbb{V}(t)$ behaves classically

$$\mathbb{D}(t) \sim \frac{k_B T}{m} t, \qquad t \to 0, \ T \to \infty. \tag{4.86b}$$

In the classical limit $\hbar \to 0$, physically interpreted as $t \gg \hbar/2k_B T$, the non-ballistic quantum behavior (4.84b) becomes ballistic, $\mathbb{X}(t) \sim t\sqrt{k_B T/m}$, while the quantum instantaneous velocity (4.85b) tends to the time-independent instantaneous velocity $\mathbb{V}(t) \sim \sqrt{k_B T/m}$, whose measuring has recently borne out the energy equipartition theorem of statistical mechanics [232].

Moreover, it is predicted that the quantum instantaneous velocity (4.85b) can break down the equipartition theorem at quantum time scales $t < \hbar/2k_B T$. At room temperature, for example, in Eq. (4.84b) the classical ballistic regime turns to be valid for times $t \gtrsim 10^{-14}$ s while the quantum non-ballistic one comes about at times $t < 10^{-14}$ s. At very short time scales $t \ll 10^{-14}$, Eq. (4.84a) predicts a non-zero displacement fluctuation $\mathbb{X}(t) \sim 10^{-17}$ m, albeit both $\mathbb{V}(t)$ and $\mathbb{D}(t)$ virtually vanish.

By way of example, let a particle of mass $m \sim 10^{-15}$ kg be immersed in a thermal bath at room temperature $T \sim 10^2$ K, the quantum non-ballistic regime (4.84b), (4.85b), and (4.86b) at short times $t \sim 10^{-17}$ s yields the following approximate numerical values: $\mathbb{X}(t) \sim 10^{-17}$ m, $\mathbb{V}(t) \sim 10^{-6}$ m s$^{-1}$, and $\mathbb{D}(t) \sim 10^{-23}$ m$^2$ s$^{-1}$. Yet, the same Eqs. (4.84b), (4.85b), and (4.86b) at $t \sim 10^{-6}$ s provide the classical ballistic regime characterized by $\mathbb{X}(t) \sim 10^{-9}$ m, $\mathbb{V}(t) \sim 10^{-3}$ m s$^{-1}$, and $\mathbb{D}(t) \sim 10^{-12}$ m$^2$ s$^{-1}$.

## 4.3. Summary

We have described the quantum Brownian motion of a free particle in terms of both Fokker-Planck and Langevin equations. Both descriptions start from the same initial condition (4.6) satisfying the Heisenberg minimum relation $\mathbb{X}(0)\mathbb{P}(0) = \hbar/2$. Nevertheless, the time-evolution of the Wigner function via the quantum Fokker-Planck equation (4.2) is constrained to the Heisenberg minimum relation (4.8) for all times $t > 0$, whereas the description based on the quantum Langevin equation (4.4) obeys the Heisenberg relation in its more general form given by Eq. (4.63).

On the one hand, the Fokker-Planck description of quantum Brownian free motion does feature the following upshots:

(a) The state of a quantum Brownian free particle is described by the time-dependent Wigner function (4.14), solution of the quantum non-Markovian Fokker-Planck equation (4.2). The consequences of its steady solution (4.15) have been investigated at both high and zero temperatures, as expected. The energy equipartition theorem of statistical mechanics is obeyed by the quantum free particle in the high-temperature regime, whereas it is violated at zero temperature. Such a prediction could be verified for both bosonic and fermionic heat baths as well as in the case of a thermal reservoir of quantum harmonic oscillators.

(b) The differentiability property of both fluctuations $\mathbb{X}(t)$ and $\mathbb{P}(t)$ at $t = 0$ leads respectively to the concepts of instantaneous velocity, $\mathbb{V}(t) = d\mathbb{X}(t)/dt$, and of environmental force, $\mathbb{F}(t) = d\mathbb{P}(t)/dt$, which are valid for all $t \geq 0$. In addition, multiplying $\mathbb{X}(t)$ by $\mathbb{V}(t)$ gives rise to the concept of quantum diffusion coefficient, i.e., $\mathbb{D}(t) = \mathbb{X}(t)\mathbb{V}(t)$.

(c) For thermal systems, both the quantum force and the instantaneous velocity can measure deviations from the energy equipartition theorem at high and low temperatures, including the $T = 0$ case. For example, measuring the thermal force (4.46a) brought about by a thermal heat bath of quantum harmonic oscillators at high temperatures can indicate deviations from the equipartition theorem at times of the order of the quantum time $t_q = \hbar/2k_B T$. This conclusion can be also reached by measuring the quantum instantaneous velocity at high-temperatures via Eq. (4.52a).

On the other hand, the Langevin description of quantum Brownian free motion does feature the following upshots:

(i) The quantum diffusive regime as $t \to \infty$ is governed by the displacement fluctuation $\mathbb{X}(t) \sim \sqrt{2\mathbb{D}(\infty)t}$ and the instantaneous velocity $\mathbb{V}(t) \sim \sqrt{\mathbb{D}(\infty)/2t}$, where $\mathbb{D}(\infty)$ is the stationary quantum diffusion constant $\mathbb{D}(\infty) \sim \mathcal{E}_\hbar/2\gamma m$. For non-thermal systems in which $\mathcal{E}_\hbar \propto \gamma\hbar$, for instance, the diffusive regime, given by Eqs. (4.70b) and (4.71b), reveals to be a universal Markovian property valid for any environments. Moreover, for thermal environments at high temperatures, whereby $\mathcal{E}_\hbar \to \mathcal{E} = k_B T$, the classical Markovian diffusive regime is regained: $\mathbb{X}(t) \sim \sqrt{2D(\infty)t}$, $\mathbb{V}(t) \sim \sqrt{D(\infty)/2t}$, with $D(\infty) \sim \mathcal{E}_\hbar/2\gamma m$.

(ii) At short times, it is predicted that the non-ballistic regime is given by Eq. (4.73) that in turn becomes ballistic at times $t \gg \hbar/2\mathcal{E}_\hbar$. For thermal systems at high temperatures, the classical ballistic regime is obtained, leading to the validity of the energy equipartition theorem. Yet, the quantum ballistic regime at low temperatures causes the violation of such a theorem. The same conclusion is reached starting from the non-ballistic regime for thermal environments at low temperatures.

# 5. Quantum Smoluchowski equation

In the classical domain the Brownian movement undergone by a particle in the absence of inertial force is described by the Langevin equation (2.61) and its corresponding non-Gaussian Kolmogorov equation in configuration space (2.62). In Sect. 5.1 we show how quantum effects on the non-inertial Brownian motion can be studied by quantizing directly the equation of motion (2.62). In the Gaussian approximation a non-Markovian quantum Smoluchowski equation in phase space is derived. Next, our quantum Smoluchowski equation is solved for the cases of a free particle and a harmonic oscillator in Sects. 5.2 and 5.3, respectively, for both thermal and non-thermal environments. The thermal systems (heat bath of quantum harmonic oscillators, of fermions, and of bosons) are treated at zero and high temperatures. Lastly, a discussion on some quantum Smoluchowski equations existing in the literature is presented in Sect. 5.4.

## 5.1. Quantizing the Kolmogorov equation in configuration space

In this section we wish to show how quantum effects on the non-inertial Brownian motion can be studied by quantizing directly the equation of motion (2.62). To this end, let $\chi_1 = \chi(x_1, t)$ and $\chi_2 = \chi(x_2, t)$ be two distinct solutions of Eq. (2.62) given respectively at points $x_1$ and at $x_2$

$$\frac{\partial \chi(x_1, t)}{\partial t} = \sum_{k=1}^{\infty} \frac{(-1)^k}{k!} \frac{\partial^k}{\partial x_1^k} [\overline{A}_k(x_1, t) \chi(x_1, t)], \qquad (5.1a)$$

$$\frac{\partial \chi(x_2, t)}{\partial t} = \sum_{i=1}^{\infty} \frac{(-1)^k}{k!} \frac{\partial^k}{\partial x_2^k} [\overline{A}_k(x_2, t) \chi(x_2, t)], \qquad (5.1b)$$

where $\overline{A}_k(x_1, t)$ and $\overline{A}_k(x_2, t)$ are coefficients associated with solutions $\chi(x_1, t)$ and $\chi(x_2, t)$, respectively. Multiplying (5.1a) and (5.1b) by $\chi(x_2, t)$ and $\chi(x_1, t)$, respectively, and then adding the resulting equations we arrive at

$$\frac{\partial \xi(x_1, x_2, t)}{\partial t} = \sum_{k=1}^{\infty} \frac{(-1)^k}{k!} \left[ \frac{\partial^k}{\partial x_1^k} \overline{A}_k(x_1, t) + \frac{\partial^k}{\partial x_2^k} \overline{A}_k(x_2, t) \right] \xi(x_1, x_2, t), \qquad (5.2)$$

where $\xi(x_1, x_2, t) = \chi(x_1, t) \chi(x_2, t)$. By quantizing via quantization conditions (3.3) and making use of the relations

$$\frac{\partial}{\partial x_1} = \frac{1}{2}\frac{\partial}{\partial x} - \frac{1}{\hbar}\frac{\partial}{\partial \eta} \quad (5.3a)$$

and

$$\frac{\partial}{\partial x_2} = \frac{1}{2}\frac{\partial}{\partial x} + \frac{1}{\hbar}\frac{\partial}{\partial \eta}, \quad (5.3b)$$

the classical equation (5.2) becomes the quantum Kolmogorov equation in configuration space $(x, \hbar\eta)$, where $x = (x_1 + x_2)/2$ and $\hbar\eta = x_2 - x_1$,

$$\frac{\partial \rho(x,\eta,t)}{\partial t} = \mathbb{L}_\hbar\, \rho(x,\eta,t) \quad (5.4)$$

the $\hbar$-dependent operator $\mathbb{L}_\hbar$ being given by

$$\mathbb{L}_\hbar = \sum_{k=1}^{\infty} \frac{(-1)^k}{k!}\left[\left(\frac{\partial}{2\partial x} - \frac{1}{\hbar}\frac{\partial}{\partial \eta}\right)^k A_k^{(\hbar)} + \left(\frac{\partial}{2\partial x} + \frac{1}{\hbar}\frac{\partial}{\partial \eta}\right)^k A_k^{(\hbar)}\left(x + \frac{\eta\hbar}{2}, t\right)\right]. \quad (5.4a)$$

## 5.1.1. Quantum Smoluchowski equation in phase space

The quantum Kolmogorov equation (5.4) exhibits non-Gaussian features in full. Yet, as far as the Gaussian approximation $|x_2 - x_1|^3 \ll 0$, or $|\eta|^3 \ll 0$, is concerned it reduces to the following quantum master equation

$$\frac{\partial \rho}{\partial t} = \frac{1}{m\gamma}\left[\frac{dV(x)}{dx} + \frac{1}{2}\left(\frac{\eta\hbar}{2}\right)^2 \frac{d^3V(x)}{dx^3}\right]\frac{\partial \rho}{\partial x} + \frac{\eta}{m\gamma}\frac{d^2V(x)}{dx^2}\frac{\partial \rho}{\partial \eta}$$
$$+ \frac{2}{m\gamma}\left[\frac{d^2V(x)}{dx^2} + \frac{1}{2}\left(\frac{\eta\hbar}{2}\right)^2 \frac{d^4V(x)}{dx^4}\right]\rho + \frac{\mathcal{E}_\hbar I(t)}{m\gamma}\left[\frac{\partial^2 \rho}{\partial x^2} + \frac{4}{\hbar^2}\frac{\partial^2 \rho}{\partial \eta^2}\right] \quad (5.5)$$

where we have used $\rho \equiv \rho(x, \hbar\eta, t)$, $\beta \equiv \gamma$ and assumed $\langle \Psi(t) \rangle = 0$. The quantum equation of motion (5.5) does correspond to the quantization of the non-Markovian Smoluchowski equation (2.66) (see Appendix H). Hence, we call it the quantum Smoluchowski equation in configuration space $(x, \hbar\eta)$. Its solution $\rho(x, \hbar\eta, t)$ can be viewed as a density matrix whose elements are given by the coordinates $x = (x_1 + x_2)/2$ and $\hbar\eta = x_2 - x_1$. The diagonal elements $x_1 = x_2$ follow the time evolution

$$\frac{\partial \rho(x,t)}{\partial t} = \frac{1}{m\gamma}\frac{\partial}{\partial x}\left[\frac{dV(x)}{dx}\rho(x,t)\right] + \frac{\mathcal{E}_\hbar I(t)}{m\gamma}\frac{\partial^2 \rho(x,t)}{\partial x^2} + \frac{1}{m\gamma}\frac{d^2V(x)}{dx^2}\rho(x,t). \quad (5.5a)$$

It is not evident that both equations (5.5) are compatible with the Heisenberg fluctuation relationship. Hence, we resort to the phase-space representation of quantum mechanics upon performing the Fourier transform (the Wigner function)

$$W(x,p,t) = \frac{1}{2\pi} \int_{-\infty}^{\infty} \rho(x,\eta,t) e^{ip\eta} d\eta \qquad (5.6)$$

on Eq. (5.5). In so doing, we arrive at the quantum Smoluchowski equation in phase space

$$\frac{\partial W(x,p,t)}{\partial t} = \frac{1}{m\gamma} \frac{dV(x)}{dx} \frac{\partial W(x,p,t)}{\partial x} - \frac{p}{m\gamma} \frac{d^2V(x)}{dx^2} \frac{\partial W(x,p,t)}{\partial p}$$
$$+ \frac{\mathcal{E}_\hbar I(t)}{m\gamma} \frac{\partial^2 W(x,p,t)}{\partial x^2} - \frac{\hbar^2}{8m\gamma} \frac{d^3V(x)}{dx^3} \frac{\partial^3 W(x,p,t)}{\partial x \partial p^2}$$
$$- \frac{\hbar^2}{4m\gamma} \frac{d^4V(x)}{dx^4} \frac{\partial^2 W(x,p,t)}{\partial p^2} + \left[ \frac{1}{m\gamma} \frac{d^2V(x)}{dx^2} - \frac{4\mathcal{E}_\hbar I(t)}{m\gamma\hbar^2} p^2 \right] W(x,p,t).$$
$$(5.7)$$

For harmonic potentials the terms proportional to $\hbar^2$ in both equations of motion (5.5) and (5.7) do not occur, while in case of anharmonic potentials they result from higher-than-second-order derivatives in $\partial V(x \pm \eta\hbar/2)/\partial \eta$ (see Appendix H).

*Initial condition.* Our quantum phase-space Smoluchowski equation (5.7) may be solved starting from the non-thermal initial condition

$$W(x,p,t=0) = \frac{1}{\pi\hbar} e^{-\left(\frac{ap^2}{m\hbar} + \frac{mx^2}{\hbar a}\right)} \qquad (5.8)$$

leading to the distribution $\mathcal{F}(x,p,t=0) = \delta(x)\delta(p)$ in the classical limit $\hbar \to 0$. The constant $a$ has dimensions of time and is assumed to be $\hbar$-independent. The Gaussian probability distribution function (5.8) fulfills the Heisenberg minimum fluctuation relationship $\mathbb{X}(0)\mathbb{P}(0) = \sqrt{\langle P^2(0)\rangle\langle X^2(0)\rangle} = \hbar/2$.

*Classical limit.* Taking into account the condition

$$p = 0, \qquad (5.9)$$

which implies from the Wigner transform (5.6) that $W(x,p=0,t) \equiv W(x,t)$, we obtain from Eq. (5.7) the following equation of motion

$$\frac{\partial W(x,t)}{\partial t} = \frac{1}{m\gamma}\frac{dV(x)}{dx}\frac{\partial W(x,t)}{\partial x} + \frac{1}{m\gamma}\frac{d^2V(x)}{dx^2}W(x,t) + \frac{\mathcal{E}_\hbar I(t)}{m\gamma}\frac{\partial^2 W(x,t)}{\partial x^2}, \quad (5.10)$$

and next taking the classical limit $\hbar \to 0$, such that

$$\lim_{\hbar \to 0} \mathcal{E}_\hbar = \mathcal{E} \quad (5.11a)$$

and

$$\lim_{\hbar \to 0} W(x,t) = \mathcal{F}(x,t), \quad (5.11b)$$

the quantum Smoluchowski equation in phase space (5.7) reduces to the classical Smoluchowski equation in configuration space (2.66):

$$\frac{\partial \mathcal{F}(x,t)}{\partial t} = \frac{1}{m\gamma}\frac{dV(x)}{dx}\frac{\partial \mathcal{F}(x,t)}{\partial x} + \frac{1}{m\gamma}\frac{d^2V(x)}{dx^2}\mathcal{F}(x,t) + \frac{\mathcal{E}}{m\gamma}I(t)\frac{\partial^2 \mathcal{F}(x,t)}{\partial x^2}. \quad (5.12)$$

It is worth noting that the quantum Smoluchowski equation in configuration space (5.10) violates the Heisenberg relation since the momentum displays no fluctuations in view of the condition (5.9). This result suggests that phase space, and not configuration space, is the suitable locus to investigate quantum Brownian motion in the absence of inertial force.

## 5.2. Free particle

### 5.2.1. Non-thermal systems

In the case of a free particle the quantum master equation (5.7) reads

$$\frac{\partial W(x,p,t)}{\partial t} = \frac{\mathcal{E}_\hbar}{m\gamma}\left(1 - e^{\frac{-t}{t_c}}\right)\frac{\partial^2 W(x,p,t)}{\partial x^2} - \frac{4\mathcal{E}_\hbar}{m\gamma\hbar^2}\left(1 - e^{\frac{-t}{t_c}}\right)p^2 W(x,p,t), \quad (5.13)$$

where we have used $I(t) = 1 - e^{-t/t_c}$. Starting from the initial condition (5.8) the time-dependent solution of Eq. (5.13) reads

$$W(x,p,t) = \frac{1}{\pi\hbar}e^{-\left[\frac{4b_\hbar(t)}{\hbar^2}p^2 + \frac{x^2}{4b_\hbar(t)}\right]} \quad (5.14)$$

where the function $b_\hbar(t)$ is given by

$$b_\hbar(t) = \frac{\hbar a}{4m} + \frac{\mathcal{E}_\hbar}{m\gamma}\left[t + t_c\left(e^{\frac{-t}{t_c}} - 1\right)\right]. \quad (5.14a)$$

Solution (5.14) leads to $\langle X(t) \rangle = \langle P(t) \rangle = 0$, $\langle X^2(t) \rangle = 2b_\hbar(t)$, and $\langle P^2(t) \rangle = \hbar^2/8b_\hbar(t)$. Accordingly, we obtain

$$\mathbb{X}(t) = \sqrt{\frac{\hbar a}{2m} + \frac{2\mathcal{E}_\hbar}{m\gamma}\left[t + t_c\left(e^{\frac{-t}{t_c}} - 1\right)\right]} \qquad (5.15)$$

and

$$\mathbb{P}(t) = \frac{\hbar}{\sqrt{(2\hbar a/m) + (8\mathcal{E}_\hbar/m\gamma)\left[t + t_c\left(e^{\frac{-t}{t_c}} - 1\right)\right]}} \qquad (5.16)$$

satisfying $\mathbb{X}(t)\mathbb{P}(t) = \hbar/2$ for all times $t \geq 0$.

The classical limit $\hbar \to 0$ of Eq. (5.15) yields

$$\mathbb{X}(t) = \sqrt{\frac{2\mathcal{E}}{m\gamma}\left[t + t_c\left(e^{\frac{-t}{t_c}} - 1\right)\right]} \qquad (5.17)$$

which is the same that Eq. (2.74) for thermal systems $\mathcal{E} = k_B T$. In contrast, the momentum fluctuation (5.16) vanishes in the classical realm, thereby standing for that the momentum is a deterministic variable. This result is consistent with the condition (5.9) for taking the classical limit.

*Long-time regime.* From Eq. (5.15) at $t \to \infty$, i.e., $t \gg t_c$, $(\gamma\hbar a/4\mathcal{E}_\hbar)$, it follows that the displacement fluctuation $\mathbb{X}(t)$ asymptotically displays the diffusive behavior

$$\mathbb{X}(t) \approx \sqrt{\frac{2\mathcal{E}_\hbar}{m\gamma}t}, \quad t \to \infty, \qquad (5.18)$$

whereas the momentum fluctuation (5.16) exhibits the following non-diffusive behavior

$$\mathbb{P}(t) \approx \sqrt{\frac{\hbar^2 m\gamma}{8\mathcal{E}_\hbar t}}, \quad t \to \infty. \qquad (5.19)$$

Quantity (5.18) as $\lim_{\hbar \to 0} \mathcal{E}_\hbar = \mathcal{E}$ turns out to be simply the Einstein law of classical diffusion, whereas momentum quantum fluctuation (5.19) vanishes, as expected.

*Short-time regime.* At short times $t \to 0$, i.e., $(t/t_c)^3 \ll 1$, Eq. (5.15) becomes non-ballistic

$$\mathbb{X}(t) \sim \sqrt{\frac{\hbar a}{2m} + \frac{\mathcal{E}_\hbar}{m\gamma t_c} t^2}, \qquad t \to 0, \tag{5.20}$$

which in turn renders ballistic at $t \gg \sqrt{a\hbar\gamma t_c / \mathcal{E}_\hbar}$:

$$\mathbb{X}(t) \sim t \sqrt{\frac{\mathcal{E}_\hbar}{m\gamma t_c}}. \tag{5.20a}$$

The classical limit of Eq. (5.20) is

$$\mathbb{X}(t) \sim t \sqrt{\frac{\mathcal{E}}{m\gamma t_c}}, \qquad t \to 0, \tag{5.21}$$

which is Eq. (2.77) for $\mathcal{E} = k_B T$.

At short times the root mean square momentum $\mathbb{P}(t) = \hbar/2\mathbb{X}(t)$ reads

$$\mathbb{P}(t) \sim \frac{\hbar}{\sqrt{(2\hbar a/m) + (4\mathcal{E}_\hbar/m\gamma t_c)t^2}}, \qquad t \to 0, \tag{5.22}$$

and

$$\mathbb{P}(t) \sim \frac{\hbar}{2t} \sqrt{\frac{m\gamma t_c}{\mathcal{E}_\hbar}} \tag{5.23}$$

at $t \gg \sqrt{\hbar a \gamma t_c / \mathcal{E}_\hbar}$.

*Instantaneous velocity and quantum force.* Due to non-Markovian effects both quantities (5.20) and (5.22) are differentiable at $t = 0$. For this reason, there exist both the instantaneous velocity

$$\mathbb{V}(t) = \frac{\mathcal{E}_\hbar}{m\gamma} \frac{\left(1 - e^{\frac{-t}{t_c}}\right)}{\sqrt{(\hbar a/2m) + (2\mathcal{E}_\hbar/m\gamma)\left[t + t_c\left(e^{\frac{-t}{t_c}} - 1\right)\right]}} \tag{5.24}$$

and the attractive quantum force

$$\mathbb{F}(t) = -\frac{4\hbar \mathcal{E}_{\hbar}}{m\gamma} \frac{\left(1 - e^{\frac{-t}{t_c}}\right)}{\left\{(2\hbar a/m) + (8\mathcal{E}_{\hbar}/m\gamma)\left[t + t_c\left(e^{\frac{-t}{t_c}} - 1\right)\right]\right\}^{3/2}}. \quad (5.25)$$

From the existence of the instantaneous velocity (5.24) it follows that the quantum diffusion coefficient $\mathbb{D}(t) = \mathbb{X}(t)\mathbb{V}(t)$, given by

$$\mathbb{D}(t) = \frac{\mathcal{E}_{\hbar}}{m\gamma}\left(1 - e^{\frac{-t}{t_c}}\right), \quad (5.26)$$

coincides with the diffusion coefficient in the quantum Smoluchowski equation (5.13).

The classical limit of Eq. (5.24) is given by

$$\mathbb{V}(t) = \sqrt{\frac{\mathcal{E}}{m}} \frac{\left(1 - e^{\frac{-t}{t_c}}\right)}{\sqrt{2\gamma\left[t + t_c\left(e^{\frac{-t}{t_c}} - 1\right)\right]}} \quad (5.27)$$

which in turn becomes identical to Eq. (2.75) as far as the thermal diffusion energy $\mathcal{E} = k_B T$ is concerned, whereas the quantity (5.25) vanishes as $\hbar \to 0$, thereby implying that effects of the quantum force show up only in the quantum realm. Lastly, the quantum diffusion coefficient (5.26) becomes $\mathbb{D}(t) = (\mathcal{E}/m\gamma)(1 - e{-t/tc}$ in the classical limit.

*Long-time regime.* At long time $t \to \infty$, i.e., $t \gg t_c, (\gamma\hbar a/4\mathcal{E}_{\hbar})$, both quantities (5.24) and (5.25) asymptotically render Markovian:

$$\mathbb{V}(t) \sim \sqrt{\frac{\mathcal{E}_{\hbar}}{2m\gamma}} t^{-1/2}, \quad t \to \infty, \quad (5.28)$$

and

$$\mathbb{F}(t) \sim -\frac{\hbar}{4}\sqrt{\frac{m\gamma}{2\mathcal{E}_{\hbar}}} t^{-3/2}, \quad t \to \infty, \quad (5.29)$$

where in this long-time regime the quantum diffusion coefficient (5.26) is steady, i.e.,

$$\mathbb{D}(\infty) = \frac{\mathcal{E}_\hbar}{m\gamma}. \qquad (5.30)$$

*Short-time regime.* On the other hand, at short times $t \to 0$, i.e., $(t/t_c)^2 \ll 1$, Eqs. (5.24) and (5.25) hold non-Markovian:

$$\mathbb{V}(t) \sim \frac{2\mathcal{E}_\hbar}{\gamma\sqrt{2m\hbar a}} \frac{t}{t_c}, \qquad t \to 0, \qquad (5.31)$$

and

$$\mathbb{F}(t) \sim -\frac{\mathcal{E}_\hbar}{2\gamma a}\sqrt{\frac{m}{\hbar a}} \frac{t}{t_c}, \qquad t \to 0. \qquad (5.32)$$

The diffusion coefficient (5.26) also remains non-Markovian

$$\mathbb{D}(t) \sim \mathbb{D}(\infty) \frac{t}{t_c}, \qquad t \to 0. \qquad (5.33)$$

### 5.2.2. Thermal systems

So far, we have studied the quantum non-inertial Brownian motion of a free particle immersed in a generic environment $\mathcal{E}_\hbar$. In order to provide physical significance to our theoretical predictions, Eqs. (5.24) and (5.25), we now turn to consider at short and long times non-Markovian effects on a quantum free Brownian particle immersed in thermal environments at zero and high temperatures.

*Heat bath of quantum harmonic oscillators.* The quantum diffusion energy $\mathcal{E}_\hbar$ due to a heat bath of quantum harmonic oscillators is given by expression (3.7). At zero temperature, we obtain from Eqs. (5.28) and (5.29) at long time the following Markovian quantities

$$\mathbb{V}(t) \sim \sqrt{\frac{\omega\hbar}{4m\gamma}} t^{-1/2}, \qquad t \to \infty \text{ and } T = 0, \qquad (5.34)$$

and

$$\mathbb{F}(t) \sim -\frac{1}{4}\sqrt{\frac{\hbar m\gamma}{\omega}} t^{-3/2}, \qquad t \to \infty \text{ and } T = 0, \qquad (5.35)$$

whereas at short time we find the non-Markovian behavior

$$\mathbb{V}(t) \sim \frac{\omega}{\gamma}\sqrt{\frac{\hbar}{2ma}\frac{t}{t_c}}, \qquad t \to 0 \text{ and } T = 0, \qquad (5.36)$$

$$\mathbb{F}(t) \sim -\frac{\omega}{4\gamma a}\sqrt{\frac{\hbar m}{a}\frac{t}{t_c}}, \qquad t \to 0 \text{ and } T = 0. \qquad (5.37)$$

*Heat bath of fermions.* According to Eq. (3.8a) the quantum diffusion energy in a fermionic heat bath at $T = 0$ is given by $\mathcal{E}_\hbar = 3k_B T_F/5$. Therefore, at long time $t \to \infty$, Eqs. (5.28) and (5.29) read respectively

$$\mathbb{V}(t) \sim \sqrt{\frac{3k_B T_F}{10m\gamma}} t^{-1/2}, \qquad t \to \infty \text{ and } T = 0, \qquad (5.38)$$

and

$$\mathbb{F}(t) \sim -\frac{\hbar}{4}\sqrt{\frac{5m\gamma}{6k_B T_F}} t^{-3/2}, \qquad t \to \infty \text{ and } T = 0, \qquad (5.39)$$

while at short times they become

$$\mathbb{V}(t) \sim \frac{6k_B T_F}{5\gamma\sqrt{2m\hbar a}}\frac{t}{t_c}, \qquad t \to 0 \text{ and } T = 0, \qquad (5.40)$$

and

$$\mathbb{F}(t) \sim -\frac{3k_B T_F}{10\gamma a}\sqrt{\frac{m}{a\hbar}\frac{t}{t_c}}, \qquad t \to 0 \text{ and } T = 0. \qquad (5.41)$$

*Heat bath of bosons.* According to Eq. (3.12) the quantum diffusion energy in a bosonic heat bath at $T = 0$ is given by $\mathcal{E}_\hbar = 0.77 k_B T_{BE}$. Therefore, at long time $t \to \infty$, Eqs. (5.28) and (5.29) turn out to be given respectively by

$$\mathbb{V}(t) \sim \sqrt{\frac{0.77 k_B T_{BE}}{2m\gamma}} t^{-1/2}, \qquad t \to \infty \text{ and } T = 0, \qquad (5.42)$$

and

$$\mathbb{F}(t) \sim -\frac{\hbar}{4}\sqrt{\frac{m\gamma}{1.54 k_B T_{BE}}} t^{-3/2}, \qquad t \to \infty \text{ and } T = 0. \qquad (5.43)$$

At short times we have

$$\mathbb{V}(t) \sim \frac{0.77 k_B T_{\text{BE}}}{\gamma} \sqrt{\frac{2}{m\hbar a}} \frac{t}{t_c}, \qquad t \to 0 \text{ and } T = 0, \qquad (5.44)$$

and

$$\mathbb{F}(t) \sim -\frac{0.77 k_B T_{\text{BE}}}{2\gamma a} \sqrt{\frac{m}{\hbar a}} \frac{t}{t_c}, \qquad t \to 0 \text{ and } T = 0. \qquad (5.45)$$

*High temperatures.* At high temperatures $T \to \infty$, i.e., $T \gg \hbar\omega/2k_B, T_F, T_{\text{BE}}$, we obtain in the long-time regime the results

$$\mathbb{V}(t) \sim \sqrt{\frac{k_B T}{2m\gamma}} t^{-1/2}, \qquad t \to \infty \text{ and } T \to \infty, \qquad (5.46)$$

and

$$\mathbb{F}(t) \sim -\frac{\hbar}{4} \sqrt{\frac{m\gamma}{2k_B T}} t^{-3/2}, \qquad t \to \infty \text{ and } T \to \infty. \qquad (5.47)$$

Notice that the instantaneous velocity (5.46) is a classical effect because it is independent of the Planck constant $\hbar$, whereas the thermal force (5.47) holds as a quantum effect.

On the other hand, at short times we find that $\mathbb{V}(t)$ and $\mathbb{F}(t)$ hold quantum effects at high temperatures, i.e.,

$$\mathbb{V}(t) \sim \frac{2k_B T}{\gamma\sqrt{2m\hbar a}} \frac{t}{t_c}, \qquad t \to 0 \text{ and } T \to \infty, \qquad (5.48)$$

and

$$\mathbb{F}(t) \sim -\frac{k_B T}{2\gamma a} \sqrt{\frac{m}{\hbar a}} \frac{t}{t_c}, \qquad t \to 0 \text{ and } T \to \infty. \qquad (5.49)$$

### 5.3. Harmonic oscillator

### 5.3.1. Non-thermal systems

The quantum Smoluchowski equation (5.7), with $I(t) = 1 - e^{-t/t_c}$, describing a harmonic oscillator $V(x) = kx^2/2$, is given by

$$\frac{\partial W(x,p,t)}{\partial t} = \frac{kx}{m\gamma}\frac{\partial W(x,p,t)}{\partial x} - \frac{kp}{m\gamma}\frac{\partial W(x,p,t)}{\partial p} + D_x(t)\frac{\partial^2 W(x,p,t)}{\partial x^2}$$
$$+ \left[\frac{k}{m\gamma} - \frac{4D_x(t)}{\hbar^2}p^2\right]W(x,p,t). \tag{5.50}$$

where $D_x(t)$ is the time-dependent diffusion coefficient

$$D_x(t) = \frac{\mathcal{E}_\hbar}{m\gamma}\left(1 - e^{\frac{-t}{t_c}}\right). \tag{5.50a}$$

*The time-dependent solution.* Starting from the initial condition (5.8) a time-dependent solution to Eq. (5.50) reads

$$W(x,p,t) = \frac{1}{\pi\hbar}e^{-\left[\frac{p^2}{\hbar^2}B_\hbar(t) + \frac{x^2}{B_\hbar(t)}\right]} \tag{5.51}$$

where the function $B_\hbar(t)$, given by

$$B_\hbar(t) = \frac{\hbar a}{m}e^{\frac{-2kt}{m\gamma}} + \frac{2\mathcal{E}_\hbar}{k}\left[1 - e^{\frac{-2kt}{m\gamma}} + \frac{2kt_c}{2kt_c - m\gamma}\left(e^{\frac{-2kt}{m\gamma}} - e^{\frac{-t}{t_c}}\right)\right], \tag{5.51a}$$

is expressed in terms of the time evolution $t$, the correlation time $t_c$, and the relaxation time $t_r = m\gamma/2k$. Notice that in the limit $k \to 0$ the function $B_\hbar(t)$ reduces to $4b_\hbar(t)$, $b_\hbar(t)$ being given by Eq. (5.14a).

Solution (5.51) yields $\langle X(t)\rangle = \langle P(t)\rangle = 0$, $\langle X^2(t)\rangle = B_\hbar(t)/2$ and $\langle P^2(t)\rangle = \hbar^2/2B_\hbar(t)$. So we have the following fluctuations

$$\mathbb{X}(t) = \sqrt{\frac{B_\hbar(t)}{2}}, \tag{5.52}$$

$$\mathbb{P}(t) = \frac{\hbar}{\sqrt{2B_\hbar(t)}}, \tag{5.53}$$

both satisfying the Heisenberg constraint $\mathbb{X}(t)\mathbb{P}(t) = \hbar/2$. The fluctuation (5.52) generates the instantaneous velocity

$$\mathbb{V}(t) = \frac{1}{\sqrt{2B_\hbar(t)}}\left[\frac{2\mathcal{E}_\hbar}{(2kt_c - m\gamma)}\left(e^{\frac{-t}{t_c}} - e^{\frac{-2kt}{m\gamma}}\right) - \frac{\hbar a k}{\gamma m^2}e^{\frac{-2kt}{m\gamma}}\right], \tag{5.54}$$

which is well defined at $t = 0$, i.e.,

$$\mathbb{V}(0) = -\frac{k}{m\gamma}\sqrt{\frac{a\hbar}{2m}}, \tag{5.54a}$$

while differentiating the root mean square momentum (5.53) yields the non-thermal force

$$\mathbb{F}(t) = \frac{-\hbar}{B_\hbar(t)\sqrt{2B_\hbar(t)}}\left[\frac{2\mathcal{E}_\hbar}{(2kt_c - m\gamma)}\left(e^{\frac{-t}{t_c}} - e^{\frac{-2kt}{m\gamma}}\right) - \frac{\hbar ak}{\gamma m^2}e^{\frac{-2kt}{m\gamma}}\right]. \tag{5.55}$$

which is also well defined at $t = 0$:

$$\mathbb{F}(0) = \frac{k\hbar}{\gamma}\sqrt{\frac{\hbar}{2am}}. \tag{5.55a}$$

The existence of both quantities (5.54a) and (5.55a) implies that the root mean square displacement (5.52) and the root mean square momentum (5.53) are said to be differentiable functions for all $t \geq 0$.

Multiplying Eq. (5.54) by Eq. (5.52) leads to the following quantum diffusion coefficient, $\mathbb{D}(t) = \mathbb{X}(t)\mathbb{V}(t)$,

$$\mathbb{D}(t) = \frac{\mathcal{E}_\hbar}{(2kt_c - m\gamma)}\left(e^{\frac{-t}{t_c}} - e^{\frac{-2kt}{m\gamma}}\right) - \frac{\hbar ak}{2\gamma m^2}e^{\frac{-2kt}{m\gamma}} \tag{5.56}$$

which coincides with the diffusion coefficient of the quantum Smoluchowski equation, Eq. (5.48a), only in the free particle limit $k \to 0$. The diffusion coefficient (5.56) at $t = 0$ becomes

$$\mathbb{D}(0) = -\frac{\hbar ak}{2\gamma m^2} \tag{5.56a}$$

in contrast to the diffusion coefficient (5.50a) at $t = 0$ which vanishes, i.e., $D_x(0) = 0$. Notice that the upshot (5.56a) for $k < 0$ is a genuine quantum effect since it vanishes in the classical limit $\hbar \to 0$, assuming the parameter $a$ to be independent of the Planck's constant.

*Steady solution.* In the long-time regime $t \to \infty$, physically interpreted as $t \gg t_c, t_r = m\gamma/2k$, the probability distribution function (5.51) renders steady, i.e.,

$$W_{st}(x, p, \infty) = \frac{1}{\pi\hbar}e^{-\left[\frac{2\mathcal{E}_\hbar p^2}{k\hbar^2} + \frac{kx^2}{2\mathcal{E}_\hbar}\right]}, \quad k > 0, \tag{5.57}$$

while both fluctuations (5.52) and (5.53) become, respectively,

$$\mathbb{X}(\infty) = \sqrt{\frac{\mathcal{E}_\hbar}{k}} \qquad (5.58)$$

and

$$\mathbb{P}(\infty) = \frac{\hbar}{2}\sqrt{\frac{k}{\mathcal{E}_\hbar}}. \qquad (5.59)$$

In contrast, both the instantaneous velocity (5.54) and the non-thermal force (5.55) vanish, thereby meaning that they are non-equilibrium effects. The diffusion coefficient (5.54) also becomes null, i.e., $\mathbb{D}(\infty) = 0$, whereas $D_x(\infty) = \mathcal{E}_\hbar/m\gamma$.

*Classical limit.* The classical limit of the time-dependent solution (5.51), i.e., $\lim_{\hbar \to 0} W(x, p, t) = f(x, t)\delta(p)$, provides

$$f(x,t) \propto e^{\frac{-x^2}{B(t)}}, \qquad (5.60)$$

where the function $B(t)$ coming from Eq. (5.50a) as $\lim_{\hbar \to 0} \mathcal{E}_\hbar = \mathcal{E}$ reads

$$B(t) = \frac{2\mathcal{E}}{k}\left[1 - e^{\frac{-2kt}{m\gamma}} + \frac{2kt_c}{2kt_c - m\gamma}\left(e^{\frac{-2kt}{m\gamma}} - e^{\frac{-t}{t_c}}\right)\right]. \qquad (5.60a)$$

Normalizing the probability distribution function (5.60) and next taking into account $\mathcal{E} = k_B T$ and $k = m\omega^2$, we arrive at the classical solution (2.85). For these thermal systems, the Boltzmann distribution function (2.92) can be derived directly from the steady quantum Wigner function (5.60).

It is worth noticing that the root mean square momentum (5.53) vanishes in the classical limit, whereas the root mean square displacement (5.54) becomes

$$\mathbb{X}(t) = \sqrt{\frac{\mathcal{E}}{k}\left[1 - e^{\frac{-2kt}{m\gamma}} + \frac{kt_c}{2kt_c - m\gamma}\left(e^{\frac{-2kt}{m\gamma}} - e^{\frac{-t}{t_c}}\right)\right]}, \qquad (5.61)$$

thereby breaking down the Heisenberg constraint, i.e., $\mathbb{P}(t)\mathbb{X}(t) = 0$, as expected. This result implies that in the classical realm the momentum becomes a deterministic variable while the displacement holds a stochastic one. In other words, the momentum variable is said to be eliminated from the classical probabilistic description of the Brownian harmonic oscillator in the absence of inertial force.

Further, while the quantum force (5.55) vanishes in the classical realm, characterized by both limits $\hbar \to 0$ and $\mathcal{E}_\hbar \to \mathcal{E}$, the instantaneous velocity (5.54) and the diffusion coefficient (5.56) become respectively

$$\mathbb{V}(t) = \frac{2\mathcal{E}}{(2kt_c - m\gamma)\sqrt{2B(t)}} \left( e^{\frac{-t}{t_c}} - e^{\frac{-2kt}{m\gamma}} \right), \qquad (5.62)$$

and

$$\mathbb{D}(t) = \frac{\mathcal{E}}{(2kt_c - m\gamma)} \left( e^{\frac{-t}{t_c}} - e^{\frac{-2kt}{m\gamma}} \right). \qquad (5.63)$$

In Eq. (5.62), the function $B(t)$ is given by Eq. (5.60a).

### 5.3.2. Thermal systems

*Heat bath of quantum harmonic oscillators at $T = 0$.* As far as a heat bath of harmonic oscillators at zero temperature is concerned, the diffusion energy is $\mathcal{E}_\hbar = \hbar\omega/2$ and, acccordingly, Eqs. (5.54), (5.55), and (5.56) at short time $t \to 0$, i.e., $t^2/(m\gamma/2k)^2 \ll 1$, become respectively

$$\mathbb{V}(t) \sim \sqrt{\frac{\hbar}{2\gamma a(m\gamma - 2kt)}} \left[ \left( \frac{\omega}{t_c} + \frac{ak^2}{\gamma m^2} \right) t - \frac{ak}{m} \right], \qquad t \to 0 \text{ and } T = 0, \quad (5.64)$$

$$\mathbb{F}(t) \sim -\frac{m^2}{a(m\gamma - 2kt)} \sqrt{\frac{\hbar\gamma}{a(m\gamma - 2kt)}} \left[ \left( \frac{\omega}{t_c} + \frac{ak^2}{\gamma m^2} \right) t - \frac{ak}{\gamma m} \right], t \to 0 \text{ and } T = 0$$

$$(5.65)$$

and

$$\mathbb{D}(t) \sim \frac{\hbar}{2m\gamma} \left[ \left( \frac{\omega}{t_c} + \frac{2ak^2}{\gamma m^2} \right) t - \frac{ak}{m} \right], \qquad t \to 0 \text{ and } T = 0. \qquad (5.66)$$

Note that the above quantum-mechanical quantities are non-Markovian at $t \neq 0$.

*Fermionic heat bath at $T = 0$.* In the case of a fermionic thermal reservoir at $T = 0$ the diffusion energy is given by $\mathcal{E}_\hbar = 3\epsilon_F/5$. So, Eqs. . (5.54), (5.55), and (5.56) read respectively

$$\mathbb{V}(t) \sim \frac{2}{\sqrt{2\gamma\hbar a(m\gamma - 2kt)}} \left[ \left( \frac{3\epsilon_F}{5t_c} + \frac{\hbar ak^2}{2\gamma m^2} \right) t - \frac{\hbar ak}{2m} \right], t \to 0 \text{ and } T = 0, \quad (5.67)$$

$$\mathbb{F}(t) \sim \frac{-2m^2}{a(m\gamma - 2kt)}\sqrt{\frac{\gamma}{2\hbar a(m\gamma - 2kt)}}\left[\left(\frac{3\epsilon_F}{5t_c} + \frac{\hbar a k^2}{2\gamma m^2}\right)t - \frac{\hbar a k}{2\gamma m}\right], t \to 0 \text{ and } T = 0$$

(5.68)

and

$$\mathbb{D}(t) \sim \left(\frac{3\epsilon_F}{5m\gamma t_c} + \frac{\hbar a k^2}{\gamma^2 m^3}\right)t - \frac{\hbar a k}{2\gamma m^2}, t \to 0 \text{ and } T = 0. \qquad (5.69)$$

*Bosonic heat bath at $T = 0$.* In the case of a bosonic thermal reservoir at $T = 0$ the diffusion energy is given by $\mathcal{E}_\hbar = 0.77 k_B T_{\text{BE}}$. So, Eqs. (5.54), (5.55), and (5.56) read respectively

$$\mathbb{V}(t) \sim \frac{2}{\sqrt{2\gamma\hbar a(m\gamma - 2kt)}}\left[\left(\frac{0.77 k_B T_{\text{BE}}}{t_c} + \frac{\hbar a k^2}{2\gamma m^2}\right)t - \frac{\hbar a k}{2m}\right], t \to 0 \text{ and } T = 0,$$

(5.70)

$$\mathbb{F}(t) \sim \frac{-2m^2}{a(m\gamma - 2kt)}\sqrt{\frac{\gamma}{2\hbar a(m\gamma - 2kt)}}\left[\left(\frac{0.77 k_B T_{\text{BE}}}{t_c} + \frac{\hbar a k^2}{2\gamma m^2}\right)t - \frac{\hbar a k}{2\gamma m}\right],$$

$$t \to 0 \text{ and } T = 0, \qquad (5.71)$$

and

$$\mathbb{D}(t) \sim \left(\frac{0.77 k_B T_{\text{BE}}}{m\gamma t_c} + \frac{\hbar a k^2}{\gamma^2 m^3}\right)t - \frac{\hbar a k}{2\gamma m^2}, t \to 0 \text{ and } T = 0. \qquad (5.72)$$

*Heat baths at high temperatures.* At high temperatures and short times we find the following quantum effects

$$\mathbb{V}(t) \sim \frac{2}{\sqrt{2\gamma\hbar a(m\gamma - 2kt)}}\left[\left(\frac{k_B T}{t_c} + \frac{\hbar a k^2}{2\gamma m^2}\right)t - \frac{\hbar a k}{2m}\right], \quad t \to 0, T \to \infty, \quad (5.73)$$

$$\mathbb{F}(t) = \frac{-2m^2}{a(m\gamma - 2kt)}\sqrt{\frac{\gamma}{2\hbar a(m\gamma - 2kt)}}\left[\left(\frac{k_B T}{t_c} + \frac{\hbar a k^2}{2\gamma m^2}\right)t - \frac{\hbar a k}{2\gamma m}\right], t \to 0, T \to \infty,$$

(5.74)

$$\mathbb{D}(t) \sim \left( \frac{k_B T}{m\gamma t_c} + \frac{\hbar a k^2}{\gamma^2 m^3} \right) t - \frac{\hbar a k}{2\gamma m^2}, \qquad t \to 0, \qquad T \to \infty. \tag{5.75}$$

## 5.4. Summary and Discussion

Starting directly from the non-Markovian, non-Gaussian Kolmogorov equation in configuration space, Eq. (2.62), describing the non-inertial Brownian motion, we have derived a non-Markovian, non-Gaussian quantum Kolmogorov equation in configuration space, Eq. (5.4), which in turn reduces to the quantum master equation (5.5) in the Gaussian approximation for both diagonal and non-diagonal elements of the matrix density. Next, a non-Markovian quantum Smoluchowski equation in phase space has been derived, Eq. (5.7), and solved for both cases of a free particle and a harmonic oscillator immersed as much in thermal as non-thermal environments. Our quantum Smoluchowski equations (5.5) and (5.7) are valid for both harmonic and nonharmonic potentials even in the Gaussian approximation.

Our quantum Smoluchowski equation (QSE) in phase space (5.7) proves to be a consistent way to investigate the quantum Brownian motion in the absence of inertial force because

- as $\hbar \to 0$ it correctly reduces to the classical Smoluchowski equation in configuration space;
- as far as a free particle and a harmonic oscillator are concerned we have shown explicitly that the solutions to our QSE does not violate the Heisenberg fluctuation relations;
- our non-Markovian QSE is valid for all temperatures $T \geq 0$ in the cases of thermal systems (reservoirs of quantum harmonic oscillators, of bosons, and of fermions).

Further, our QSE furnishes some theoretical predictions: the mathematical property of differentiability or analyticity of the quantum Brownian trajectories give rise to both concepts of instantaneous velocity $\mathbb{V}(t) = d\mathbb{X}(t)/dt$ and generalized force $\mathbb{F}(t) = d\mathbb{P}(t)/dt$, where $\mathbb{X}(t)$ and $\mathbb{P}(t)$ are the root mean square displacement and the root mean square momentum, respectively. Physically, for thermal systems it is predicted that the equipartition theorem is violated for all times $t \geq 0$ and temperatures $T \geq 0$.

Eventually, we want to emphasize that the derivation of our quantum Smoluchowski equation in phase space (5.7) is based on a non-Hamiltonian quantization method, whereby we make use of neither the canonical quantization scheme nor the path integral formalism. Furthermore, our QSE (5.7) describing a free particle does not coincide with the classical Smoluchowski equation.

We now turn to compare our findings with some quantum Smoluchowski equations existing in the literature.

*Ankerhold's quantum Smoluchowski equations.* On the basis of the Caldeira-Leggett Hamiltonian system-plus-reservoir model [133,167,187] in tandem with the path integral formalism, Ankerhold and co-workers have attempted to derive a consistent quantum Smoluchowski equation (QSE) in configuration space for the diagonal elements of the matrix density [379-386] as well in phase space (for a free particle) [380,388]. Further works [188,387-390], however, have disclosed some drawbacks underlying the QSE in configuration space, for it may violate the second law of thermodynamics, for example. A highly questionable result is that such Ankerhold's QSE for the case of a free quantum Brownian particle coincides with the classical Smoluchowski equation. In addition, both kinds of Ankerhold's QSE are Markovian and not valid at zero temperature, for it is argued that non-Markovian effects are intractable since they are determined by a time nonlocal damping kernel [385]. In comparison to our non-Markovian QSEs, Eqs. (5.5) and (5.7), restricting to the Markovian character and non-zero temperature regime $T > 0$ seems to be indeed a strong limitation underlying the path integral approach to quantum Brownian motion.

Furthermore, Coffey et al. [389,390] have shown that the stationary solution of Ankerhold's QSE in the high temperature limit is different from the true Wigner equilibrium distribution in configuration space.

Also, Ankerhold's approach provides no derivation of the time evolution of the non-diagonal elements of the so-called density matrix, for it is claimed that they are "*strongly suppressed during the time evolution*" (see e.g. Refs. [383,386]). In contrast, we have derived our master equation (5.5) for the off-elements $x_1 \neq x_2$. According to our Eq. (5.5a), even the diagonal elements do not behave classically.

*Indian group's quantum Smoluchowski equations.* Another Hamiltonian system-plus-reservoir account aiming at to derive non-Markovian quantum Smoluchowski equations has been developed in Refs. [391-394]. This c-number approach [394] to quantum Brownian motion does not depend on the path integral formalism but it is based on the canonical quantization. For the case of a quantum free Brownian particle, although the QSE in configuration space differs from the classical Smoluchowski equation due to the presence of a quantum diffusion coefficient that is valid at $T = 0$, the so-termed QSE in phase space (see Eq. (17) in Ref. [391]) is simply identical to the QSE in configuration space. Nevertheless, according to our Smoluchowski equations (5.5) and (5.7) such singular result does not occur at all.

*Coffey's quantum Smoluchowski equation.* Coffey and co-workers [390,395-399] have obtained a Markovian QSE in configuration space valid only in the semi-classical region of quantum mechanics, that is, in the high-temperature and weak-coupling limits. They used a heuristic method for the determination of the effective position-dependent diffusion coefficient. Such a method is based on Brinkman's hierarchy expressed in terms of the time evolution for the Wigner function obtained from the Caldeira-Leggett Hamiltonian system-plus-reservoir model. According to Maier and Ankerhold [386], the Coffey and co-workers' semiclassical method is not a consistent way of obtaining higher-order quantum corrections to the QSE. Moreover, it is worth highlighting that although the Coffey's semiclassical QSE is very similar but not identical to Ankerhold's QSE, both quantum-mechanical equations of motion are plagued with the same disease, namely, they coincide with the classical Smoluchowski equation for the case of a free Brownian motion [190,400]. They predict therefore Einstein's diffusion law $\mathbb{X}(t) = \sqrt{2Dt}$ for a quantum free Brownian particle. In contrast, in the semiclassical realm of quantum mechanics, characterized by $\mathcal{E}_\hbar \to k_B T$, our approach predicts that Eqs. (5.15) and (5.16) for a quantum free particle read respectively

$$\mathbb{X}(t) = \sqrt{\frac{\hbar a}{2m} + \frac{2k_B T}{m\gamma}\left[t + t_c\left(e^{\frac{-t}{t_c}} - 1\right)\right]} \qquad (5.76)$$

and

$$\mathbb{P}(t) = \frac{\hbar}{\sqrt{(2\hbar a/m) + (8k_B T/m\gamma)\left[t + t_c\left(e^{\frac{-t}{t_c}} - 1\right)\right]}}. \qquad (5.77)$$

Notice that only at long times the position fluctuation (5.76) behaviors classically according to Einstein's diffusion law, i.e., $\mathbb{X}(t) \approx \sqrt{(2k_B T/m\gamma)t}$, while the momentum fluctuation (5.77) $\mathbb{P}(t) \approx \hbar/\sqrt{(8k_B T/m\gamma)t}$ holds its quantum nature, thereby fulfilling the Heisenberg relation $\mathbb{X}(t)\mathbb{P}(t) \approx \left[\sqrt{(2k_B T/m\gamma)t}\right]\left[\hbar/\sqrt{8k_BT/m\gamma t}\right] = \hbar/2$. Therefore, the non-inertial quantum motion of a free Brownian particle cannot be viewed as a classical motion.

*Vacchini's quantum Smoluchowski equation.* Starting from the quantum linear Boltzmann equation, Vacchini [172] has derived a QSE describing a free quantum particle at high temperature (a diffusion equation containing a diffusion coefficient with quantum correction). Such quantum dynamics is distinct from a classical free particle. Such an approach is however restricted to the case of a free quantum Brownian particle.

*Tsekov's nonlinear quantum Smoluchowski equation.* Lastly, Tsekov [400-405] has arrived at a nonlinear version of QSE in configuration space without referring to any quantization process. He starts with a Hamiltonian system-plus-reservoir model described by a nonlinear Schrödinger equation in the Madelung representation of quantum mechanics and then obtains a QSE for a free Brownian particle different from the classical Smoluchowski equation, too. His nonlinear approach to QSE is valid at zero temperature.

# 6. Quantum anomalous diffusion

In Sect. 2.3 we have examined the anomalous behavior of a Brownian particle in the classical domain. We now wish to investigate quantum effects on anomalous diffusion in the presence and absence of inertial force for thermal and non-thermal systems. For thermal heat baths quantum anomalous are studied at zero and high temperatures.

## 6.1. Quantum anomalous diffusion in the presence of inertial force

### 6.1.1. Non-thermal systems

The quantum master equation (4.1), with $I(t) = [\lambda(t/t_c)^{\lambda-1} + 1 - e^{-t/t_c}]$ instead of $I(t) = 1 - e^{-t/t_c}$, describing a free particle is expressed in terms of the Wigner function (4.3) by the following quantum Fokker-Planck equation in phase space

$$\frac{\partial W}{\partial t} = -\frac{p}{m}\frac{\partial W}{\partial x} + 2\gamma \frac{\partial}{\partial p} pW + 2\gamma m \mathcal{E}_\hbar \left(\lambda \frac{t^{\lambda-1}}{t_c^{\lambda-1}} + 1 - e^{\frac{-t}{t_c}}\right)\frac{\partial^2 W}{\partial p^2}, \quad (6.1)$$

which is defined for $\lambda > 1$ (see Appendix D). The initial condition for solving Eq. (6.1) is deemed to be the function (4.6) obeying $\langle X^2(0)\rangle\langle P^2(0)\rangle = \hbar^2/4$. Furthermore, assuming $W \equiv W(x,p,t)$ to be given by Eq. (4.7), i.e., $W(x,p,t) = \mathcal{W}(p,t)\mathcal{G}(x,t)$, such that it satisfies the minimal Heisenberg relationship (4.8) in the form

$$\langle X^2(t)\rangle\langle P^2(t)\rangle = \frac{\hbar^2}{4}, \quad (6.2)$$

for all time $t > 0$, we can derive the quantum Rayleigh equation

$$\frac{\partial \mathcal{W}(p,t)}{\partial t} = 2\gamma \frac{\partial}{\partial p}[p\mathcal{W}(p,t)] + 2\gamma m \mathcal{E}_\hbar \left(\lambda \frac{t^{\lambda-1}}{t_c^{\lambda-1}} + 1 - e^{\frac{-t}{t_c}}\right)\frac{\partial^2 \mathcal{W}(p,t)}{\partial p^2}, \quad (6.3)$$

whose solution at long times $t \to \infty$ yields the non-steady behavior according to the values of $\lambda$ as follows.

- For $\lambda = 2$,

$$\langle P^2(t)\rangle \sim 2m\mathcal{E}_\hbar \frac{t}{t_c}, \quad t \to \infty; \quad (6.4a)$$

- For $\lambda = 3$,

$$\langle P^2(t)\rangle \sim 3m\mathcal{E}_\hbar \frac{t^2}{t_c^2}, \quad t\to\infty; \tag{6.4b}$$

- For $\lambda = 4$,

$$\langle P^2(t)\rangle \sim 4m\mathcal{E}_\hbar \frac{t^3}{t_c^3}, \quad t\to\infty. \tag{6.4c}$$

Generalizing such results leads to the expression

$$\langle P^2(t)\rangle \sim \lambda m\mathcal{E}_\hbar \frac{t^{\lambda-1}}{t_c^{\lambda-1}}, \quad t\to\infty. \tag{6.5}$$

As a consequence of the minimal Heisenberg relationship (6.2), we obtain the following mean square displacement at long time

$$\langle X^2(t)\rangle \sim \frac{\hbar^2 t_c^{\lambda-1}}{4\lambda m \mathcal{E}_\hbar t^{\lambda-1}}, \quad t\to\infty. \tag{6.6}$$

Notice that the anomalous behavior (6.5) reduces in the classical limit $\lim_{\hbar\to 0}\mathcal{E}_\hbar = \mathcal{E}$ to Eq. *2.115*, whereas the behavior *6.6* holds a non-classical effect, for it vanishes as $\hbar \to 0$. This result implies that the position shows up as a deterministic quantity in the classical realm.

### 6.1.2. Thermal systems at zero temperature

In the case of a heat bath of quantum harmonic oscillators the Brownian free particle's quantum diffusion energy $\mathcal{E}_\hbar$ is given by Eq. (3.7), i.e., $\mathcal{E}_\hbar = (\omega\hbar/2)\coth\omega\hbar/2k_BT$. At zero temperature, Eqs. *6.5* and *6.6* read respectively

$$\langle P^2(t)\rangle \sim \lambda m \frac{\hbar\omega}{2}\frac{t^{\lambda-1}}{t_c^{\lambda-1}}, \quad t\to\infty,\ T=0, \tag{6.5a}$$

and

$$\langle X^2(t)\rangle \sim \frac{\hbar t_c^{\lambda-1}}{2\lambda m\omega t^{\lambda-1}}, \quad t\to\infty,\ T=0. \tag{6.5b}$$

For a fermionic thermal reservoir the quantum diffusion energy at low temperatures is given by Eq. (3.8a) which becomes $\mathcal{E}_\hbar = (3/5)\,k_B T_F$ at zero temperature. Hence, we find

$$\langle P^2(t)\rangle \sim \lambda m \frac{3k_B T_F}{5} \frac{t^{\lambda-1}}{t_c^{\lambda-1}}, \qquad t \to \infty, \; T = 0, \qquad (6.6a)$$

and

$$\langle X^2(t)\rangle \sim \frac{5\hbar^2 t_c^{\lambda-1}}{12\lambda m k_B T_F t^{\lambda-1}}, \qquad t \to \infty, \; T = 0. \qquad (6.6b)$$

Although the anomalous behavior of the mean square displacement (6.6b) occurs in the absence of quantum effects, because $T_F \propto \hbar^2$, the mean square momentum (6.6a) exhibits quantum effects, so that the Heisenberg relation (6.2) still holds valid at long time.

As far as a bosonic heat bath is concerned the quantum diffusion energy is given by Eq. (3.12) at temperatures below or equal to the Bose-Einstein temperature $T_{\text{BE}}$, so that at $T = 0$ it reads $\mathcal{E}_\hbar = 0.77 k_B T_{\text{BE}}$. So, Eqs. (6.5) and (6.6) become respectively

$$\langle P^2(t)\rangle \sim 0.77 \lambda m k_B T_{\text{BE}} \frac{t^{\lambda-1}}{t_c^{\lambda-1}}, \qquad t \to \infty, \qquad T = 0, \qquad (6.7a)$$

and

$$\langle X^2(t)\rangle \sim \frac{\hbar^2 t_c^{\lambda-1}}{3.08 \lambda m k_B T_{\text{BE}} t^{\lambda-1}}, \qquad t \to \infty, \; T = 0. \qquad (6.7b)$$

According to Eq. (3.11a) the Bose-Einstein temperature $T_{\text{BE}}$ is directly proportional to $\hbar^2$, i.e., $T_{\text{BE}} \propto \hbar^2$. Hence, there exists no quantum effect on the anomalous behavior of the mean square displacement (6.7b), whereas the mean square momentum (6.7a) displays quantum effects.

*Thermal systems at high temperatures.* At high temperatures, i.e., $T \gg \omega\hbar/2k_B, T_F, T_{\text{BE}}$, the quantum diffusion energy $\mathcal{E}_\hbar$ tends to the classical thermal energy $\mathcal{E} \propto k_B T$. Accordingly, we have

$$\langle P^2(t)\rangle \sim \lambda m k_B T \frac{t^{\lambda-1}}{t_c^{\lambda-1}}, \qquad t \to \infty, \; T \to \infty, \qquad (6.8a)$$

$$\langle X^2(t)\rangle \sim \frac{\hbar^2 t_c^{\lambda-1}}{4\lambda m k_B T t^{\lambda-1}}, \qquad t \to \infty, \; T \to \infty. \qquad (6.8b)$$

Although the mean square momentum (6.8a) does not depend on the Planck constant, the Heisenberg relation $\langle P^2(t)\rangle\langle X^2(t)\rangle \sim \hbar^2/4$ holds valid owing to quantum effects on the mean square displacement (6.8b).

## 6.2. Quantum anomalous diffusion in the absence of inertial force

### 6.2.1. Non-thermal systems

In order to investigate the anomalous behavior of a quantum free particle in the absence of inertial force, we start with the quantum Smoluchowski equation (5.7) for $V(x) = 0$, with the correlational function $I(t) = \left[\lambda(t/t_c)^{\lambda-1} + 1 - e^{-t/t_c}\right]$, i.e.,

$$\frac{\partial W}{\partial t} = \frac{\mathcal{E}_\hbar}{m\gamma}\left(\lambda \frac{t^{\lambda-1}}{t_c^{\lambda-1}} + 1 - e^{\frac{-t}{t_c}}\right)\frac{\partial^2 W}{\partial x^2} - \frac{4\mathcal{E}_\hbar}{m\gamma\hbar^2}\left(\lambda \frac{t^{\lambda-1}}{t_c^{\lambda-1}} + 1 - e^{\frac{-t}{t_c}}\right)p^2 W \qquad (6.9)$$

the solution $W \equiv W(x,p,t)$ of which leads to

$$\langle X^2(t)\rangle \sim \frac{2\mathcal{E}_\hbar}{m\gamma}\frac{t^\lambda}{t_c^{\lambda-1}}, \qquad t \to \infty, \qquad (6.10a)$$

and

$$\langle P^2(t)\rangle \sim \frac{t_c^{\lambda-1} m\gamma \hbar^2}{8\mathcal{E}_\hbar t^\lambda}, \qquad t \to \infty, \qquad (6.10b)$$

both satisfying the Heisenberg relationship (6.2) at $t \to \infty$.

In the classical limit $\hbar \to 0$, it is expected that $\mathcal{E}_\hbar$ goes to $\mathcal{E}$. Accordingly, Eq. (6.10a) reduces to Eq. (2.102), whereas Eq. (6.10b) vanishes, standing for that the momentum of a free Brownian particle is a deterministic variable in the classical domain.

### 6.2.2. Thermal systems at zero temperature

Let us consider a heat bath of quantum harmonic oscillators at zero temperature, so that $\mathcal{E}_\hbar = \omega\hbar/2$. Both the mean square displacement (6.10a) and the mean square momentum (6.10b) turn out to be respectively

$$\langle X^2(t)\rangle \sim \frac{\omega\hbar}{m\gamma}\frac{t^\lambda}{t_c^{\lambda-1}}, \qquad t \to \infty, \; T = 0, \qquad (6.11a)$$

and

$$\langle P^2(t)\rangle \sim \frac{t_c^{\lambda-1} m\gamma\hbar}{4\omega t^\lambda}, \qquad t\to\infty,\ T=0. \tag{6.11b}$$

For a fermionic thermal reservoir at zero temperature, we have $\mathcal{E}_\hbar = (3/5)k_B T_F$. Hence,

$$\langle X^2(t)\rangle \sim \frac{6k_B T_F}{5m\gamma}\frac{t^\lambda}{t_c^{\lambda-1}}, \qquad t\to\infty,\ T=0, \tag{6.12a}$$

and

$$\langle P^2(t)\rangle \sim \frac{5t_c^{\lambda-1} m\gamma\hbar^2}{24k_B T_F t^\lambda}, \qquad t\to\infty,\qquad T=0. \tag{6.12b}$$

Since $T_F \propto \hbar^2$ the mean square momentum (6.12b) takes place in the absence of quantum effects while the anomalous behavior of the mean square displacement (6.12a) is under quantum effects, so that the Heisenberg relation (6.2) holds valid at long time.

In the case of a bosonic heat bath at $T=0$, we find $\mathcal{E}_\hbar = 0.77 k_B T_{BE}$. So Eqs. (6.10) become respectively

$$\langle X^2(t)\rangle \sim \frac{1.54 k_B T_{BE}}{m\gamma}\frac{t^\lambda}{t_c^{\lambda-1}}, \qquad t\to\infty,\ T=0, \tag{6.13a}$$

and

$$\langle P^2(t)\rangle \sim \frac{t_c^{\lambda-1} m\gamma\hbar^2}{6.61 k_B T_{BE} t^\lambda}, \qquad t\to\infty,\qquad T=0. \tag{6.13b}$$

Due to the fact that $T_{BE} \propto \hbar^2$ there exist no quantum effects on the mean square momentum (6.13b), whereas the anomalous behavior of the mean square displacement (6.13a) holds its quantum nature.

*Thermal systems at high temperatures.* At high temperatures, $\mathcal{E} \propto k_B T$. For this reason, we have

$$\langle X^2(t)\rangle \sim \frac{2k_B T}{m\gamma}\frac{t^\lambda}{t_c^{\lambda-1}}, \qquad t\to\infty,\qquad T\to\infty, \tag{6.14a}$$

and

$$\langle P^2(t)\rangle \sim \frac{t_c^{\lambda-1} m\gamma\hbar^2}{8k_B T t^\lambda}, \qquad t\to\infty,\qquad T\to\infty. \tag{6.14b}$$

Notice that the mean square displacement (6.14a) is the same one given by Eq. (2.102), for $\mathcal{E} = k_B T$. In the quantum realm both quantities (6.14) fulfill the Heisenberg relation $\langle P^2(t)\rangle\langle X^2(t)\rangle \sim \hbar^2/4$.

The classical limit $\hbar \to 0$ predicts no fluctuations on the mean square momentum (6.14b), meaning that the momentum renders deterministic in the classical domain.

# 7. Quantum tunneling: Non-Markov and Markov effects

Tunneling is a genuine quantum phenomenon in which a particle can penetrate and in most cases pass through a potential barrier, which is assumed to be higher than the kinetic energy of the particle [406]. In isolated systems, the tunneling probability can be calculated on the basis of the Schrödinger equation, the path integral method, the Heisenberg equations, or the Wigner representation of quantum mechanics [406]. Furthermore, there is a description of quantum tunneling in terms of density matrix for calculating escape rates rather than probabilities [385].

In the context of open quantum systems in which there is no Schrödinger function describing a quantum Brownian particle, the tunneling phenomenon is commonly investigated via the quantum Kramers rate on the basis of a system-plus-reservoir model Hamiltonian in tandem with the path integral formalism [139,239,385,407,408] or in terms of c-numbers [409-416]. In addition, quantum Kramers escape rate is calculated within the Wigner picture of quantum mechanics but without necessarily resorting to a model Hamiltonian [389,390,395,396,399].

In the sequel we intend to look at quantum tunneling phenomenon in presence as well as in absence of inertial force in Sect. 7.1 and Sect. 7.2, respectively. More specifically, we reckon with both Markovian and non-Markovian effects on the quantum tunneling for thermal and non-thermal open systems. Thermal systems are studied at zero and high temperatures.

## 7.1. Quantum tunneling in the presence of inertial force

### 7.1.1. Markovian effects

*Non-thermal systems.* Performing the Wigner transform [376,389]

$$\mathcal{W}(x,p,t) = \frac{1}{2\pi} \int_{-\infty}^{\infty} \rho\left(x + \frac{\eta\hbar}{2}, x - \frac{\eta\hbar}{2}, t\right) e^{-ip\eta} d\eta \qquad (7.1)$$

leads our quantum master equation (3.5), with $\beta \equiv 2\gamma$, to the following non-Markovian Fokker-Planck equation in quantum phase space

$$\frac{\partial \mathcal{W}}{\partial t} = -\frac{p}{m}\frac{\partial \mathcal{W}}{\partial x} + \left[\frac{dV(x)}{dx}\frac{\partial}{\partial p} + \hat{\mathcal{O}}\right]\mathcal{W} + 2\gamma \mathcal{W} + 2\gamma p \frac{\partial \mathcal{W}}{\partial p} + 2\gamma m \mathcal{E}_\hbar I(t) \frac{\partial^2 \mathcal{W}}{\partial p^2} \qquad (7.2)$$

where $\mathcal{W} \equiv \mathcal{W}(x,p,t)$ and the operator $\hat{\mathcal{O}}$ is defined as

$$\hat{\mathcal{O}} = \sum_{n=3,5,7,\ldots}^{\infty} \frac{1}{n!}\left(\frac{i\hbar}{2}\right)^{n-1} \frac{d^n V(x)}{dx^n} \frac{\partial^n}{\partial p^n}. \tag{7.2a}$$

In addition, we suppose a steady solution of Eq. (7.2) to be in the form

$$W_{st}(x,p) \propto e^{-\left[\frac{V(x)}{\mathcal{E}_\hbar} + \frac{p^2}{2m\mathcal{E}_\hbar}\right]}, \tag{7.3}$$

where the potential energy $V(x)$ is a quadratic function:

$$V(x) = k\frac{x^2}{2}. \tag{7.3a}$$

Assuming the barrier top to be located at point $x_b$, while the two bottom wells are at $x_a$ and $x_c$, with $x_a < x_b$, an expansion of the potential function (7.3a) in a Taylor series around the point $x_b$ yields $V(x) \sim V(x_b) - (m\omega_b^2/2)(x - x_b)^2$, where we have used $k = -m\omega_b^2$, the quantity $\omega_b$ denoting the particle's oscillation frequency over the potential barrier. Accordingly, the quantum Fokker-Planck equation (7.2) in the steady regime $t \to \infty$ reads

$$-\frac{p}{m}\frac{\partial W}{\partial x} - m\omega_b^2(x - x_b)\frac{\partial W}{\partial p} + 2\gamma W + 2\gamma p\frac{\partial W}{\partial p} + 2\gamma m\mathcal{E}_\hbar \frac{\partial^2 W}{\partial p^2} = 0. \tag{7.4}$$

We now introduce the non-equilibrium steady function

$$\mathcal{W}(x,p) = \mathcal{W}'(x,p) W_{st}(x,p), \tag{7.5}$$

$W_{st}(x,p)$ being the steady function (7.3) over the potential barrier at $x_b$, i.e.,

$$W_{st}(x,p) = e^{\frac{-V(x_b)}{\mathcal{E}_\hbar}} e^{-\left[\frac{p^2}{2m\mathcal{E}_\hbar} - \frac{m\omega_b^2(x-x_b)^2}{2\mathcal{E}_\hbar}\right]}. \tag{7.6}$$

So, inserting Eq. (7.5) into Eq. (7.4) leads to the following stationary equation for $\mathcal{W}' \equiv \mathcal{W}'(x,p)$:

$$-\frac{p}{m}\frac{\partial \mathcal{W}'}{\partial x} - [2\gamma p + m\omega_b^2(x - x_b)]\frac{\partial \mathcal{W}'}{\partial p} + 2\gamma m\mathcal{E}_\hbar \frac{\partial^2 \mathcal{W}'}{\partial p^2} = 0. \tag{7.7}$$

Upon changing the variable, i.e., $\xi = p - a(x - x_b)$, we arrive at

$$\frac{d^2 \mathcal{W}'}{d\xi^2} = -\mathcal{B}\xi \frac{d\mathcal{W}'}{d\xi}, \tag{7.8}$$

with

$$B = \frac{1}{2\gamma m \mathcal{E}_\hbar}\left(\frac{a}{m} - 2\gamma\right). \tag{7.8a}$$

The solution of Eq. (7.8) is given by

$$\mathcal{W}'(\xi) = \sqrt{\frac{B}{2\pi}} \int_{-\infty}^{\xi} e^{-\frac{B}{2}\xi^2} d\xi, \tag{7.9}$$

provided that

$$a = m\gamma + m\sqrt{\gamma^2 + \omega_b^2}. \tag{7.10}$$

Making use of the function (7.5) with Eqs. (7.6) and (7.9), the probability current $J_b = (1/m)\int_{-\infty}^{\infty} \mathcal{W}(x = x_b, p, t) p\, dp$ reads

$$J_b = \frac{\mathcal{E}_\hbar}{\omega_b}\left(\sqrt{\gamma^2 + \omega_b^2} - \gamma\right) e^{\frac{-V(x_b)}{\mathcal{E}_\hbar}}, \tag{7.11}$$

whereas from the steady function (7.3) at the bottom well $x_a$, whereby $V(x) \sim V(x_a) + (m\omega_a^2/2)(x - x_a)^2$, i.e.,

$$W_{st}(x,p) = e^{\frac{-V(x_a)}{\mathcal{E}_\hbar}} e^{-\left[\frac{p^2}{2m\mathcal{E}_\hbar} + \frac{m\omega_a^2(x-x_a)^2}{2\mathcal{E}_\hbar}\right]}, \tag{7.12}$$

the number of particles, given by $N_a = \int_{-\infty}^{\infty}\int_{-\infty}^{\infty} W_{st}(x,p)\,dxdp$, is calculated as

$$N_a = \frac{2\pi\mathcal{E}_\hbar}{\omega_a} e^{\frac{-V(x_a)}{\mathcal{E}_\hbar}}. \tag{7.13}$$

On taking both quantities (7.11) and (7.13) into account, we find the Markovian quantum Kramers rate, $\Gamma_\hbar = J_b/N_a$, as

$$\Gamma_\hbar = \frac{\omega_a}{2\pi\omega_b}\left(\sqrt{\gamma^2 + \omega_b^2} - \gamma\right) e^{\frac{-[V(x_b)-V(x_a)]}{\mathcal{E}_\hbar}}. \tag{7.14}$$

For weak friction $\gamma \ll \omega_b$, we obtain from Eq. (7.14)

$$\Gamma_\hbar = \frac{\omega_a}{2\pi} e^{\frac{-[V(x_b)-V(x_a)]}{\mathcal{E}_\hbar}}, \quad \gamma \ll \omega_b, \tag{7.14a}$$

whereas in the strong friction case, i.e., $\gamma \gg \omega_b$, the quantum Kramers rates (7.14) becomes

$$\Gamma_\hbar = \frac{\omega_a \omega_b}{2\pi\gamma} e^{\frac{-[V(x_b)-V(x_a)]}{\mathcal{E}_\hbar}}, \quad \gamma \gg \omega_b. \qquad (7.14b)$$

As an example of non-thermal system let us consider $\mathcal{E}_\hbar = \hbar\gamma$. The quantum Kramers rate (7.14) then reads

$$\Gamma_\hbar = \frac{\omega_a}{2\pi\omega_b}\left(\sqrt{\gamma^2 + \omega_b^2} - \gamma\right) e^{\frac{-[V(x_b)-V(x_a)]}{\hbar\gamma}}. \qquad (7.15)$$

Notice that effects of the Planck constant $\hbar$ on the quantum escape rate turn up only through the quantum diffusion energy $\mathcal{E}_\hbar$ present in the exponential factor in Eqs. (7.14) and (7.15).

The quantum rate (7.15) does vanish in the classical limit $\hbar \to 0$, meaning that it is a pure quantum effect without any classical analogue. Yet, if it is expected that $\mathcal{E}_\hbar$ turns out to be given by $\mathcal{E}$ as $\hbar \to 0$, then the following classical rates are obtained from Eq. (7.14)

$$\Gamma = \frac{\omega_a}{2\pi\omega_b}\left(\sqrt{\gamma^2 + \omega_b^2} - \gamma\right) e^{\frac{-[V(x_b)-V(x_a)]}{\mathcal{E}}}, \qquad (7.16)$$

$$\Gamma = \frac{\omega_a}{2\pi} e^{\frac{-[V(x_b)-V(x_a)]}{\mathcal{E}}}, \quad \gamma \ll \omega_b, \qquad (7.16a)$$

and

$$\Gamma = \frac{\omega_a \omega_b}{2\pi\gamma} e^{\frac{-[V(x_b)-V(x_a)]}{\mathcal{E}}}, \quad \gamma \gg \omega_b. \qquad (7.16b)$$

For thermal systems, i.e., $\mathcal{E} = k_B T$, Eqs. (7.16) are the classical escape rates found out by Kramers [217].

### Thermal systems at zero temperature

*(a) Heat bath of quantum harmonic oscillators.* As far as a thermal reservoir of quantum harmonic oscillators is concerned the Brownian particle's quantum diffusion energy $\mathcal{E}_\hbar$ is given by Eq. (3.7), i.e., $\mathcal{E}_\hbar = (\omega_a \hbar/2) \coth(\omega_a \hbar/2k_B T)$, the angular frequency being assumed to be the same as $\omega_a$, inside the potential well around the point $a$. The quantum Kramers rate (7.14) then reads

$$\Gamma_\hbar = \frac{\omega_a}{2\pi\omega_b}\left(\sqrt{\gamma^2 + \omega_b^2} - \gamma\right) e^{\frac{-2[V(x_b)-V(x_a)]}{\omega_a \hbar \coth\left(\frac{\omega_a \hbar}{2k_B T}\right)}}, \qquad (7.17)$$

becoming at $T = 0$

$$\Gamma_\hbar = \frac{\omega_a}{2\pi\omega_b}\left(\sqrt{\gamma^2 + \omega_b^2} - \gamma\right) e^{\frac{-2[V(x_b)-V(x_a)]}{\omega_a \hbar}}, \qquad T = 0. \qquad (7.17a)$$

For the weak friction case, we find

$$\Gamma_\hbar = \frac{\omega_a}{2\pi} e^{\frac{-2[V(x_b)-V(x_a)]}{\omega_a \hbar \coth\left(\frac{\omega_a \hbar}{2k_B T}\right)}}, \qquad \gamma \ll \omega_b, \qquad (7.18)$$

which reduces at zero temperature to

$$\Gamma_\hbar = \frac{\omega_a}{2\pi} e^{\frac{-2[V(x_b)-V(x_a)]}{\omega_a \hbar}}, \qquad T = 0. \qquad (7.18a)$$

Notice that the quantum rates (7.18) take place in the absence of friction. In contrast, in the strong friction regime we obtain

$$\Gamma_\hbar = \frac{\omega_a \omega_b}{2\pi\gamma} e^{\frac{-2[V(x_b)-V(x_a)]}{\omega_a \hbar \coth\left(\frac{\omega_a \hbar}{2k_B T}\right)}}, \qquad \gamma \gg \omega_b, \qquad (7.19)$$

and

$$\Gamma_\hbar = \frac{\omega_a \omega_b}{2\pi\gamma} e^{\frac{-2[V(x_b)-V(x_a)]}{\omega_a \hbar}}, \qquad T = 0. \qquad (7.19a)$$

Results (7.17) and (7.18) were found out by Faria et al. in the context of stochastic electrodynamics [417].

Upshots (7.17), (7.18), and (7.19) should be contrasted with Chadhuri et al.'s quantum Kramers rates [418] which are not valid at $T = 0$, since they are calculated in the semiclassical regime of quantum mechanics. Moreover, their semiclassical effects on quantum Kramers rate vanish for harmonic potential. In other words, such a semiclassical escape rates rely on the nonlinearity of the cubic potential.

*(b) Fermionic heat bath.* On the condition that the quantum Brownian motion takes place in presence of a fermionic heat bath, the quantum Kramers rate (7.14) at low temperatures reads

$$\Gamma_\hbar = \frac{\omega_a}{2\pi\omega_b}\left(\sqrt{\gamma^2 + \omega_b^2} - \gamma\right) e^{\frac{-5[V(x_b)-V(x_a)]}{3k_B T_F\left(1+\frac{5\pi^2 T^2}{12\,T_F^2}\right)}}, \qquad (7.19)$$

where we have used the quantum diffusion energy $\mathcal{E}_\hbar$ given by Eq. (3.8a). At $T = 0$, quantum effects on Eq. (7.19) turn out to be determined by the Fermi energy $\epsilon_F = k_B T_F = (\hbar^2/2m)(6\pi^2/g)^{2/3}(N/V)^{2/3}$, i.e.,

$$\Gamma_\hbar = \frac{\omega_a}{2\pi\omega_b}\left(\sqrt{\gamma^2 + \omega_b^2} - \gamma\right)e^{\frac{-5[V(x_b)-V(x_a)]}{3\epsilon_F}}, \quad T = 0. \quad (7.19a)$$

*(c) Bosonic heat bath.* If the quantum Brownian particle is immersed in an ideal boson gas, the quantum Kramers rate (7.14) is then given by

$$\Gamma_\hbar = \frac{\omega_a}{2\pi\omega_b}\left(\sqrt{\gamma^2 + \omega_b^2} - \gamma\right)e^{\frac{-[V(x_b)-V(x_a)]}{0.77k_B\left(\sqrt{T^5/T_{BE}^5}+T_{BE}\sqrt{1-(T^3/T_{BE}^3)}\right)}}, \quad (7.20)$$

where we have used the Deeney-O'Leary energy as the quantum diffusion energy, i.e., $\mathcal{E}_\hbar = 0.77k_B\left(\sqrt{T^5/T_{BE}^5} + T_{BE}\sqrt{1-(T^3/T_{BE}^3)}\right)$, valid for $T \leq T_{BE}$. At zero temperature, Eq. (7.20) reduces to

$$\Gamma_\hbar = \frac{\omega_a}{2\pi\omega_b}\left(\sqrt{\gamma^2 + \omega_b^2} - \gamma\right)e^{\frac{-[V(x_b)-V(x_a)]}{0.77k_B T_{BE}}}, \quad T = 0, \quad (7.20a)$$

$T_{BE}$ being the Bose-Einstein temperature $T_{BE} = g(2\pi\hbar^2/mk_B)(1/2.612)^{2/3}(N/V^{2/3}$.

At very high temperatures the three kinds of quantum thermal systems leads to $\mathcal{E}_\hbar \propto \mathcal{E} = k_B T$. The quantum Kramers escape rate is therefore given by expression (7.16).

### 7.1.2. Non-Markovian effects

*Non-thermal systems.* In order to investigate non-Markov effects on quantum escape rate, we suppose the non-steady function $\mathcal{W}(x,p,t)$ to be factorized as

$$\mathcal{W}(x,p,t) = e^{2\gamma t}\mathcal{W}(x,p), \quad t < \infty, \quad (7.21)$$

so that the equation of motion (7.2) turns out to be time-independent

$$-\frac{p}{m}\frac{\partial \mathcal{W}}{\partial x} + \left[\frac{dV(x)}{dx}\frac{\partial}{\partial p} + \hat{\mathcal{O}}\right]\mathcal{W} + 2\gamma p\frac{\partial \mathcal{W}}{\partial p} + 2\gamma m\mathcal{E}_\hbar I(\alpha)\frac{\partial^2 \mathcal{W}}{\partial p^2} = 0, \quad (7.22)$$

$\hat{\mathcal{O}}$ being given by Eq.(7.2a). The constant $I(\alpha)$ is determined by the expression (see Appendix D)

$$I(\alpha) = \lambda\alpha^{\lambda-1} + 1 - e^{-\alpha}, \qquad (7.22a)$$

where $\alpha = t/t_c$ is a dimensionless parameter accounting for non-Markov effects and $\lambda$ a parameter associated with the anomalous diffusion.

Expanding the potential function $V(x)$ around the barrier at point $x_b$, i.e., $V(x) \sim V(x_b) - (m\omega_b^2/2)(x - x_b)^2$, and introducing the variable $\xi = p - a(x - x_b)$, Eq. (7.22) becomes

$$\frac{d^2\mathcal{W}(\xi)}{d\xi^2} = -\mathcal{A}\xi\frac{d\mathcal{W}(\xi)}{d\xi}, \qquad (7.23)$$

whose solution is

$$\mathcal{W}(\xi) = \sqrt{\frac{\mathcal{A}}{2\pi}} \int_{-\infty}^{\xi} e^{-\frac{\mathcal{A}}{2}\xi^2} d\xi, \qquad (7.24)$$

with

$$\mathcal{A} = \frac{\gamma + \sqrt{\gamma^2 + \omega_b^2}}{2\gamma m\mathcal{E}_\hbar I(\alpha)}. \qquad (7.25)$$

Making use of the non-steady function (7.21), with Eq. (7.24), we construct the following time-dependent probability distribution function

$$W(x,p,t) = e^{2\gamma t}\sqrt{\frac{\mathcal{A}}{2\pi}} e^{\frac{-1}{\mathcal{E}_\hbar}\left[\frac{p^2}{2m}+V(x_b)-\frac{m\omega_b^2}{2}(x-x_b)^2\right]} \int_{-\infty}^{\xi=p-a(x-x_b)} e^{-\frac{\mathcal{A}}{2}\xi^2} d\xi, \qquad (7.26)$$

where the stationary function (7.6) appears at $t = 0$ and $\xi \to \infty$. Inserting Eq. (7.26) into the probability current $J_b = (1/m)\int_{-\infty}^{\infty} W(x = x_b, p, t) p\, dp$ leads to

$$J_b = e^{2\gamma t}\mathcal{E}_\hbar e^{\frac{-V(x_b)}{\mathcal{E}_\hbar}}\sqrt{\frac{m\mathcal{E}_\hbar\mathcal{A}}{1 + m\mathcal{E}_\hbar\mathcal{A}}}. \qquad (7.27)$$

On the other hand, the number of particles at the bottom well $x_a$, i.e., $N_a = \int_{-\infty}^{\infty}\int_{-\infty}^{\infty} W_{st}(x,p)\,dxdp$, where $W_{st}(x,p)$ is given by Eq. (7.12), reads

$$N_a = \frac{2\pi\mathcal{E}_\hbar}{\omega_a} e^{\frac{-V(x_a)}{\mathcal{E}_\hbar}}. \qquad (7.28)$$

Consequently, the non-Markovian quantum Kramers escape rate $\Gamma_\hbar(t) = J_b(t)/N_a$ becomes

$$\Gamma_\hbar(t) = \Gamma_\hbar(0) e^{2\gamma t} \sqrt{\frac{\gamma + \sqrt{\gamma^2 + \omega_b^2}}{2\gamma I(\alpha) + \gamma + \sqrt{\gamma^2 + \omega_b^2}}}, \quad t < \infty, \quad (7.29)$$

where $\Gamma(0)$ is the Markovian friction-independent quantum rate at $t = 0$:

$$\Gamma_\hbar(0) = \frac{\omega_a}{2\pi} e^{\frac{-[V(x_b)-V(x_a)]}{\mathcal{E}_\hbar}}, \quad t = 0. \quad (7.29a)$$

Non-Markov effects on the quantum escape rate (7.29) are due to the constant $I(\alpha)$, given by Eq. (7.22a), for $\alpha > 0$. In the case $\lambda = 0$, i.e., $I(\alpha) = 1 - e^{-\alpha}$, the non-Markovian rate (7.29) corresponds to the normal quantum diffusion, whereas for $\lambda > 1$ it is related to the anomalous quantum diffusion.

## Thermal systems

*(a) Heat bath of quantum harmonic oscillators.* For a thermal reservoir of quantum harmonic oscillators the time-dependent quantum escape rate (7.29) is expressed in terms of $\mathcal{E}_\hbar = (\omega_a \hbar/2) \coth(\omega_a \hbar/2 k_B T)$, which at zero temperature becomes

$$\Gamma_\hbar(t) = \frac{\omega_a}{2\pi} e^{2\gamma t} \sqrt{\frac{\gamma + \sqrt{\gamma^2 + \omega_b^2}}{2\gamma I(\alpha) + \gamma + \sqrt{\gamma^2 + \omega_b^2}}} e^{\frac{-2[V(x_b)-V(x_a)]}{\omega_a \hbar}}, \quad T = 0. \quad (7.30)$$

*(b) Fermionic heat bath.* In the case of a fermionic heat bath Eq. (7.29) is given in terms of $\mathcal{E}_\hbar = (3/5) k_B T_F [1 + (5\pi^2/12)(T^2/T_F^2)]$. At $T = 0$, the fermionic Kramers rate reads

$$\Gamma_\hbar(t) = \frac{\omega_a}{2\pi} e^{2\gamma t} \sqrt{\frac{\gamma + \sqrt{\gamma^2 + \omega_b^2}}{2\gamma I(\alpha) + \gamma + \sqrt{\gamma^2 + \omega_b^2}}} e^{\frac{-5[V(x_b)-V(x_a)]}{3 k_B T_F}}, \quad T = 0. \quad (7.31)$$

*(c) Bosonic heat bath.* As the quantum Brownian particle is immersed in an ideal boson gas at low temperatures, the quantum diffusion energy is $\mathcal{E}_\hbar = 0.77 k_B \left(\sqrt{T^5/T_{BE}^5} + T_{BE}\sqrt{1 - (T^3/T_{BE}^3)}\right)$, valid for $T \leq T_{BE}$. At zero temperature the quantum Kramers rate (7.29) turns out to be expressed as

$$\Gamma_\hbar(t) = \frac{\omega_a}{2\pi} e^{2\gamma t} \sqrt{\frac{\gamma + \sqrt{\gamma^2 + \omega_b^2}}{2\gamma I(\alpha) + \gamma + \sqrt{\gamma^2 + \omega_b^2}}} e^{\frac{-[V(x_b)-V(x_a)]}{0.77 k_B T_{BE}}}, \quad T = 0. \quad (7.32)$$

At high temperatures, $\mathcal{E}_\hbar \propto k_B T$, the non-Markovian quantum escape rate (7.29), with (7.29a), reduces to the non-Markovian classical escape rate given by (2.140).

## 7.2. Quantum tunneling in the absence of inertial force

### 7.2.1. Markovian effects

*Non-thermal systems.* We start from the quantum Smoluchowski equation (5.7) in the steady regime on the potential barrier assumed to be approximately harmonic, i.e., $V(x) \sim V(x_b) - (m\omega_b^2/2)(x - x_b)^2$,

$$\frac{-\omega_b^2(x - x_b)}{\gamma}\frac{\partial W}{\partial x} + \frac{\omega_b^2 p}{\gamma}\frac{\partial W}{\partial p} + \frac{\mathcal{E}_\hbar}{m\gamma}\frac{\partial^2 W}{\partial x^2} - \left[\frac{\omega_b^2}{\gamma} + \frac{4\mathcal{E}_\hbar}{m\gamma\hbar^2}p^2\right]W = 0, \quad (7.33)$$

whose equilibrium steady solution may be expressed as (see Eq. (5.57))

$$\mathcal{W}_{st}(x,p) = e^{\frac{-V(x_b)}{\mathcal{E}_\hbar}} e^{-\left[\frac{2\mathcal{E}_\hbar p^2}{m\omega_b^2 \hbar^2} - \frac{m\omega_b^2(x-x_b)^2}{2\mathcal{E}_\hbar}\right]}. \quad (7.34)$$

With the help of Eq. (7.34) we now introduce the non-equilibrium steady function

$$\mathcal{W}(x,p) = \mathcal{W}'(x,p)\mathcal{W}_{st}(x,p), \quad (7.35)$$

so that Eq. (7.33) changes into

$$\frac{\omega_b^2(x - x_b)}{\gamma}\frac{\partial \mathcal{W}'(x,p)}{\partial x} + \frac{\omega_b^2 p}{\gamma}\frac{\partial \mathcal{W}'(x,p)}{\partial p} + \frac{\mathcal{E}_\hbar}{m\gamma}\frac{\partial^2 \mathcal{W}'(x,p)}{\partial x^2} = 0. \quad (7.36)$$

Changing the variable

$$\xi = a(x - x_b) - p \quad (7.37)$$

leads to

$$\frac{d^2 \mathcal{W}'(\xi)}{d\xi^2} = -\frac{m\omega_b^2}{a^2 \mathcal{E}_\hbar}\xi \frac{d\mathcal{W}'(\xi)}{d\xi} \quad (7.38)$$

the solution of which is

$$\mathcal{W}'(\xi) = \frac{\omega_b}{a}\sqrt{\frac{m}{2\pi\mathcal{E}_\hbar}}\int_{-\infty}^{\xi} e^{\frac{-m\omega_b^2\xi^2}{2a^2\mathcal{E}_\hbar}}\,d\xi. \tag{7.39}$$

So the non-equilibrium function (7.35), with (7.34) and (7.39), reads

$$\mathcal{W}(x,p) = \sqrt{\frac{m\omega_b^2}{2\pi a^2 \mathcal{E}_\hbar}}\, e^{\frac{-V(x_b)}{\mathcal{E}_\hbar}} e^{-\left[\frac{2\mathcal{E}_\hbar p^2}{m\omega_b^2\hbar^2} - \frac{m\omega_b^2(x-x_b)^2}{2\mathcal{E}_\hbar}\right]} \int_{-\infty}^{\xi} e^{\frac{-m\omega_b^2\xi^2}{2a^2\mathcal{E}_\hbar}}\,d\xi. \tag{7.40}$$

The probability current $J_b$ on the potential barrier is given by

$$J_b = -\frac{\omega_b^4 \hbar^3 m}{4\mathcal{E}_\hbar\sqrt{m^2\omega_b^4\hbar^2 - 4a^2\mathcal{E}_\hbar^2}}\, e^{\frac{-V(x_b)}{\mathcal{E}_\hbar}}, \tag{7.41}$$

while the number of particles $n_a$ at bottom well is

$$n_a = \hbar\pi e^{\frac{-V(x_a)}{\mathcal{E}_\hbar}} \tag{7.42}$$

whose calculation is based on the equilibrium function

$$W_{eq}(x,p) = e^{\frac{-V(x_a)}{\mathcal{E}_\hbar}} e^{-\left[\frac{2\mathcal{E}_\hbar p^2}{m\omega_a^2\hbar^2} + \frac{m\omega_a^2(x-x_a)^2}{2\mathcal{E}_\hbar}\right]}. \tag{7.43}$$

Accordingly, the quantum Kramers escape rate becomes

$$\Gamma_\hbar = \frac{|J_b|}{n_a} = \frac{\omega_b^4 m \hbar^2}{4\pi\mathcal{E}_\hbar}\, \frac{e^{\frac{-[V(x_b)-V(x_a)]}{\mathcal{E}_\hbar}}}{\sqrt{m^2\omega_b^4\hbar^2 - 4a^2\mathcal{E}_\hbar^2}}. \tag{7.44}$$

Notice that the undetermined parameter $a$ introduced in the linear transformation (7.37) displays dimension of mass per time and satisfies the condition

$$0 \leq a < \frac{m\hbar\omega_b^2}{2\mathcal{E}_\hbar}. \tag{7.44a}$$

For $a = 0$, the quantum escape rate (7.44) reads

$$\Gamma_\hbar = \frac{\hbar\omega_b^2}{4\pi\mathcal{E}_\hbar}\, e^{\frac{-[V(x_b)-V(x_a)]}{\mathcal{E}_\hbar}}. \tag{7.45}$$

In the case of a non-thermal system, given, for example, by $\mathcal{E}_\hbar = \hbar\gamma$, the escape rates (7.44) and (7.45) read respectively

$$\Gamma_\hbar = \frac{\omega_b^4 m}{4\pi\gamma} \frac{e^{\frac{-[V(x_b)-V(x_a)]}{\hbar\gamma}}}{\sqrt{m^2\omega_b^4 - 4a^2\gamma^2}}, \tag{7.46}$$

and

$$\Gamma_\hbar = \frac{\omega_b^2}{4\pi\gamma} e^{\frac{-[V(x_b)-V(x_a)]}{\hbar\gamma}}. \tag{7.47}$$

Therefore, quantum tunneling takes place in the presence of friction. Yet, in the case of thermal systems for which $\mathcal{E}_\hbar$ does not depend on friction, Eq. (7.45) predicts a frictionless tunneling, a kind of a superfluidity phenomenon. Superfluidity also occurs if $a = m\omega_b/2$, for instance, is put in Eq. (7.44). So we have

$$\Gamma_\hbar = \frac{\omega_b^3 \hbar^2}{4\pi\mathcal{E}_\hbar} \frac{e^{\frac{-[V(x_b)-V(x_a)]}{\mathcal{E}_\hbar}}}{\sqrt{\omega_b^2 \hbar^2 - \mathcal{E}_\hbar^2}}, \tag{7.48}$$

on the condition that $\mathcal{E}_\hbar < \hbar\omega_b$.

**Thermal systems**

*(a) Heat bath of quantum harmonic oscillators.* For a thermal reservoir of quantum harmonic oscillators the non-dissipative quantum Kramers escape rates (7.45) and (7.48) are expressed in terms of $\mathcal{E}_\hbar = (\omega_a \hbar/2) \coth(\omega_a \hbar/2k_B T)$. Eq. (7.45) is then valid for all temperatures $T \geq 0$, whereas Eq. (7.48) holds valid at low temperatures $\coth(\omega_a \hbar/2k_B T) < 2$ alone, including zero temperature.

At zero temperature, Eq. (7.45) becomes

$$\Gamma_\hbar = \frac{1}{2\pi} \frac{\omega_b^2}{\omega_a} e^{\frac{-2[V(x_b)-V(x_a)]}{\omega_a \hbar}}, \quad T = 0, \tag{7.49}$$

while the escape rate (7.48) at zero temperature reads

$$\Gamma_\hbar = \frac{\omega_b^3}{\pi\omega_a\sqrt{4\omega_b^2 - \omega_a^2}} e^{\frac{-2[V(x_b)-V(x_a)]}{\omega_a \hbar}}, \quad T = 0. \tag{7.50}$$

*(b) Fermionic heat bath.* On the condition that the quantum Brownian motion takes place in presence of a fermionic heat bath, the quantum Kramers rate (7.45) at low temperatures reads

$$\Gamma_\hbar = \frac{5\hbar\omega_b^2}{12\pi k_B T_\text{F}\left(1+\frac{5\pi^2}{12}\frac{T^2}{T_\text{F}^2}\right)} e^{\frac{-5[V(x_b)-V(x_a)]}{3k_B T_\text{F}\left(1+\frac{5\pi^2 T^2}{12\,T_\text{F}^2}\right)}}. \tag{7.51}$$

which reduces at $T=0$ to

$$\Gamma_\hbar = \frac{5\hbar\omega_b^2}{12\pi k_B T_\text{F}} e^{\frac{-5[V(x_b)-V(x_a)]}{3k_B T_\text{F}}}, \quad T=0. \tag{7.52}$$

The quantum escape rate (7.48) turns out to be

$$\Gamma_\hbar = \frac{5\omega_b^3\hbar^2}{12\pi k_B T_\text{F}\left(1+\frac{5\pi^2}{12}\frac{T^2}{T_\text{F}^2}\right)} \frac{e^{\frac{-5[V(x_b)-V(x_a)]}{3k_B T_\text{F}\left(1+\frac{5\pi^2 T^2}{12\,T_\text{F}^2}\right)}}}{\sqrt{\omega_b^2\hbar^2 - \left[\frac{3}{5}k_B T_\text{F}\left(1+\frac{5\pi^2}{12}\frac{T^2}{T_\text{F}^2}\right)\right]^2}} \tag{7.53}$$

which becomes at $T=0$

$$\Gamma_\hbar = \frac{5\omega_b^3\hbar^2}{12\pi k_B T_\text{F}} \frac{e^{\frac{-5[V(x_b)-V(x_a)]}{3k_B T_\text{F}}}}{\sqrt{\omega_b^2\hbar^2 - \left(\frac{3}{5}k_B T_\text{F}\right)^2}}, \quad T=0, \tag{7.54}$$

on the condition that $\omega_b > 3k_B T_\text{F}/5\hbar$.

*(c)Bosonic heat bath.* If the quantum Brownian particle is immersed in an ideal boson gas, the dissipative quantum Kramers rate (7.45) is expressed in terms of $\mathcal{E}_\hbar = 0.77 k_B \left(\sqrt{(T/T_\text{BE})^5} + T_\text{BE}\sqrt{1-(T/T_\text{BE})^3}\right)$, which at $T=0$ it becomes

$$\Gamma_\hbar = \frac{\hbar\omega_b^2}{3.08\pi k_B T_\text{BE}} e^{\frac{-[V(x_b)-V(x_a)]}{0.77 k_B T_\text{BE}}}, \quad T=0, \tag{7.55}$$

The non-dissipative quantum Kramers rate (7.48) at zero temperature reads

$$\Gamma_\hbar = \frac{\omega_b^3\hbar^2}{3.08\pi k_B T_\text{BE}} \frac{e^{\frac{-[V(x_b)-V(x_a)]}{0.77 k_B T_\text{BE}}}}{\sqrt{\omega_b^2\hbar^2 - (0.77 k_B T_\text{BE})^2}}, \quad T=0. \tag{7.56}$$

on the condition that $\omega_b > 0.77 k_B T_\text{BE}/\hbar$.

At very high temperatures, the quantum rate (7.45) turns out to be given by

$$\Gamma_\hbar = \frac{\hbar\omega_b^2}{4\pi k_B T} e^{\frac{-[V(x_b)-V(x_a)]}{k_B T}}, \qquad T \to \infty. \tag{7.57}$$

Therefore, for thermal environments the quantum tunneling process happens in the absence of damping for all temperatures $T \geq 0$. These findings corroborate our previous result found in Ref. [188] according to which frictionless quantum tunneling has been predicted to occur at low temperatures, including $T = 0$.

### 7.2.2. Non-Markovian effects

*Non-thermal systems.* If we assume the parameter $a$ in Eq. (7.44) to be given by $a = m/2t_c$, $t_c > 0$ being a sort of correlation time, then the following non-Markovian quantum Kramers escape rate can be obtained

$$\Gamma_\hbar = \frac{t_c \omega_b^4 \hbar^2}{4\pi \mathcal{E}_\hbar} \frac{e^{\frac{-[V(x_b)-V(x_a)]}{\mathcal{E}_\hbar}}}{\sqrt{t_c^2 \omega_b^4 \hbar^2 - \mathcal{E}_\hbar^2}}, \tag{7.58}$$

provided that

$$t_c > \frac{\mathcal{E}_\hbar}{\hbar \omega_b^2}. \tag{7.58a}$$

In the case of a non-thermal environment characterized by $\mathcal{E}_\hbar = \hbar\gamma$, Eq. (7.58) reads

$$\Gamma_\hbar = \frac{t_c \omega_b^4}{4\pi\gamma} \frac{e^{\frac{-[V(x_b)-V(x_a)]}{\hbar\gamma}}}{\sqrt{t_c^2 \omega_b^4 - \gamma^2}}, \tag{7.59}$$

valid for $t_c > \gamma/\omega_b^2$.

*Thermal systems.* The quantum Kramers escape rate (7.58) in the presence of a thermal reservoir comprised of a set of harmonic oscillators is given by

$$\Gamma_\hbar = \frac{t_c \omega_b^4 \hbar^2}{4\pi \left(\frac{\omega_a \hbar}{2}\right) \coth\left(\frac{\omega_a \hbar}{2k_B T}\right)} \frac{e^{\frac{-[V(x_b)-V(x_a)]}{\left(\frac{\omega_a \hbar}{2}\right) \coth\left(\frac{\omega_a \hbar}{2k_B T}\right)}}}{\sqrt{t_c^2 \omega_b^4 \hbar^2 - \left[\left(\frac{\omega_a \hbar}{2}\right) \coth\left(\frac{\omega_a \hbar}{2k_B T}\right)\right]^2}}, \tag{7.60}$$

the correlation time $t_c$ satisfying the validity condition

$$t_c > \frac{\omega_a}{2\omega_b^2}\coth\left(\frac{\omega_a\hbar}{2k_BT}\right). \qquad (7.60a)$$

At zero temperature Eq. (7.60) reduces to

$$\Gamma_\hbar = \frac{t_c\omega_b^4\, e^{\frac{-2[V(x_b)-V(x_a)]}{\omega_a\hbar}}}{\pi\omega_a\sqrt{4t_c^2\omega_b^4 - \omega_a^2}}, \qquad T = 0, \qquad (7.61)$$

provided that $t_c > \omega_a/2\omega_b^2$.

The quantum escape rate (7.58) in the presence of a fermionic heat bath at low temperature is

$$\Gamma_\hbar = \frac{5t_c\omega_b^4\hbar^2}{12\pi k_B T_F\left(1+\frac{5\pi^2}{12}\frac{T^2}{T_F^2}\right)}\frac{e^{\frac{-5[V(x_b)-V(x_a)]}{3k_BT_F\left(1+\frac{5\pi^2 T^2}{12\,T_F^2}\right)}}}{\sqrt{t_c^2\omega_b^4\hbar^2 - \left[\frac{3}{5}k_BT_F\left(1+\frac{5\pi^2}{12}\frac{T^2}{T_F^2}\right)\right]^2}}, \qquad (7.62)$$

At $T = 0$, it becomes

$$\Gamma_\hbar = \frac{5t_c\omega_b^4\hbar^2}{12\pi k_B T_F}\frac{e^{\frac{-5[V(x_b)-V(x_a)]}{3k_BT_F}}}{\sqrt{t_c^2\omega_b^4\hbar^2 - \left(\frac{3}{5}k_BT_F\right)^2}}, \qquad T = 0, \qquad (7.62a)$$

valid for $t_c > 3k_BT_F/5\hbar\omega_b^2$.

The quantum escape rate (7.58) in the presence of a bosonic heat bath at zero temperature reads

$$\Gamma_\hbar = \frac{t_c\omega_b^4\hbar^2}{3.08\pi k_B T_{BE}}\frac{e^{\frac{-[V(x_b)-V(x_a)]}{0.77 k_B T_{BE}}}}{\sqrt{t_c^2\omega_b^4\hbar^2 - (0.77 k_B T_{BE})^2}}, \qquad T = 0, \qquad (7.63)$$

valid for $t_c > 0.77 k_B T_{BE}/\hbar\omega_b^2$.

At high temperatures, Eq. (7.58) becomes

$$\Gamma_\hbar = \frac{t_c\omega_b^4\hbar^2}{4\pi k_B T}\frac{e^{\frac{-[V(x_b)-V(x_a)]}{k_B T}}}{\sqrt{t_c^2\omega_b^4\hbar^2 - (k_B T)^2}}, \qquad T \to \infty, \qquad (7.64)$$

valid for $t_c > k_BT/\hbar\omega_b^2$,

In summary, as far as thermal systems are concerned the non-Markovian quantum rate (7.58) also predicts frictionless quantum tunneling at zero and high temperatures.

## 7.3. Summary and discussion

Our quantum Fokker-Planck equation (7.2) predicts dissipative quantum tunneling in the presence of inertial force for thermal systems at low temperatures, including zero temperature, reckoning with non-Markov and Markov effects on the quantum Brownian particle.

In contrast, as far as quantum thermal environments (a heat bath of quantum harmonic oscillators as well as bosonic and fermionic baths) are concerned, our quantum Smoluchowski (7.33) predicts dissipationless quantum tunnelling in the absence of inertial force for all $T \geq 0$, taking into account both non-Markov and Markov effects. Such a result corroborates our previous findings in Ref. [188], missunderstood nevertheless by Ankerhold et al. [419].

# 8. Summary and outlook

As pointed out in Chapter 1 of this work, realistic physical systems are never found isolated from its environment. Since Galileo and Newton nevertheless the idea of isolated system became the most fundamental metaphysical concept underlying physics [3]. That is, only by isolating a given physical system from the world surrounding it, neglecting and abstracting some features taken as secondary, such as air resistance or friction, can we come up with the true mathematical laws behind the motion of bodies. Newton's laws are a well-known example of such a process of mathematizing physical processes. Experimentally, such mathematical laws are never exactly born out, for physical bodies generally move in media where some kind of friction is present. The major breakthrough is that in the ground of such a mathematical idealization physical systems are subject to a experimentation whose control leads to the approximate verification or confirmation of the exact mathematical laws. Thus on the basis of the concept of isolated system, Galileo and Newton established the mathematical-experimental foundation of modern physics.

Besides being deterministic, isolated systems are conservative, meaning that the forces acting on them comply with the principle of conservation of the mechanical energy. Accordingly, it is claimed that in experimental situations, for instance, non-isolated systems (e.g., dissipative systems) arise from our inability or ignorance about the details of all conservative forces acting on the system.

Moreover, a quite relevant development from a formal point of view was that conservative Newtonian systems could be alternatively described by Hamilton's equations. Historically, the Hamiltonian formalism was the starting point for quantizing classical systems. So, the Schrödinger equation could come into being.

Afterwards, the Hamiltonian methodology was extended to investigating non-isolated systems. First, in 1941 Caldirola [420] attempted to quantizing dissipative system without fluctuations within a Hamiltonian structure. Later, the first attempts to study the quantum physics of an open system, that is, a dissipative system with fluctuations (e.g., a quantum Brownian particle), were carried out by Prigogine, Toda [421,423], Magalinskii [277], and George [160] via the canonical quantization following also the tenets of the Hamiltonian formalism, according to which the quantum Brownian motion of a particle appears to result from a total Hamiltonian function $H_T$, assumed to be separated as $H_T = H_S + H_E + H_I$, where $H_S$ is associated with the open system (the Brownian particle), $H_E$ with the environment, and $H_I$ with the interaction between them. It is worth highlighting that as a consequence of the Hamiltonian approach to quantum open systems both phenomena of dissipation and fluctuation do not ensue as intrinsic physical properties. According to Ankerhold [385], it is claimed that

> *the underlying idea of this separation [of the total Hamiltonian] is that dissipation is not something intrinsic, but something that is born out of our ignorance to focus on a small subsystem only*.

The Hamiltonian methodology revealed to be a powerful mathematical superstructure encompassing classical and quantum mechanics of both isolated and open systems. In our monograph [187], the process of mathematizing physical systems via the Hamiltonian methodology we have called it the *Hamiltonization of physics*. By contrast, our non-Hamiltonian approach to (classical and quantum) open systems developed elsewhere [181-191] and reviewed in the present work assumes that both dissipation and fluctuation are physical processes intrinsic to a Brownian particle. In contrast to Ankerhold's above quotation, it is therefore utterly unacceptable to uphold the view that our ignorance can account for any physical phenomena, such as quantum Brownian motion.

Our non-Hamiltonian methodology for open systems could be illustrated by the following pictorial scheme:

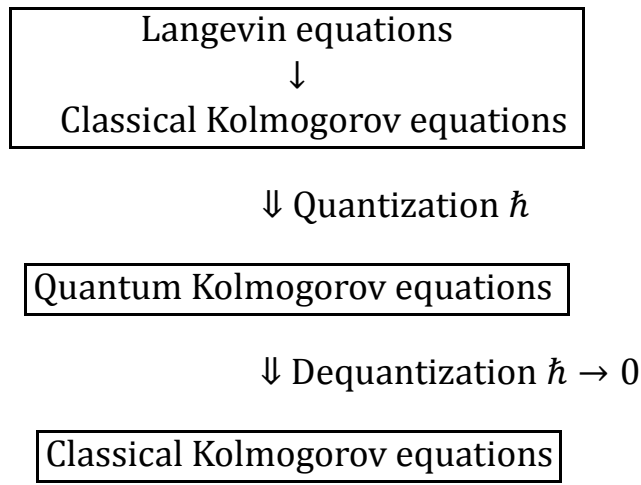

We start from Langevin equations giving rise to the corresponding classical Kolmogorov equations that display non-Markovian, non-Gaussian and non-linear effects on a Brownian particle. Kolmogorov's equations reduce to non-Markovian Fokker-Planck equations as far as the Gaussian approximation is concerned. Next, we quantize classical Brownian motion by bringing Planck's constant $\hbar$ directly in Kolmogorov equations through a linear change of coordinates. This Hamiltonian-independent quantization method underpins our dynamical-quantization approach to quantum open systems. So we arrive at the quantum Kolmogorov equations exhibiting non-Markovian, non-Gaussian and non-linear effects on a Brownian particle immersed in a generic environment. The classical limit or the dequantization $\hbar \to 0$ of the quantum Brownian motion is taken into account, too.

We now turn to summarize the main epistemological implications underlying our non-Hamiltonian methodology, pointing out the physical meaning of our dynamical-quantization approach to quantum open systems.

Initially, in Chapter 2 we have worked on our non-Markovian approach to normal and anomalous Brownian motion in the classical domain. Here our main finding has been that non-Markov effects are naturally introduced owing to the existence of a correlational function $I(t)$ that renders the diffusion coefficient time-dependent. We have shown that non-Markov effects account for the differentiability property of the Brownian trajectories, so implying the existence of two physical quantities: A force and an instantaneous velocity, that is, the derivative of the root mean square momentum and the root mean square displacement, respectively. We emphasize that we have predicted the breakdown of the energy equipartition of statistical mechanics at short times in the following physical situations:

- In the presence of inertial forces in momentum space via the existence of the concept of thermal force, given by Eq. (2.47);
- In the absence of inertial forces in configuration space through our instantaneous velocity $\mathbb{V}(t) \sim \sqrt{k_B T/m\gamma t_c}$, which is valid for a free particle, Eq. (2.78), and a harmonic oscillator, Eq. (2.94);
- In the non-inertial anomalous diffusion of a free particle for the cases in which $\lambda \geq 2$, on account of the existence of the instantaneous velocities (2.104), as well as in the anomalous diffusion of a harmonic oscillator as described by the mean square displacements (2.106b), (2.107b), and (2.108b);
- In the inertial anomalous diffusion for $\lambda = 2, 3, 4$ due to the existence of the forces (2.116).

In Chapter 3 quantum effects were investigated by directly quantizing the stochastic classical dynamics described by Fokker-Planck equations. On the basis of this dynamical-quantization approach to quantum open systems, we have derived a non-Markovian quantum master equation, given by Eq. (3.5), which is valid for both thermal and non-thermal environments. In the case of thermal systems quantum Brownian motion may occur in presence of a heat bath of quantum harmonic oscillators, of bosons, or of fermions for any temperatures $T \geq 0$.

Chapter 4 dealt with the quantum Brownian motion of a free particle according to both Fokker-Planck and Langevin descriptions. For thermal systems both descriptions lead to the violation of the energy equipartition theorem of statistical mechanics in the following cases:

- At zero temperature in the steady regime as $t \to \infty$ via Eqs. (4.20), (4.22), and (4.24), for instance;
- At zero temperature for all time $t \geq 0$ through Eqs. (4.46b), (4.47b), and (4.49);
- At high temperatures via Eq. (4.46a) or Eq. (4.85b) for times of the order of the quantum time $t_q = \hbar/2k_B T$, a kind of correlation time according to Eq. (4.40).

Such theoretical predictions could be experimentally confirmed through the measurement of both the quantum force and the quantum instantaneous velocity whose existence proves that the quantum Brownian trajectories are differentiable.

Chapter 5 in turn addressed quantum Brownian motion in the absence of inertial force. A non-Markovian quantum Smoluchowski equation in phase space, Eq. (5.7), has been derived and solved for a free particle and a harmonic oscillator in the case of both thermal and non-thermal systems. For general environments the differentiability of the quantum trajectories leads to the concepts of quantum instantaneous velocity and quantum force. For thermal systems it was predicted that the equipartition theorem can be violated for all times $t \geq 0$ and $T \geq 0$.

Chapter 6 approached quantum anomalous diffusion of a free Brownian particle in presence and absence of inertial force for thermal and non-thermal baths. On the one hand, in presence of inertial force, for fermionic and bosonic heat baths the anomalous behavior of the mean square displacement takes place classically, that is, it is $\hbar$-independent at zero temperature. In contrast, at high temperatures it is the mean square momentum that behaves classically. On the other hand, in absence of inertial force our non-Markovian Smoluchowski equation has predicted quantum effects on the mean square displacement for both fermionic and bosonic heat baths at zero temperature. In this case, it is the mean square momentum that behaves classically. Yet, at high temperatures displacement fluctuations are classical while momentum fluctuations are under quantum effects.

Lastly, in Chapter 7 on the basis of our non-Markovian quantum Smoluchowski equation we have predicted a kind of dissipationless quantum tunneling in the following cases: For all $T \geq 0$ and in the low-temperatures regime, including $T = 0$.

From a physical point of view we believe that our theoretical predictions can be subject to experimental confirmation, thereby going far beyond the predictions based on the Hamiltonian methodology.

The ontological implications of our dynamical-quantization approach to quantum open systems can be summed up as follows.

- Our contribution to open system theory has been to provide an ontological status to quantum Brownian motion. That is, quantum motion *is* really Brownian. Deterministic quantum motion is only a special case.
- Our dynamical-quantization approach to quantum open systems has revealed that the von Neumann function is more fundamental than the Schrödinger function for describing quantum physical systems, for open systems are not described in terms of $\psi$. As long as the environment can be neglected we obtain an isolated system as a special case to be described by the Schrödinger equation. Even in this case, the Schrödinger function $\psi$ should play a secondary role in the theoretical framework of quantum physics.
- As a consequence of Galileo's and Newton's establishment of the mathematical-experimental foundation of modern science based on the concept of isolated system, probability in classical physics appears to be a concept of epistemic nature only. It is common to associate it with our ignorance or inaccuracy to specify or predict the initial conditions of a system of many isolated particles, such as in statistical mechanics[34] [423,424]. In quantum physics probability also displays an epistemological character, for it is associated with statistical records according to Born's probabilistic rule. In contrast, underlying our non-Hamiltonian methodology for quantum open systems, the probability concept shows up as an ontological feature characterizing Brownian motion and fades out as far as the environment can be left out.
- Idealizing mathematically physical phenomena as open systems allows for abandoning the concept of isolated system in accordance with some suggestions made by Bohr and Heisenberg as well as unveiling the ontological status of physical systems without resorting to any measurement apparatus in accord with Einstein's philosophical viewpoint. In addition, our dynamical-quantization approach to quantum open systems makes some theoretical predictions that could be experimentally confirmed.

---

[34]In contrast, Einstein imagined the task of physics as being to justify the epistemologically objective importance of the probability concept from a theory constructed ontologically on the grounds of a theory of isolated systems. The ergodic theory, in which statistical properties are meant to be derived from the dynamics of isolated systems, was a model for Einstein [197,210].

- According to our non-Hamiltonian quantization method, the theory of quantum Brownian motion can be ontologically formulated without appealing in any way to a dualistic framework of matter (wave vs. particle).
- Both classical and quantum Brownian movements are differentiable, thereby implying that the concept of trajectory exists as an intrinsic or ontological attribute for both quantum and classical particles. In other words, quantum Brownian particles as well as Schrödinger particles actually follow a certain quantum trajectory.

Lastly, we would like to mention some forthcoming works.

- As a future work we would like to extend our non-Hamiltonian approach to Brownian motion to the relativistic regime. Our previous articles on bosons and fermions in relativistic quantum phase space [425,426] as well as the recent review paper on relativistic Brownian motion [242] will be our starting points.
- More recently, quantum information and quantum computation are guided by the dream of isolating and controlling quantum-mechanical systems against environmental influences so as to make use of some properties, such as the superposition principle and entanglement [159,371]. Taming quantum Brownian motion may be a precondition for attaining such a goal.
- To investigate no-Gaussian effects on the quantum Brownian motion on the basis of our quantum Kolmogorov equations (3.4) and (5.4).
- The concept of quantum force given by the differentiability of the root mean square momentum could be looked upon as a physical criterion for measuring non-Markovianity. It is worth comparing this criterion with those existing in the literature on quantum open systems (see Ref. [190], and references therein).

# Appendix A. The Einstein approach to Markovian Brownian motion

From the physical viewpoint, the first mathematization of Brownian motion was achieved by Einstein in 1905 [194] by building up a theoretical model in which the environment (the liquid particles, for instance) acts on the Brownian particle in a probabilistic fashion. In this Einstein picture the position or displacement of the Brownian particle is represented by a stochastic variable $X \equiv X(t)$ whose realizations $x \equiv x(t)$ are arranged according to a determined probability distribution function $\mathcal{F}(x,t) \geq 0$, the time parameter $t \geq 0$ being a non-random quantity. Then Einstein came up with a time evolution for $\mathcal{F}(x,t)$ describing the possible positions $x$ of a (non-inertial) free Brownian particle immersed in a generic fluid (the so-called Fick diffusion equation[35])

$$\frac{\partial \mathcal{F}(x,t)}{\partial t} = D \frac{\partial^2 \mathcal{F}(x,t)}{\partial x^2}, \qquad (A.1)$$

where $D \geq 0$ with dimensions of $[\text{length}^2 \times \text{time}^{-1}]$ is a non-random constant responsible for diffusing the probability density function $\mathcal{F}(x,t)$ in configuration space. Hence, $D$ is called diffusion constant.

The random displacement $X(t)$ of the Brownian particle is physically characterized by its root mean square (RMS)

$$\mathbb{X}(t) = \sqrt{\langle X^2(t) \rangle - \langle X(t) \rangle^2}, \qquad (A.2)$$

which measures the fluctuation of $X(t)$ about its average value $\langle X(t) \rangle$ calculated with the aid of the solution $\mathcal{F}(x,t)$ to Eq. (A.1) according to

$$\langle X(t) \rangle = \int_{-\infty}^{\infty} x(t) \mathcal{F}(x,t) dx. \qquad (A.3)$$

The quantity

$$\langle X^2(t) \rangle = \int_{-\infty}^{\infty} x^2(t) \mathcal{F}(x,t) dx \qquad (A.4)$$

in Eq. (A.2) is called the mean square displacement (MSD).

---

[35]In 1855, Fick [427] posited the diffusion equation for the particle concentration in analogy to Fourier's law for heat conduction [211,428,429].

*Deriving the diffusion equation.* In order to derive the diffusion equation (A.1), Einstein [194] posited the motion of the Brownian particle as a translation transformation from the point $x_1 = x + \Delta x$ at $t_1 = t$ to the point $x_2 = x$ at $t_2 = t + \tau$. That movement evolves at a distance

$$|\Delta x| = |x_1 - x_2| \tag{A.5}$$

in a time interval

$$\tau = t_2 - t_1, \tag{A.6}$$

through the following integral equation[36]

$$\mathcal{F}(x, t + \tau) = \int_{-\infty}^{\infty} \mathcal{F}(x + \Delta x, t)\varphi(\Delta x)d(\Delta x), \tag{A.7}$$

where the time-independent function $\varphi(\Delta x)$ accounting for the transition from $\mathcal{F}(x_1, t_1) \equiv \mathcal{F}(x + \Delta x, t)$ to $\mathcal{F}(x_2, t_2) \equiv \mathcal{F}(x, t + \tau)$ complies with the normalization condition

$$\int_{-\infty}^{\infty} \varphi(\Delta x)d(\Delta x) = 1, \tag{A.8}$$

as far as $\mathcal{F}(x, t + \tau) = \mathcal{F}(x + \eta, t) = 1$ is concerned, as well as the symmetry property (parity)

$$\varphi(\Delta x) = \varphi(-\Delta x) \tag{A.9}$$

introduced due to the concept of distance (A.5).

Assuming both distribution functions $\mathcal{F}(x, t + \tau)$ and $\mathcal{F}(x + \Delta x, t)$ in Eq. (A.7) to be expanded in a Taylor series about $\tau$ and $\Delta x$, respectively, one obtains[37]

$$\sum_{k=0}^{\infty} \frac{\tau^k}{k!} \frac{\partial^k \mathcal{F}(x,t)}{\partial t^k} = \int_{-\infty}^{\infty} \sum_{k=0}^{\infty} \frac{(\Delta x)^k}{k!} \frac{\partial^k \mathcal{F}(x,t)}{\partial \eta^k} \varphi(\eta)d\eta. \tag{A.10}$$

Taking into account the following infinitesimality conditions

---

[36]An integral equation like Eq. (A.7) was originally put forward by Louis Bachelier in 1900 [205] in the context of the theory of financial speculation [211,431]. Such an integral equations are also known as Chapman-Kolmogorov equations [120,207,220,229,240,431].

[37]The term on the right side of Eq. (A.10) is known as the Kramers-Moyal expansion [120,207,220,229,240,431], albeit Kramers [217] himself had employed explicitly no expansion at all. In truth, it was Einstein [194] who implicitly just introduced such an expansion.

$$\tau^3 \ll 1 \tag{A.11}$$

and[38]

$$|\Delta x|^3 = |x_1 - x_2|^3 \ll 1, \tag{A.12}$$

Eq. (A.10) can be truncated in the form of the generalized diffusion equation[39] (the so-called telegraph equation)

$$\frac{\tau}{2}\frac{\partial^2 \mathcal{F}(x,t)}{\partial t^2} + \frac{\partial \mathcal{F}(x,t)}{\partial t} = D(\tau)\frac{\partial^2 \mathcal{F}(x,t)}{\partial x^2}, \tag{A.13}$$

after formally identifying the diffusion coefficient $D(\tau)$ in the following way

$$D(\tau) \equiv \frac{\overline{(\Delta X)^2}}{2\tau}, \tag{A.14}$$

where the mean square of the stochastic variable $\Delta X \equiv X_2 - X_1$ is given by

$$\overline{(\Delta X)^2} = \int_{-\infty}^{\infty} (\Delta x)^2 \varphi(\Delta x) d(\Delta x). \tag{A.15}$$

As $\tau$ approaches zero, i.e., $\tau \to 0$, the telegraph equation (A.13) does reduce to the Fickian diffusion equation (A.1), $D(\tau)$ becoming the $\tau$-independent diffusion constant[40] $D$ (see also Ref. [207], p. 55)

$$D \equiv \frac{1}{2}\lim_{\tau \to 0}\frac{\overline{(\Delta X)^2}}{\tau}. \tag{A.16}$$

According to Einstein [194], the time parameter $\tau$, present in the integral equation (A.7) and then used for deriving the diffusion equation (A.1), is to be interpreted as a time interval "*very small with respect of a given time interval of observation, but large enough to ensure that the motions of the Brownian particle at*

---

[38]Einstein [194] used the parity property (A.9) for justifying partly condition (A.12) which indeed is related to the Gaussian property for truncating the right side of Eq. (A.10). Yet, he did not provide any physical reason for discarding non-Gaussian contributions. Later, Pawula (see [227,228]) proved that there exists no non-Gaussian approximation to the so-called Kramers-Moyal expansion complying with the positivity of the probability distribution function.

[39]In his 1905 paper on Brownian motion, Einstein did not write down explicitly the telegraph equation (A.13).

[40]Einstein [194], in fact, identified inappropriately $D$ with the expression $(1/2\tau)\int_{-\infty}^{\infty}\eta^2\varphi(\eta)d\eta$, thereby overlooking its $\tau$-dependence. Deriving the diffusion equation (A.1) from the telegraph equation (A.14) seems therefore to eschew that Einstein's slight oversight!

*two consecutive intervals of time τ can be treated as independent of each other*"[41]. That Einstein's operational statement as well as the time-independent diffusion constant (A.16) seem to have been the first seeds of Markovianity assumption underlying the physical interpretation of the elusive parameter $\tau$ posited by Einstein in Eq. (A.7)[42].

In addition, the picture of Brownian motion based on the Bachelier-Einstein integral equation (A.7) along with both Gaussianity conditions (A.11) and (A.12) had been employed by Einstein himself [199], Smoluchowski [432,433], Fokker [434], Planck [435], Fürth [231,436], Ornstein [214], Klein [216], Chapmann [437], Uhlenbeck and Ornstein [218], Kolmogorov [196,202], Kramers [217], Chandrasekhar [193], Wang and Uhlembeck [219] and many others, as starting point for describing Brownian motion in several physical contexts thus generalizing the diffusion equation (A.1) by means of the so-called Fokker-Planck equations [120,149,207,220,229,240].

Moreover, Markovianity property ingrained in the integral equation (A.7) has been deemed to be the pivotal assumption underpinning the theory of Brownian motion [120,149,207,220,229,240,439].

*The diffusion constant.* The physics underlying the Brownian movement turns up in determining the transport coefficient $D$ in the diffusion equation (A.1). On the ground of both the van't Hoff law for the osmotic pressure exerted by the Brownian particle on a thermal reservoir (a liquid, for instance) and the Stokes law for the mobility of the particle, Einstein [194,196,210,211,439-442] derived the following expression for the diffusion constant[43]

$$D = \frac{k_B T}{\beta m}, \qquad (A.17)$$

---

[41]"*Wir führen ein Zeitinterval τ in die Betrachtung ein, welches sehr klein sei gegen die beochactbaren Zeitintervalle, aber noch so gross, dass die in zwei aufeinanderfolgenden Zeitintervallen τ von einem Teilchen ausgeführten Bewegungen als voneinander unabhängige auflassen sind.*" [194]. The physical reason underlying the infinitesimal character of $\tau$ had been quite a controversial issue: Fürth [442], for instance, stressed that the introduction of such a time interval $\tau$ is the weak point of Einstein's theory of Brownian motion because there is no theoretical justification for the independence of the motion in this interval (see also [187,211,246,429,439,440,443]).

[42]According to Chandrasekhar [193], p. 31, the integral equation (A.7) *per se* sets up a Markovianity condition in sense that the Brownian particle "*depends only on the instantaneous values of its physical parameters and is entirely independent of its whole previous history*".

[43]Sutherland [206] also derived Eq. (A.17), hence it is dubbed the Sutherland-Einstein diffusion constant [210,211]. Eq. (A.17) is also known as the Stokes-Einstein diffusion constant [207,429].

whereby the thermal energy $\mathcal{E} \equiv k_B T$, responsible for activating the Brownian particle through an osmotic force, comes from the heat bath at thermodynamic equilibrium characterized by the Boltzmann constant $k_B$ and the temperature $T$, whereas the viscosity coefficient $\alpha \equiv \beta m$ arises from the Stokes viscous force accounting for damping the motion of the Brownian particle of mass $m$ at a relaxation timescale $t_r \equiv \beta^{-1}$, where $\beta$ is the frictional constant.

*Solving the diffusion equation.* Starting from the deterministic initial condition $\mathcal{F}(x, t = 0) = \delta(x)$ and using the fluctuation-dissipation relation (A.17), the solution to Eq. (A.1) reads

$$\mathcal{F}(x,t) = \sqrt{\frac{m\beta}{4\pi k_B T t}} e^{\frac{-m\beta x^2}{4 k_B T t}}, \tag{A.18}$$

providing both the average displacement

$$\langle X(t) \rangle = 0 \tag{A.19}$$

and the mean square displacement (MSD)

$$\langle X^2(t) \rangle = \frac{2 k_B T}{m\beta} t. \tag{A.20}$$

The position fluctuation of the free Brownian particle is then represented by the following Einstein's root mean square displacement (RMSD)

$$\mathbb{X}(t) = \sqrt{\frac{2 k_B T}{m\beta} t}, \tag{A.21}$$

the differentiability of which yields the instantaneous velocity $\mathbb{V}(t) \equiv d\mathbb{X}(t)/dt$

$$\mathbb{V}(t) = \sqrt{\frac{k_B T}{2 m \beta t}}, \tag{A.22}$$

which in turn blows up[44] at short times $t \to 0$. In other words, the displacement fluctuation $\mathbb{X}(t)$ is not a differentiable quantity at $t = 0$. Einstein accounted for such a feature resorting to the so-called Markovian assumption [199]:

---

[44]Such singular short-time behavior of the instantaneous velocity (A.22) is operationally explained by Einstein himself as follows: "*Since an observer operating with definite means of observation in a definite manner can never perceive the actual path traversed in an arbitrarily small time, a certain mean velocity will always appear to him as an [infinite] instantaneous velocity. But it is clear that the velocity ascertained*

*We have implicitly assumed in our development that the [stochastic] processes during the time t are to be considered as phenomena independent of the processes in the time immediately preceding. Yet this assumption renders inapplicable as far as increasingly tiny times t are concerned.*[45]

In brief, because the position fluctuation (A. 21) is non-differentiable at $t = 0$ the concept of instantaneous velocity of a non-inertial free particle cannot exist in the Einstein picture of Brownian motion. Only Eq. (A. 21) could be expected to have physical significance.

On the ground of the Einsteinian mathematization of Brownian motion, Perrin [444,445] developed experiments to verify the Einstein's MSD law, Eq. (A. 20),[46] pointing out a close analogy between the reality of atoms and the non-analytic character of the Brownian trajectories (curves without tangents) [211,439]:

*One may be tempted to define an 'average velocity of agitation' by following a particle as accurately as possible. But such evaluations are* grossly wrong. *The apparent average velocity varies crazily in magnitude and direction. It is easy to see that in practice the notion of tangent is meaningless for such curves.* (Perrin [444], Mandelbrot's translation [446], p. 12.)

Moreover, Perrin emphasized that

*It is curves with derivatives that are now the exceptions; or, if one prefers the geometrical language, curves with no tangent at any point become the rule, while the familiar regular curves become some kind of curiosities, doubtless interesting, but still very special*" (Perrin in Brush [439], p. 32).

Conceptually, the mathematical property of the non-analyticity of Einstein's Brownian paths, given by Eq. (A. 21), or rather, the non-existence of the instantaneous velocity (A. 22), raises the question about the very physical reality of the random trajectory of a Brownian particle. Put it differently, do Brownian paths actually exist in nature or do they play merely an elusive or heuristic role useful to assess the atomic nature of matter, for instance?[47]

---

*thus corresponds to no objective property of the motion under investigation – at least, if the theory corresponds to the facts*" [200].

[45]"*[weil] wir in unserer Entwicklung implizite angenohmen haben, dass der Vorgang während der Zeit als von dem Vorgänge in den unmitttelbar vorangehenden Zeiten unabhängiges Ereignis aufzufassen sei. Diese Annahme trifft aber um so weniger zu, jener kleiner die Zeiten t gewählt werden*" [199].

[46]On the basis of Eq. (A. 20) Perrin [439,444,445] was able to measure the Boltzmann constant (or Avogadro's number) by investigating the motion of Brownian particles with radius of the order of $10^{-6}$ m, thereby upholding the evidence for the hypothesis of the atomic structure of matter. In 1926, Perrin was awarded the Nobel Prize for his experimental works [447].

Lastly, even though not pointed out by Einstein himself, it is worth stressing that multiplying $\mathbb{X}(t)$ by $\mathbb{V}(t)$ results in a physically well defined quantity, namely, the diffusion coefficient $\mathbb{D}(t) \equiv \mathbb{X}(t)\mathbb{V}(t)$ that in turn is in general a time dependent quantity. In the case of Eqs. (A.21) and (A.22), nevertheless, $\mathbb{D}(t)$ becomes the diffusion constant (A.17), albeit $\mathbb{V}(t)$ is not well-defined at $t = 0$.

---

[47]Although the Brownian motion of a particle immersed in a thermal reservoir allows for calculating the Boltzmann constant, thereby unveiling the atomic structure intrinsic to matter, the very concept of non-differentiable trajectory does not seem to be an ontological feature underlying the Brownian movement: For Chandrasekhar, *"The perpetual motion of the Brownian particles is maintained by fluctuations in the collisions with the molecules of the surrounding fluid. Under normal conditions, in a liquid, a Brownian particle will suffer about $10^{21}$ collisions per second and this is so frequent that we cannot really speak of separate collisions. Also, since each collision can be thought of as producing a kink in the path of the particle, it follows that we cannot hope to follow the path in any detail—indeed, to our senses the details of the path are impossible fine"* [193]. Moreover, according to von Plato *"the motion observable through a microscope is not the same as the true motion of the particle. For it experiences in normal circumstances up to some $10^{20}$ collisions a second, so that one sees average effects of theses collisions"* [197], p. 128.

# Appendix B. The Langevin approach to Markovian Brownian motion

Rather than focusing on the time evolution of probability distribution functions, the Langevin approach [195] to Brownian movement deals with the dynamics of random variables [193,233,448-455]. The position $X(t)$ of a Brownian particle of mass $m$ is assumed to be a solution, for example, of the stochastic differential equation (the so-called Langevin equation [120,149,207,220,229,240,448])

$$m\frac{d^2 X(t)}{dt^2} = -\frac{dV(X)}{dX} - \beta m \frac{dX(t)}{dt} + b\Psi(t), \quad (B.1)$$

in which the term $md^2 X(t)/dt^2$ denotes an inertial force[48] offsetting a conservative force, $F_c \equiv -dV(X)/dX$, derived from the Brownian particle's potential energy $V \equiv V(X)$, and two kinds of environmental forces: a linearly velocity-dependent dissipative force, $F_d \equiv -\beta m dX/dt$, and a fluctuating force, $L(t) \equiv b\Psi(t)$, dubbed Langevin's force.

*The Langevin method* [195]. For the case of a free particle, $V(X) = 0$, multiplying Eq. (B.1) by $X(t)$ and averaging the resulting equation over the joint probability distribution function $F_{X\Psi}(x, \psi, t)$ lead to

$$\frac{m}{2}\frac{d^2 \langle X^2(t) \rangle}{dt^2} = m\langle v^2(t) \rangle + \frac{\beta m}{2}\frac{d\langle X^2(t) \rangle}{dt} + b\langle X(t)\Psi(t) \rangle, \quad (B.2)$$

$v(t)$ being the stochastic velocity $v(t) \equiv dX(t)/dt$ of the Brownian particle. Supposing the Brownian particle's kinetic energy $m\langle v^2(t) \rangle/2$ to obey the energy equipartition for all time $t \geq 0$, i.e.,

$$\frac{m\langle v^2(t) \rangle}{2} = \frac{k_B T}{2}, \quad (B.3)$$

where $T$ denotes the medium temperature, it can be shown that $X(t)$ and $\Psi(t)$ are statistically uncorrelated, i.e., $\langle X(t)\Psi(t) \rangle = \langle X(t)\rangle\langle\Psi(t)\rangle = 0$ [213]. So, one obtains from Eq. (B.2) the following differential equation for the mean square displacement $\langle X^2(t) \rangle$

$$m\frac{d^2 \langle X^2(t) \rangle}{dt^2} = -\beta m \frac{d\langle X^2(t) \rangle}{dt} + 2k_B T. \quad (B.4)$$

---

[48]In contrast to Einstein's approach to Brownian motion [194], the Langevin one [195] reckons with inertial effects on Brownian motion due to the term $md^2 X(t)/dt^2$ in Eq. (B.1).

In short, Langevin's method[49] [195] for describing the free Brownian motion consists in changing the stochastic differential equation (B.1) for the random variable $X(t)$ into the deterministic differential equation (B.4) for the mean square displacement $\langle X^2(t) \rangle$, because $X(t)$ and the Langevin force $L(t) = b\Psi(t)$ are uncorrelated[50]: $\langle X(t)\Psi(t) \rangle = 0$.

Solution to Eq. (B.4) is given by the mean square displacement[51]

$$\langle X^2(t) \rangle = \langle X^2(0) \rangle + \frac{2k_B T}{\beta m} t - \frac{C}{\beta}(e^{-\beta t} - 1), \tag{B.5}$$

$\langle X^2(0) \rangle$ denoting the mean square displacement at $t = 0$ and $C$ an integration constant. Assuming $\langle X(0) \rangle = 0$, it follows from Eq. (B.5) that the root mean square displacement reads

$$\mathbb{X}(t) = \sqrt{\langle X^2(0) \rangle + \frac{2k_B T}{\beta m} t - \frac{C}{\beta}(e^{-\beta t} - 1)}, \tag{B.6}$$

the derivative of which provides the instantaneous velocity

$$\mathbb{V}(t) = \frac{1}{2\mathbb{X}(t)} \left( \frac{2k_B T}{\beta m} + Ce^{-\beta t} \right) \tag{B.7}$$

that in turn leads to the time-dependent diffusion coefficient

$$\mathbb{D}(t) \equiv \mathbb{X}(t)\mathbb{V}(t) = \frac{k_B T}{\beta m} + \frac{C}{2} e^{-\beta t}. \tag{B.8}$$

Further, on the condition that Eq. (B.8) vanishes at $t = 0$, i.e., $\mathbb{D}(0) = 0$, one obtains

$$C = -\frac{2k_B T}{\beta m}. \tag{B.9}$$

---

[49]The Langevin method of converting the stochastic differential equation (B.1) for the random variable $X(t)$ into a deterministic differential equation for the mean square displacement $\langle X^2(t) \rangle$ is valid only for linear systems, i.e., for potentials of the form $V(X) = a_1 + a_2 X + a_3 X^2$, where $a_1$, $a_2$ and $a_3$ are non-random constants.

[50]Naqvi [213] has recently raised some objections against Langevin's assumption $\langle X(t)\Psi(t) \rangle = \langle X(t) \rangle \langle \Psi(t) \rangle = 0$.

[51] At $t \gg 1/\beta$, the mean square displacement (B.5) does approximate to Einstein's result given Eq. (A.20), i.e., $\langle X^2(t) \rangle \sim (2k_B T/\beta m)t$. That upshot allowed Langevin [195] to identify $k_B T/\beta m$ with the diffusion constant $D$.

Thus, according to the Langevin approach the Brownian dynamics of an inertial free particle turns out to be characterized by the Ornstein-Fürth RMSD[52] [214,231]

$$\mathbb{X}(t) = \sqrt{\langle X^2(0)\rangle + \frac{2k_B T}{\beta m}t + \frac{2k_B T}{\beta^2 m}(e^{-\beta t} - 1)}. \qquad (B.10)$$

At long timescales $t \to \infty$, i.e., $t \gg 1/\beta$, Eq. (B.10) leads to the Einstein diffusive regime $\mathbb{X}(t) \sim \sqrt{2(k_B T/\beta m)t}$, the instantaneous velocity (B.7) in turn approaches $\mathbb{V}(t) \sim \sqrt{(k_B T/\beta m)(1/2t)}$, whereas the diffusion coefficient (B.8), with Eq. (B.9), becomes the Sutherland-Einstein diffusion constant $\mathbb{D}(\infty) = (k_B T/\beta m)$.

On the other hand, at short times $t \to 0$, i.e., $t \ll 1/\beta$, assuming the initial condition to have no fluctuations, such that $\langle X^2(0)\rangle = \langle X(0)\rangle = 0$, the Ornstein-Fürth RMSD (B.10) renders ballistic, i.e.,

$$\mathbb{X}(t) \sim v_{\text{th}} t, \qquad (B.11)$$

the instantaneous velocity (B.7) in turn goes to

$$\mathbb{V}(t) \sim v_{\text{th}}, \qquad (B.12)$$

while the diffusion coefficient turns out to be

$$\mathbb{D}(t) \sim v_{\text{th}}^2 t, \qquad (B.13)$$

where $v_{\text{th}}$ denotes the thermal velocity $v_{\text{th}} = \sqrt{k_B T/m}$ coming from the energy equipartition theorem (B.3).

Since both the displacement (B.11) and the instantaneous velocity (B.12) do not depend on the friction coefficient $\beta$, the particle evolves for the span of time $0 \le t \ll 1/\beta$ before any collisions occur. Hence, the frictionless inertial free Brownian motion at short times is said to be ballistic, that is, the free Brownian particle moves in straight line with constant speed given by the equipartition energy, before collisions with bath particles slow it down and randomize its motion [456-460].

---

[52]Originally, but without making use of the condition $\mathbb{D}(0) = 0$, Ornstein [214] and Fürth [231], derived the mean square displacement, $\langle X^2(t)\rangle = \langle X^2(0)\rangle + (2k_B T/\beta m)t + (2k_B T/\beta^2 m)(e^{-\beta t} - 1)$ (see also Ref. [213]).

Experimental verifications of theoretical predictions based on the Ornstein-Fürth formula (B. 10) have been reported through the measurement of the transition from ballistic to diffusive regime of Brownian motion of micrometer-sized particles in rarified gases [460-463]. Moreover, Li et al. [232] have recently measured the instantaneous velocity of a Brownian particle held in air by an optical tweezer, thereby verifying through Eq. (B.12) the validity of the energy equipartition theorem at short times.

*The inertial free Brownian motion in momentum space.* The differentiability of the position fluctuation (B. 10) implies the existence of the stochastic velocity $dX(t)/dt$, or the stochastic momentum $P(t) = mdX(t)/dt$, present in the Langevin equation (B. 1). It remains to examine if there exists the acceleration $d^2X/dt^2$, or the random inertial force $md^2X/dt^2$. We now turn to answer that question within the Markovian Langevin framework.

Notice that the Langevin equation (B. 1), for $V = 0$, may be written down in terms of the stochastic momentum $P(t) = mdX(t)/dt$ as

$$\frac{dP(t)}{dt} = -\beta P(t) + b\Psi(t), \tag{B.14}$$

whose formal solution is the following

$$P(t) = P(0)e^{-\beta t} + b\int_0^t e^{-\beta(t-s)}\Psi(s)ds. \tag{B.15}$$

The mean square momentum then reads

$$\langle P^2(t)\rangle = \langle P^2(0)\rangle e^{-2\beta t} + b^2 e^{-2\beta t} \int_0^t\int_0^t e^{\beta(s+s')}\langle\Psi(s)\Psi(s')\rangle dsds'. \tag{B.16}$$

In addition, making use of the Markovian property

$$\langle\Psi(s)\Psi(s')\rangle = \delta(s-s'), \tag{B.17}$$

Eq. (B. 16) changes into

$$\langle P^2(t)\rangle = \langle P^2(0)\rangle e^{-2\beta t} + \frac{b^2}{2\beta}\left(1 - e^{-2\beta t}\right). \tag{B.18}$$

Moreover, assuming that the energy equipartition is valid in the steady regime, i.e., $\langle P^2(\infty)\rangle = mk_BT$, the following fluctuation-dissipation is obtained from Eq. (B. 18)

$$b = \sqrt{2\beta m k_B T}.\qquad(\mathrm{B.}\,19)$$

For sharp initial condition, i.e., $\langle P^2(0)\rangle = \langle P(0)\rangle = 0$, and $\langle \Psi(t)\rangle = 0$, the root mean square momentum $\mathbb{P}(t) \equiv \sqrt{\langle P^2(t)\rangle - \langle P(t)\rangle^2}$, with the upshot (B. 19), reads

$$\mathbb{P}(t) = \sqrt{m k_B T(1 - e^{-2\beta t})}.\qquad(\mathrm{B.}\,20)$$

It is readily to check that Eq. (B. 20) is nondifferentiable at $t \to 0$, so implying that the stochastic differential equation (B. 14), or the Langevin equation (B. 1), is devoid of any mathematical significance because the force $dP(t)/dt$, or the acceleration $d^2X(t)/dt^2$, cannot exist. To overcome this hurdle, Doob [220,229,233,240,438] in 1942 pointed out that the Langevin equation (B.14) in the form

$$dP(t) = -\beta P(t)dt + dB(t),\qquad(\mathrm{B.}\,21)$$

where $dB(t) = b\Psi(t)dt$, should be rigorously interpreted not as a differential equation but as the Wiener integral equation [229,451]

$$\int_{s=0}^{t} h(s)dP(s) = -\beta \int_{0}^{t} h(s)P(s)ds + \int_{s=0}^{t} h(s)dB(s),\qquad(\mathrm{B.}\,22)$$

where $h(s)$ is a continuous function[53].

Summarizing, in the context of the inertial Brownian motion the property of non-differentiability of Eq. (B. 20) at $t \to 0$ implies the mathematical inexistence of the concept of force $dP(t)/dt$, or acceleration $d^2X(t)/dt^2$, albeit the concept of velocity $dX(t)/dt$, or momentum $P(t) = mdX(t)/dt$, is mathematically well defined due to the differentiability of the Ornstein-Fürth process (B. 10).

---

[53]Similarly, as far as inertial effects are neglected in the Langevin equation (B. 1) the resulting stochastic differential equation $dX(t)/dt = (b/\beta m)\Psi(t)$ has no mathematical significance, unless it is interpreted as an integral equation according to Doob's interpretation based on Wiener integrals [220,229,233,240,451].

## Appendix C. Kolmogorov equations

In this Appendix we wish to show how a given stochastic differential equation gives rise to a Kolmogorov equation which in turn reduces to a Fokker-Planck equation in the Gaussian approximation. First, we take up the case of one random variable and then the case of two variables [192].

We suppose the dynamics of the stochastic process $\Phi = \Phi(t)$ to be governed by the ordinary differential equation

$$\frac{d\Phi(t)}{dt} = \mathcal{K}(\Phi, t). \tag{C.1}$$

To find out the time evolution of the probability distribution function $\mathcal{F}(\varphi, t)$ expressed in terms of the realizations $\varphi$ of the random variable $\Phi$, we closely follow Stratonovich's procedure [192,204]. We resort to the definition of conditional probability density given by

$$W(\varphi', t'|\varphi, t) = \frac{f(\varphi', t'; \varphi, t)}{\mathcal{F}(\varphi, t)}, \tag{C.2}$$

where $f(\varphi, t; \varphi', t')$ is the joint probability density function of $\Phi$ at different times $t$ and $t'$ with $\varphi' \equiv \varphi(t')$ and $\varphi \equiv \varphi(t)$.

From (C.2), we arrive at the Bachelier-Einstein integral equation [194,205]

$$\mathcal{F}(\varphi', t') = \int_{-\infty}^{\infty} W(\varphi', t'|\varphi, t)\mathcal{F}(\varphi, t)d\varphi, \tag{C.3}$$

after using the Kolmogorov compatibility condition [202]

$$\int_{-\infty}^{\infty} f(\varphi', t'; \varphi, t)d\varphi = \mathcal{F}(\varphi', t'). \tag{C.4}$$

The characteristic function for the increment $\Delta\Phi \equiv \Phi(t') - \Phi(t)$ is expressed in terms of the conditional probability density (C.2) as

$$\langle e^{iu\Delta\Phi} \rangle = \int_{-\infty}^{\infty} e^{iu\Delta\varphi} W(\varphi', t'|\varphi, t)d\varphi, \tag{C.5}$$

the inverse of which is

$$W(\varphi', t'|\varphi, t) = \frac{1}{2\pi} \int_{-\infty}^{\infty} e^{-iu\Delta\varphi} \langle e^{iu\Delta\Phi} \rangle du. \tag{C.6}$$

Now, using the expansion

$$\langle e^{iu\Delta\Phi} \rangle = \sum_{s=0}^{\infty} \frac{(iu)^s}{s!} \langle (\Delta\Phi)^s \rangle, \tag{C.7}$$

Eq. (C.6) turns out to be given by

$$W(\varphi', t'|\varphi, t) = \sum_{s=0}^{\infty} \frac{(-1)^s}{s!} \langle (\Delta\Phi)^s \rangle \frac{\partial^s}{\partial \varphi^s} \delta(\varphi - \varphi'). \tag{C.8}$$

Inserting (C.8) into (C.3) and dividing the resulting equation by $\epsilon$, we obtain

$$\frac{\mathcal{F}(\varphi', t') - \mathcal{F}(\varphi', t)}{\epsilon} = \sum_{s=1}^{\infty} \frac{(-1)^s}{s!} \frac{\partial^s}{\partial \varphi^s} \left\{ \frac{\langle (\Delta\Phi)^s \rangle}{\epsilon} \mathcal{F}(\varphi, t) \right\}, \tag{C.9}$$

with $t' = t + \epsilon$ and $\varphi' \equiv \varphi(t + \epsilon)$. The procedure of taking the limit $\epsilon \to 0$ in both sides of (C.9) leads to the Kolmogorov equation

$$\frac{\partial \mathcal{F}(\varphi, t)}{\partial t} = \mathbb{K}\mathcal{F}(\varphi, t), \tag{C.10}$$

where the Kolmogorovian operator $\mathbb{K}$ acts upon the probability distribution function $\mathcal{F}(\varphi, t)$ according to

$$\mathbb{K}\mathcal{F}(\varphi, t) = \sum_{s=1}^{\infty} \frac{(-1)^s}{s!} \frac{\partial^s}{\partial \varphi^s} \left[ B^{(s)}(\varphi, t) \mathcal{F}(\varphi, t) \right], \tag{C.11}$$

with the coefficients $B^{(s)}(\varphi, t)$, given by

$$B^{(s)}(\varphi, t) = \lim_{\epsilon \to 0} \frac{\langle (\Delta\Phi)^s \rangle}{\epsilon}, \tag{C.12}$$

calculated from the stochastic differential equation (C.1) in the following integral form

$$\Delta\Phi \equiv \Phi(t + \epsilon) - \Phi(t) = \int_{t}^{t+\epsilon} \mathcal{K}(\Phi, t') dt', \tag{C.13}$$

after averaging $(\Delta\Phi)^s$ over a given conditional probability density $W(\varphi', t'|\varphi, t)$ according to Eq. (C.5).

Summing up, we have set out a general scheme for deriving from the stochastic differential equation (C.1) the Kolmogorov equation (C.10) which reckons with non-Gaussian features on account of the presence of the $s$th moment of the increment $\Delta\Phi$, i.e., $\langle(\Delta\Phi)^s\rangle$.

According to Pawula's theorem [227,228], there exists no non-Gaussian approximation to the non-Gaussian Kolmogorov equation (C.10) in compliance with the positivity of $\mathcal{F}(\varphi, t)$. Hence, in the Gaussian approximation Eq. (C.10) reads

$$\frac{\partial \mathcal{F}_\Phi(\varphi, t)}{\partial t} = -\frac{\partial}{\partial \varphi}\left[B^{(1)}(\varphi, t)\mathcal{F}_\Phi(\varphi, t)\right] + \frac{1}{2}\frac{\partial^2}{\partial \varphi^2}\left[B^{(2)}(\varphi, t)\mathcal{F}_\Phi(\varphi, t)\right]. \quad (C.14)$$

In the physics literature, the class of Gaussian stochastic differential equation (C.1) is known as Langevin equation (e. g., Eqs. (2.49) and (2.61)), whereas the corresponding Gaussian Kolmogorov equation (C.14) is dubbed Fokker-Planck equation in configuration space or in momentum space, for example, the non-Markovian Smoluchowski equation (2.66) and the non-Markovian Rayleigh equation (2.37), respectively.

For the case of two random variables $\Pi$ and $\Phi$, the set of stochastic differential equations, given by

$$\frac{d\Pi(t)}{dt} = \mathcal{K}_1(\Pi, \Phi, t), \quad (C.15a)$$

$$\frac{d\Phi(t)}{dt} = \mathcal{K}_2(\Pi, \Phi, t), \quad (C.15b)$$

generates the following phase space Kolmogorov equation for the joint probability distribution $\mathcal{F}_{\Pi\Phi}(\pi, \varphi, t)$

$$\frac{\partial \mathcal{F}_{\Pi\Phi}(\pi, \varphi, t)}{\partial t} = \sum_{s=1}^{\infty}\sum_{r=0}^{s}\frac{(-1)^s}{r!(s-r)!}\frac{\partial^s}{\partial \pi^{s-r}\partial \varphi^r}\left[B^{(s-r,r)}(\pi, \varphi, t)\mathcal{F}_{\Pi\Phi}(\pi, \varphi, t)\right], (C.16)$$

with

$$B^{(s-r,r)}(\pi, \varphi, t) = \lim_{\epsilon \to 0}\left[\frac{\langle(\Delta\Pi)^{s-r}(\Delta\Phi)^r\rangle}{\epsilon}\right]. \quad (C.17)$$

The increments $\Delta\Pi$ and $\Delta\Phi$ are evaluated from (C.15a) and (C.15b) in the integral form

$$\Delta\Phi \equiv \Phi(t+\epsilon) - \Phi(t) = \int_{t}^{t+\epsilon} \mathcal{K}_1(\Pi,\Phi,t)dt \tag{C.18}$$

and

$$\Delta\Pi \equiv \Pi(t+\epsilon) - \Pi(t) = \int_{t}^{t+\epsilon} \mathcal{K}_2(\Pi,\Phi,t)dt. \tag{C.19}$$

In the Gaussian approximation the phase space Kolmogorov equation (C.16) reduces to the following Fokker-Planck equation in phase space

$$\frac{\partial \mathcal{F}_{\Pi\Phi}(\pi,\varphi,t)}{\partial t} = \sum_{s=1}^{2}\sum_{r=0}^{s}\frac{(-1)^s}{r!(s-r)!}\frac{\partial^s}{\partial \pi^{s-r}\partial \varphi^r}\left[B^{(s-r,r)}(\pi,\varphi,t)\mathcal{F}_{\Pi\Phi}(\pi,\varphi,t)\right]. \tag{C.20}$$

Both Langevin equations (2.7) in phase space are physical examples of equations of motion (C.15), while the non-Markovian Klein-Kramers equation (2.33) is a special case of phase space Fokker-Planck equation (C.20).

In Ref. [192] we have found out explicitly the coefficients $B^{(s)}(\varphi,t)$ in (C.14) for the cases of the non-Markovian Rayleigh equation (2.37) and the non-Markovian Smoluchowski equation (2.66), as well as the coefficients $B^{(s-r,r)}(\pi,\varphi,t)$ concerning the non-Markovian Klein-Kramers equation (2.33).

# Appendix D. The correlational function $I(t)$

The correlational function $I(t)$ is defined in Eq. (2.24) as

$$I(t) = \lim_{\epsilon \to 0} \frac{1}{\epsilon} \int_t^{t+\epsilon} \int_t^{t+\epsilon} \langle \Psi(t')\Psi(t'') \rangle dt'dt'' \geq 0. \tag{D.1}$$

If the autocorrelation function $\langle \Psi(t')\Psi(t'') \rangle$ is given by

$$\langle \Psi(t')\Psi(t'') \rangle = \left( \lambda \frac{t'^{\lambda-1}}{t_c^{\lambda-1}} + 1 - e^{\frac{-(t'+t'')}{2t_c}} \right) \delta(t' - t''), \tag{D.2}$$

where $t_c$ is deemed to be the correlation time of $\Psi(t)$ at times $t'$ and $t''$, then it follows that

$$I(t) = \lambda \frac{t^{\lambda-1}}{t_c^{\lambda-1}} + 1 - e^{\frac{-t}{t_c}}. \tag{D.3}$$

For $\lambda > 1$, Eq. (D.3) is defined at $t = 0$, i.e., $I(0) = 0$, while in the range $0 < \lambda < 1$ it is not defined at short times $t \to 0$.

For $\lambda = 0$, Eq. (D.3) becomes

$$I(t) = 1 - e^{\frac{-t}{t_c}} \tag{D.4}$$

whose Markovian regime is characterized by the steady behavior $I(\infty) = 1$ at long times $t \to \infty$, i.e., $t \gg t_c$, or by $I(t) = 1$ as $t_c \to 0$, i.e., $t \ll t_c$. In addition, the correlational function (D.4) vanishes at $t = 0$: $I(0) = 0$.

Eq. (D.3) displays the following asymptotic behavior at long times $t \to \infty$:

$$I(t) \sim \lambda \frac{t^{\lambda-1}}{t_c^{\lambda-1}}, \tag{D.5}$$

as long as $\lambda > 1$. The autocorrelation function $\langle \Psi(t')\Psi(t'') \rangle$, Eq. (D.2), for $\lambda = 0$, defines a sort of nonwhite noise which in the Markovian limit $t_c \to 0$ changes into the so-called white noise

$$\langle \Psi(t')\Psi(t'') \rangle = \delta(t' - t''), \tag{D.6}$$

meaning that the stochastic function $\Psi(t)$ is delta-correlated. It has been argued that the Markov property (D.4) is a highly idealized feature [187,220-224], because the physical interaction between the Brownian particle and the thermal bath actually takes places for a finite correlation time $t_c \neq 0$. Hence, our auto-correlation function (D.2) seems to be a more realistic feature underlying the statistical behavior of the Langevin stochastic force.

In addition, it is worth highlighting that the form of the autocorrelation function $\langle \Psi(t')\Psi(t'') \rangle$ is not a directly observed effect. Analytically, the functional form $I(t) = 1 - e^{-t/t_c}$, for instance, is suitable to solving the non-Markovian Fokker-Planck equations as well as examining non-Markovian effects on physically measurable quantities (see Chapter 2).

Moreover, it is worth stressing that imposing the condition $I(t) = 1$ implies Markovianity for all time $t$. Yet this special case turns up as a sufficient but not necessary condition for Markovianity property.

Therefore, we wish to point out that our non-Markovian approach to Brownian motion is based on the correlational function $I(t)$ the introduction of which naturally rises in the Fokker-Planck equations by being built from the corresponding Langevin equation. This fact has been overlooked by the centennial literature on Brownian motion theory.

# Appendix E. General solution of the non-Markovian diffusion equation

Let us consider the non-Markovian Smoluchowski equation (2.68) for a free Brownian particle, i.e.,

$$\frac{\partial \mathcal{F}(x,t)}{\partial t} = \frac{k_B T}{m\gamma} I(t) \frac{\partial^2 \mathcal{F}(x,t)}{\partial x^2}, \tag{E.1}$$

the solution of which reads

$$f(x,t) = \sqrt{\frac{\gamma m}{4\pi k_B T \mathcal{J}(t)}} e^{\frac{-\gamma m x^2}{4 k_B T \mathcal{J}(t)}}, \tag{E.2}$$

where the time-dependent function $\mathcal{J}(t)$ is given by

$$\mathcal{J}(t) = \int_0^t I(t) dt. \tag{E.3}$$

Solution (E.2) yields $\langle X(t) \rangle = 0$ and

$$\langle X^2(t) \rangle = \frac{2 k_B T}{\gamma m} \mathcal{J}(t). \tag{E.4}$$

So, the root mean square displacement, $\mathbb{X}(t) = \sqrt{\langle X^2(t) \rangle - \langle X(t) \rangle^2}$, reads

$$\mathbb{X}(t) = \sqrt{\frac{2 k_B T}{\gamma m} \mathcal{J}(t)}. \tag{E.5}$$

By evaluating the instantaneous velocity, $\mathbb{V}(t) = d\mathbb{X}(t)/dt$, we obtain

$$\mathbb{V}(t) = \sqrt{\frac{2 k_B T}{\gamma m}} \frac{d}{dt} \sqrt{\mathcal{J}(t)}. \tag{E.6}$$

Therefore, the differentiability property of the quantity (E.5) does depend on the convergence of Eq. (E.6) at short times, including $t = 0$.

The Markovian case $I(t) = 1$ gives rise to $\mathcal{J}(t) = t$ that, in turn, makes $\mathbb{V}(t)$ to diverge at $t = 0$. In consequence, $\mathbb{X}(t)$ is said to be a non-differentiable or non-analytic function at $t = 0$. This property characterizes the Einstein's Brownian particle [194] (see also Appendix A). On the contrary, the non-Markovian case $I(t) = 1 - e^{-t/t_c}$ leads to $\mathcal{J}(t) \sim t^2/2t_c$ at short times, so that $\lim_{t \to 0}(d\sqrt{\mathcal{J}(t)}/dt) \sim 1/\sqrt{2t_c}$. In this case, the paths of our Brownian particle are differentiable [192].

## Appendix F. Deriving Schrödinger equations

As far as the environment can be neglected, i.e., $\gamma = 0$, our quantum master equation (3.4) turns out to describe a quantum isolated system given by the von Neumann equation

$$i\hbar \frac{\partial \rho(x_1, x_2, t)}{\partial t} = \left[ V(x_1, t) - V(x_2, t) - \frac{\hbar^2}{2m}\left(\frac{\partial^2}{\partial x_1^2} - \frac{\partial^2}{\partial x_2^2}\right) \right] \rho(x_1, x_2, t), \quad (F.1)$$

where $x_1$ and $x_2$ are our quantization conditions (3.3)

$$x_1 = x + \frac{\eta\hbar}{2}, \quad (F.2)$$

$$x_2 = x - \frac{\eta\hbar}{2}. \quad (F.3)$$

Both variables $x$ and $\eta$ do not rely on Planck's constant $\hbar$, that is, they are classical quantities. Besides, assuming the von Neumann function $\rho(x_1, x_2, t)$ in Eq. (F.1) to be factorized as

$$\rho(x_1, x_2, t) = \psi(x_1, t)\psi^*(x_2, t), \quad (F.4)$$

where $\psi^*(x_2, t)$ denotes the complex conjugate of $\psi$ at point $x_2$, we obtain the pair of Schrödinger's equations

$$i\hbar \frac{\partial \psi(x_1, t)}{\partial t} = V(x_1, t)\, \psi(x_1, t) - \frac{\hbar^2}{2m}\frac{\partial^2 \psi(x_1, t)}{\partial x_1^2} \quad (F.5)$$

and

$$-i\hbar \frac{\partial \psi^*(x_2, t)}{\partial t} = V(x_2, t)\, \psi^*(x_2, t) - \frac{\hbar^2}{2m}\frac{\partial^2 \psi^*(x_2, t)}{\partial x_2^2}. \quad (F.6)$$

If equations of motion (F.5) and (F.6) are interpreted as evolutions forward and backward in time, respectively, then our non-Hamiltonian quantization conditions (F.2) and (F.3) justify the Schwinger's Lagrangian interpretation of quantum mechanics of isolated systems [27]: "*imagine that the positive and negative senses of time development are governed by different dynamics*". In this Schwinger's picture, Blasone et al. [28] have been able to explain the two-slit experiment whereby the difference between the two motions accounts for quantum interference whereas the motion of the same point $x_1 = x_2 \equiv x$ provides the classical behavior (no interference pattern).

In terms of the von Neumann equation (F.1), the classical limit $x_1 = x_2 \equiv x$ can be interpreted as yielding the time-independent probability distribution

$$\rho(x) = |\psi(x)|^2, \tag{F.7}$$

to be appropriately interpreted as a classical probability distribution. Quantum effects arise actually for $x_1 \neq x_2$ in the quantal configuration space.

Lastly, we wish to point out that in our approach a quantum Brownian particle (the quantum open system) does not exhibits interference effects since the von Neumann function $\rho(x_1, x_2, t)$ in our quantum master equation (3.4) cannot be factorized, i.e.,

$$\rho(x_1, x_2, t) \neq \psi(x_1, t)\psi^*(x_2, t). \tag{F.8}$$

# Appendix G. The environment's physics

Quantum Brownian motion is the result of the influence of a given environment or medium (e.g., a quantum fluid), comprised of $N$ particles, on the movement of a tagged particle. More precisely, the quantum diffusion energy $\mathcal{E}_\hbar$ arising from the quantization of the classical diffusion energy $\mathcal{E}$, defined in Eq. (2.25a), is to be identified with the medium's internal energy $U$ per particle, i.e.,

$$\mathcal{E}_\hbar = \frac{U}{N}. \tag{G.1}$$

How can we then derive such environmental energy? For thermal systems, the environment is viewed as a quantum thermal reservoir or a heat bath, and hence its temperature-dependent internal energy $U \equiv U(T; \hbar)$ can be found out on the basis of statistical thermodynamics according to three sorts of quantum statistics[54] [351,464-470]: The Maxwell-Boltzmann statistics, the Fermi-Dirac statistics [471,472], and the Boson-Einstein statistics [473-475].

*Maxwell-Boltzmann systems.* Let us assume that the environment can be devised as a heat bath comprising of a set of $N$ quantum harmonic oscillators having the same oscillation frequency $\omega$ in thermal equilibrium at temperature $T$. This system obeys the Maxwell-Boltzmann statistics according to which the harmonic oscillators are distinguishable from each other. The internal energy $U$ of this Maxwell-Boltzmann system is given by $U = N\overline{E}$, where $\overline{E} = (\omega\hbar/2)\coth(\omega\hbar/2k_BT)$ is the mean energy of such oscillators [351]. Accordingly, the quantum diffusion energy (G.1) of a Brownian particle immersed in such oscillators heat bath reads

$$\mathcal{E}_\hbar = \frac{\omega\hbar}{2}\coth\left(\frac{\omega\hbar}{2k_BT}\right), \tag{G.2}$$

$k_B$ being dubbed Boltzmann's constant and $\coth(y) \equiv (e^{2y} + 1)/(e^{2y} - 1)$, with $y = \omega\hbar/2k_BT$. The $\hbar$-dependent energy, given by

$$\mathcal{E}_\hbar = \frac{\omega\hbar}{2}, \tag{G.2a}$$

corresponds to the zero point diffusion energy coming from the heat bath in the quantum limit at zero temperature, $T = 0$, while the $\hbar$-independent energy, i.e.,

$$\mathcal{E} = k_BT, \tag{G.2b}$$

---

[54]Although the expression *quantum statistics* could be looked upon as a "misnomer" [464], such expression is employed here for pointing out the fact that the Maxwell-Boltzmann, Fermi-Dirac, and Boson-Einstein statistics are reliant on Planck's constant $\hbar$.

denotes the thermal energy of the heat bath in the classical limit $\hbar \to 0$ at high temperatures, $T \gg \omega\hbar/2k_B$.

*Fermi-Dirac systems.* Let the environment be an ideal quantum gas of $N$ identical particles each of mass $m$, inside a volume $V$ at temperature $T$, and obeying the Fermi-Dirac statistics [471,472]. A tagged particle immersed in this gas undergoes quantum Brownian motion owing to the diffusion energy (G.1) given by

$$\mathcal{E}_\hbar = \frac{U_{\text{FD}}}{N}, \tag{G.3}$$

where $U_{\text{FD}}$ is the internal energy of the Fermi-Dirac gas [351,377,468-470]

$$U_{\text{FD}} = \frac{gVm^{3/2}}{\sqrt{2}\pi^2\hbar^3} \int_0^\infty \frac{\varepsilon^{3/2}d\varepsilon}{z^{-1}e^{\frac{\varepsilon}{k_BT}} + 1}. \tag{G.3a}$$

The quantity $g = 2s + 1$ is termed the degeneracy factor of the half-integral spin states, whereas $z = e^{\mu/k_BT}$ is called the fugacity of the gas expressed in terms of its chemical potential $\mu$.

At low temperatures $T \to 0$, the Fermi-Dirac internal energy (G.3a) becomes $U_{\text{FD}} \sim (3/5)Nk_BT_F[1 + (5\pi^2/12)(T^2/T_F^2)]$ and accordingly the diffusion energy (G.3) of a quantum Brownian particle reads

$$\mathcal{E}_\hbar = \frac{3}{5}k_BT_F\left(1 + \frac{5\pi^2}{12}\frac{T^2}{T_F^2}\right), \quad T \to 0, \tag{G.4}$$

becoming at zero temperature

$$\mathcal{E}_\hbar = \frac{3}{5}k_BT_F, \quad T = 0. \tag{G.4a}$$

$T_F$ is the Fermi temperature defined by

$$T_F = \frac{\epsilon_F}{k_B}, \tag{G.5}$$

$\epsilon_F$ being the Fermi energy given by

$$\epsilon_F = \frac{\hbar^2}{2m}\left(\frac{6\pi^2}{g}\right)^{2/3}\left(\frac{N}{V}\right)^{2/3}. \tag{G.5a}$$

The Fermi energy $\epsilon_F$ is interpreted as the single-particle energy corresponding to the Fermi momentum $p_F = \hbar(6\pi^2/g)^{1/3}(N/V)^{2/3}$ via the relation $\epsilon_F = p_F^2/2m$, geometrically viewed as the radius of the Fermi sphere, within which the momentum states are occupied [468].

For $T \ll T_F$, the fermionic gas is said to be nearly degenerate, rendering completely degenerate at $T = 0$, the so-called ground state whose internal energy is $U_{FD} = (3/5)Nk_B T_F$ [470]. Therefore, in a completely degenerate gas the quantum Brownian motion of a particle comes about due to the Fermi energy, i.e., $\mathcal{E}_\hbar = (3/5)\epsilon_F$.

*Bose-Einstein systems.* If the environment is assumed to be an ideal Boson gas of $N$ identical particles contained in a volume $V$ at temperature $T$, satisfying the Bose-Einstein statistics [473-475], then the quantum diffusion energy of a Brownian particle turns out to be given by

$$\mathcal{E}_\hbar = \frac{U_{BE}}{N}, \tag{G.6}$$

$U_{BE}$ being the internal energy of the bosonic gas [351,377,468-470]

$$U_{BE} = \frac{gVm^{3/2}}{\sqrt{2}\pi^2\hbar^3} \int_0^\infty \frac{\varepsilon^{3/2} d\varepsilon}{z^{-1}e^{\frac{\varepsilon}{k_B T}} - 1}, \tag{G.6a}$$

where $g = 2s + 1$ is the degeneracy factor of the integral spin states. Bose-Einstein systems are characterized by the following temperature regimes [352,353]

$$U_{BE} = \frac{3}{2}Nk_B T \frac{g_{5/2}(z)}{g_{3/2}(z)}, \qquad T > T_{BE}, \tag{G.7a}$$

$$U_{BE} = \frac{3}{2}Nk_B T \frac{g_{5/2}(1)}{g_{3/2}(1)} = 0.77 Nk_B T_{BE}, \qquad T = T_{BE}, \tag{G.7b}$$

and

$$U_{BE} = 0.77 Nk_B \left( \sqrt{\frac{T^5}{T_{BE}^3}} + T_{BE}\sqrt{1 - \frac{T^3}{T_{BE}^3}} \right), \qquad T < T_{BE}, \tag{G.7c}$$

where $T_{BE}$ is dubbed the Bose-Einstein temperature defined as

$$T_{\text{BE}} = g\frac{2\pi\hbar^2}{mk_B}\left(\frac{1}{g_{3/2}(1)}\frac{N}{V}\right)^{2/3}. \tag{G.8}$$

In Eq. (G.7a), $g_{3/2}(z)$ and $g_{5/2}(z)$ are special cases of Bose-Einstein functions defined in the general form as

$$g_{\nu+1}(z) = \frac{1}{(k_B T)^{\nu+1}\Gamma(\nu+1)}\int_0^\infty \frac{\varepsilon^\nu d\varepsilon}{z^{-1}e^{\frac{\varepsilon}{k_B T}} - 1}, \tag{G.9}$$

expressed in terms of $z = e^{\mu/k_B T}$ and the gamma function defined as $\Gamma(\nu+1) = \nu\int_0^\infty dx e^{-x}x^{\nu-1}$, with $\text{Re}\,\nu > 0$ [469]. In Eq. (G.7b) we have used the approximate values $g_{5/2}(1) \sim 2.612$ and $g_{3/2}(1) \sim 1.341$.

It had been predicted that the particles at temperatures $T < T_{\text{BE}}$ undergo a Bose-Einstein condensation [473-475]. Especially, according to Eq. (G.7c) the gas in the ground state at $T = 0$ displays a non-null internal energy given by $U_{\text{FD}}(0) = 0.77 N k_B T_{\text{BE}}$. Only recently, the existence of this zero point energy of the Boson gas compatible with the Heisenberg principle has been reported by Deeney and O'Leary [352,353] on the assumption that the gas particles do not lose all of their kinetic energy on dropping into the ground state at temperatures below $T_{\text{BE}}$. These particles just move and hence contribute to the internal energy of the Boson gas. In short, the ground state is not a state of zero energy. Accordingly, particles in the Bose-Einstein condensate make contribution to the total energy of the Bose-Einstein gas owing to the presence in Eq. (G.7c) of the Deeney-O'Leary energy $U(T) = 0.77 N k_B T_{\text{BE}}\sqrt{1 - (T/T_{\text{BE}})^3}$, which is interpreted as a sort of ordered kinetic energy since the particles in the condensate have the same energy $\sim 0.77 k_B T_{\text{BE}}$ [352,353]. Such a Deeney-O'Leary energy accounts for the quantum Brownian motion during the Bose-Einstein condensation in view of the following diffusion energy

$$\mathcal{E}_\hbar = 0.77 k_B\left(\sqrt{\frac{T^5}{T_{\text{BE}}^3}} + T_{\text{BE}}\sqrt{1 - \frac{T^3}{T_{\text{BE}}^3}}\right), \quad T < T_{\text{BE}}. \tag{G.10}$$

# Appendix H. Quantization of the non-Markovian Smoluchowski equation

In order to quantize the Brownian motion in the absence of inertial forces, we start with the non-Markovian Smoluchowski equation (2.66), given by

$$\frac{\partial \mathcal{F}(x,t)}{\partial t} = \frac{1}{m\gamma}\frac{dV(x)}{dx}\frac{\partial \mathcal{F}(x,t)}{\partial x} + \frac{1}{m\gamma}\frac{d^2V(x)}{dx^2}\mathcal{F}(x,t) + \frac{\mathcal{E}}{m\gamma}I(t)\frac{\partial^2 \mathcal{F}(x,t)}{\partial x^2}. \quad (H.1)$$

Now let $\chi_1 = \chi(x_1,t)$ and $\chi_2 = \chi(x_2,t)$ be two distinct solutions of Eq. (H.1) given respectively at points $x_1$ and at $x_2$

$$\frac{\partial \chi_1}{\partial t} = \frac{1}{m\gamma}\frac{dV(x_1)}{dx_1}\frac{\partial \chi_1}{\partial x_1} + \frac{1}{m\gamma}\frac{d^2V(x_1)}{dx_1^2}\chi_1 + \frac{2\mathcal{E}}{m\gamma}I(t)\frac{\partial^2 \chi_1}{\partial x_1^2} \quad (H.2a)$$

and

$$\frac{\partial \chi_2}{\partial t} = \frac{1}{m\gamma}\frac{dV(x_2)}{dx_2}\frac{\partial \chi_2}{\partial x_2} + \frac{1}{m\gamma}\frac{d^2V(x_2)}{dx_2^2}\chi_2 + \frac{2\mathcal{E}}{m\gamma}I(t)\frac{\partial^2 \chi_2}{\partial x_2^2}. \quad (H.2b)$$

Notice that both solutions $\chi_1$ and $\chi_2$ evolve with the diffusion coefficient $D(t) = 2\mathcal{E}I(t)/m\gamma$ in contrast to the solution $\mathcal{F}(x,t)$ of (H.1).

Multiplying (H.2a) and (H.2b) by $\chi(x_2,t)$ and $\chi(x_1,t)$, respectively,

$$\chi_2\frac{\partial \chi_1}{\partial t} = \frac{1}{m\gamma}\frac{dV(x_1)}{dx_1}\chi_2\frac{\partial \chi_1}{\partial x_1} + \frac{1}{m\gamma}\frac{d^2V(x_1)}{dx_1^2}\chi_2\chi_1 + \frac{2\mathcal{E}}{m\gamma}I(t)\chi_2\frac{\partial^2 \chi_1}{\partial x_1^2} \quad (H.3a)$$

and

$$\chi_1\frac{\partial \chi_2}{\partial t} = \frac{1}{m\gamma}\frac{dV(x_2)}{dx_2}\chi_1\frac{\partial \chi_2}{\partial x_2} + \frac{1}{m\gamma}\frac{d^2V(x_2)}{dx_2^2}\chi_1\chi_2 + \frac{2\mathcal{E}}{m\gamma}I(t)\chi_1\frac{\partial^2 \chi_2}{\partial x_2^2} \quad (H.3b)$$

and then adding the resulting equations, we arrive at

$$\frac{\partial \xi}{\partial t} = \frac{1}{m\gamma}\left[\frac{dV(x_1)}{dx_1}\frac{\partial \xi}{\partial x_1} + \frac{dV(x_2)}{dx_2}\frac{\partial \xi}{\partial x_2}\right] + \frac{1}{m\gamma}\left[\frac{d^2V(x_1)}{dx_1^2} + \frac{d^2V(x_2)}{dx_2^2}\right]\xi$$
$$+ \frac{2\mathcal{E}I(t)}{m\gamma}\left[\frac{\partial^2 \xi}{\partial x_1^2} + \frac{\partial^2 \xi}{\partial x_2^2}\right], \quad (H.4)$$

where $\xi \equiv \xi(x_1,x_2,t) = \chi(x_1,t)\,\chi(x_2,t)$.

We quantize the equation of motion (H.4) by means of the change of variables

$$x_1 = x - \frac{\eta\hbar}{2}, \tag{H.5a}$$

$$x_2 = x + \frac{\eta\hbar}{2}, \tag{H.5b}$$

and the quantization of the diffusion energy

$$\mathcal{E} \to \mathcal{E}_\hbar. \tag{H.6}$$

By making use of the relations

$$\frac{\partial}{\partial x_1} = \frac{1}{2}\frac{\partial}{\partial x} - \frac{1}{\hbar}\frac{\partial}{\partial \eta}, \tag{H.7a}$$

$$\frac{\partial}{\partial x_2} = \frac{1}{2}\frac{\partial}{\partial x} + \frac{1}{\hbar}\frac{\partial}{\partial \eta}, \tag{H.7b}$$

and

$$\frac{\partial^2}{\partial x_1^2} + \frac{\partial^2}{\partial x_2^2} = \frac{1}{2}\frac{\partial^2}{\partial x^2} + \frac{2}{\hbar^2}\frac{\partial^2}{\partial \eta^2}, \tag{H.7c}$$

expanding the potentials $V(x_1)$ and $V(x_2)$ according to

$$V\left(x \pm \frac{\eta\hbar}{2}\right) = \sum_{i=0}^{\infty} \frac{1}{i!}\left(\frac{\eta\hbar}{2}\right)^i \frac{d^i V(x)}{dx^i}, \tag{H.8}$$

as well as neglecting the terms $O(\eta^3)$, we arrive at the following quantum master equation in configuration space $(x, \hbar\eta)$

$$\frac{\partial \rho}{\partial t} = \frac{1}{m\gamma}\left[\frac{dV(x)}{dx} + \frac{1}{2}\left(\frac{\eta\hbar}{2}\right)^2 \frac{d^3 V(x)}{dx^3}\right]\frac{\partial \rho}{\partial x} + \frac{\eta}{m\gamma}\frac{d^2 V(x)}{dx^2}\frac{\partial \rho}{\partial \eta}$$
$$+ \frac{2}{m\gamma}\left[\frac{d^2 V(x)}{dx^2} + \frac{1}{2}\left(\frac{\eta\hbar}{2}\right)^2 \frac{d^4 V(x)}{dx^4}\right]\rho + \frac{\mathcal{E}_\hbar I(t)}{m\gamma}\left[\frac{\partial^2 \rho}{\partial x^2} + \frac{4}{\hbar^2}\frac{\partial^2 \rho}{\partial \eta^2}\right], \tag{H.9}$$

the time-dependent quantum-mechanical function $\rho \equiv \rho(x, \hbar\eta, t)$ being expressed in terms of the coordinates $x$ and $\hbar\eta$, given by $x = (x_1 + x_2)/2$ and $\eta\hbar = x_2 - x_1$, respectively.

The Fourier transform (the Wigner function)

$$W(x, p, t) = \frac{1}{2\pi}\int_{-\infty}^{\infty} \rho(x, \eta, t) e^{ip\eta} d\eta \tag{H.10}$$

leads the quantum master equation (H.9) to the quantum Smoluchowski equation in phase space

$$\frac{\partial W(x,p,t)}{\partial t} = \frac{1}{m\gamma}\frac{dV(x)}{dx}\frac{\partial W(x,p,t)}{\partial x} - \frac{p}{m\gamma}\frac{d^2V(x)}{dx^2}\frac{\partial W(x,p,t)}{\partial p}$$
$$+ \frac{\mathcal{E}_\hbar I(t)}{m\gamma}\frac{\partial^2 W(x,p,t)}{\partial x^2} - \frac{\hbar^2}{8m\gamma}\frac{d^3V(x)}{dx^3}\frac{\partial^3 W(x,p,t)}{\partial x \partial p^2}$$
$$- \frac{\hbar^2}{4m\gamma}\frac{d^4V(x)}{dx^4}\frac{\partial^2 W(x,p,t)}{\partial p^2} + \left[\frac{1}{m\gamma}\frac{d^2V(x)}{dx^2} - \frac{4\mathcal{E}_\hbar I(t)}{m\gamma\hbar^2}p^2\right]W(x,p,t)$$

(H.11)

whose classical limit $\hbar \to 0$ is the non-Markovian Smoluchowski equation (H.1) as long as

$$\lim_{\hbar \to 0} \mathcal{E}_\hbar = \mathcal{E} \qquad \text{(H.12a)}$$

and

$$\lim_{\hbar \to 0} W(x, p = 0, t) = \mathcal{F}(x,t). \qquad \text{(H.12b)}$$

# References

[1] C. F. von Weizsäcker, Grosse Physiker: von Aristoteles bis Werner Heisenberg, Carl Hanser, München, 1999.

[2] R. Dugas, A History of Mechanics, Routledge and Kegan-Paul, London, 1957.

[3] M. Heidegger, Die Frage nach dem Ding, Third Edition, Max Niemeyer, Tübingen, 1987.

[4] A. Koyré, Galileo and Plato, Journal of the History of Ideas 4 (4) (1943) 400-428.

[5] A. Koyré, Galileo and the scientific revolution of the seventeenth century, Philosophical Review 52 (4) (1943) 333-348.

[6] A. Einstein, L. Infeld, The Evolution of Physics, Cambridge University Press, Cambridge, 1938.

[7] M. Jammer, Concepts of Force: A Study in the Foundations of Dynamics, Cambridge University Press, Cambridge, 1957.

[8] S. Chandrasekhar, Newton's Principia for the Common Reader, Oxford University Press, Oxford, 1995.

[9] C. Lanczos, The Variational Principles of Mechanics, University of Toronto Press, Toronto, 1949.

[10] R. Santilli, Foundations of Theoretical Mechanics: The Inverse Problem in Newtonian Mechanics, Springer, New York, 1978.

[11] A. Einstein, The Meaning of Relativity, Princeton University Press, Princeton, 1922.

[12] L. de Broglie, Recherches sur la théorie de quanta, Annales de Physiques 10 (3) (1925) 22-128.

[13] M. Born, Problems of Atomic Dynamics, Massachusetts Institute of Technology, Cambridge, 1926.

[14] M. Born, The Mechanics of the Atom, Bell and Sons, London, 1927.

[15] P. A. M. Dirac, The Principles of Quantum Mechanics, Oxford University Press, Oxford, 1930.

[16] W. Heisenberg, The Physical Principles of the Quantum Theory, University of Chicago Press, Chicago, 1930.

[17] L. de Broglie, An Introduction to the Study of Wave Mechanics, Methuen, London, 1930.

[18] W. Pauli, Die allgemeinen Prinzipien der Wellenmechanik, Springer, Berlin, 1933.

[19] L. Pauling, E. B. Wilson Jr., Introduction to Quantum Mechanics, McGraw-Hill, New York, 1935.

[20] M. Jammer, Conceptual Developments of Quantum Mechanics, McGraw-Hill, New York, 1966.

[21] W. Yourgrau, S. Mandelstam, Variational Principles in Dynamics and Quantum Theory, Dover, New York, 1968.

[22] W. Heisenberg, Über quantentheoretische Umdeutung kinematischer und mechanischer Beziehungen, Zeitschrift für Physik 33 (1) (1925) 879-893.

[23] E. Schrödinger, Quantisierung als Eigenwertproblem, Annalen der Physik 79 (1926) 361-376.

[24] P. A. M. Dirac, The fundamental equations of quantum mechanics, Proceedings of the Royal Society A 109 (1926) 642-653.

[25] R. P. Feynman, Space-time approach to non-relativistic quantum mechanics, Reviews of Modern Physics 20 (1948) 367-387.

[26] A. Eddington, The Nature of the Physical World, MacMillan, Cambridge, 1928.

[27] J. Schwinger, Brownian motion of a quantum oscillator, Journal of Mathematical Physics 2 (1961) 407-432.

[28] M. Blasone, Y N. Srivastava, G. Vitiello, A. Widom, Phase coherence in quantum Brownian motion, Annals of Physics 267 (1998) 61-74.

[29] M. Born, Zur Quantenmechanik der Stossvorgänge, Zeitschrift für Physik 37 (12) (1926) 863-867.

[30] M. Born, Zur Wellenmechanik der Stoßvorgänge, Nachrichten von der Gesellschaft der Wissenschaften zu Göttingen, Mathematisch-Physikalische Klasse, Göttingen (1926) 146-160.

[31] M. Born, The statistical interpretation of quantum mechanics, Nobel Lecture, 1954. 863

[32] M. Bitbol, Schrödinger's Philosophy of Quantum Mechanics, Kluwer, Dordrecht, 1996.

[33] F. Laloë, Do We Really Understand Quantum Mechanics?, Cambridge University Press, Cambridge, 2012.

[34] M. Jammer, The Philosophy of Quantum Mechanics: The Interpretations of Quantum Mechanics in Historical Perspective, Wiley, New York, 1974.

[35] B. d'Espagnat, Conceptual Foundations of Quantum Mechanics, Perseus Books Publishing, Second Edition, Reading, Massachusetts, 1999.

[36] D. Home, Conceptual Foundations of Quantum Physics: An Overview from Modern Perspectives, Springer, New York, 1997.

[37] G. Auletta, Foundations and Interpretation of Quantum Mechanics, World Scientific, Singapore, 2001.

[38] W. Heisenberg, Über den anschaulichen Inhalt der quantentheoretischen Kinematik und Mechanik, Zeitschrift für Physik 43 (3-4) (1927) 172-198.

[39] N. Bohr, The quantum postulate and the recent development of atomic theory, Nature 14 (1928) 580-590.

[40] N. Bohr, Discussion with Einstein on epistemological problems in atomic physics, in: P. A. Schilpp, (Ed.), Albert Einstein: Philosopher-Scientist, The Library of Living Philosophers, Evanston (1949), pp. 200-241.

[41] N. Bohr, Atomic Theory and the Description of Nature, Cambridge University Press, Cambridge, 1934.

[42] D. Bohm, Quantum Theory, Prentice-Hall, New York, 1951.

[43] J. von Neumann, Mathematische Grundlagen der Quantenmechanik, Springer, Berlin, 1932. English translation: Mathematical Foundations of Quantum Mechanics, Princeton University Press, Princeton, 1955.

[44] A. Petersen, The philosophy of Niels Bohr, Bulletin of the Atomic Scientist 19 (1963) 8-14.

[45] J. A. Wheeler, W. H. Zurek (Eds.), Quantum Theory and Measurement, Princeton University Press, Princeton, 1983.

[46] N. Bohr, On the notions of causality and complementarity, Dialectica 2 (1948) 312-319.

[47] M. Katsumori, Niels Bohr's Complementarity: Its Structure, History, and Intersections with Hermeneutics and Deconstruction, Springer, Dordrecht, 2011.

[48] A. Plotnitsky, Niels Bohr and Complementarity: An Introduction, Springer, New York, 2013.

[49] H. Krips, Measurement in quantum theory, Stanford Encyclopedia of Philosophy. http://seop.illc.uva.nl/archives/fall2008/entries/qt-measurement/

[50] A. Einstein, B. Podolsky, N. Rosen, Can quantum-mechanical description of physical reality be complete?, Physical Review 47 (1935) 777-780.

[51] A. Einstein, Quanten-mechanik und Wirklichkeit, Dialectica 2 (3-4) (1948) 320-324.

[52] A. Einstein, Elementare Überlegungen zur Interpretation der Grundlagen der Quanten-Mechanik. In: Scientific Papers Presented to Max Born on his Retirement from the Tait Chair of Natural Philosophy in the University of Edinburgh, Oliver and Boyd, Edinburgh, 1953, pp. 33-40.

[53] D. W. Belousek, Einstein's 1927 unpublished hidden-variable theory: Its background, context and significance, Studies in History and Philosophy of Science B: Studies in History and Philosophy of Modern Physics 27 (4) (1996) 437-461.

[54] P. Holland, What's wrong with Einstein's 1927 hidden-variable interpretation of quantum mechanics?, Foundations of Physics 35 (2) (2005) 177-196.

[55] A. Einstein, Physik und Realität, Journal of The Franklin Institute: Engineering and Applied Mathematics 221 (3) (1936) 313-347.

[56] A. Einstein, Autobiographical notes, in: P. A. Schilpp (Ed.), Albert Einstein: Philosopher-Scientist, The Library of Living Philosophers, Evanston (1949), pp. 200-241.

[57] L. Ballentine, Einstein's interpretation of quantum mechanics, American Journal of Physics 40 (1972) 1763-1771.

[58] L. Ballentine, Quantum Mechanics: A Modern Development, World Scientific, New Jersey, 1998.

[59] M. Bunge, Survey of the interpretations of quantum mechanics, American Journal of Physics 24 (4) (1956) 272-286.

[60] L. de Broglie, L'interprétation de la mécanique ondulatoire, Journal de Physique et Radium 20 (12) (1959) 963-979.

[61] L. de Broglie, The reinterpretation of wave mechanics, Foundations of Physics 1 (1) (1970) 5-15.

[62] L. de Broglie, Interpretation of quantum mechanics by the double solution theory, Annales de la Fondation Louis de Broglie 12 (4) (1987) 1-23.

[63] C. Bialobrzeski, L'interpretation concrète de la mécanique quantique, Revue de Métaphysique et de Morale 41 (1) (1934) 83-103.

[64] D. Bohm, A suggested interpretation of the quantum theory in terms of 'hidden' variables. I and II, Physical Review 85 (2) (1952) 166-193.

[65] T. Takabayasi, Remarks on the formulation of quantum mechanics with classical pictures and on relations between linear scalar fields and hydrodynamical fields, Progress of Theoretical Physics 9 (3) (1953) 187-222.

[66] L. de Broglie. Non-linear Wave Mechanics: A Causal Interpretation, Elsevier, Amsterdam, 1960.

[67] D. Bohm, B. J. Hiley, P. N. Kaloyerou, An ontological basis for the quantum theory Physics Reports 144 (6) (1987) 321-375.

[68] D. Bohm, B. J. Hiley, The Undivided Universe: An Ontological Interpretation of Quantum Theory, Routledge, London, 1993.

[69] D. Bohm, B. J. Hiley, Statistical mechanics and the ontological interpretation, Foundations of Physics 26 (6) (1996) 823-846.

[70] P. R. Holland, The Quantum Theory of Motion: An Account of the de Broglie-Bohm Causal Interpretation of Quantum Mechanics, Cambridge University Press, Cambridge, 1993.

[71] J. T. Cushing, A. Fine, S. Goldstein (Eds.), Bohmian Mechanics and Quantum Theory: An Appraisal, Springer, Dordrecht, 1996.

[72] P. Riggs, Quantum Causality: Conceptual Issues in the Causal Theory of Quantum Mechanics, Springer, Berlin.

[73] D. Dürr, Bohmsche Mechanik als Grundlage der Quantenmechanik, Springer, Berlin, 2001.

[74] D. Dürr, S. Teufel, Bohmian Mechanics: The Physics and Mathematics of Quantum Theory, Springer, Dordrecht, 2009.

[74] D. Dürr, S. Goldstein, N. Zanghì. Quantum Physics Without Quantum Philosophy, Springer, Berlin, 2013.

[75] A. Ney, D. Z. Albert (Eds.), The Wave Function: Essays on the Metaphysics of Quantum Mechanics, Oxford University Press, Oxford, 2013.

[76] M. Cini, J. M. Lévy-Leblond (Eds.), Quantum Theory Without Reduction, Hilger, London, 1990.

[77] D. Dürr, S. Goldstein, N. Zanghi, Bohmian mechanics and the meaning of the wave function, in: R. S. Cohen, M. Horne, J. J. Stachel (Eds.), Experimental Metaphysics: Quantum Mechanical Studies for Abner Shimony, Volume One, 1996, pp. 25-38.

[78] J. Bricmont, What is the meaning of the wave function?, in: J.-M. Frere, M. Henneaux, A. Sevrin, Ph. Spindel (Eds.), Fundamental interactions: from symmetries to black holes, Université Libre de Bruxelles, Belgium, 1999, pp. 53-67.

[79] Y. Aharonov, J. Anandan, L. Vaidman, Meaning of the wave function, Physical Review A 47 (6) (1993) 4616-4626.

[80] C. Rovelli, Comment on ``Meaning of the wave function'', Physical Review A 50 (3) (1994) 2788- 2792.

[81] W. G. Unruh, Reality and measurement of the wave function, Physical Review A 50 (1) (1994) 882-887.

[82] J. Uffink, How to protect the interpretation of the wave function against protective measurements, Physical Review A 60 (5) (1999) 3474 -3481.

[83] Y. Aharonov, D. Rohrlich, Quantum Paradoxes: Quantum Theory for the Perplexed, Wiley-VCH, Weinheim, 2005.

[84] S. Gao, On Uffink's criticism of protective measurements, Studies in History and Philosophy of Modern Physics 44 (2013) 513-518.

[85] J. Uffink, Reply to Gao's "On Uffink's criticism of protective measurements", Studies in History and Philosophy of Modern Physics 44 (2013) 519-523.

[86] S. Gao (Ed.), Protective Measurement and Quantum Reality: Towards a New Understanding of Quantum Mechanics, Cambridge University Press, Cambridge, 2014.

[87] C. A. Fuchs, A. Peres. Quantum theory needs no 'interpretation', Physics Today 53 (3) (2000) 70-71.

[88] E. Dennis, T. Norsen, Quantum theory: Interpretation cannot be avoided. arXiv: 0408178.

[89] L. Hardy, Quantum ontological excess baggage, Studies in History and Philosophy of Modern Physics 35 (2004) 267-276.

[90] T. T. Yong, Failure of ontological excess baggage as a criterion of the ontic approaches to quantum theory, Studies in History and Philosophy of Modern Physics 41 (2010) 318-321.

[91] A. I. M. Rae, Quantum Mechanics, Fourth Edition, Institute of Physics, Bristol, 2002.

[92] A. I. M. Rae, Quantum Physics, Second Edition, Cambridge University Press, Cambridge, 2004.

[93] S. Gao, Meaning of the wave function, International Journal of Quantum Chemistry 111 (2011) 4124-4138.

[94] H. D. Zeh, The wave function: It or Bit?, in: J. D. Barrow, P. C. W. Davies, C. L. Harper Jr., (Eds.), Science and Ultimate Reality, Cambridge University Press, Cambridge, 2004, pp. 103-120.

[95] M. F. Pusey, J. Barret, T. Rudolph, On the reality of the quantum state, Nature Physics 8 (2012) 475-478.

[96] D. Z. Albert, Wave function realism, in: A. Ney, D. Z. Albert (Eds.), The Wave Function: Essays on the Metaphysics of Quantum Mechanics, Oxford University Press, Oxford, 2013, pp. 52-57.

[97] S. French, Wither wave function realism? in: A. Ney, D. Z. Albert (Eds.), The Wave Function: Essays on the Metaphysics of Quantum Mechanics, Oxford University Press, Oxford, 2013, pp. 76-90.

[98] S. Goldstein, N, Zanghì, Reality and the role of wave functions in quantum theory, in: A. Ney, D. Z. Albert (Eds.), The Wave Function: Essays on the Metaphysics of Quantum Mechanics, Oxford University Press, Oxford, 2013, pp. 91-109.

[99] T. Maudlin, The nature of the quantum state, in: A. Ney, D. Z. Albert (Eds.), The Wave Function: Essays on the Metaphysics of Quantum Mechanics, Oxford University Press, Oxford, 2013, pp. 126-153.

[100] A. Ney, Ontological reduction and the wave function ontology, in: A. Ney, D. Z. Albert (Eds.), The Wave Function: Essays on the Metaphysics of Quantum Mechanics, Oxford University Press, Oxford, 2013, pp. 168-183.

[101] M. Dorato, F. Laudisa, Realism and instrumentalism about the wave function: how should we choose?, in: S. Gao (Ed.), Protective Measurement and Quantum Reality: Towards a New Understanding of Quantum Mechanics, Cambridge University Press, Cambridge, 2014, pp. 119-134.

[102] M. Schlosshauer, T. V. B. Claringbold, Entanglement, scaling, and the meaning of the wave function in protective measurement, in: S. Gao (Ed.), Protective Measurement and Quantum Reality: Towards a New Understanding of Quantum Mechanics, Cambridge University Press, Cambridge, 2014, pp.180-194.

[103] V. Lam, Protective measurement and the nature of the wave function within the primitive ontology approach, in: S. Gao (Ed.), Protective Measurement and Quantum Reality: Towards a New Understanding of Quantum Mechanics, Cambridge University Press, Cambridge, 2014, pp. 195-210.

[104] S. Gao, Reality and meaning of the wave function, in: S. Gao (Ed.), Protective Measurement and Quantum Reality: Towards a New Understanding of Quantum Mechanics, Cambridge University Press, Cambridge, 2014, pp. 211-229.

[105] F. J. Tipler. Quantum nonlocality does not exist, Proceedings of the National Academy of Sciences of the United States of America 111 (31) (2014) 11281-11286.

[106] A. A. Grib, W. R. Rodrigues Jr., Nonlocality in Quantum Physics, Springer, New York, 1999.

[107] A. O. Bolivar, The Bohm quantum potential and the classical limit of quantum mechanics, Canadian Journal of Physics 81 (2003) 971-976.

[108] I. Prigogine, The formulation of classical and quantum mechanics for nonintegrable systems, International Journal of Quantum Chemistry 53 (1995) 105-118.

[109] T. Petrosky, I. Prigogine, The Liouville space extension of quantum mechanics, resonances, instability and irreversibility, Advances in Chemical Physics 99 (1997) 1-120.

[110] D. I. Blokhintsev, Classical statistical physics and quantum mechanics, Soviet Physics Uspekhi 20 (8) (1977) 683-690.

[111] M. Cini, Quantum mechanics without waves: A generalization of classical statistical mechanics, Annals of Physics 273 (1) (1999) 99-113.

[112] M. Cini, How real are quantons?, International Journal of Modern Physics B 18 (4-5) (2004) 565-574.

[113] M. Cini, The physical nature of wave-particle duality, in: M. R. Pahlavani (Ed.), Theoretical Concepts of Quantum Mechanics, InTech, Rijeka, 2012.

[114] M. Cini, Is time real?, in: S. Alberio, P. Blanchard (Eds.), Direction of Time, Springer, Cham, 2014.

[115] R. Balian, From Microphysics to Macrophysics: Methods and Applications of Statistical Physics, Second Edition, Springer, Berlin, 2007.

[116] D. Bohm, B. Hiley, On a quantum algebraic approach to a generalized phase space, Foundations of Physics 11 (3-4) (1981) 179-203.

[117] B. Fain, Irreversibilities in Quantum Mechanics, Kluwer, New York, 2002.

[118] G. G. Emch, G. L. Sewell, Nonequilibrium statistical mechanics of open systems, Journal of Mathematical Physics 9 (6) (1968) 946-958.

[119] L. D. Landau, E. M. Lifshitz, Quantum mechanics: Non-relativistic Theory, Third Edition, Pergamon Press, Oxford, 1977.

[120] N. G. van Kampen, Stochastic Processes in Physics and Chemistry, Third Edition, North Holland, 2007.

[121] J. von Neumann, Wahrscheinlichkeitstheoretischer Aufbau der Quantenmechanik, Göttinger Nachrichten 1 (1927) 245-272.

[122] L. D. Landau, Das Dämpfungsproblem in der Wellenmechanik, Zeitschrift für Physik 45 (1927) 430-441.

[123] U. Fano, Description of states in quantum mechanics by density matrix and operator techniques, Reviews of Modern Physics 29 (1957) 74-93.

[124] R. K. Wangsness, F. Bloch, The dynamical theory of nuclear induction, Physical Review 89 (4) (1953) 728-739.

[125] H. Haken, Laser Theory, Springer, Berlin, 1970.

[126] A. Kossakowski, On quantum statistical mechanics of non-Hamiltonian systems, Reports on Mathematical Physics 3 (4) (1972) 247-274.

[127] V. Gorini, A. Kossakowski, E. C. G. Sudarshan, Completely positive semigroups of N-level systems, Journal of Mathematical Physics 17 (5) (1976) 821-825.

[128] G. Lindblad, Brownian motion of a quantum harmonic oscillator, Reports on Mathematical Physics 10 (3) 393-406.

[129] G. Lindblad, On the generators of quantum dynamical semigroups, Communications on Mathematical Physics 48 (1976) 119-130.

[130] E. B. Davies, Quantum Theory of Open Systems, Academic Press, New York, 1976.

[131] V. Gorini, A. Frigerio, M. Verri, A. Kossakowski, E. C. G. Sudarshan, Properties of quantum markovian master equations, Reports on Mathematical Physics 13 (1978) 149-173.

[132] H. Spohn, Kinetics equations from Hamiltonian dynamics, Reviews of Modern Physics 53 (3) (1980) 569-615.

[133] H.-P. Breuer, F. Petruccione, The Theory of Open Quantum Systems, Oxford University Press, Oxford, 2002.

[134] R. Alicki, K. Lendi, Quantum Dynamical Semigroups and Applications, Lecture Notes in Physics 717, Springer, Berlin, 2007.

[135] Á. Rivas, S. F. Huelga, Open Quantum Systems: An Introduction, Springer, Heidelberg, 2012.

[136] K. Blum, Density Matrix Theory and Applications, Third Edition, Springer, Berlin, 2012.

[137] H. -P. Breuer, Non-Markovian generalization of the Lindblad theory of open quantum systems, Physical Review A 75 (2007) 022103.

[138] D. Kohen, C. C. Marston, D. J. Tannor, Phase space approach to theories of quantum dissipation, Journal of Chemical Physics 107 (13) (1997) 5236-5253.

[139] U. Weiss, Quantum Dissipative Systems, third edition, World Scientific, Singapore, 2008.

[140] A. G. Redfield, On the theory of relaxation processes, IBM Journal of Research and Development 1 (1957) 19-31.

[141] A. G. Redfield, On the theory of relaxation processes, Advances in Magnetic Resonance 1 (1965) 1-32.

[142] B. Cowan, Nuclear Magnetic Resonance and Relaxation, Cambridge University Press, Cambridge, 1997.

[143] T. Prosen, B. Zunkovic, Exact solution of Markovian master equations for quadratic Fermi systems, New Journal of Physics **12** (2010) 025016.

[144] G. S. Agarwal, Brownian motion of a quantum oscillator, Physical Review A 4 (2) (1971) 739-747.

[145] G. S. Agarwal, Master equation methods in quantum optics, in: E. Wolf (ed.), Progress in Optics, Volume 11, North-Holland, Amsterdam, 1973.

[146] W. H. Louisell, Quantum Statistical Properties of Radiation, Wiley, New York, 1973.

[147] F. Haake, Statistical treatment of open systems by generalized master equations, in: Quantum Statistics in Optics and Solid-State Physics, Springer Tracts in Modern Physics, Springer, New York, 1973, pp. 98-166.

[148] M. Sargent, M. O. Scully, W. E. Lamb, Laser Physics, Addison-Wesley, New York, 1974.

[149] K. Lindenberg, B. J. West, The Nonequilibrium Statistical Mechanics of Open and Closed Systems, VCH, New York, 1990.

[150] H. Carmichael, An Open Systems Approach to Quantum Optics, Springer, Berlin, 1993.

[151] H. Carmichael, Statistical Methods in Quantum Optics 1: Master Equations and Fokker-Planck Equations, Springer, Berlin, 1999.

[152] H. Carmichael, Statistical Methods in Quantum Optics 2: Non-Classical Fields, Springer, Berlin, 2008.

[153] D. F. Walls, G. J. Milburn, Quantum Optics, Second Edition, Springer, 2008.

[154] C. W. Gardiner, P. Zoller, Quantum Noise: A Handbook of Markovian and Non-Markovian Quantum Stochastic Methods, Second Edition Enlarged, Springer, Berlin, 2000.

[155] R. R. Puri, Mathematical Methods of Quantum Optics, Springer, Berlin, 2001.

[156] C. Cohen-Tannoudji, J. Dupont-Roc, G. Grynberg, Atom-Photon Interactions: Basic Processes and Applications, Wiley, New York, 2004.

[157] A. Nitzan, Chemical Dynamics in Condensed Phases: Relaxation, Transfer, and Reactions in Condensed Molecular Systems, Oxford University Press, New York, 2006.

[158] V. May, O. Kuhn, Charge and Energy Transfer Dynamics in Molecular Systems, Wiley-VCH, Weinheim, 2011.

[159] A. O. Caldeira, An Introduction to Macroscopic Quantum Phenomena and Quantum Dissipation, Cambridge University Press, Cambridge, 2014.

[160] C. George, Mouvement brownien d'un oscillateur quantique, Physica 26 (1960) 453-477.

[161] I. Prigogine, R. Balescu, Sur la theorie moleculaire du mouvement brownien, Physica 23 (1957) 555-568.

[162] J. Schwinger, Brownian motion of a quantum oscillator, Journal of Mathematical Physics 2 (3) (1961) 407-432.

[163] R. P. Feynman, F. L. Vernon Jr., The theory of a general quantum system interacting with a linear dissipative system, Annals of Physics 24 (1963) 118-173.

[164] S. Attal, A. Joye, C.-A. Pillet (Eds.), Open Quantum Systems I: The Hamiltonian Approach, Springer, Berlin, 2006.

[165] S. Nakajima, On quantum theory of transport phenomena, Progress on Theoretical Physics 20 (1958) 948-959.

[166] R. Zwanzig, Ensemble method in the theory of irreversibility, Journal of Chemical Physics 33 (1960) 1338-1341.

[167] A. O. Caldeira, A. J. Leggett, Path integral approach to quantum Brownian motion, Physica A 121 (1983) 587-616.

[168] L. Diòsi, Quantum master equation of a particle in a gas environment, Europhysics Letters, 30 (2) (1995) 63-68.

[169] L. Diòsi, Quantum linear Boltzmann equation with finite intercollision time, Physical Review A 80 (2009) 064104.

[170] B. Vacchini, Completely positive quantum dissipation, Physical Review Letters 84 (7) (2000) 1374-1377.

[171] B. Vacchini, Test particle in a quantum gas, Physical Review E 63 (2001) 066115.

[172] B. Vacchini, Non-Abelian linear Boltzmann equation and quantum correction to Kramers and Smoluchowski equation, Physical Review E 66 (2002) 027107.

[173] B. Vacchini, K. Hornberger, Quantum linear Boltzmann equation, Physics Reports 478 (4-6) (2009) 71-120.

[174] S. M. Barnett, D. Cresser, Quantum theory of friction, Physical Review A **72** (2005) 022107.

[175] F. Petruccione, B. Vacchini, Quantum description of Einstein's Brownian motion, Physical Review E 71 (2005) 046134.

[176] K. Hornberger, Master equation for a quantum particle in a gas, Physical Review Letters 97 (2006) 060601.

[177] J. J. Halliwell, Two derivations of the master equation of quantum Brownian motion, Journal of Physics A: Mathematical and Theoretical 40 (2007) 3067-3080.

[178] K. Hornberger, B. Vacchini, Monitoring derivation of the quantum linear Boltzmann equation, Physical Review A **77** (2008) 022112.

[179] B. Vacchini, F. Petruccione, Kinetic description of quantum Brownian motion, European Physical Journal, Special Topics 159 (2008) 135-141.

[180] I. Kamleitner, J. Cresser, Quantum position diffusion and its implications for the quantum linear Boltzmann equation, Physical Review A 81 (2010) 012107.

[181] A. O. Bolivar, Quantization of non-Hamiltonian physical systems, Physical Review A 58 (1998) 4330-4335.

[182] A. O. Bolivar, Representação de Wigner da Mecânica Clássica, Quantização e Limite Clássico, PhD Thesis, Centro Brasileiro de Pesquisas Físicas, Rio de Janeiro, 2000 (in Portuguese).

[183] A. O. Bolivar, Quantization and classical limit of a linearly damped particle, a van der Pol system and a Duffing system, Random Operators and Stochastic Equations 9 (2001) 275-286.

[184] A. O. Bolivar, The Wigner representation of classical mechanics, quantization and classical limit, Physica A 301 (2001) 219-240.

[185] A. O. Bolivar, Quantization of the anomalous Brownian motion, Physics Letters A 307 (2003) 229-232.

[186] A. O. Bolivar, Dynamical quantization and classical limit, Canadian Journal of Physics 81 (2003) 663-673.

[187] A. O. Bolivar, Quantum-Classical Correspondence: Dynamical Quantization and the Classical Limit, Springer, Berlin, 2004.

[188] A. O. Bolivar, Quantum tunneling at zero temperature in the strong friction regime, Physical Review Letters 94 (2005) 026807.

[189] A. O. Bolivar, Non-equilibrium effects upon the non-Markovian Caldeira-Leggett quantum master equation, Annals of Physics 326 (2011) 1354-1367.

[190] A. O. Bolivar, The dynamical-quantization approach to open quantum systems, Annals of Physics 327 (2012) 705-732.

[191] A. O. Bolivar, Generalization of the classical Kramers rate for non-Markovian open systems out of equilibrium, Journal of Mathematical Physics 49 (2008) 01330.

[192] A. O. Bolivar, Non-Markovian effects on the Brownian motion of a free particle, Physica A 390 (2011) 3095-3107.

[193] S. Chandrasekhar, Stochastic problems in physics and astronomy, Reviews of Modern Physics 15 (1943) 1-89.

[194] A. Einstein, Über die von der molekularkinetischen Theorie der Wärme geforderte Bewegung von in ruhenden Flüssigkeiten suspendierten Teilchen, Annalen der Physik 17 (1905) 549-560.

[195] P. Langevin, Sur la theorie du mouvement brownien, Comptes Rendus de l'Académie des Sciences Paris 146 (1908) 530-533.

[196] A. Kolmogorov, Grundbegriffe der Wahrscheinlichkeitsrechnung, Springer, Berlin, 1933.

[197] J. von Plato, Creating Modern Probability: Its Mathematics, Physics, and Philosophy in Historical Perspective, Cambridge University Press, Cambridge, 1994.

[198] A. Einstein, Eine neue Bestimmung der Moleküldimensionen, Annalen der Physik 19 (1906) 289-306.

[199] A. Einstein, Zur Theorie der Brownschen Bewegung, Annalen der Physik 19 (1906) 371-381.

[200] A. Einstein, Theoretische Bemerkungen über die Brownsche Bewegung, Zeitschrift für Elektrochemie und angewandte physikalische Chemie 13 (1907) 41-42.

[201] A. Einstein, Elementare Theorie der Brownschen Bewegung, Zeitschrift für Elektrochemie 14 (1908) 235-239.

[202] A. Kolmogorov, Über die analytischen Methoden in der Wahrscheinlichkeitsrechnung, Mathematische Annalen 104 (1931) 414-458.

[203] A. Kolmogorov, Zur Umkehrkeit der statistischen Naturgesetze, Mathematische Annalen 113 (1937) 766-772.

[204] R. L. Stratonovich, Topics in the Theory of Random Noise, Vol. 1, Gordon and Breach, New York, 1963.

[205] L. Bachelier, Théorie de la spéculation, Annales Scientifiques de l'École Normale Supérieure 17 (3) (1900) 21-86.

[206] W. Sutherland, Dynamical theory of diffusion for non-electrolytes and the molecular mass of albumin, Philosophical Magazine 9 (6) (1905) 781-785.

[207] R. M. Mazo, Brownian Motion: Fluctuations, Dynamics and Applications, Oxford University Press, Oxford, 2002.

[208] R. Kubo, The fluctuation-dissipation theorem, Reports on Progress of Physics 29 (1966) 255-284.

[209] R. Kubo, Brownian motion and nonequilibrium statistical mechanics, Science 233 (1986) 330-334.

[210] A. Pais, Subtle is the Lord: The Science and the Life of Albert Einstein, Oxford University Press, New York, 1982.

[211] B. Duplantier, Brownian Motion, "Diverse and Undulating", in Einstein 1905-2005: Poincaré Seminar 2005, pp. 201-293, Birkhäuser, Basel, 2006.

[212] U. M. B. Marconi, A. Puglisi, L. Rondoni, A. Vulpiani, Fluctuation–dissipation: Response theory in statistical physics, Physics Reports 461 (2008) 111-195.

[213] K. R. Naqvi. The origin of the Langevin equation and the calculation of the mean squared displacement: Let's set the record straight. arXiv: 0502141.

[214] L. S. Ornstein, On the Brownian motion, Proceedings of the Royal Academy of Amsterdam 21 (I) (1919) 96-108.

[215] L. S. Ornstein, W. R. van Wijk, On the derivation of distribution functions in problems of Brownian motion, Physica 1 (7-12) (1934) 235-254.

[216] O. Klein, Zur statistische Theorie der Suspensionen und Lösungen, Arkiv för Matematik, Astronomi och Fysik 16 (5) (1922) 1-51.

[217] H. A. Kramers, Brownian motion in a field of force and the diffusion model of chemical reactions, Physica 7 (4) (1940) 284-304.

[218] G. E. Uhlenbeck, L. S. Ornstein, On the theory of the Brownian motion, Physical Review 36 (1930) 823-841.

[219] M. C. Wang, G. E. Uhlenbeck, On the theory of the Brownian motion II, Reviews of Modern Physics 17 (1945) 323-342.

[220] H. Risken, The Fokker-Planck Equation: Methods of Solution and Applications, Second Edition, Springer, Berlin, 1989.

[221] P. Hänggi, P. Jung, Colored noise in dynamical systems, Advances on Chemical Physics 89 (1995) 239-326.

[222] R. Zwanzig, Nonequilibrium Statistical Mechanics, Oxford University Press, New York, 2001.

[223] J. Luczka, Non-Markovian stchastic processes: Colored noise, Chaos 15 (2005) 026107.

[224] N. G. van Kampen, Remarks on non-Markov processes, Brazilian Journal of Physics 28 (2) (1998) 90-96.

[225] I. Oppenheim, K. E. Shuler, Master equations and Markov processes, Physical Review 138 (4B) (1965) 1007-1011.

[226] L. D. Landau, E. M. Lifschitz, Mechanics, Third Edition, Butterworth-Heinemann, Oxford, 1976.

[227] R. F. Pawula, Approximation of the linear Boltzmann equation by the Fokker-Planck equation, Physical Review 162 (1) (1967) 186-188.

[228] R. F. Pawula, Generalizations and Extensions of the Fokker-Planck-Kolmogorov, IEEE Transactions on Information Theory 13 (1) (1967) 33-41.

[229] W. T. Coffey, Y. P. Kalmykov, J. T. Waldron, The Langevin Equation: with Applications to Stochastic Problems in Physics, Chemistry and Electrical Engineering, Second Edition, World Scientific, Singapore, 2004.

[230] L. Ferrari, Particles dispersed in a dilute gas: Limits of validity of the Langevin equation, Chemical Physics 336 (2007) 27-35. Erratum: Chemical Physics 337 (2007) 177.

[231] R. Fürth, Einige Untersuchungen über Brownsche Bewegung an einem Einzelteilchen, Annalen der Physik 358 (11) (1917) 177-213.

[232] T. Li, S. Kheifets, D. Medellin, M. G. Raizen, Measurement of the instantaneous velocity of a Brownian particle, Science 328 (2010) 1673-1675.

[233] J. L. Doob, The Brownian movement and stochastic equations, Annals of Mathematics 43 (2) (1942) 351-369.

[234] M. von Smoluchowski, Über Brownsche Molekularbewegung unter Einwirkung ausserer Kdifte und deren Zusammenhang mit der verallgemeinerten Diffusionsgleichung, Annalen Physik 48 (1915) 1103-1112.

[235] J-P. Bouchaud, A. Georges, Anomalous diffusion in disordered media: Statistical mechanisms, models and physical applications, Physics Reports 195 (4-5) (1990) 127-293.

[236] R. Metzler, J. Klafter, The random walk's guide to anomalous diffusion: A fractional dynamics approach, Physics Reports 339 (2000) 1-77.

[237] D. H. Zanette, Statistical thermodynamical foundations of anomalous diffusion, Brazilian Journal of Physics 29 (1) (1999) 108-124.

[238] S. A. Trigger, Anomalous diffusion in velocity space, Physics Letters A 374 (2009) 134-138.

[239] P. Hänggi, P. Talkner, M. Borkovec, Reaction-rate theory: fifty years after Kramers, Reviews of Modern Physics 62 (2) (1990) 251-341.

[240] C. W. Gardiner, Handbook of Stochastic Methods: For Physics, Chemistry, and the Natural Sciences, Third Edition, Springer, Berlin, 2004.

[241] E. Pollak, P. Talkner, Reaction rate theory: What it was, where is it today, and where is it going? Chaos 15 (2005) 026116.

[242] J. Dunkel, P. Hänggi, Relativistic Brownian motion, Physics Reports 471 (2009) 1-73.

[243] M. S. Green, Brownian motion in a gas of noninteracting molecules, Journal of Chemical Physics 19 (8) (1951) 1036-1046.

[244] J. Logan, M. Kac, Fluctuations and the Boltzmann equation I, Physical Review A 13 (1) (1976) 458-470.

[245] L. Bocquet, J. Piasecki, Microscopic derivation of non-Markovian thermalization of a Brownian particle, Journal of Statistical Physics 87 (5-6) (1997) 1005-1035.

[246] J. G. Kirkwood, The statistical mechanical of transport processes, Journal of Chemical Physics 14 (3) (1946) 180-201.

[247] J. Ross, Statistical mechanical theory of transport processes. IX. Contribution to the theory of Brownian motion, Journal of Chemical Physics 24 (2) (1956) 375-380.

[248] V. B. Magalinskij, Ja. P. Terletskij, On the statisticaf theory of nonequilibrium processes, Annalen der Physik 460 (5-6) (1960) 296-307.

[249] I. Prigogine, Non-equilibrium Statistical Mechanics, Wiley, New York, 1960.

[250] R. Kubo, M. Toda, N. Hashitsume, Nonequilibrium Statistical Mechanics, Springer, Berlin, 1985.

[251] P. Español, Statistical mechanics of coarse-graining, in: M. Karttunen I. Vattulainen A. Lukkarinen (Eds.), Novel Methods in Soft Matter Simulations, Vol. 640 of Lecture Notes in Physics, Springer, Berlin, 2004, pp. 69-115.

[252] P. Castiglione, M. Falcioni, A. Lesne, A. Vulpiani, Chaos and Coarse Graining in Statistical Mechanics, Cambridge University Press, Cambridge, 2008.

[253] H. Mori, Transport, collective motion, and Brownian motion, Progress of Theoretical Physics 33 (1965) 423-455.

[254] R. I. Cukier, J. M. Deustsch, Microscopic theory of Brownian motion: The multiple-time-scale point of view, Physical Review 177 (1) (1969) 240-244.

[255] L. Boquet, From a stochastic to a microscopic approach to Brownian motion, Acta Physica Polonica 29 (6) (1998) 1551-1564.

[256] J. Lebowitz, E. Rubin, Dynamical study of Brownian motion, Physical Review 131 (6) (1963) 2381-2306.

[257] P. Resibois, H. T. Davis, Transport equation of a Brownian particle in an external field, Physica 30 (1964) 1077-1091.

[258] J. Lebowitz, P. Resibois, Microscopic theory of Brownian motion in an oscillating field-connection with macroscopic theory, Physical Review 139 (4A) (1965) 1101-1111.

[259] P. Mazur, I. Oppenheim, Molecular theory of Brownian motion, Physica 50 (2) (1970) 241-258.

[260] J. T. Hynes, R. Kapral, S. Weinberg, Microscopic theory of Brownian motion: Mori friction kernel and Langevin-equation derivation, Physica 80A (1975)105-127.

[261] P. Hänggi, Generalized Langevin equations: a useful tool for the perplexed modeler of nonequilibrium fluctuations? In: L. Schimansky-Geier, T. Pöschel (Eds.), Stochastic Dynamics, Vol. 484 of Lecture Notes in Physics, Springer, Berlin, 1997, pp. 15-22.

[262] R. Zwanzig, Nonlinear generalized Langevin equations, Journal of Statistical Physics 9 (3) (1973) 215-220.

[263] R. M. Mazo, On the theory of Brownian motion. Interaction Between Brownian Particles, Journal of Statistical Physics 1 (1) (1969) 89-99.

[264] J. Albers, J. M. Deutch, I. Oppenheim, Generalized Langevin Equations, Journal of Chemical Physics 54 (1971) 3541-3546.

[265] H. Grabert, P. Thomas, H. Hänggi, Microdynamics and time-evolution of macroscopic non-Markovian systems, Zeitschrift für Physik B 26 (1977) 389-395.

[266] H. Grabert, P. Talkner, P. Hänggi, Microdynamics and time-evolution of macroscopic non-Markovian systems II, Zeitschrift für Physik B 29 (1978) 273-280.

[267] P. Hänggi, Correlation functions and master equations of generalized non-Markovian Langevin Equations, Zeitschrift für Physik B 31 (1978) 407-416.

[268] P. Hänggi, Note on the evaluation of the memory-kernel occuring in generalized master equations, Zeitschrift für Physik B 34 (1979) 409-410.

[269] N. G. van Kampen, I. Oppenheim, Brownian motion as a problem of eliminating fast variables, Physica A 138 (1-2) (1986) 231-248.

[270] J. M. Deutch, I. Oppenheim, The concept of Brownian motion in modern statistical mechanics, Faraday Discussions of the Chemical Society 83 (1987) 1-20.

[271] J-E. Shea, I. Oppenheim, Fokker-Planck equation and Langevin equation for one Brownian particle in a nonequilibrium bath, Journal of Physical Chemistry 100 (1996) 19035-19042.

[272] M. Schwartz, R. Brustein, From Lagrangian to Brownian motion, Journal of Statistical Physics 51 (1988) 585-613.

[273] G. Frenckel, M. Schwartz, The structure of Langevin's memory kernel from Lagrangian dynamics, Europhysics Letters 50 (5) (2000) 628-634.

[274] A. V. Plyukhin, J. Schofield, Langevin equation for the Rayleigh model with finite-range interactions, Physical Review E 68 (2003) 041107.

[275] A. V. Plyukhin, J. Schofield,, Langevin equation for the extended Rayleigh model with an asymmetric bath, Physical Review E 69 (2004) 021112.

[276] A. V. Plyukhin, Generalized Fokker-Planck equation, Brownian motion, and ergodicity, Physical Review E 77 (2008) 061136.

[277] V. B. Magalinskii, Dynamical model in the theory of the Brownian motion, Soviet Physics JETP 9 (6) (1959) 1381-1382.

[278] J. M. Deutch, R. Silbey, Exact generalized Langevin equation for a particle in a harmonic lattice, Physical Review A 3 (6) (1971) 2049-2052.

[279] S. Kim, Non-Markovian irreversible behavior in a simple model, Journal of Mathematical Physics 15 (5) (1974) 578-582.

[280] A. M. Levine, M. Shapiro, E. Pollak, Hamiltonian theory for vibrational dephasing rates of small molecules in liquids, Journal of Chemical Physics 88 (3) (1988) 1959-1966.

[281] S.-B. Zhu, S. Singh, G. W. Robinson, Breakdown of the Brownian motion model in ultrafast dynamics, Physical Review A 40 (2) (1989) 1109-1115.

[282] J. R. Chaudhuri, G. Gangopadhyay, D. S. Ray, Theory of nonstationary activated rate processes: Nonexponential kinetics, Journal of Chemical Physics 109 (13) (1998) 5565-5575.

[283] J. R. Chaudhuri, S. K. Banik, B. C. Bag, D. S. Ray, Analytical and numerical investigation of escape rate for a noise driven bath, Physical Review E 63 (2001) 061111.

[284] J. R. Chaudhuri, Barik, S. K. Banik, Escape rate from a metastable state weakly interacting with a heat bath driven by external noise, Physical Review E **73** (2006) 051101.

[285] D. Boilley, Y. Lallouet, Non-Markovian diffusion over a saddle with a generalized Langevin equation, Journal of Statistical Physics 125 (2) (2006) 477-493.

[286] L. Kantorivich, Generalized Langevin equation for solids. I. Rigorous derivation and main properties, Physical Review B 78 (2008) 094304.

[287] S. Bhattacharya, S. K. Banik, S. Chattopadhyay, J. R. Chaudhuri, Time dependent current: A microscopic approach, Journal of Mathematical Physics 49 (2008) 063302.

[288] D. Segal, Thermal conduction in molecular chains: Non-Markovian effects, Journal of Chemical Physics 128 (2008) 224710.

[289] J. R. Chaudhuri, P. Chaudhury, S. Chattopadhyay, Harmonic oscillator in presence of nonequilibrium environment, Journal of Mathematical Physics 130 (2009) 234109.

[290] A. Morita, A critical investigation of the generalized Langevin equation, Chemical Physics Letters 222 (4) (1994) 399-402.

[291] I. V. L. Costa, R. Morgado, M. V. B. T. Lima, F. A. Oliveira, The fluctuation-dissipation theorem fails for fast superdiffusion, Europhysics Letters 63 (2) (2003) 173-179.

[292] J. M. Porrà, K-G. Wang, J. Masoliver, Generalized Langevin equations: Anomalous diffusion and probability distributions, Physical Review E 53 (3) (1996) 5872- 5881.

[293] K. G. Wang, M. Tokuyama, Nonequilibrium statistical description of anomalous diffusion, Physica A 265 (1999) 341-351.

[294] N. G. van Kampen, Stochastic differential equations, Physics Reports 24 (3) (1976) 171-228.

[295] A. V. Plyukhin, Does a Brownian particle equilibrate? Europhysiccs Letters 75 (1) (2006) 15-21.

[296] A. V. Plyukhin, Generalized Fokker-Planck equation, Brownian motion, and ergodicity, Physical Review E 77 (2008) 061136.

[297] V. Vladimirsky, Ya. Terletzky, On the hydrodynamic memory in the theory of Brownian motion. Translation: V. Lisy, J. Tothova, arXiv: 0410222.

[298] E. H. Hauge, A. Martin-Löf, Fluctuating hydrodynamics and Brownian motion Journal of Statistical Physics 7 (3) (1973) 259-281.

[299] J. W. Dufty, Gaussian model for fluctuation of a Brownian particle, The Physics of Fluids 17 (2) (1974) 328-333.

[300] S. A. Adelman, Fokker-Planck equations for simple non-Markovian systems, Journal of Chemical Physics 64 (1) (1976) 124-130.

[301] P. Hänggi, P. Thomas, Time evolution, correlations and linear response of non-Markov processes, Zeitschrift für Physik B (26) (1977) 85-92.

[302] R. M. Mazo, Aspects of the theory of Brownian motion. In: L. Garrido, P. Seglar, P. J. Shepherd (Eds.), Stochastic Processes in Nonequilibrium Systems, Vol. 84, of Lecture Notes in Physics, Springer, Berlin, 1978, pp. 53-81.

[303] R. F. Fox, Analysis of nonstationary, Gaussian and non-Gaussian, generalized Langevin equations using methods of multiplicative stochastic processes, Journal of Statistical Physics 16 (3) (1977) 259-279.

[304] R. F. Fox, The generalized Langevin equation with Gaussian fluctuations, Journal of Mathematical Physics 18 (12) (1977) 2331-2335.

[305] P. Hänggi, P. Talkner, On the equivalence of time-convolutionless master equations and generalized Langevin equations, Physics Letters A 68 (1) (1978) 9-11.

[306] P. Hänggi, P. Thomas, R. Grabert, P. Talkner, Note on time evolution of non-Markov processes: propagator approach, Journal of Statistical Physics 18 (2) (1978) 155-159.

[307] M. san Miguel, J. M. Sancho, A colored-noise approach to Brownian motion in position space, Journal of Statistical Physics 22 (5) (1980) 605-624.

[308] P. Grigolini, M. Ferrario, Probability Diffusion in non-Markovian, non-Gaussian Molecular Ensembles: A Theoretical Analysis and Computer Simulation, Zeitschrift für Physik B 41(2) (1981) 165-176.

[309] K. Ibuki, M. Ueno, A generalized Fokker-Planck equation treatment of inertia and non-Markovian effects on the short-time dynamics of a collision-induced reaction, Bulletin of the Chemical Society of Japan 70 (3) (1997) 543-553.

[310] P. Hänggi, P. Talkner, Non-Markov processes: The problem of the mean first passage time, Zeitschrift für Physik B 45 (1981) 79-83.

[311] L. Reichl, Translational Brownian motion in a fluid with internal degrees of freedom, Physical Review A 24 (3) (1981) 1609-1616.

[312] P. Hänggi, F. Mojtabai, Thermally activated escape rate in presence of long-time memory, Phys. Rev. A 26 (1982) 1168-1170.

[313] P. Hanggi, P. Talkner, Memory index of first-passage time: A simple measure of non-Markovian character, Physical Review Letters 51 (1983) 2242-2245. Erratum: Physical Review Letters 52 (1984) 484.

[314] V. S. Volkov, V. N. Pokrowsky, Generalized Fokker-Planck equation for non-Markovian processes, Journal of Mathematical Physics 24 (2) (1983) 267-270.

[315] A. Hernandez-Machado, M. San Miguel, Dynamical properties of non-Markovian stochastic differential equations, Journal of Mathematical Physics 25 (4) (1984) 1066-1075.

[316] P. Hänggi, Path integral solutions for non-Markovian processes, Zeitschrift für Physik B 75 (1989) 275-281.

[317] K. Kawasaki, T. Kawakatsu, Projector formalism of generalized Brownian motion theory applied to dissipative and noisy systems, Journal of Statistical Physics 67 (3-4) (1992) 795-811.

[318] A. Fulinski, Non-Markovian noise, Physical Review 50 (4) E (1994) 2668-2681.

[319] J. Luczka, P. Hänggi, A. Gadomski, Non-Markovian process driven by quadratic noise: Kramers-Moyal expansion and Fokker-Planck modeling Physical Review E 51 (4) (1995) 2933-2938.

[320] V. S. Volkov, A. I. Leonov, Non-Markovian Brownian motion in a viscoelastic fluid, Journal of Chemical Physics 104 (15) (1996) 5922-5931.

[321] S. K. Banik, J. R. Chaudhuri, D. S. Ray, The generalized Kramers theory for nonequilibrium open one-dimensional systems, Journal of Chemical Physics 112 (19) (2000) 8330-8337.

[322] I. M. Sokolov, Solutions of a class of non-Markovian Fokker-Planck equations, Physical Review E 66 (2002) 041101.

[323] C. C. Martens, Qualitative dynamics of generalized Langevin equations and the theory of chemical reaction rates, Journal of Chemical Physics 116 (6) (2002) 2516-2528.

[324] A. V. Mokshin, R. M. Yulmetyev, P. Hänggi, Diffusion processes and memory effects, New Journal of Physics **7** (9) (2005) 1-10.

[325] J-D. Bao, P. Hänggi, Y-Z. Zhuo, Non-Markovian Brownian dynamics and nonergodicity, Physical Review E 72 (2005) 061107.

[326] J-D. Bao, Y-L. Song, Q. Ji, Y-Z. Zhuo, Harmonic velocity noise: Non-Markovian features of noise-driven systems at long times, Physical Review E 72 (2005) 011113.

[327] K. Zabrocki, S. Tatur, S. Trimper, R. Mahnke, Relationship between a non-Markovian process and Fokker-Planck equation, Physics Letters A 359 (2006) 349-356.

[328] J. R. Chaudhuri, S. Chattopadhyay, S. K. Banik, Generalization of the escape rate from a metastable state driven by external cross-correlated noise processes, Physical Review E 76 (2007) 021125.

[329] D. O. Soares-Pinto, W. A. M. Morgado, Exact time-average distribution for a stationary non-Markovian massive Brownian particle coupled to two heat baths, Physical Review E 77 (2008) 011103.

[330] R. L. S. Farias, R. O. Ramos, L. A, da Silva, Stochastic Langevin equations: Markovian and non-Markovian dynamics, Physical Review E 80 (2009) 031143.

[331] J. Tóthová, V. Lisý, Hydrodynamic Memory in the Motion of Charged Brownian Particles across the Magnetic Field, Acta Physica Polonica A 118 (5) (2010) 1051-1053.

[332] L. F. Richardson, Atmospheric diffusion shown on a distance-neighbour graph. Proceedings of the Royal Society of London A 110 (756) (1926) 709-737.

[333] B. J. Alder, T. E. Wainright, Velocity autocorrelations for hard spheres, Physical Review Letters 18 (23) (1967) 988-900.

[334] B. J. Alder, T. E. Wainright, Decay of the velocity autocorrelation function, Physical Review A 1 (1) (1970) 18-21.

[335] R. F. Fox, Long-time tails and diffusion, Physical Review A 27 (6) (1983) 3216- 3233.

[336] K. G. Wang, Long-time-correlation effects and biased anomalous diffusion, Physical Review A 45 (2) (1992) 833-837.

[337] J. Heinrichs, Probability distributions for second-order processes driven by Gaussian noise, Physical Review E 47 (5) (1993) 3007-3012.

[338] J. Masoliver, K. G. Wang, Free inertial processes driven by Gaussian noise: Probability distributions, anomalous diffusion, and fractal behavior, Physical Review E 51 (4) (1995) 2987-2995.

[339] J. M. Porrà, K-G. Wang, J. Masoliver, Generalized Langevin equations: Anomalous diffusion and probability distributions, Physical Review E 53 (6) (1996) 5872-5881.

[340] K-G. Wang, M. Tokuyama, Nonequilibrium statistical description of anomalous diffusion, Physica A 265 (1999) 341-351.

[341] J. Tothova, V. Lisy, A. V. Zatovsky, Long-time tails in the dynamics of Rouse polymers, Journal of Chemical Physics 119 (24) (2003) 13135-13137.

[342] J. Tothova, V. Lisy, A. V. Zatovsky, Long-time dynamics of Rouse-Zimm polymers in dilute solutions, Journal of Chemical Physics 121 (21) (2004) 10699-10706.

[343] J-.D. Bao, Anomalous transport in unbound and ratchet potentials, Physical Review E 69 (2004) 016124.

[344] K. L. Lü, J-D. Bao, Numerical simulation of generalized Langevin equation with arbitrary correlated noise, Physical Review E **72** (2005) 067701.

[345] T. D. Franck, Nonlinear Fokker-Planck Equations: Fundamentals and Applications, Springer, Berlin, 2005.

[346] P. Olla, L. Pignagnoli, Local evolution equations for non-Markovian processes, Physics Letters A 350 (2006) 51-55.

[347] A. Mura, M.S. Taqqu, F. Mainardi, Non-Markovian diffusion equations and processes, Physica A 387 (2008) 5033-5064.

[348] C. Tsallis, Introduction to Nonextensive Statistical Mechanics: Approaching a Complex World, Springer, New York, 2009.

[349] J. Tóthova, G. Vasziová, L. Glod, V. Lisý, Langevin theory of anomalous Brownian motion made simple, European Journal of Physics 32 (2011) 645-655.

[350] A. Einstein, O. Stern, Einige Argumente für die Annahme einer molekularen Agitation beim absoluten Nullpunkt, Annalen der Physik 40 (3) (1913) 551-560.

[351] R. Tolman, The Principles of Statistical Mechanics, Clarendon, Oxford, 1938.

[352] F. A. Deeney, J. P. O'Leary, A note on the zero point energy of an ideal boson gas, European Journal of Physics 32 (2011) L19-L24.

[353] F. A. Deeney, J. P. O'Leary, The internal energy and thermodynamic behaviour of a boson gas below the Bose-Einstein temperature, Physics Letters A 375 (2011) 1637-1639.

[354] R. P. Feynman, The Principle of Least Action in Quantum Mechanics, in: L. M. Brown (Ed.), Feynman's Thesis: A New Approach to Quantum Theory, World Scientific, Singapore, 2005.

[355] R. P. Feynman, A. R. Hibbs, Quantum Mechanics and Path Integrals, McGraw-Hill, New York, 1965.

[356] V. Ambegaokar, Quantum Brownian motion and its classical limit, Berichte der Bunsengesellschaft für physikalische Chemie 95 (3) (1991) 400-404.

[357] L. Diòsi, On high-temperature markovian equation for quantum Brownian motion, Europhysics Leters 22 (1) (1993) 1-3.

[358] L. Diòsi, Caldeira-Leggett master equation and medium temperatures, Physica A 199 (1993) 517-526.

[359] W. J. Munro and C. W. Gardiner, Non-rotating-wave master equation, Physical Review A 53 (4) (1996) 2633-2640.

[360] A. Tameshtit, J. E. Sipe, Physical Review Letters 77 (13) (1996) 2600-2603.

[361] R. Karrlein, H. Grabert, Exact time evolution and master equations for the damped harmonic oscillator, Physical Review E 55 (1) (1997) 153-164.

[362] S. Gao, Lindblad approach to quantum dynamics of open systems, Physical Review B 57 (8) (1998) 4509- 4517.

[363] S. M. Barnett, J. Jeffers, J. D. Cresser, From measurements to quantum friction, Journal of Physics: Condensed Matter 18 (2006) S401-S410.

[364] A. O. Caldeira, H. A. Cerdeira, R. Ramaswamy, Limits of weak damping of a quantum harmonic oscillator, Physical Review A 40 (1989) 3438-3440.

[365] A. O. Caldeira, A. J. Leggett, Path integral approach to quantum Brownian motion, Physica A 121 (1983) 587-616.

[366] A. Săndulescu, H. Scutaru, Open quantum systems and the damping of collective modes in deep inelastic collisions, Annals of Physics 173 (2) (1987) 277-317.

[367] E. Stefanescu, W. Scheid, A. Sandulescu, Non-Markovian master equation for a system of Fermions interacting with an electromagnetic field, Annals of Physics 323 (5) (2008) 1168-1190.

[368] J. G. Peixoto de Faria, M. C. Nemes, Phenomenological criteria for the validity of quantum Markovian equations, Journal of Physics A: Mathematical and General 31 (1998) 7095-7103.

[369] R. C. de Berrêdo, J. G. P. de Faria, F. Camargo, M. C. Nemes, H. E. Borges, K. M. Fonseca Romero, A. F. R. de Toledo Piza, A. N. Salgueiro, On the physical content of Lindblad form master equations, Physica Scripta 57 (1998) 533-534.

[370] C. H. Fleming, A. Roura, B. L. Hu, Exact analytical solutions to the master equation of quantum Brownian motion for a general environment, Annals of Physics 326 (2011) 1207-1258.

[371] M. Schlosshauer, Decoherence and the Quantum to Classical Transition, Springer, Berlin, 2007.

[372] F. Guinea, Friction and particle-hole pairs, Physical Review Letters 53 (13) (1984) 1268-1271.

[373] P. Hedegard, A. O. Caldeira, Static properties of a particle coupled to a fermionic environment, Physical Review B 53 (1) (1987) 106-114.

[374] P. Hedegard, A. O. Caldeira, Quantum dynamics of a particle in a fermionic environment, Physica Scripta 35 (1987) 609-622.

[375] A. H. Castro Neto, A. O. Caldeira, Motion of heavy particles coupled to fermionic and bosonic environments in one dimension, Physical Review B 52 (6) (1995) 4198-4208.

[376] E. P. Wigner, On the quantum correction for thermodynamic equilibrium, Physical Review 40 (1932) 749-759.

[377] K. Huang, Introduction to Statistical Physics, Taylor and Francis, London, 2001.

[378] M. J. Biercuk, H. Uys, J. W. Britton, A. P. VanDevender, J. J. Bollinger, Ultrasensitive force and displacement detection using trapped ions, Nature Nanotechnology 5 (9) (2010) 646-650.

[379] P. Pechukas, J. Ankerhold, H. Grabert, Quantum Smoluchowski equation, Annalen der Physik 9 (9-10) (2000) 794-803.

[380] J. Ankerhold, Quantum dynamics with strong friction: the quantum Smoluchowski equation and beyond, Acta Physica Polonica B 34 (7) (2003) 3569-3579.

[381] J. Ankerhold, Overdamped quantum phase diffusion and charging effects in Josephson junctions, Europhysics Letters 67 (2) (2004) 280-286.

[382] J. Ankerhold, H. Lehle, Low temperature electron transfer in strongly condensed phase. Jouran of Chemical Physics 120 (3) (2004) 1436-1449.

[383] J. Ankerhold, H. Grabert, P. Pechukas, Quantum Brownian motion with large friction, Chaos 15 (2005) 026106.

[384] J. Łuczka, R. Rudnicki, P. Hänggi, The diffusion in the quantum Smoluchowski equation Physica A 351 (2005) 60-68.

[385] J. Ankerhold, Quantum Tunneling in Complex Systems: The Semiclassical Approach, Springer, Berlin, 2007.

[386] S. A. Maier, J. Ankerhold, Quantum Smoluchowski equation: A systematic study, Physical Review E 81 (2010) 021107.

[387] L. Machura, M. Kostur, P. Hänggi, P. Talkner, J. Łuczka, Consistent description of quantum Brownian motors operating at strong friction, Physical Review E 70 (2004) 031107.

[388] A. Ankerhold, Dynamics of dissipative quantum systems: From path integrals to master equations, in E. Benatti, R. Floreanini (Eds.), Irreversible Quantum Dynamics, vol. 622 of Lecture Notes in Physics, Springer, Berlin, 2003, pp. 165-178.

[389] W. T. Coffey, Yu P. Kalmykov, S. V. Titov, B. P. Mulligan, Wigner function approach to the quantum Brownian motion of a particle in a potential, Physical Chemistry Chemical Physics 9 (2007) 3361-3382.

[390] W. T. Coffey, Yu P. Kalmykov, S. V. Titov, B. P. Mulligan, Semiclassical Klein-Kramers and Smoluchowski equations for the Brownian motion of a particle in an external potential, Journal of Physics A: Mathematical and Theoretical 40 (2007) F91-F98.

[391] S. K. Banik, B. C. Bag, D. S. Ray, Generalized quantum Fokker-Planck, diffusion, and Smoluchowski equations with true probability distribution functions, Physical Review E 65 (2002) 051106.

[392] D. Banerjee, B. C. Bag, S. K. Banik, D. S. Ray, Quantum Smoluchowski equation: escape from a metastable state, Physica A 318 (2003) 6-13.

[393] S. Bhattacharya, P. Chaudhury, S. Chattopadhyay, J. R. Chaudhuri, Phase induced current in presence of nonequilibrium bath: A quantum approach, Journal of Chemical Physics 129 (2008) 124708.

[394] D. Barik, D. Banerjee, D. S. Ray, Quantum Brownian Motion in c-Numbers: Theory and Applications, Nova Science Publishers, New York, 2009.

[395] W. T. Coffey, Yu. P. Kalmykov, S. V. Titov, B. P. Mulligan, Semiclassical master equation in Wigner's phase space applied to Brownian motion in a periodic potential, Physical Review E 75 (2007) 041117.

[396] W. T. Coffey, Yu. P. Kalmykov, S. V. Titov, L. Cleary, Smoluchowski equation approach for quantum Brownian motion in a tilted periodic potential, Physical Review E 78 (2008) 031114.

[397] L. Cleary, W. T. Coffey, Yu. P. Kalmykov, S. V. Titov, Semiclassical treatment of a Brownian ratchet using the quantum Smoluchowski equation, Physical Review E 80 (2009) 051106.

[398] W. T. Coffey, Yu. P. Kalmykov, S. V. Titov, L. Cleary, Quantum effects in the Brownian motion of a particle in a double well potential in the overdamped limit, Journal of Chemical Physics 131 (2009) 084101.

[399] W. T. Coffey, Yu. P. Kalmykov, S. V. Titov, L. Cleary, Smoluchowski equation approach for quantum Brownian motion in a tilted periodic potential, Physical Review E 78 (2008) 031114.

[400] R. Tsekov, Comment on 'Semiclassical Klein-Kramers and Smoluchowski equations for the Brownian motion of a particle in an external potential', Journal of Physics A: Mathematical and Theoretical 40 (2007) 10945-10947.

[401] R. Tsekov, Dissipation in quantum systems, Journal of Physics A: Mathematical and Theoretical 28 (1995) L557-L561.

[402] R. Tsekov, Nonlinear theory of quantum Brownian motion, International Journal of Theoretical Physics 48 (2009) 85-94.

[403] R. Tsekov, Thermo-quantum diffusion, International Journal of Theoretical Physics 48 (2009) 630-636.

[404] R. Tsekov, Towards nonlinear quantum Fokker-Planck equations, International Journal of Theoretical Physics 48 (2009) 1431-1435.

[405] R. Tsekov, Quantum diffusion, Physica Scripta 83 (2011) 035004.

[406] M. Razavy, Quantum Theory of Tunneling, World Scientific, Singapore, 2003.

[407] P. Hänggi, Escape from a Metastable State, Journal of Statistical Physics 42 (1-2) (1986) 105-148. Addendum and Erratum: Journal of Statistical Physics 44 (5-6) (1986) 1003-1004.

[408] V. I. Mel'nikov. The Kramers problem: Fifty years of development, Physics Reports 299 (1-2) (1991 ) 1-71.

[409] D. Banerjee, B. C. Bag, S. K. Banik, D. S. Ray, Generalized quantum Fokker-Planck, diffusion, and Smoluchowski equations with true probability distribution functions, Physical Review E (65) 051106.

[410] D. Banerjee, B. C. Bag, S. K. Banik, D. S. Ray, Approach to quantum Kramers' equation and barrier crossing dynamics, Physical Review E (65) (2002) 021109.

[411] D. Banerjee, B. C. Bag, S. K. Banik, D. S. Ray, Quantum Smoluchowski equation: Escape from a metastable state, Physica A 318 (2003) 6-13.

[412] D. Barik, S. K. Banik, D. S. Ray, Quantum phase-space function formulation of reactive flux theory, Journal of Chemical Physics 119 (2) (2003) 680-695.

[413] D. Barik, D. S. Ray, Quantum equilibrium factor in the decay of a metastable state at low temperature, Journal of Statistical Mechanics: Theory and Experiment (2006) 06001.

[414] S. Bhattacharya, S. Chattopadhyay, J. R. Chaudhuri, Investigation of noise-induced escape rate: A quantum mechanical approach, Journal of Statistical Physics 136 (2009) 733-750.

[415] P. Ghosh, A. Shit, S. Chattopadhyay, J. R. Chaudhuri, A semiclassical approach to explore the bistable kinetics of a Brownian particle in a nonequilibrium environment, Journal of Statistical Mechanics: Theory and Experiment (2011) 02026.

[416] A. Shit, S. Chattopadhyay, J. R. Chaudhuri, Towards an understanding of escape rate and state dependent diffusion for a quantum dissipative system, Chemical Physics 386 (2011) 56-72.

[417] A. J. Faria, H. M. França, R. C. Sponchiado, Tunneling as a classical escape rate induced by the vacuum zero-point radiation, Foundations of Physics 36 (2) (2006) 307-320.

[418] J. R. Chaudhuri, B. C. Bag, D. S. Ray, A semiclassical approach to the Kramers' problem, Journal of Chemical Physics 111 (24) (1999) 10852-10858.

[419] J. Ankerhold, H. Grabert, P. Pechukas, Comment on ''Quantum Tunneling at Zero Temperature in the Strong Friction Regime'', Physical Review Letters 95 (2005) 079801.

[420] P. Caldirola, Forze non conservative nella meccanica quantistica, Nuovo Cimento 18 (9) (1941) 393-400.

[421] I. Prigogine, M. Toda, On the irreversible processes in quantum mechanics, Molecular Physics 1 (1958) 48-62.

[422] M. Toda, On the theory of the Brownian motion, Journal of Physical Society of Japan 13 (11) (1958) 1266-1280.

[423] J. Bricmont, Science of chaos, or chaos in science?, in: P. R. Gross, N. Levitt, M. W. Lewis (Eds.), The Flight from Science and Reason, Annals of the New York Academy of Sciences 775, New York, 1996, pp. 131-175.

[424] J. Bricmont, Bayes, Boltzmann and Bohm: probabilities in physics, in: J. Bricmont, D. Dürr, M. C. Gallavoti, G. C. Ghirardi, P. Petruccione, N. Zanghi (Eds.), Chance in Physics: Foundations and Perspectives, Lectures Notes in Physics 574, Springer, Berlin, 2001, pp. 3-21.

[425] A. O. Bolivar, Classical limit of fermions in phase space, Journal of Mathematical Physics 42 (9) (2001) 4020-4030.

[426] A. O. Bolivar, Classical limit of bosons in phase space, Physica A 315 (2002) 601-615.

[427] A. Fick, Über Diffusion, Poggendorff's Annalen der Physik und Chemie 94 (1855) 59-86. Reprinted in Journal of Membrane Science 100 (1995) 33-38.

[428] J. Philibert, One and a half century of diffusion: Fick, Einstein, before and beyond, Diffusion Fundamentals 2 (2005) 1-10.

[429] E. Frey, K. Kroy, Brownian motion: a paradigm of soft matter and biological physics, Annalen der Physik 14 (1-3) (2005) 20-50.

[430] D. Hobson, A survey of mathematical finance, Proceedings of the Royal of the Society of London A 460 (2004) 3369-3401.

[431] L. E. Reichl, A Modern Course in Statistical Physics, Second Edition, Wiley, New York, 1998.

[432] M. von Smoluchowski, Einige Beispiele Brown'scher Molekularbewegung unter Einfluss äusserer Kräfte, Bulletin International de l'Académie des Sciences de Cracovie, Classe des Sciences Mathématiques et Naturelles A (1913) 418-434.

[433] M. von Smoluchowski, Drei Vorträge über Diffusion, Brownsche Bewegung, Physikalische Zeitschrift 17 (1916) 557-571, and 587-599.

[434] A. D. Fokker, Die mittlere Energie rotierender elektrischer Dipole im Strahlungsfeld, Annalen der Physik 43 (1914) 810-820.

[435] M. Planck, Über einen Satz der statistischen Dynamik und seine Erweiterung in der Quantentheorie, Sitzungsberichte der Königlich Preussischen Akademie der Wissenschaften, Sitzung der physikalische-mathematischen Klasse 1(1917) 324-341.

[436] R. Fürth, Die Brownsche Bewegung bei Berücksichtgung einer Persistenz der Bewegungsrichtung. Mit Anwendungen auf die Bewegung lebender Infusorien, Zeitschrift für Physik 2 (3) (1920) 244-256.

[437] S. Chapman, On the Brownian displacements and thermal diffusion of grains suspended in a non-uniform fluid, Proceedings of the Royal Society of London A 119 (1928) 34-54.

[438] E. Nelson, Dynamical Theories of Brownian Motion, Second Edition, Princeton University Press, Princeton, 1967.

[439] S. G. Brush, A history of random processes I. Brownian movement from Brown to Perrin, Archive for History of Exact Sciences 5 (1) (1968) 1-36.

[440] D. J. Shaw, Introduction to Colloid and Surface Chemistry, Fourth Edition, Butterworth-Heinemann, Oxford, 1992.

[441] G. Inzelt, Einstein and the osmotic theory, Journal of Solid State Electrochemistry 10 (2006) 1008-1011.

[442] R. Fürth, Comments, in: A. Einstein, Untersuchungen über die Theorie der Brownschen Bewegung, Ostwalds Klassiker der Exakten Wissenschaften, Vol. 199, Third Edition, Deutsch, Frankfurt am Main, 1997.

[443] M. A. Olivares-Robles, L. S. Garcia-Colín, Mesoscopic derivation of hyperbolic transport equations, Physical Review E 50 (1994) 2451-2457.

[444] J. Perrin, Mouvement brownien et réalité moléculaire, Annales de Chimie et de Physique 18 (1909) 5-114.

[445] J. Perrin, Atoms, Iniversity of California, New York, 1916.

[446] B. B. Mandelbrot, The Fractal Geometry of Nature, Freeman, New York, 1983.

[447] J. B. Perrin, Discontinuous structure of matter, Nobel Lecture. http://nobelprize.org/nobel_prizes/physics/laureates/1926/perrin-lecture.html.

[448] G. L. de Haas-Lorentz, Die Brownsche Bewegung und einige verwandte Erscheinungen, Springer, Wiesbaden, 1913.

[449] J. D. van der Waals Jr., On the theory of the Brownian movement, Proceedings of the Royal Acadamy of Amsterdam 20 (1918) 1254-1271.

[450] L. S. Ornstein, H. C. Burger, On the theory of the Brownian motion, Proceedings of the Royal Acadamy of Amsterdam 21 (1919) 922-931.

[451] N. Wiener, The average of an analytic functional and the Brownian movement, Proceedings of the National Academy of Science 7 (1921) 294-298.

[452] L. S. Ornstein, Zur Theorie der Brownschen Bewegung für Systeme, worin mehrere Temperaturen vorkommen, Zeitschrift für Physik 41 (4-5) (1927) 848-856.

[453] L. S. Ornstein, W. R. van Wijk. On the derivation of distribution functions in problems of Brownian motion, Physica 1 (1) (1934) 235-254.

[454] J. M. W. Milatz, L. S. Ornstein, Properties of the fortuitous force in the Einstein-Langevin equation, Physica 7 (8) (1940) 793-801.

[455] S. Chandrasekhar, Brownian motion, dynamical friction and stellar dynamics, Reviews of Modern Physics 21 (3) (1949) 383-388.

[456] D. A. Weitz, D. J. Pine, P. N. Pusey, R. J. A. Tough, Nondiffusive Brownian motion studied by diffusing-wave spectroscopy 63 (16) (1989) 1747-1750.

[457] P. N. Pusey, Brownian motion goes ballistic, Science 332 (2011) 802-803.

[458] B. Lukic, S. Jeney, C. Tischer, A. J. Kulik, L. Forró, E.-L. Florin, Direct observation of nondiffusive motion of a Brownian particle, Physical Review Letters 95 (2005) 160601.

[459] J. Blum, Astrophysical microgravity experiments with dust particles, Microgravity, Sciences and Technology 22 (4) (2010) 517-527.

[460] R. Huang, I. Chavez, K. M. Taute, B. Lukic, S. Jeney, M. G. Raizen, E-L. Florin, Direct observation of the full transition from ballistic to diffusive Brownian motion in a liquid, Nature Physics **7** (2011) 576-580.

[461] J. Blum, G. Wurm, S. Kempf, T. Henning, The Brownian motion of dust particles in the solar nebula: An experimental approach to the problem of pre-planetary dust aggregation, Icarus 124 (1996) 141-151.

[462] J. Blum, G. Wurm, S. Kempf, T. Poppe, H. Klahr, T. Kozasa, M. Rott, T. Henning, J. Dorschner, R. Schräpler, H. U. Keller, W. J. Markiewicz, I. Mann, B. A. S. Gustafson, F. Giovane, D. Neuhaus, H. Fechtig, E. Grün, B. Feuerbacher, H. Kochan, L. Ratke, A. El Goresy, G. Morfill, S. J. Weidenschilling, G. Schwehm, K. Metzler, W.-H. Ip, Growth and form of planetary seedlings: results from a microgravity aggregation experiment, Physical Review Letters 85 (12) (2000) 2426-2429.

[463] J. Blum, S. Bruns, D. Rademacher, A. Voss, B. Willenberg, M. Krause, Measurement of the translational and rotational Brownian motion of ndividual particles in a rarefied gas, Physical Review Letters 97 (2006) 230601.

[464] D. ter Haar, Elements of Statistical Mechanics, Third Edition, Butterworth-Heinemann, Oxford, London, 1995.

[465] E. Schrödinger, Statistical Thermodynamics, Cambridge University Press, Cambridge, 1948.

[466] T. L. Hill, An Introduction to Statistical Thermodynamics, Addison-Wesley, Reading, 1960.

[467] L. D. Landau, E. M. Lifshitz, Statistical Physics, Third Edition, Part 1, Pergamon, Oxford, 1980.

[468] K. Huang, Statistical Mechanics, Wiley, New York, 1987.

[469] S. R. A. Salinas, Introduction to Statistical Physics, Springer, New York, 2001.

[470] F. Schwabl, Statistical Mechanics, Second Edition, Springer, Berlin, 2006.

[471] E. Fermi, Zur Quantelung des idealen einatomigen Gases, Zeitschrift für Physik 36 (11-12) (1926) 902-912.

[472] P. A. M. Dirac, On the theory of quantum mechanics, Proceedings of the Royal Society A 112 (762) (1926) 661-677.

[473] N. Bose, Plancks Gesetz und Lichtquantenhypothese, Zeitschrift für Physik 26 (1924) 178-181.

[474] A. Einstein, Quantentheorie des einatomigen idealen Gases, Sitzungsberichte der Preussischen Akademie der Wissenschaft (1924) 261-267.

[475] A. Einstein, Quantentheorie des einatomigen idealen Gases, Sitzungsberichte der Preussischen Akademie der Wissenschaft (1925) 3-14.

[476] J-G. Li, J. Zou, B. Shao, Factorization law for entanglement evolution of two qubits in non-Markovian pure dephasing channels, Physics Letters A 375 (2011) 2300-2304.

[477] R. Huang, I. Chavez, K. M. Taute, B. Lukic, S. Jeney, M. G. Raizen, and E-L. Florin, Direct observation of the full transition from ballistic to diffusive Brownian Motion in a liquid, Nature Physics 7 (7) (2011) 576-580.